%% file: review.tex
\documentclass[preprint,sort&compress]{elsarticle}
\usepackage{graphicx}
\usepackage{amssymb}
\usepackage{epstopdf}
\usepackage{slashed}
\usepackage{amsmath}
\usepackage{bm}
\usepackage{hyperref}
\newcommand{\Tr}{{\rm Tr}}

\newcommand\mn[1]{\mathcal{#1}}
\def \nfd {n_\mathrm{F}}
\def \nbe {n_\mathrm{B}}
\def \nn{\nonumber}
\def\q{{\bm q}}
\def\p{{\bm p}}
\def \mm{m_\infty^2}
\renewcommand \Im {{\rm Im} }
\renewcommand \Re {{\rm Re} }
\def\sumint{\hbox{$\sum$}\!\!\!\!\!\!\!\int}

\newcommand{\msbar}{{\overline{\mbox{\rm MS}}}}
\newcommand{\lambdamsbar}{{\Lambda_{\overline{\rm MS}}}}

\usepackage[usenames,dvipsnames]{color}
\definecolor{darkblue}{RGB}{0,0,196}

\begin{document}

%%%%%%%%%%%%%%%%%%%%%%%%%%%%%%%%%%%%%%%%%%%%%%%%%%%%%%%%%%%%%%%%%%%%%%%%%%%%%%%%
% Frontmatter
%%%%%%%%%%%%%%%%%%%%%%%%%%%%%%%%%%%%%%%%%%%%%%%%%%%%%%%%%%%%%%%%%%%%%%%%%%%%%%%%

\begin{frontmatter}

\title{Perturbative Thermal QCD: Formalism and Applications}
\author[jg]{Jacopo Ghiglieri}
\author[ak]{Aleksi Kurkela}
\author[m]{Michael Strickland}
\author[av]{Aleksi Vuorinen\corref{cor1}}

\address[jg]{SUBATECH,  Universit\'e  de  Nantes,  IMT  Atlantique,  IN2P3/CNRS,\\
4  rue  Alfred  Kastler,  44307  Nantes  cedex  3,  France}
\address[ak]{Theoretical Physics Department, CERN, CH-1211 Gen\`eve 23, Switzerland, and \\Faculty of Science and Technology, University of Stavanger, 4036 Stavanger,
Norway}
\address[m]{Physics Department, Kent State University, OH 44242, United States}
\address[av]{Department of Physics and Helsinki Institute of Physics,
FI-00014 University of Helsinki, Finland}

\cortext[cor1]{Corresponding author}

\begin{abstract}
In this review article, we discuss the current status and future prospects of perturbation theory as a means of studying the equilibrium thermodynamic and near-equilibrium transport properties of deconfined QCD matter. We begin with a brief introduction to the general topic, after which we review in some detail the foundations and modern techniques of the real- and imaginary-time formalisms of thermal field theory, covering e.g.~the different bases used in the real-time formalism and the resummations required to deal with soft and collinear contributions. After this, we discuss the current status of applications of these techniques, including topics such as electromagnetic rates, transport coefficients, jet quenching, heavy quarks and quarkonia, and the Equations of State of hot quark-gluon plasma as well as cold and dense quark matter. Finally, we conclude with our view of the future directions of the field, i.e.~how we anticipate perturbative calculations to contribute to our collective understanding of strongly interacting matter in the coming years.
\end{abstract}

\date{}

%\preprint{asd}


\end{frontmatter}

%%%%%%%%%%%%%%%%%%%%%%%%%%%%%%%%%%%%%%%%%%%%%%%%%%%%%%%%%%%%%%%%%%%%%%%%%%%%%%%%
% Main body
%%%%%%%%%%%%%%%%%%%%%%%%%%%%%%%%%%%%%%%%%%%%%%%%%%%%%%%%%%%%%%%%%%%%%%%%%%%%%%%%

\tableofcontents
\input{introduction}
\input{qftIV}

\input{realtimeformalism}

\input{realtimeresults}
\input{imagtimeformalism}
\input{imagtimeresults}
\input{futuredirections}

%%%%%
%Acknos
%%%%%
\section{Acknowledgments}
\label{sec:acknowledgments}

We would like to thank Nora Brambilla, Mikko Laine, and Guy Moore for detailed and very useful comments on our manuscript. The work of MS has been supported by U.S. Department of Energy, Office of Science, Office of Nuclear Physics under award no.~DE-SC0013470, and that of AV by the European Research Council, grant no.~725369, and by the Academy of Finland, grant no.~1322507.

%%%%%%%%%%%%%%%%%%%%%%%%%%%%%%%%%%%%%%%%%%%%%%%%%%%%%%%%%%%%%%%%%%%%%%%%%%%%%%%%
% Appendices
%%%%%%%%%%%%%%%%%%%%%%%%%%%%%%%%%%%%%%%%%%%%%%%%%%%%%%%%%%%%%%%%%%%%%%%%%%%%%%%%

\appendix
\input{appendix-gauge}
%\input{appendix-conventions}

%%%%%%%%%%%%%%%%%%%%%%%%%%%%%%%%%%%%%%%%%%%%%%%%%%%%%%%%%%%%%%%%%%%%%%%%%%%%%%%%
% Bibliography
%%%%%%%%%%%%%%%%%%%%%%%%%%%%%%%%%%%%%%%%%%%%%%%%%%%%%%%%%%%%%%%%%%%%%%%%%%%%%%%%

\bibliographystyle{elsarticle-num}
\bibliography{review}

\end{document}

%% file: introduction.tex
% !TEX root = review.tex

\section{Introduction}

The need to quantitatively understand extended quantum field theory systems is abundant in particle and nuclear physics as well as in cosmology. Depending on the details of the system under consideration, a number of different computational methods, including nonperturbative ones, may be applicable. However, there is only one theoretical framework that is both based on first principles and maximally versatile in the sense that it is applicable even to real-time quantities and can accommodate nonzero densities. This is perturbation theory, or more generally a class of weak-coupling methods based on expanding the functional integrals that define different physical quantities in powers (and logarithms) of a coupling constant. The purpose of the review article at hand is to introduce the most important aspects of modern thermal perturbation theory, including its theoretical foundations and recent milestone results, within Quantum Chromodynamics or QCD. For compactness, we limit our discussion to thermal equilibrium, but consider a variety of different quantities and settings. In particular, we discuss both bulk thermodynamic and real-time observables, and cover both the realms of high temperatures, relevant for heavy-ion physics, and high densities, needed in the description of cold quark matter, possibly present inside (some) neutron star cores.

A recurrent issue in perturbative calculations is the need to perform resummations of diagrams of all loop orders to reach a result valid to a given power of the strong coupling constant $g$. Also, additional resummations are often necessary to remedy poorly converging results. These issues are typically related to contributions from soft collective excitations, associated with momentum scales such as $gT$, $g^2T$, or $g\mu$, or with almost collinear (small-angle) scatterings. Such contributions often lead to infrared (IR) divergences in naive, or unresummed, perturbation theory, making their first appearance at different orders depending on the quantity in question. What all these  excitations have in common is that their consistent  inclusion in a weak coupling calculation requires first a proper identification of the IR sensitive degrees of freedom, and then the development of some type of an effective description for them. How such identifications are made and the effective theories or resummations constructed in practical calculations is one of the leading themes of this article. %[[AK: I'm not sure if I understand the sentence here]]

There exist two largely complementary formulations of perturbative thermal field theory, dubbed the `real time' and `imaginary time' formalisms. Despite their seemingly different starting points, they are equivalent and share many of the same features regarding, e.g.~IR sensitivity. Which one is the more practical tool of the two depends on the observable. As explained in some detail in section~\ref{sec:IV}, the real-time formalism is amenable to describing systems even outside thermal equilibrium, and is well-suited for the determination of inherently Minkowskian quantities, such as spectral functions and particle production rates. The basics of this formalism, the resummations of soft and collinear modes within it, and a selection of its most prominent recent applications are discussed in sections~\ref{sec:realtimept}, \ref{sec_soft_collinear} and \ref{sec:realresults}, respectively. We note that due to the technically rather involved nature of this formalism and the unavailability of modern textbooks on the subject, section \ref{sec_soft_collinear} is likely the most involved one in our review.

In comparison, the imaginary-time formalism, which is formulated assuming thermal equilibrium from the outset, is considerably more straightforward to follow. Its development relies on a formal analogy between the definition of the Boltzmann operator $e^{-\hat{H}/T}$ appearing in thermal expectation values and the time-evolution operator of zero-temperature quantum field theory. This formalism is particularly well suited for the determination of bulk thermodynamic quantities which are time independent and thus inherently "Euclidean" in nature, though an analytic continuation often allows one to address Minkowskian quantities as well. The imaginary time formalism is covered in sections~\ref{sec:imagtimeform} and \ref{sec:imagtimeresults} of our review, beginning again from the basic formalism and subsequently moving on to recent highlight results. In both parts, particular emphasis will be given to a comparison of the two leading schemes used for resumming IR contributions to physical quantities, dubbed Hard Thermal Loops perturbation theory (HTLpt) and Dimensional Reduction (DR).

Before commencing with the article, we note that our discussion naturally owes a lot to a number of existing textbooks and review articles on thermal field theory. From textbooks, we should mention the the three classics by Joseph Kapusta \cite{Kapusta:1989tk}, Kapusta and Charles Gale \cite{Kapusta:2006pm}, and Michel Le Bellac \cite{Bellac:2011kqa}, applying respectively the imaginary  \cite{Kapusta:1989tk,Kapusta:2006pm} and real time \cite{Bellac:2011kqa} formalisms, as well as a more recent book by Mikko Laine and Aleksi Vuorinen, concentrating on practical perturbative computations mostly in the imaginary time formalism \cite{Mikko}. Among review articles, we owe gratitude to both \cite{Blaizot:2001nr} and \cite{Kraemmer:2003gd}, which respectively concentrate on the Hard Thermal Loop framework and perturbative thermal field theory in general. While both of these excellent articles have significant overlap with our review, we feel that an update has been due for some time. The reason for this is related to recent advances on one hand in high-order perturbative calculations both within the real and imaginary time formalisms, and on the other hand to conceptual advances in how to optimally handle the IR sensitive degrees of freedom. 

%The structure of our article is as follows: We begin with a modern take on the basics of statistical quantum field theory in section~\ref{sec:IV}. [[fix what follows]] After this, our discussion bifurcates into the real- and imaginary-time formalisms, which are successively developed in sections 3 and 4. This is followed by a review of a selection of important recent results from both formalisms, given in sections 5 and 6. After this, we finally conclude in section 7 with a brief outlook of the future of perturbative calculations in thermal quantum field theories.

Our notational choices will be specified in detail when necessary, but we list the most important elements here. Euclidean momenta are denoted by $K=(\omega_n,\mathbf{k})$, while Minkowskian ones read $\mn{K}=(k^0,\mathbf{k})$, with the Minkowskian metric following the $-+++$ convention.  Finally, throughout our review, we will be working in natural units, in which both the reduced Planck constant $\hbar$ and the speed of light $c$ are set to unity.

%% file: qftIV.tex
% !TEX root = review.tex

\section{Thermal quantum field theory as initial value problem \label{sec:IV}}
In ordinary zero-temperature quantum field theory, one is typically interested in $S$-matrix elements 
related through the Lehmann-Symanzik-Zimmer\-mann (LSZ) reduction (see e.g.~\cite{Peskin}) to vacuum 
expectation values of a set of time-ordered operators ${\rm T} [\hat{\mathcal{O}}]$.
These operators $\mathcal{\hat O}$ act on the vacuum state $| \Omega \rangle$ to create the \emph{asymptotic states} 
that correspond to particles infinitely far away in the past or future.

When developing the formalism of statistical field theory, two differences arise. First, one needs to account for \emph{statistical fluctuations}.
From the quantum-mechanical point of view, the vacuum state is a pure state. As such,
a simple expectation value constructed from pure states accounts only for quantum-mechanical
fluctuations. However in a medium, there may be also statistical fluctuations arising from
having only limited, macroscopic information of  the state of the system at some specified
moment $t_0$ (in a suitable frame). If the medium is described by states $| i \rangle$ with probabilities
$p_i(t_0)$ at $t_0$, an expectation value combining both statistical and quantum fluctuations
is given by
\begin{equation}
	\langle \hat{\mathcal{O}}(t_0) \rangle  \equiv \sum_i p_i(t_0) \langle i | \hat{\mathcal{O}} | i \rangle,
\end{equation}
or equivalently
\begin{equation}
\langle \hat{\mathcal{O}}(t_0) \rangle=\Tr\, \hat{\rho}(t_0) \hat{\mathcal{O}}(t_0), \quad \hat{\rho}(t_0) \equiv \sum_i p_i(t_0) | i \rangle \langle i |,
\end{equation}
where $\hat{\rho}$ is the \emph{density operator}.
The states $|i\rangle$ may be any complete set of states and do not need to form an orthonormal basis. 
% The evaluation of the Hilbert space trace in the left equation, on the other hand, naturally does require an orthonormal basis.

In order to evaluate the expectation value at a later time $t_1$, the density
operator needs to be evolved to the later time by the application of the time-translation operator\textcolor{red}{, }
$\hat{\rho}(t_1) = U(t_1,t_0) \hat{ \rho}(t_0) U(t_0,t_1)$. Then the expectation value of the
operator $\mathcal{\hat O}$ at the later time $t_1$ reads
\begin{equation}
	\langle \hat{\mathcal{O}} (t_1)\rangle = \Tr \,\hat{\rho}(t_1) \hat{\mathcal{O}}(t_1)
	= \Tr \, U(t_1,t_0) \hat\rho(t_0) U(t_0,t_1) \hat{\mathcal{O}}(t_1)  \,.
\end{equation}
Writing the expression in the field basis, we obtain\footnote{In this section, we consider a generic bosonic field $\phi$. This field can be thought of as a single component of the $A_\mu$ gauge field, as the discussion does not depend explicitly on the spin of the field. }
% For complications arising from gauge fixing see \ref{app_horror}. \redflag{Check this!}}
\begin{equation}
	\label{Oatt1}
\langle \hat{\mathcal{O}} (t_1)\rangle= \sum_{i,j,k,l} \langle \phi_i | U(t_1,t_0)  | \phi_j \rangle \,  {\rho}_{jk}\,  \langle \phi_k | U(t_0,t_1) | \phi_l \rangle \,{\mathcal{O}}_{li}\,,
\end{equation}
where $\rho_{jk}\equiv\langle \phi_j\vert\hat\rho\vert\phi_k\rangle$ and similarly for $\mathcal{O}_{li}$.
The matrix elements of the evolution operator can be given in terms of the path integral representation
\begin{align}
	\label{forwardev}
 \langle \phi_i | U(t_1,t_0)  | \phi_j \rangle  &=   \langle \phi_i | e^{-i \hat{H} (t_1 - t_0)}  | \phi_j \rangle = \int_{\phi_1(t_0) = \phi_j}^{\phi_1(t_1) = \phi_i}\mathcal{D}\phi_1(t) e^{i S(\phi_1) },\\
 \label{backwardev}
  \langle \phi_k | U(t_0,t_1)  | \phi_l \rangle & =   \langle \phi_k | e^{-i \hat{H} (t_0 - t_1)}  | \phi_l \rangle = \int_{\phi_2 (t_0)=\phi_k}^{\phi_2(t_1) = \phi_l}\mathcal{D}\phi_2(t) e^{ - i S(\phi_2)},
\end{align}
where $\hat{H}$ is the Hamiltonian of the system and
the second equation follows from unitarity. Hence Eq.~\eqref{Oatt1} becomes
\begin{align}
\langle \hat{\mathcal{O}} (t_1)\rangle= \sum_{i,j,k,l} \int_{\phi_1(t_0) = \phi_j }^{\phi_1(t_1)=\phi_i} \mathcal{D}\phi_1(t) \int_{\phi_2(t_0) = \phi_k}^{\phi_2(t_1) = \phi_l}\mathcal{D}\phi_2(t) e^{i S(\phi_1)-i S(\phi_2)} {\rho}_{jk} {\mathcal{O}}_{li} \,.
\label{eq:bc}
\end{align}
The labels $1$ and $2$ should not be confused with the $ijkl$ ones. While the latter
refer to the field configurations of the corresponding states $|i\rangle,\ldots$, the former
are only used to label the time evolution of the ket ($1$) and bra ($2$), which by the unitarity
of the evolution operator lead to the relative minus sign in the exponent of the action. This labeling
is oftentimes called \emph{doubling of the degrees of freedom}. It represents a crucial point of time-dependent
statistical field theory and we will explore its details and physical implications in Sec.~\ref{sec:realtimept}.

In the special case of thermal equilibrium in the grand canonical ensemble, the density operator
 takes the form
\begin{align}
	\hat{\rho}_{\rm eq} = \frac{1}{Z}e^{-\beta (\hat{H}-\mu_i \hat{N}_i )},\qquad
	Z=\mathrm{Tr}\,e^{-\beta (\hat{H}-\mu_i \hat{N}_i )},
\end{align}
where %$\hat{H}$ is the Hamiltonian and
$\mu_i$ and $\hat{N}_i$ correspond to the chemical potentials
and associated number operators for possible conserved charges that commute with each other and with the Hamiltonian. In the case of QCD, these are,
e.g., quark numbers of flavor $f$
\begin{align}
\hat{N}_f = \int d^3 x \, \bar{q}_f(x) \gamma^0 q_f(x).
\end{align}
The normalization constant $Z$, the partition function, enforces $\langle 1 \rangle =1$.

As can be readily verified, the equilibrium form
of the density matrix resembles that of an evolution operator with a time argument of $- i \beta$. This allows one to write also the density matrix in the path integral form
\begin{align}
({\rho_{\rm eq}})_{jk} \equiv \langle \phi_j | \hat{\rho}_{eq} | \phi_k \rangle = \frac{1}{Z}\int_{\phi_E(t_0) = \phi_k}^{\phi_E( t_0 - i \beta) = \pm \phi_j} \mathcal{D}\phi_E \,  e^{- S_E(\phi_E)}\,,\label{eq:pif}
\end{align}
where $S_E$ is the \emph{Euclidean action}, $S_E=\int_0^\beta d\tau\, L_E$, or in 
the presence of nonzero chemical potentials $S_E=\int_0^\beta d\tau\, (L_E-\mu_f N_f) $.
%(\redflag{JACOPO: is $\mu_f$ consistent with other notation?}). 
The field at $t=t_0-i\beta$ is equal to $\pm \phi_j$, with the upper sign enforcing
a \emph{periodic boundary condition} for bosons and the lower one an \emph{antiperiodic
boundary condition} for fermions (see e.g.~\cite{Kapusta:2006pm} for a careful derivation of both boundary conditions). 

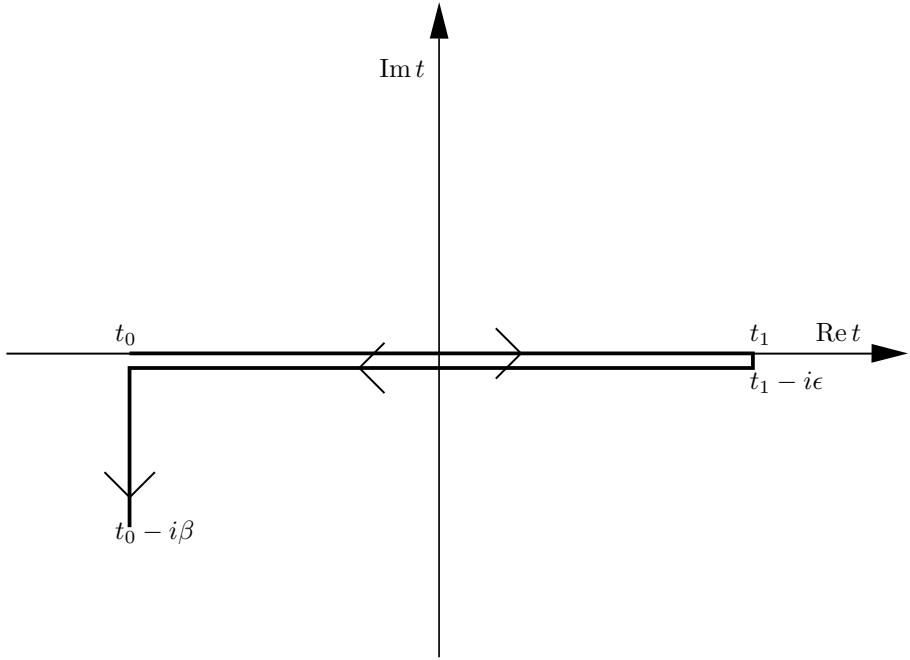
\begin{figure}[t!]
	\begin{center}
		\includegraphics[width=12cm]{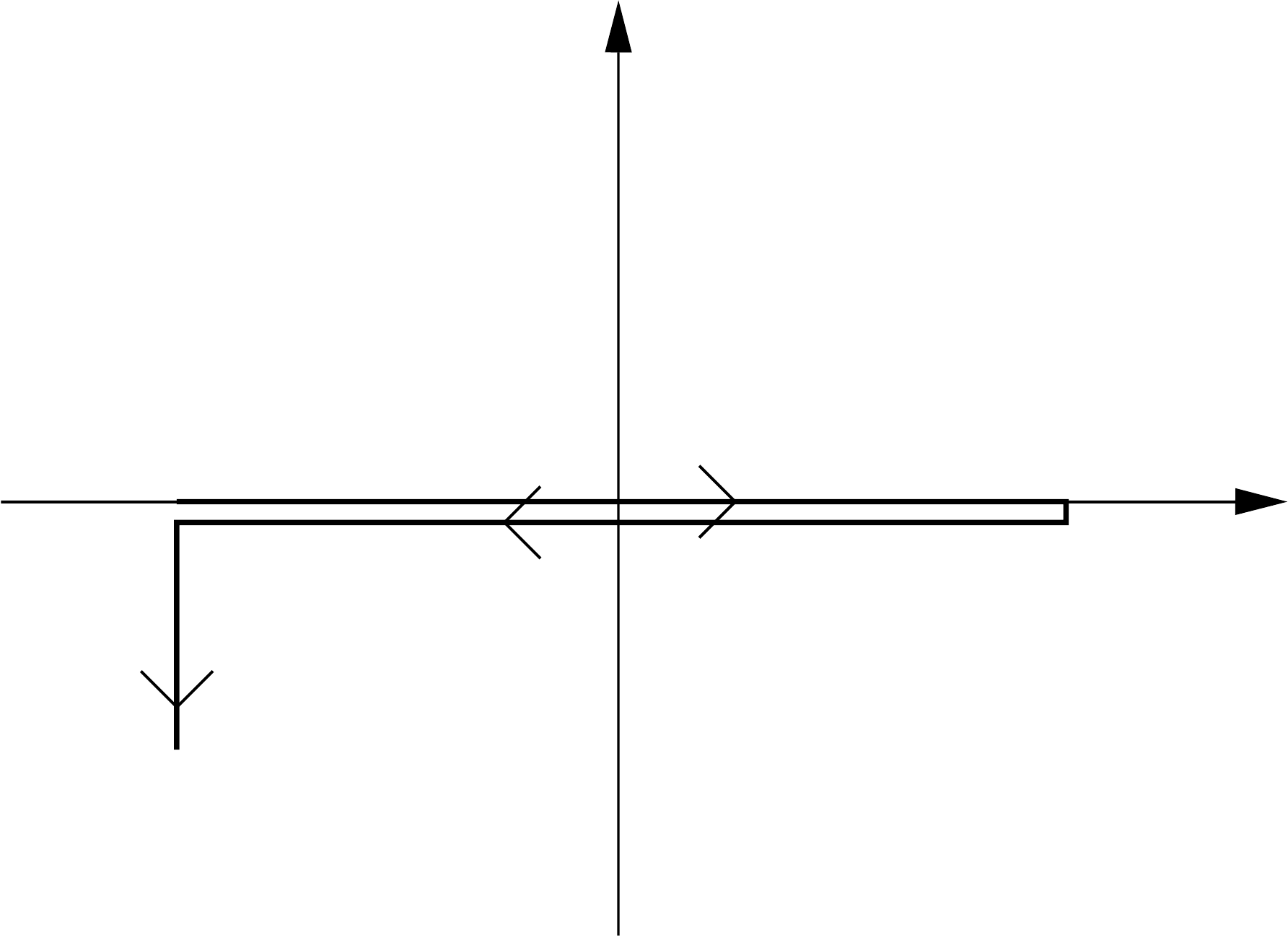}
		\put(-300,120){$t_0$}
		\put(-300,45){$t_0-i\beta$}
		\put(-200,220){$\mathrm{Im}\,t$}
		\put(-35,120){$\mathrm{Re}\,t$}
		\put(-60,120){$t_1$}
		\put(-60,102){$t_1-i\epsilon$}
	\end{center}
	\caption{The Schwinger--Keldysh contour on the complex $t$-plane.}
	\label{fig_sk}
\end{figure}

It is clear that the equilibrium density operator commutes with the Hamiltonian, $[\hat\rho_{eq},\hat{H}] = 0$, and
is thus time-translation invariant. Therefore, in equilibrium, the initial time $t_0$ is completely arbitrary, and for an operator local in time we may simply choose $t_0=t_1$.
In this case,  the $\mathcal{D}\phi_1(t)$ and $\mathcal{D}\phi_2(t)$ integrals
disappear and we are left with only
the Euclidean branch of the path integral. This purely Euclidean path integral will be the starting point
of our discussion in Sec.~\ref{sec:imagtimeform}. On the other hand, in case of operators separated in (real)
time, such as $\mathcal{\hat{O}}=\mathcal{\hat{O}}_i(t_1)\mathcal{\hat{O}}_j(t_2)$ with $t_1< t_2$,
the action of the first operator $\mathcal{O}_i(t_1)$ on the density operator creates a
\emph{non-equilibrium} state characterized by a new density matrix
$\hat{\rho}(t_1)=\hat{\rho}_{\rm eq} \mathcal{\hat{O}}_i(t_1)$. This new density operator no
longer commutes with the Hamiltonian and therefore the integrals
over the real branches no longer
trivialize. The contour formed by the two real branches and the imaginary
one is called the \emph{Schwinger--Keldysh contour}
\cite{Schwinger:1960qe,Keldysh:1964ud}, and is depicted in Fig.~\ref{fig_sk} (see also \cite{Bellac:2011kqa} for a more pedagogical introduction).

The second important difference with ordinary $T=0$ QFT is that a thermal medium induces
random interactions which, in turn, do not preserve any state. Therefore, one cannot separate
\`a la LSZ the far-away asymptotics from the space-time region where the interactions take place.
Hence the observables of interest are not the $S$-matrix elements or the associated
time-ordered expectation values, which, as we remarked previously, are the ones relevant in vacuum perturbation theory.
In a medium, on the other hand, operator ordering plays a
much more enhanced role: at nonzero temperatures and/or densities, most observables of interest depend either on the forward or backward Wightman functions, describing physical correlations in the medium,
or on retarded and advanced functions, describing causation in medium.
For bosons,\footnote{Note that we do not display the spatial coordinates or possible color, Lorentz or spin indices of the fields in these definitions. It is understood that the indices correspond to those of the fields at the given time arguments; for example, ${D^<}_{\mu \nu}^{ab} (t_0,t_1; {\bf x}_0, {\bf x}_1) = \langle A_\nu^b(t_1,{\bf x}_1) A_\mu^a(t_0,{\bf x}_0)\rangle$.} the Wightman functions read
\begin{align}
D^>(t_1,t_0) &= \langle \phi(t_1) \phi(t_0) \rangle, \\
D^<(t_1,t_0) &= \langle \phi(t_0) \phi(t_1) \rangle,
\end{align}
whereas the retarded and advanced correlators are
\begin{align}
	\label{bosonretpos}
D^R(t_1,t_0) & = \theta(t_1-t_0) \rho_B(t_1,t_0),\\
	\label{bosonadvpos}
D^A(t_1,t_0) & = -\theta(t_0-t_1) \rho_B(t_1,t_0),
\end{align}
which are written in terms of the spectral function
\begin{align}
	\label{defrho}
\rho_B(t_1,t_0) = \langle [ \phi(t_1), \phi(t_0) ]\rangle.
\end{align}
For a fermionic field $\psi$, the corresponding expressions on the other hand read
\begin{align}
S^>(t_1,t_0) & = \langle \psi(t_1) \overline \psi(t_0) \rangle,\\
S^<(t_1,t_0) & =  - \langle  \overline \psi(t_0)  \psi(t_1)\rangle, \\
\rho_F(t_1,t_0) & = \langle  \{ \psi(t_1), \overline \psi(t_0) \} \rangle,  \\
S^R(t_1,t_0) & = \theta(t_1-t_0 )\rho_F(t_1,t_0), \\
S^A(t_1,t_0) & =  - \theta(t_0-t_1 )\rho_F(t_1,t_0) .
\end{align}
Our definitions of the different correlation functions are chosen in such a way that we may write
\begin{align}
	\label{rhoboson}
\rho_B(t_1,t_0) &= D^R(t_1,t_0) - D^A(t_1,t_0) = D^>(t_1,t_0) - D^<(t_1,t_0), \\
\rho_F(t_1,t_0) &= S^R(t_1,t_0) - S^A(t_1,t_0) = S^>(t_1,t_0) - S^<(t_1,t_0).
\end{align}

In simple terms, the significance of the different correlators defined above can be summarized as follows: the Wightman function measures \emph{correlation}, whereas the retarded function measures \emph{causation}. That is, the Wightman function between firetrucks and fires is non-zero, whereas the retarded function between them vanishes, as firetrucks are often found around fires but the trucks do not cause them.

With a generic density operator $\hat\rho$, three of the above five correlators are independent.\footnote{As is clearly seen in position space
from Eqs.~\eqref{bosonretpos} and \eqref{bosonadvpos},
knowledge of $\rho$ determines the retarded and advanced correlators.}
However, in equilibrium even these functions are related to each other through the fluctuation-dissipation theorem, known in this context as the Kubo-Martin-Schwinger (KMS) relation \cite{Kubo:1957mj,Martin:1959jp}. As discussed above, in thermal equilibrium the functions depend on the difference of the two times, $t\equiv t_1 - t_0$.  The Wightman functions $D^>(t)$ and $D^<(t)$ are strictly analytic inside the bands $-\beta < \Im (t) < 0$ and  $0 < \Im (t) < \beta$, respectively (see for instance \cite{Bellac:2011kqa, Mikko}). This is seen particularly clearly by writing the forward Wightman function in its (normal-ordered) spectral representation
\begin{align}
D^{>}(t_1,t_0)= \frac{1}{Z} \sum_{m,n} e^{- \beta E_n} e^{- i E_n(t_1 - t_0)} e^{ i E_m(t_0 - t_1)} | \langle n | \hat{\phi}(0)| m \rangle|^2.
\end{align}
Now, assuming that the convergence of the sum is governed by the exponentials, it is clear that the sum is absolutely convergent, and therefore the resulting function analytic, for $-\beta < \Im (t) < 0$.\footnote{The zero-temperature limit of this statement is equivalent with the well known vacuum field theory result that the forward (backward) Wightman function has support only for positive (negative) frequencies. \label{footnote:pos_omega}}

Using the cyclicity of the trace, the exponential form of the thermal density operator, and the commutation relations of the conserved charge, the two Wightman functions can be related to each other via
\begin{align}
	\label{kmsposboson}
D^>(t ) & = D^<(t + i \beta ),  \\
	\label{kmsposfermion}
S^>(t ) & =  - e^{-\beta \mu}S^{<}(t+i \beta)\,,
\end{align}
where, due to our interest in QCD, we have omitted the possibility of assigning
a chemical potential to bosons.

In momentum space\footnote{According to our conventions, the retarded function is related to the spectral function via
$\rho_{B}(\omega) = 2\Re D^{R}(\omega)$, and $\rho_F(\omega) = 2\Re S^{R}(\omega)$} the above relations take a particularly useful form,
\begin{align}
D^>(\omega) \equiv & \int dt e^{i \omega t}  D^> (t) = e^{\beta \omega} D^<(\omega), \\
S^>(\omega) \equiv & \int dt e^{i \omega t}  S^> (t) = -e^{\beta (\omega-\mu)} S^<(\omega),
\end{align}
or equivalently in terms of the Wightman and spectral functions
\begin{align}
	\label{kmsmombosonless}
n_B(\omega)\rho_B(\omega) & = D^<(\omega), \\
	\label{kmsmombosongreat}
(1+n_B(\omega))\rho_B(\omega) & = D^>(\omega), \\
	\label{kmsmomfermionless}
n_F(\omega-\mu )\rho_F(\omega) & = -S^<(\omega),
%\rho(\omega)  = n_B=  2 \Re D^R(\omega)
\end{align}
where $n_B(\omega) = (e^{\beta \omega}-1)^{-1}$ and $n_F(\omega) = (e^{\beta \omega}+1)^{-1}$ are the Bose--Einstein and Fermi--Dirac distributions, respectively.

%% file: realtimeformalism.tex
% !TEX root = review.tex

\newcommand{\x}{\mathbf{x}}
\newcommand{\form}{\rm form}
\def\k{{\bm k}}

\def\Biggdlangle{\Bigg\langle\!\!\!\Bigg\langle}
\def\Biggdrangle{\Bigg\rangle\!\!\!\Bigg\rangle}
\def\biggdlangle{\bigg\langle\!\!\!\bigg\langle}
\def\biggdrangle{\bigg\rangle\!\!\!\bigg\rangle}
\def\Bigdlangle{\Big\langle\!\!\Big\langle}
\def\Bigdrangle{\Big\rangle\!\!\Big\rangle}

\section{Real time formalism}
\label{sec:realtimept}
In this section, we go on to explore in detail the implications
of the Schwinger--Keldysh contour on thermal expectation values,
illustrating general methods without a specific focus on QCD.
In Sec.~\ref{sub_allyourbases}, we review the most commonly
used bases for the fields on that contour. while in
Sec.~\ref{sub_self_energies} we explain how one of the most important objects in perturbation theory,
the self energy, behaves under such bases. This is of particular relevance for the resummations that 
will be introduced in Sec.~\ref{sec_soft_collinear}.
Sec.~\ref{sub_cutrules} is then dedicated to expounding the structure of
cutting rules at finite $T$ and $\mu$, which are of great relevance e.g.~for calculations of thermal production rates, as we will again show later
in Sec.~\ref{sec_soft_collinear}. After this, we contrast the finite temperature
and density theory with the behavior encountered at $T=\mu=0$, showing how it arises as a limiting case in Sec.~\ref{sub_vacuum}.
In Sec.~\ref{sub_classical} we finally explore a different limiting case of high occupation numbers, 
where the quantum thermal field theory approaches a classical field theory.

\subsection{Field bases for the Schwinger--Keldysh contour}
\label{sub_allyourbases}
Our introduction of the Schwinger--Keldysh contour in Sec.~\ref{sec:IV}
mentioned the ``standard'' basis for the so-called ``doubling of degrees of freedom''.
These are the ``1'' and ``2'' fields of the ``1/2'' basis, which we shall cover below
in Sec.~\ref{sub_12}. However, this is neither the only possible basis nor an optimal one, depending on the problem at hand. Indeed, in Sec.~\ref{sec_rabasis}
we will introduce a second basis, the ``$r/a$'' basis, which has
two advantages. From the physical standpoint, it makes the connection to the causal 
structure of amplitudes more explicit, as we show in Sec.~\ref{sub_ra_causal}. From
the computational standpoint, the vertices and the matrix structure of the propagators %on the other hand
become simpler than in the 1/2 basis. Finally, it is possible to derive
rules for the effective Hard Thermal Loop theory within this basis that are again
physically connected to causality and well-suited for computations, as we will
show in Sec.~\ref{sub_heur_htl}.

In summary, we feel that the practitioner of real-time perturbative calculations
should have both bases firmly in her toolbox and be ready to apply the better suited
one to the problem at hand; for instance, in Sec.~\ref{sub_res_heavy} we will display an example
where one is interested in a time-ordered correlator, so that the 1/2 basis is superior,
while in Sec.~\ref{sub_heur_htl} we will see how the fermionic Hard Thermal Loop is derived
rather easily in the $r/a$ basis.

\subsubsection{The 1/2 basis}
\label{sub_12}
In the previous Section, we introduced the Schwinger--Keldysh integral
in  Eq.~\eqref{eq:bc} and drew the corresponding contour in
Fig.~\ref{fig_sk}. We now set on  developing its perturbative expansion. To this end, a generating functional is commonly introduced by
generalizing the field-doubled path integrals introduced in the previous Section,
\begin{equation}
Z[J_1,J_2] = \int \mathcal{D}\phi_E e^{-S_E(\phi_E)}  \int \mathcal{D} \phi_1 \mathcal{D}  \phi_2 e^{i S(\phi_1)-i S(\phi_2)- \int d^4 x (J_1(x)\phi_1(x) - J_2(x) \phi_2(x)) }.
\label{genfunc}
\end{equation}
Generic operators $\mathcal{\hat{O}}$ can be  introduced in the usual way,
by taking appropriate derivatives of the generating functional with respect to the sources
$J_i$. The complications
introduced by gauge fields and fermions, i.e.~gauge fixing and
Grassmann variables, are thoroughly covered in textbooks such as \cite{Bellac:2011kqa}, so we will not consider these subtleties further here.

The perturbative series is constructed by separating the free part of the action,  quadratic in fields,
\begin{equation}
{ \bf D}_{ij} =  \frac{\delta}{ \delta J_i} \frac{\delta}{ \delta J_j}  Z[J_1,J_2]\bigg\vert_{J=0}\,,
\end{equation}
from the interaction part $S_I$. In this expansion, both the propagators and the vertices
are matrices in Schwinger--Keldysh indices. From the form of the generating functional, it should
not come as a surprise that the diagonal entries of the propagators are the time- and anti-time-ordered Feynman propagators,
\begin{align}
	\label{tord}
D^F(t_1,t_0) &= \theta(t_1-t_0)\langle \phi(t_1)\phi(t_0)\rangle + \theta(t_0-t_1)\langle \phi(t_0)\phi(t_1)\rangle, \\
D^{\bar{F}}(t_1,t_0) &= \theta(t_0-t_1)\langle \phi(t_1)\phi(t_0)\rangle + \theta(t_1-t_0)\langle \phi(t_0)\phi(t_1)\rangle,
\label{antitord}
\end{align}
 whereas the off-diagonal terms
are the forward and backward Wightman functions\footnote{We use boldface letters to
identify the propagator matrix, but drop the spacetime or four-momentum dependence, as
these equations are valid both in position and momentum space.}
\begin{equation}
{ \bf D} =
\left(
\begin{array}{cc}
\langle \phi_1 \phi_1 \rangle & \langle \phi_1 \phi_2 \rangle \\
\langle \phi_2 \phi_1 \rangle & \langle \phi_2 \phi_2 \rangle
\end{array}
\right)
=
\left(
\begin{array}{cc}
D^F & D^< \\
D^> & D^{\bar{F}}
\end{array}
\right).
\end{equation}
Through the definition of (anti-)time-ordering and the Wightman and retarded correlators
together with their relation to the spectral function, Eqs.~\eqref{kmsmombosonless} and \eqref{kmsmombosongreat},
the momentum-space forms of Eqs.~\eqref{tord} and \eqref{antitord} become
\begin{align}
	\label{tordom}
D^F(\omega,k) &= \frac12\big[D_R(\omega,k)+D_A(\omega,k)\big]+\left(\frac12+n_\mathrm{B}(\omega)\right)\rho(\omega,k), \\
D^{\bar F}(\omega,k) &= -\frac12\big[D_R(\omega,k)+D_A(\omega,k)\big]+\left(\frac12+n_\mathrm{B}(\omega)\right)\rho(\omega,k) .
\label{antitordom}
\end{align}

\begin{figure}[t!]
\begin{center}
\includegraphics[width=0.5 \textwidth]{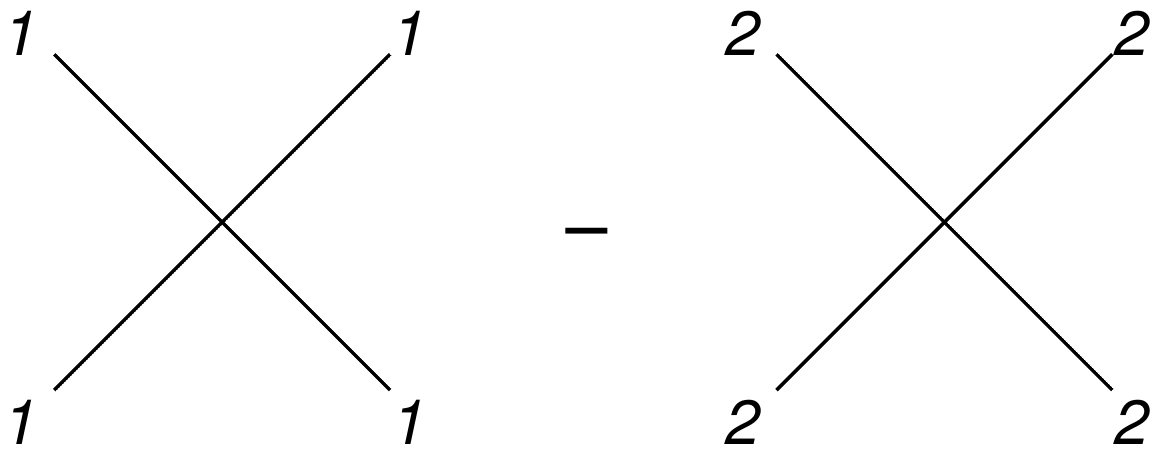}
\end{center}
\caption{The two vertices appearing in the example featuring the 1/2 basis.The vertex with type 2 fields comes with a relative minus sign because of the different signs of the actions in Eq.~\ref{genfunc}.}
\label{fig:phi4_12}
\end{figure}

As the actions $S(\phi_1)$ and $S(\phi_2)$ do not mix fields with indices 1 and 2, the vertices
have their usual vacuum field theory form with the minor modification
that all the lines in the vertex carry the Schwinger--Keldysh index 1 or 2, and that the
vertices with index 2 come with an extra minus sign, as shown in Fig.~\ref{fig:phi4_12}.

\subsubsection{The $r/a$ basis}
\label{sec_rabasis}
Instead of using the basis of 1 and 2 fields, as we have done so far, it is oftentimes convenient to introduce a
second basis. To this end, we define \cite{Keldysh:1964ud,Chou:1984es}
\begin{align}
\phi_r \equiv \frac{1}{2}(\phi_1+\phi_2) \quad \quad \phi_a \equiv \phi_1 -\phi_2,
%\\
%J_r = \frac{1}{2}(J_1+J_2) \quad \quad J_a = J_1 -J_2,
\end{align}
which we call the $r/a$ basis. In this basis, the propagator matrix reads
\begin{align}
{ \bf D} =
\left(
\begin{array}{cc}
\langle \phi_r \phi_r \rangle & \langle \phi_r \phi_a \rangle \\
\langle \phi_a \phi_r \rangle & \langle \phi_a \phi_a \rangle
\end{array}
\right)
=
\left(
\begin{array}{cc}
D^{rr} & D^R \\
D^A & 0
\end{array}
\right) \quad \quad \textrm{in $r/a$ basis,}
\end{align}
where we have  for convenience defined the symmetric $rr$-propagator
\begin{align}
	\label{kmsrr}
D^{rr} = \frac{1}{2}(D^>+D^< ).
\end{align}
The propagator between two $a$ fields is identically zero to all orders, in and
out of equilibrium, due to the
$\theta$-functions in the definitions of the different correlation functions.

\begin{figure}[t!]
\begin{center}
\includegraphics[width=0.26 \textwidth]{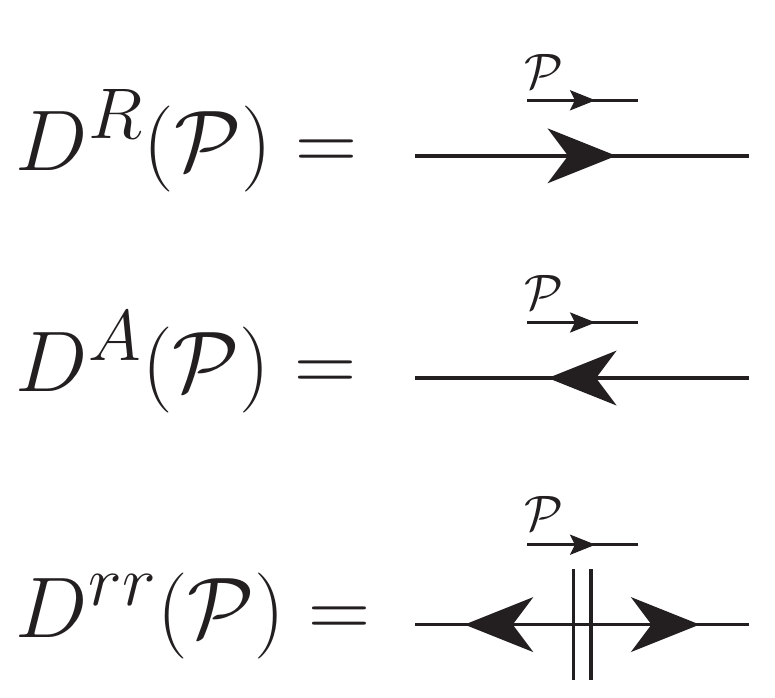}
\end{center}
\caption{Graphical representation of the propagators in the $r/a$ basis.
	The large arrows mark the direction of causation, whereas the small arrows on top
	indicate the flow of momentum. For the symmetric propagator $D^{rr}$ the direction
of the momentum does not matter for bosons. When drawing diagrams for gluons, we will use wiggly
lines instead of straight lines. For fermions, the fermionic flow must be aligned with
 momentum flow (not with causation).}
\label{fig:props_ra}
\end{figure}

In this basis, the vertices have an odd number of $a$ indices. This is so because
in the 1/2 basis the vertex with index 1 and the vertex with index 2 come with opposite
signs. For the interaction part $S_I$ of the action this gives
\begin{equation}
S_I(\phi_1)-S_I(\phi_2) = S_I\left(\phi_r+\frac{1}{2}\phi_a\right)-S_I\left(\phi_r-\frac{1}{2}\phi_a\right) .
\end{equation}
If there are no occurrences of $\phi_a$ in the two contributions, they cancel exactly. This is also the case
if there is an even number of $a$ fields. For example, consider a quartic term $\frac{1}{4!}\phi^4$;
the combined actions of fields
1 and 2 are proportional to
\begin{equation}
S_I(\phi_1)-S_I(\phi_2)\propto \frac{1}{4!}\left( \phi_1^4-\phi_2^4 \right) =  \frac{1}{2^2}\frac{1}{3!}\phi_a^3 \phi_r + \frac{1}{3!} \phi_r^3 \phi_a.
\label{eq:phi4}
\end{equation}
As can be seen from the above example, we have chosen the normalization of the $\phi_r$ and $\phi_a$ to be
such that for vertices with exactly one $a$-field, the symmetry factor $(3!)$ is reproduced correctly.
For vertices with more than one $a$-field, there is an extra factor of $1/2$ for each additional external $a$ line.

\begin{figure}[t!]
\begin{center}
\includegraphics[width=0.5 \textwidth]{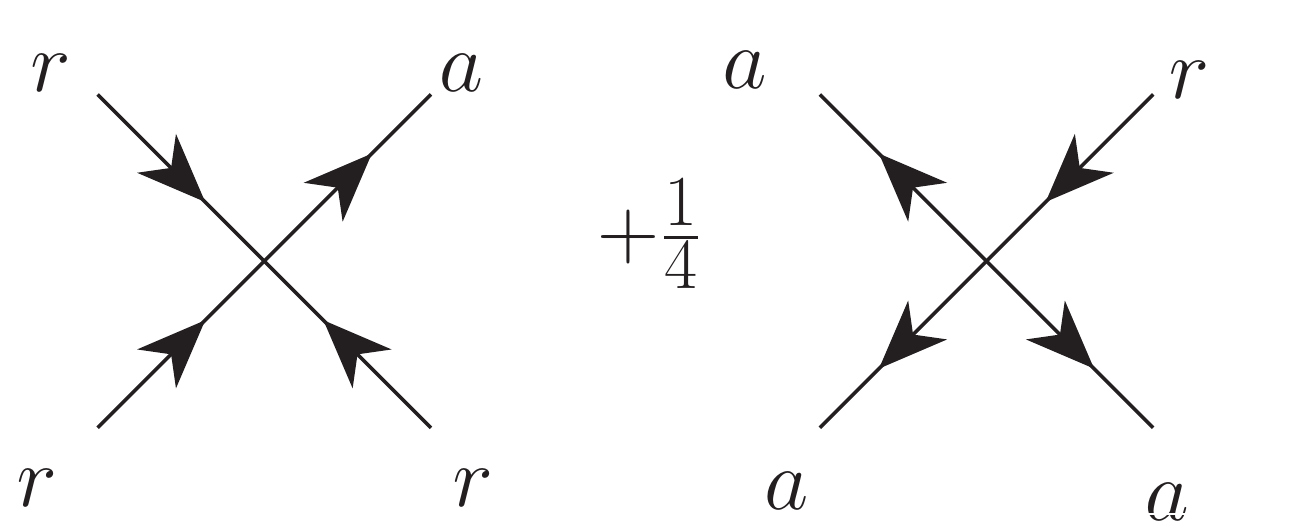}
\end{center}
\caption{Graphical representation of the vertices appearing in the example discussed in the $r/a$ basis.}
\label{fig:phi4_ra}
\end{figure}

The $r/a$ basis lends itself to a diagrammatic representation that is particularly intuitive \cite{CaronHuot:2007nw}.
Recall from response theory
that the retarded propagator $D^R(t_1,t_0)$ measures the response
of a field $\langle \phi( t_1 ,\x) \rangle$ \emph{caused} by a current $J(t_0,\x')$,
so that
\begin{align}
\label{eq:retarded}
\delta \langle \phi( t_1 ,\x) \rangle = -  i\int d^4 \x' D^R(t_1,\x;t_0,\x') J(t_0,\x'),
\end{align}
where  $\delta \langle \phi( t_1 ,\x) \rangle $ is the difference between the expectation value 
in the presence and in the absence of the source $J$. Therefore, we will use the notation
of \cite{CaronHuot:2007nw}, where retarded propagators are drawn as arrows that depict the 
\emph{flow of causation} from $t_0$ to $t_1$; see Fig.~\ref{fig:props_ra}. Similarly, we draw the 
advanced propagator as an arrow from $t_1$ to $t_0$. In vertices we draw arrows pointing out for $a$ 
fields and arrows pointing in for $r$ fields; see Fig.~\ref{fig:phi4_ra}.

The symmetric functions measure instead the \emph{correlation} between two fields.
This correlation may be either due to quantum fluctuations or to statistical fluctuations in the past, but
either way these fluctuations trace back to the density matrix at the time the system
was initialized as $\hat\rho(t_0)$, i.e.~both lines in the propagator are sourced by $\hat\rho(t_0)$ and
therefore we draw them as ``cut'' lines, where the cut is to be thought of as tracing back to $\hat\rho(t_0)$.

Before continuing our illustration of the advantages of this basis, we note that we shall present a detailed,
pedagogical calculation of the quark self energy in Sec.~\ref{sub_heur_htl}.

\subsubsection{The $r/a$ basis and causality}
\label{sub_ra_causal}
\begin{figure}[t!]
\begin{center}
\includegraphics[width=0.5 \textwidth]{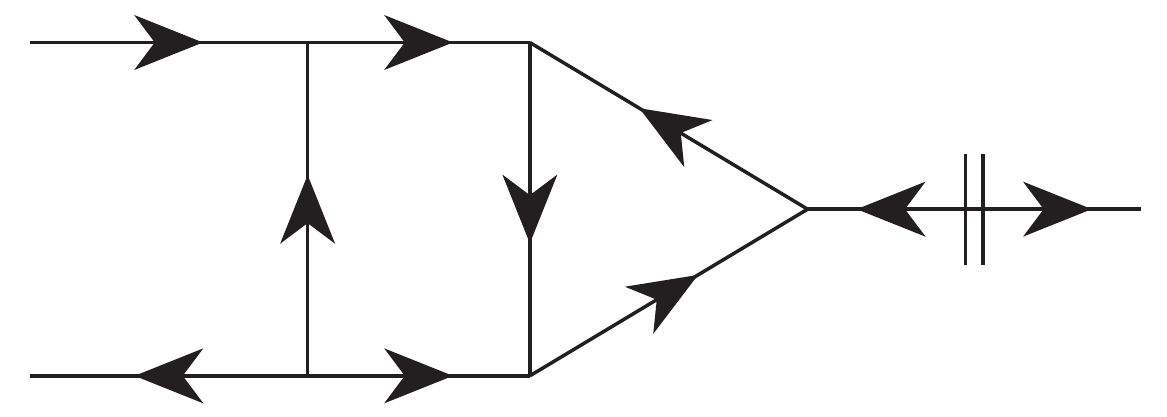}
%\put(-97,60){$t_0$}
\put(-97,60){$t_1$}
\put(-56,37){$t_0$}
\end{center}
\caption{An example of a diagram that is identically zero because it  contains a closed loop of causation.}
\label{fig:causation}
\end{figure}

One major advantage of the $r/a$ basis is its straightforward relation to causality,
which dictates that there can be no closed loops formed from advanced
or retarded propagators only. This is most
simply observed in the time domain: should we have a closed loop of causation as shown in
Figure~\ref{fig:causation},  we must have
 both a (product of) retarded and advanced propagators
  connecting the vertices at $t_0$ and $t_1$. However,
because of the step functions in the
definitions of the advanced and retarded propagators, these have support
only for $t_0 -t_1 >0$ (for advanced) or $t_0 -t_1  < 0$ (for retarded)
and one of them is necessarily zero. Thus, any diagram with a closed loop of flow of
causality is necessarily zero,
 as depicted in Fig.~\ref{fig:causation}. As the figure makes clear, these loops are clearly
 identified in the $r/a$ formalism as a succession of simple arrows in the same direction. They
 are then easily discarded when drawing all possible $r/a$ assignments.

\subsection{Self-energies and amputated diagrams in the $r/a$ formalism}
\label{sub_self_energies}
It is oftentimes useful to consider diagrams which have had the propagators
of the external legs amputated, including in particular the case of the amputated two-point function,
the self energy. We denote the amputated diagrams by $\Pi$. In the $r/a$ formalism,
the amputated diagrams carry indices as well. By convention we choose the indices so
that the amputated diagram carries those indices that appear on the near side of
the bare propagator $D_{(0)}$ which is removed.

The amputated diagrams are related to expectation values of the currents
conjugate to the amputated fields. In particular,
\begin{align}
\Pi^{aa}(t_1,t_0) &=  -i \frac{1}{2}\langle \{J(t_1),J(t_0)\}\rangle\\
\Pi^{ra}(t_1,t_0) &\equiv  \Pi^{A}(t_1,t_0)  = i \theta(t_0- t_1) \langle [J(t_1),J(t_0)]\rangle \\
\Pi^{ar}(t_1,t_0) &\equiv \Pi^{R}(t_1,t_0)   =  -i \theta(t_1- t_0) \langle [J(t_1),J(t_0)]\rangle 
\label{eq:46}\\
\Pi^{rr}(t_1,t_0) &= 0
\end{align}
and similarly 
\begin{align}
\Pi^{>}(t_1,t_0)  = -\Pi^{21} = -i \langle J(t_1)J(t_0)\rangle  \\
\Pi^{<}(t_1,t_0)  = -\Pi^{12}= -i \langle J(t_0)J(t_1)\rangle
\end{align}

\begin{figure}[t!]
	\begin{center}
		\includegraphics[width=12.3cm]{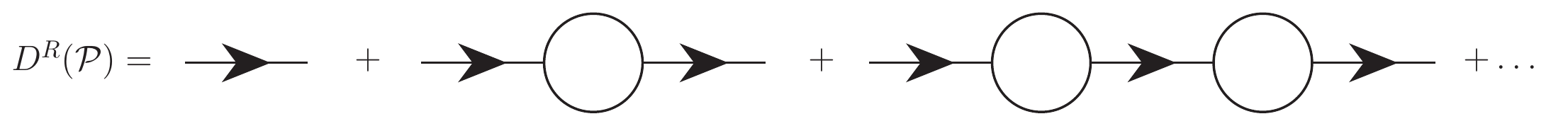}
	\end{center}
	\caption{Resummation for the retarded propagator. All self energies and bare propagators
between the self energies are retarded. This is so because, on one hand, if one or more of the
bare propagators were to be an $rr$ propagator $D^{rr}$, at least one of the self energies
would need to be $\Pi^{rr}$, which vanishes identically. On the other hand, if one or more of
the self energies were $\Pi^{aa}$, at least one of the propagators would need to be a
vanishing $D^{aa}$.}
	\label{fig_gret}
\end{figure}

That the propagator between two $a$-fields vanishes identically translates directly to the
vanishing of the $rr$ self energy $\Pi^{rr}(\mathcal{P})=0$.
This leads to a particularly simple form for the Dyson-Schwinger equation relating
the retarded and advanced self energies to the corresponding propagators
(see fig.~\ref{fig_gret})
\begin{equation}
D^{R/A}(\mathcal{P})  =
\frac{ 1 }{[D^{R/A}_{(0)}(\mathcal{P})]^{-1} + i \Pi^{R/A}(\mathcal{P})}.
\label{retresum}
\end{equation}

\begin{figure}[t!]
	\begin{center}
		\includegraphics[width=12cm]{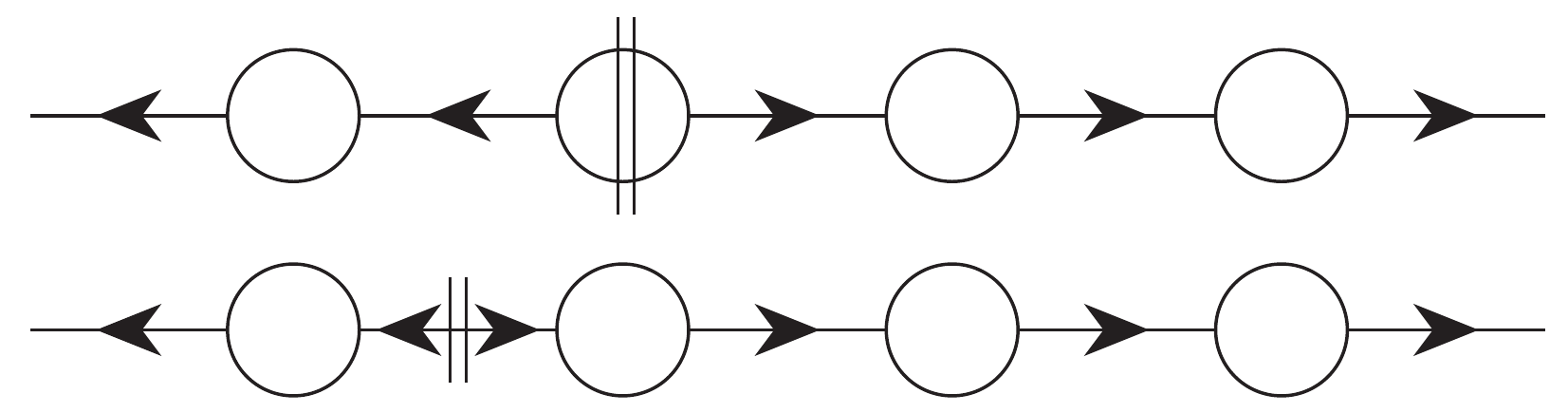}
	\end{center}
	\caption{Resummation for the $rr$ propagator. The cut self energy stands for $\Pi^{aa}$. In order to arrive to an $rr$ propagator, the flow of causation needs to be flipped exactly once by either $D^{rr}$ or $\Pi^{aa}$. The first and second diagrams corresponds to the first and second terms in Eq.~(\ref{eq:Drr_SD}), respectively. That is, the correlation in the fields can be induced either by a statistical fluctuation in the currents ($\Pi^{aa}$) or in the fields themselves ($D^{rr}$) at an earlier time. }
	\label{fig_grr}
\end{figure}

The expression for the remaining $D^{rr}$ is non-trivial, as
the cut that changes the causality flow may either appear in the
self energy or in the propagator connecting two self energies, as depicted
in Fig.~\ref{fig_grr}. Thus,
\begin{align}
D^{rr}(\mathcal{P}) =
&- D^{ra}(\mathcal{P})\,i \Pi^{aa}(\mathcal{P})\, D^{ar}(\mathcal{P}) \nonumber \\
&+ \left[  D^{ra} (\mathcal{P}) (D^{ra}_{(0)}(\mathcal{P}))^{-1} \right] D^{rr}_{(0)}(\mathcal{P})\left[  (D^{ar}_{(0)}(\mathcal{P}))^{-1}  D^{ar} (\mathcal{P})\right]. \label{eq:Drr_SD}
\end{align}
It can be easily shown that a similar relation holds also for the forward and
backward Wightman self energies,
\begin{align}
D^{>,<}(\mathcal{P}) =
&- D^{ra}(\mathcal{P}) \,i \Pi^{>,<}(\mathcal{P})\, D^{ar}(\mathcal{P}) \nonumber \\
&+ \left[  D^{ra} (\mathcal{P}) (D^{ra}_{(0)}(\mathcal{P}))^{-1} \right] D^{>,<}_{(0)}(\mathcal{P})\left[  (D^{ar}_{(0)}(\mathcal{P}))^{-1}  D^{ar} (\mathcal{P})\right].
\end{align}
Finally, we recall that the KMS conditions that we introduced for the
connected two-point functions (the propagators) in
Eqs.~\eqref{kmsmombosonless}-\eqref{kmsmomfermionless} apply in equilibrium
to the amputated function as well. One then has, for instance
\begin{equation}
	\label{kmspiaa}
	\Pi^{aa}(\mn{P})=\left(\frac12\pm n(p^0)\right)\left(\Pi^R(\mn{P})-\Pi^A(\mn{P})
	\right),
\end{equation}
and similarly for $\Pi^>$ and $\Pi^<$.

For higher $n$-point correlation functions, the assignments with only one $r$ index and rest $a$'s (i.e.  $\langle \phi^r \phi^a \phi^a\ldots\rangle$) correspond to fully retarded functions  (see e.g. \cite{Chou:1984es,Blaizot:2001nr,CaronHuot:2007nw})
 \begin{align}
% 	\nonumber
 \delta \langle \phi(y_0) \rangle  
 = \sum_{n=1}^{\infty} \frac{(-i)^n}{n!}\int d^4 y_1 d^4 y_2\ldots d^4y_n
 &D^{raa\ldots}(y_0;y_1,y_2,\ldots,y_n) \nonumber \\ 
 &\times J(y_1)J(y_2)\ldots J(y_n).
 \label{manyfields}
 \end{align}
corresponding to a linear response of an operator $\phi(y_0)$ to multiple currents $J(y_i)$ in analogy with Eq.~(\ref{eq:retarded}). It can be shown that all retarded/advanced $n$-point functions
can be obtained by analytical continuation from Euclidean correlation functions. This continuation is, however, non-trivial because of the presence of multiple frequencies, leading to multiple ways for how the continuation from Euclidean to real frequencies can be performed, depending on the signs of the frequencies of the individual lines \cite{Evans:1991ky}.

\subsection{In-medium generalization of the Cutkosky rules}
\label{sub_cutrules}

\begin{figure}[t!]
	\begin{center}
		\includegraphics[width=10cm]{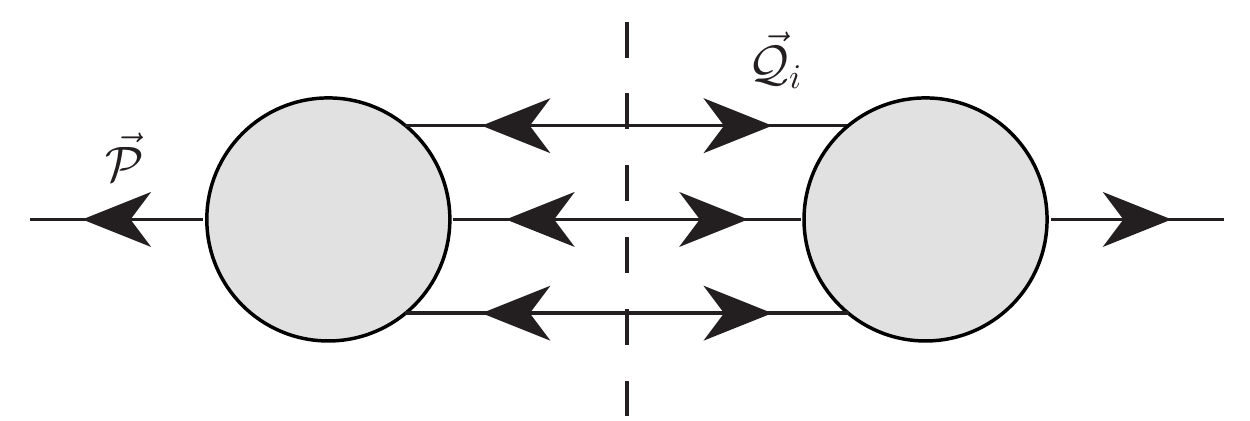}
	\end{center}
	\caption{Graphical representation of a possible cut in the evaluation
	of the cutting rule in Eq.~\eqref{simonrule}. The blobs represent
	the fully retarded amplitudes and the cut lines are replaced
	by the Wightman propagators $D^>(\mathcal{Q}_i)$.}
	\label{fig_cutrule}
\end{figure}

Similarly to the vacuum Cutkosky rules, there is an in-medium expression
for the imaginary part of the time-ordered self energy in terms of a sum over squared
amplitudes \cite{Kobes:1985kc,Kobes:1986za}. However, as argued earlier, the time-ordered
propagator is of limited use in medium and does not have a straightforward physical interpretation.
Hence, it is not surprising that the cutting rule written in terms of the time ordered
self energy becomes rather baroque and often cumbersome to use. There are several
reformulations of the rule in different Schwinger--Keldysh bases (see
\cite{Chou:1984es,Guerin:1993ik,Gelis:1997zv}), but the version by 
Caron-Huot \cite{Caron-Huot:2007zhp} in the $r/a$ basis simplifies
it significantly and provides a straightforward physical interpretation. According to it, we have
\begin{align}
	\nonumber
\Pi^>(\mathcal{P}) = & \sum_n \frac{1}{n!}\left(\prod_n\int\frac{d^4\mn{Q}_n}{(2\pi)^4}\right)
(2\pi)^4
\delta^4(\mathcal{Q}_1+\ldots+\mathcal{Q}_n -\mathcal{P})\\
& \times
 \mathcal{M}_{ar...r}(\mathcal{P};\mathcal{Q}_1,\ldots,\mathcal{Q}_n)
 \mathcal{M}_{ar...r}(-\mathcal{P};-\mathcal{Q}_1,\ldots,-\mathcal{Q}_n) \nn\\
 & \times
  D^>(\mathcal{Q}_1)\ldots D^>(\mathcal{Q}_n),
\label{simonrule}
\end{align}
where the sum runs over all possible cuts, and the cut lines are replaced by the $D^>(\mathcal{Q}_i)$
propagators, with momenta assigned from left to right.  We have not shown explicitly
any internal indices on the cut lines (color, spin, etc.), which are 
assumed to be summed over.
All cut propagators are furthermore
to be considered to be attached to neighboring vertices as $r$ fields
and the external lines as $a$ fields. Hence, one transparently sees the
physical picture: a sum over all  \emph{fully retarded} squared amplitudes 
which are the appropriate finite-temperature generalization of ``matrix elements'',
multiplying the forward Wightman propagators.\footnote{\label{amputated_foot} The
argument just seen in Sec.~\ref{sub_self_energies} refers to the correlators  --- see in particular
Eq.~\eqref{manyfields} and the discussion preceding it ---
for which it is true that those with a single $r$ field are fully retarded/advanced. For amputated amplitudes,
it then follows that those with a single $a$ leg are fully retarded/advanced, hence the fully retarded
label for $\mathcal{M}_{ar...r}$.} To fix  conventions, we define
$i\mathcal{M}$ as %correspond to 
the fully retarded  amputated Feynman diagrams with outgoing momentum $\mn{P}$ at the $a$
vertex and incoming momenta $\mn{Q}_i$ at the $r$ vertices.
 An example cut is depicted
in Fig.~\ref{fig_cutrule}, and a pedagogic application of the rules to thermal photon production will be presented in Sec.~\ref{sec_soft_collinear}.

\subsection{From the 1/2 basis to vacuum field theory}
\label{sub_vacuum}
\begin{figure}
\begin{center}
\includegraphics[width=0.8\textwidth]{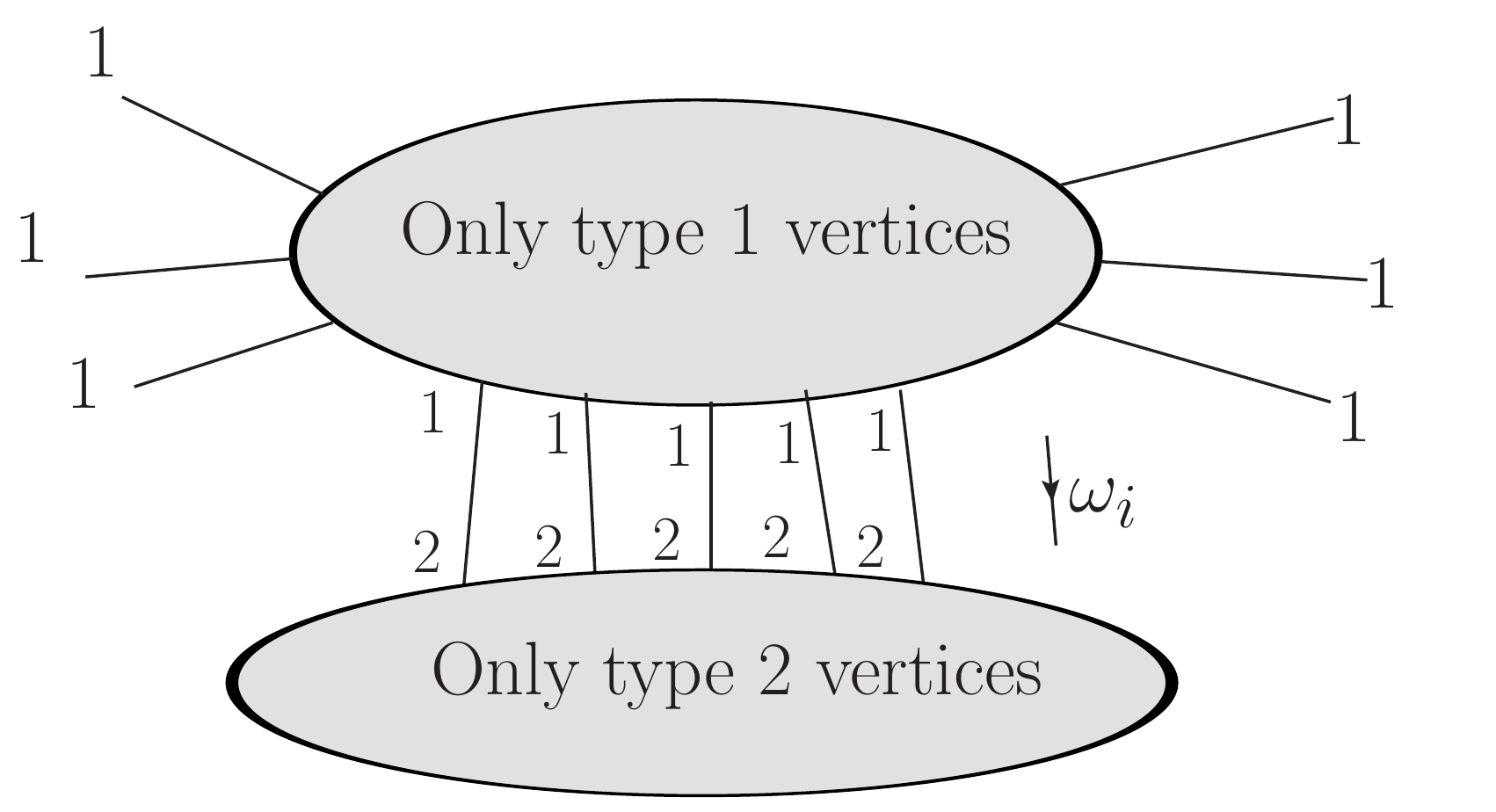}
\end{center}
\caption{A generic diagram with with only type 1 fields as external lines. The diagram is
organized such that all the vertices and propagators with index 1 are in the upper ellipse,
while the type 2 vertices and propagators live in the lower ellipse. The two regions are
connected by Wightman functions $D^>(\omega_i)$ (the direction of momentum flow is indicated
by the small arrow). Conservation of energy implies that the frequencies $\omega_i$ that
flow between the two regions must sum up to zero, $\sum_i \omega_i=0$, and therefore some of
them must be negative. In vacuum, the Wightman function $D^>(\omega_i)$ has support only for
positive frequencies, and therefore such diagrams containing type 2 fields vanish.}
\label{only_1_fields}
\end{figure}
One might wonder, how is it that in zero-temperature field theory one manages with only a
single set of fields, whereas in the statistical theory a field doubling is necessary. While
the reason for this has been already explained in Sec.~\ref{sec:IV}, it is amusing
the see how this happens diagrammatically. If we are satisfied with computing only
time-ordered correlation functions, as one usually is in vacuum field theory, then the
only correlation functions we need are those between any number of $\phi_1$ fields.
Then we may separate the diagram to parts where there are only vertices with fields
$\phi_1$ and to parts that contain only fields $\phi_2$, that arise from the loops
within the diagram. This is depicted in figure \ref{only_1_fields}. Now, these two parts
of the diagrams are connected with a number of $\langle \phi_2 \phi_1 \rangle$ propagators.
In the frequency domain, as there are no external $\phi_2$ lines in the diagram, the sum of
the frequencies appearing in the Wightman functions $D^{>}(\omega_i)$ must add up to zero,
$\sum_i \omega_i = 0$. In vacuum, the forward Wightman function has support only for positive
frequencies (see Footnote \ref{footnote:pos_omega}). Therefore, at least one of the lines
must have a negative frequency running through it, causing the
diagrams containing type 2 fields to give a vanishing contribution
when computing time-ordered correlation functions. In the presence of a medium, however,
there is no reason for the forward Wightman functions to be zero for negative frequencies.
Indeed, even when computing time-ordered correlation functions, the type-2 fields contribute
to the diagrams.

\subsection{Relation to classical field theory}
\label{sub_classical}
We conclude this section by reviewing how classical field theory arises as a limit of the full
quantum theory in the limit where the fields are strong, or equivalently the occupation numbers
 large. In equilibrium this corresponds to the infrared bosonic modes for which $\omega \ll T$. 
Following the discussion in \cite{Mueller:2002gd} (see also e.g.~\cite{Bodeker:1995pp, Bodeker:1996wb,  Greiner:1996dx,Aarts:1996qi, Aarts:1997kp,Mathieu:2014aba}), we rewrite the ``horizontal''
part of the Schwinger--Keldysh generating functional in Eq.~\ref{genfunc} in
the $r/a$ basis, i.e.
\begin{align}
\int \mathcal{D}\phi_r\int \mathcal{D}\phi_a e^{i S(\phi_r + \frac{1}{2}\phi_a)-i S(\phi_r - \frac{1}{2}\phi_a)}\,.
\end{align}
As discussed earlier, the difference of the two actions contains only terms with
an odd number of $\phi_a$-fields.
If it is the case that there is a scale hierarchy between the $\phi_a$-fields and the
$\phi_r$-fields, then the leading-order term in the expansion in $\phi_a$ fields is a
linear function of $\phi_a$, and the integral over $\phi_a$ can be explicitly performed.
To quantify when the condition is fulfilled,
consider that in equilibrium $\langle \phi^r(-\omega) \phi^r(\omega) \rangle
\sim (1/2 + n_{B}(\omega))\rho_{B}(\omega)$.
For $\omega \ll T$, the  bosonic
distribution function $n_B(\omega)\approx T/\omega$ is parametrically larger than
the constant $1/2$, and the $\phi^r$ fields are then $\mathcal{O}(n_B^{1/2}(\omega))$.
Whenever $\omega\ll T$ and thus $n_B(\omega)\gg 1$ we speak of \emph{Bose enhancement}.
To estimate instead the size of $\phi_a$ consider then
$\langle  \phi^r(-\omega) \phi^a(\omega) \rangle$, which is the retarded correlator.
Since it does not depend on $n_B(\omega)$, the $ra$ correlator is
$\mathcal{O}(n_B^0(\omega))$. Therefore $\phi_a(\omega)$ is of order $n_B^{-1/2}(\omega)$.
Hence the approximation becomes accurate in the limit of large occupation numbers
$n_B(\omega)$, which in thermal equilibrium corresponds to $\omega \ll T$.
In the case of non-equilibrium systems, $n_B(\omega)$ is replaced with the
non-equilibrium occupation number $f(\omega)$.

For example, in  $\lambda\phi^4$ theory with the Lagrangian of Eq.~\eqref{eq:phi4}, the
leading term in the expansion in $\phi_a$ reads 
\begin{align}
S= -\int d^4 x \, \phi_a \left[  (-\partial_\mu \partial^{\mu}+m^2)\phi_r  + \frac{1}{3!} \lambda (\phi^r)^3 \right],
\end{align}
or  more generally
\begin{align}
S =  \int d^4 x \,   \left[ \phi_a  \frac{\delta \mathcal{L}(\phi^r )}{\delta \phi^r} - \partial_\mu \phi_a   \frac{\delta \mathcal{L}(\phi^r )}{\delta \partial_\mu \phi^r }  \right].
\end{align}
The integral over $\mathcal{D}\phi_a$ reduces the path integral into a delta function at all space-time points
constraining the fields to be solutions to classical equations of motion,
\begin{align}
\prod_{\x,t} 2\pi \delta \left[  \frac{\delta \mathcal{L}(\phi^r )}{\delta \phi^r} -
 \partial_\mu    \frac{\delta \mathcal{L}(\phi^r )}{\delta \partial_\mu \phi^r }  \right].
\end{align}
To extend this discussion by combining the classical fields with hard quantum fields, we
refer to \cite{Jeon:2013zga}. Unlike in the full quantum theory, the real-time evolution of the classical approximation can be numerically solved on the lattice. For some recent numerical examples see e.g. \cite{Aarts:2001yx, Laine:2009dd, Moore:2010jd,  Boguslavski:2018beu}.

\section{Soft and collinear physics in QCD}
\label{sec_soft_collinear}

\begin{figure}[t!]
	\begin{center}
		\includegraphics[width=8cm]{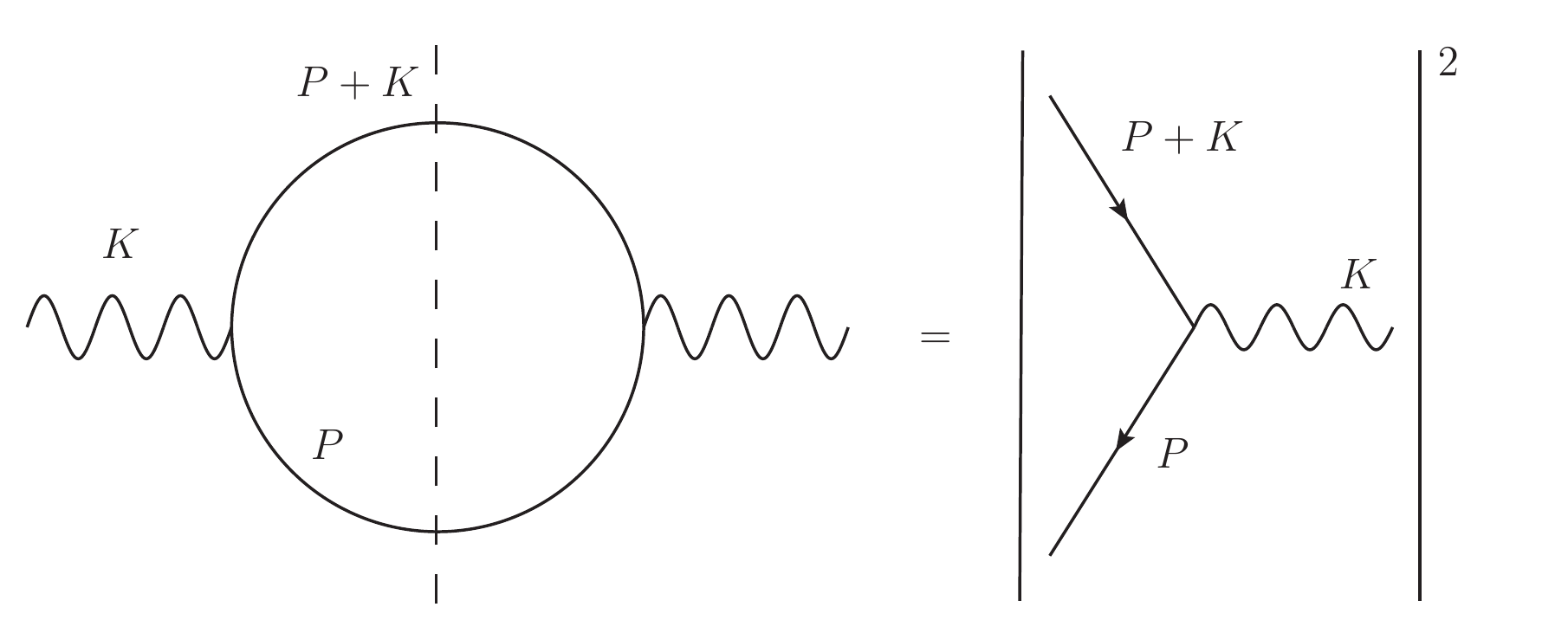}		
	\end{center}
	\caption{The zeroth-order graph for $\Pi^<$ on the left. Its cut 
	correspond to the squared amplitude for the process on the right, which vanishes
	for real photon emission. Figure taken from \cite{Ghiglieri:2014kma}.}
	\label{fig_dy}
\end{figure}

In the previous Section, we have introduced three different methods for computing
real-time correlation functions: the $1/2$ basis in Sec.~\ref{sub_12}, the $r/a$ basis
in Sec.~\ref{sec_rabasis}, and the cutting rules in Sec.~\ref{sub_cutrules}. 
Hence, performing a real-time computation might seem to boil down to finding the most 
convenient among these techniques for the problem at hand and then proceeding to 
its application. However, the naive application of the Feynman rules would in most cases result in infrared divergences. As we have mentioned in the introduction,
these in turn signal sensitivity to soft and/or collinear
regions of the phase space, where naive perturbation theory breaks down. This breakdown
corresponds to the emergence of collective effects, arising from the dynamics of the
thermal medium. The next two subsections, \ref{sub_htl_real} and \ref{sec_lpm},
will be dedicated to introducing the subtleties
of soft and collinear physics, respectively. However,
to guide the reader with a physical problem where both feature extensively, we 
now introduce a hands-on example: thermal photon production.

A thermal QCD medium can be considered weakly coupled to photons, so that the latter
are not in equilibrium and their production is a rare event. Under these assumptions,
a classic derivation \cite{McLerran:1984ay} finds that the photon emission rate
per unit volume is, at first order in $\alpha_\mathrm{em}=e^2/(4\pi)$,
\begin{equation}
	\label{photonrate}
	\frac{dN_\gamma}{d^4\mn{X}d^3k}\equiv\frac{d\Gamma_\gamma}{d^3k}=\frac{\Pi^{<}(\mn{K})}{(2\pi)^32k}
	,\qquad\Pi^<(\mn{K})=\int d^4\mn{X} e^{-i\mn{K}\cdot\mn{X}}
	\left\langle J^\mu(0)\;J_\mu(\mn{X})\right\rangle,
\end{equation}
where $\mn{K}=(k,0,0,k)$ is the photon's lightlike momentum---we assume
$k\sim T$---and the electromagnetic
current reads $ J^\mu\equiv
\sum_i^{n_f} e Q_i\overline\psi_i\gamma^\mu\psi_i $ 
for $n_f$ quarks---assumed to be massless in what follows---with electric charges $Q_i$.

Eq.~\eqref{photonrate} requires the computation of a Wightman function, $\Pi^<(\mn{K})$.
As such, the optimal technique for its evaluation lies in the cutting rules
of Sec.~\ref{sub_cutrules}, where the ``${}^<$'' version of Eq.~\eqref{simonrule} is easily
obtained by changing all occurrences of ${}^>$ to ${}^<$. 
At zeroth order in $g$, we would then have the simple one-loop diagram shown in Fig.~\ref{fig_dy}.

\begin{figure}[t!]
	\begin{center}
		\includegraphics[width=12cm]{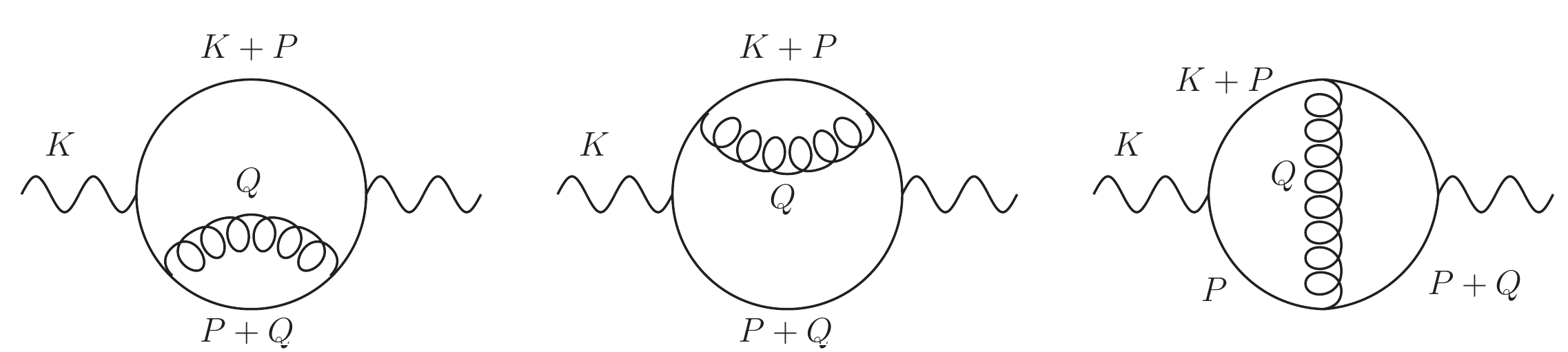}		
	\end{center}
	\caption{The first-order graphs for $\Pi^<$.
	 Figure taken from \cite{Ghiglieri:2013gia}.}
	 \label{fig_lophoton}
\end{figure}

As shown there, the cut of that diagram corresponds to the square of the tree-level
photon emission, which is well known to vanish kinematically for real photons,
which cannot be emitted from on-shell quarks. Indeed, the straightforward application
of Eq.~\eqref{simonrule} to that diagram results in
\begin{equation}
	\label{zerothphoton}
	\Pi^<_{g^0}(\mn{K})=-\int \frac{d^4\mn{P}}{(2\pi)^4}\mathrm{Tr}
	\big[(e Q \gamma^\mu)S^<(\mn{P}+\mn{K})(e Q \gamma_\mu)S^<(-\mn{P})\big],
\end{equation}
where  the retarded and advanced amplitudes 
in Eq.~\eqref{simonrule} are $\mathcal{M}_{arr}(\mn{K};\mn{P}+
\mn{K},-\mn{P})=e Q\gamma^\mu$. 
As Eq.~\eqref{kmsmomfermionless} shows, the $S^<$ propagators are proportional
to the fermion spectral density $\rho_F(\mn{P})$, $S^<(\mn{P})=-\nfd(p^0)\rho_F(\mn{P})$, 
which in the bare limit used in ordinary
perturbation theory in the interaction representation reads
$\rho_F(\mn{P})=-\slashed{\mn{P}}\epsilon(p^0)2\pi\delta(\mn{P}^2)$.\footnote{
	\label{foot_metric}
	Our convention for the Dirac algebra is slightly nonstandard, in that we
	choose $\{\gamma^\mu,\gamma^\nu\}=-2g^{\mu\nu}$. Normally (see the extensive
	discussion in App.~E of \cite{Burgess:2007zi})
	the mostly-plus metric is associated with a factor of $i$ to the $\gamma$ matrices,
	so that the anticommutator maintains a plus sign, as in the case of the 
	mostly-minus metric.
} It is then
straightforward to verify that the $d^4\mn{P}$ integration vanishes 
over the product of the two $\delta$-functions putting the two quarks on shell, 
as anticipated.

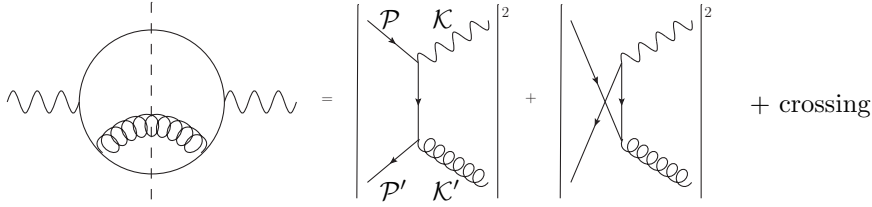
\begin{figure}[t!]
	\begin{center}
		\hspace{-2cm}\includegraphics[width=10cm]{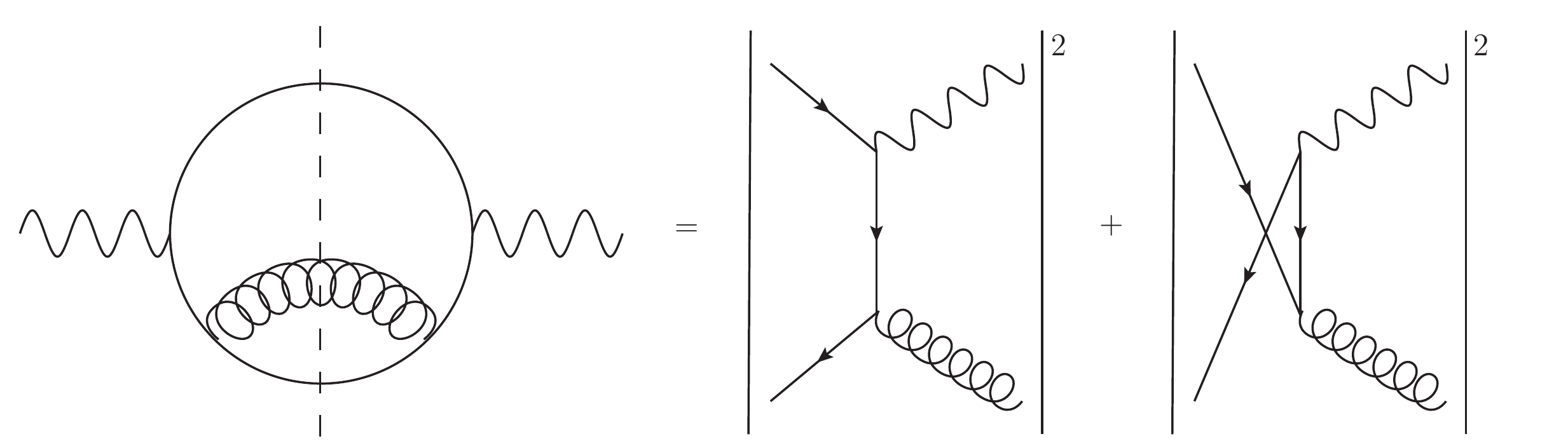}		
		\put(0,35){+ crossing}
		\put(-120,67){$\mn{K}$}
		\put(-120,2){$\mn{K}'$}
		\put(-140,67){$\mn{P}$}
		\put(-140,2){$\mn{P}'$}
	\end{center}
	\caption{The non-vanishing cut of the first diagram
	in Fig.~\ref{fig_lophoton}. The crossing, i.e.~the Compton process, is not
	shown explicitly. The cuts of the third diagram in  Fig.~\ref{fig_lophoton}
	represent the interference between the two diagrams on the right.}
	\label{fig_lo_cuts}
\end{figure}

The first contribution to photon production then needs an extra gluon to be
kinematically allowed,
and thus happens at $\mathcal{O}(g^2)$, where one encounters the diagrams of 
Fig.~\ref{fig_lophoton}. The cutting rules can now be applied to these; the class of cuts where the gluon is not
cut reproduces the $\delta$-function structure seen before and vanishes again. In other
words, they represent the \textit{interference} between the Born process of
Fig.~\ref{fig_dy} and its virtual correction. On the other hand, the cuts passing through
the gluon are kinematically allowed and correspond to the processes shown in 
Fig.~\ref{fig_lo_cuts}, i.e.~the Compton and pair annihilation processes of QCD and QED. 

A tedious but straightforward application of the cutting and Feynman rules leads to \cite{Kapusta:1991qp,Baier:1991em,Arnold:2001ms}
\begin{align}
	\label{totnaive}
\Pi_{g^2\,\mathrm{naive}}^<(\mn{K})\equiv&	
\Pi^<(\mn{K})_\mathrm{Compton}+\Pi^<(\mn{K})_\mathrm{annih},\\
\Pi^<(\mn{K})_\mathrm{Compton}=&e^2\sum_{i=1}^{n_f}Q_i^2
\int\frac{d^3p\,d^3p'\, d^3k'}{(2\pi)^9\,8\,p\,p'\,k'}(2\pi)^4\delta^{(4)}(\mn{P}+\mn{P'}-
\mn{K}-\mn{K}')	\nn\\
\label{comptonnaive}
\times&16d_F C_F g^2\left[\frac{-s}{t}+\frac{-t}{s}\right]\nfd(p)\nbe(p')(1-\nfd(k'))\,,\\
\Pi^<(\mn{K})_\mathrm{annih}=&e^2\sum_{i=1}^{n_f}Q_i^2
\int\frac{d^3p\,d^3p'\, d^3k'}{(2\pi)^9\,8\,p\,p'\,k'}(2\pi)^4\delta^{(4)}(\mn{P}+\mn{P'}-
\mn{K}-\mn{K}')	\nn\\
\label{annihnaive}
\times&8d_F C_F g^2\left[\frac{u}{t}+\frac{t}{u}\right]\nfd(p)\nfd(p')(1+\nbe(k'))\, ,
\end{align}
where, $s$, $t$, and $u$ stand for the usual Mandelstam variables. Upon taking the cuts, the momenta have been shifted into those
of Fig.~\ref{fig_lo_cuts}.  This makes particularly transparent the connection to kinetic
theory: noting how the terms in square brackets are nothing
but the matrix elements squared for these processes, we see that
 we have recovered the \emph{gain term} of a Boltzmann equation for photon
production in the case where the photon's distribution $f_\k$ is negligible,
which is precisely the approximation underlying the derivation of Eq.~\eqref{photonrate}.
Under this approximation, the \emph{loss term} vanishes entirely.

We refer to \cite{Arnold:2001ms} for technical details of the
 evaluation of Eqs.~\eqref{comptonnaive} and \eqref{annihnaive}. What we wish
 to emphasize here is instead that the phase space integrations for both processes
 span the regions $t\to 0$ and $u\to 0$, giving rise to a logarithmic IR
 divergence, which is precisely what we anticipated. It
 signals the breakdown of this naive perturbative expansion and the need
 for a proper handling of the region of \textit{soft momenta}, 
 $t=-(\mn{P}-\mn{K})^2\sim g^2T^2$, where Hard Thermal Loop
 resummation describes the emergence of collective effects. We will devote 
 Sec.~\ref{sub_htl_real} to its illustration, where we  derive in 
 detail the HTL-resummed soft contribution in Eq.~\eqref{softphotonsumrulefinal}.

 \begin{figure}[t!]
	\begin{center}
		\includegraphics[width=8cm]{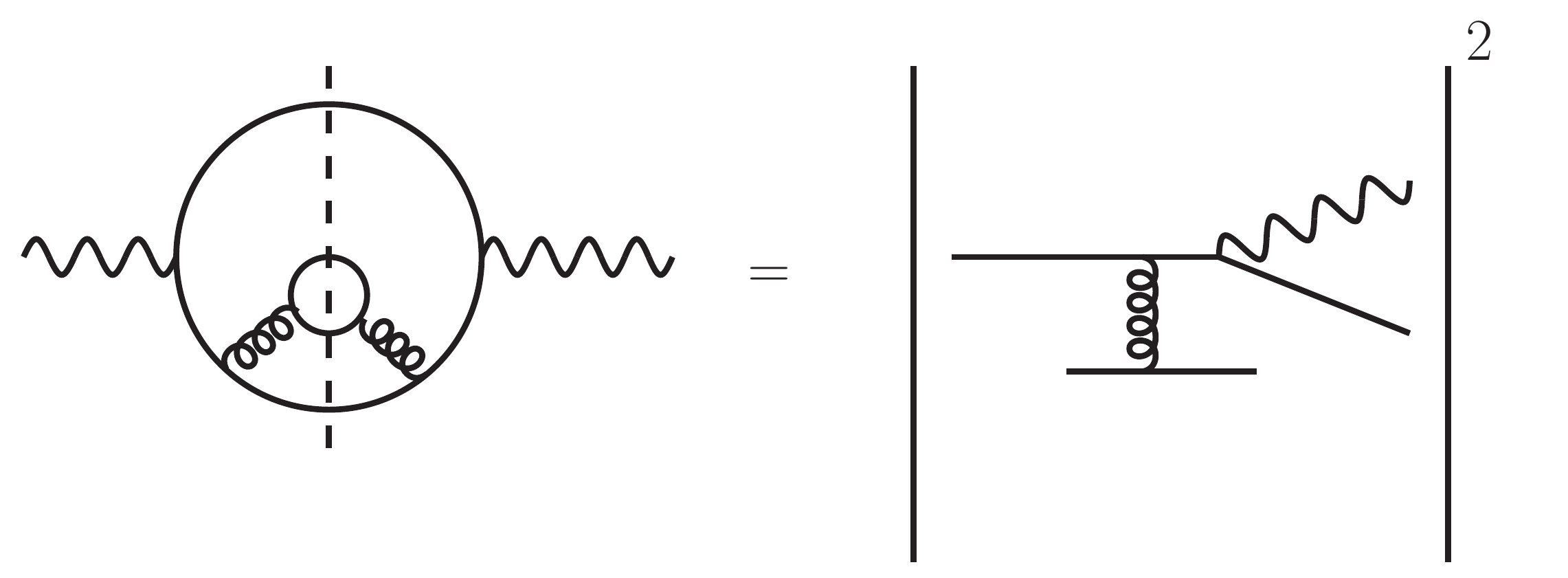}
	\end{center}
	\caption{A higher-order process that can be collinearly enhanced, with its
	cut process shown on the right. The crossing into the annihilation diagram is 
	not shown explicitly. Figure taken from~\cite{Ghiglieri:2014kma}.}
	\label{fig_photon_collinear}
\end{figure} 

This sensitivity to soft exchanges does not represent the only breakdown of the naive expansion
for thermal photon production. Indeed, it turns out there is another leading
order contribution; this stems from a proper handling of \textit{collinear physics}.
To see this, let us look at the contribution depicted in Fig.~\ref{fig_photon_collinear}: naively, it is suppressed by 
$g^2$ with respect to Fig.~\ref{fig_lophoton}. However, it was realized in 
\cite{Aurenche:1998nw} that a collinear enhancement boosts a slice of the
phase space region in these types of diagrams to leading order. This happens
when the gluon momentum is small, so that the virtual quark coupling
to the gluon on one side and the photon on the other is only slightly off-shell, 
$\mn{P}^2\sim g^2 T^2$, enhancing its propagator  by the inverse of that. Furthermore,
kinematics constrain the outgoing photon and quark to be \emph{collinear},
$\mn{P}\cdot\mn{K}\sim g^2T^2$. This small opening angle between $\k$ and $\p$
implies a long \emph{formation time} $\tau_\mathrm{form}\approx p/(2\mn{P}\cdot\mn{K})$. This is nothing
but the time it takes the wave packets of the outgoing photon and quark to separate. As we shall see
in great detail in the next sections, the interactions of the quarks
with soft gluons are so frequent that many such scatterings will overlap
during a single formation time, so that their quantum-mechanical interference
needs to be accounted for, in what is called Landau-Pomeranchuk-Migdal (LPM) resummation. It will be the core 
of Sec.~\ref{sec_lpm}. 

\begin{figure}[t!]
	\begin{center}
		\includegraphics[width=10cm]{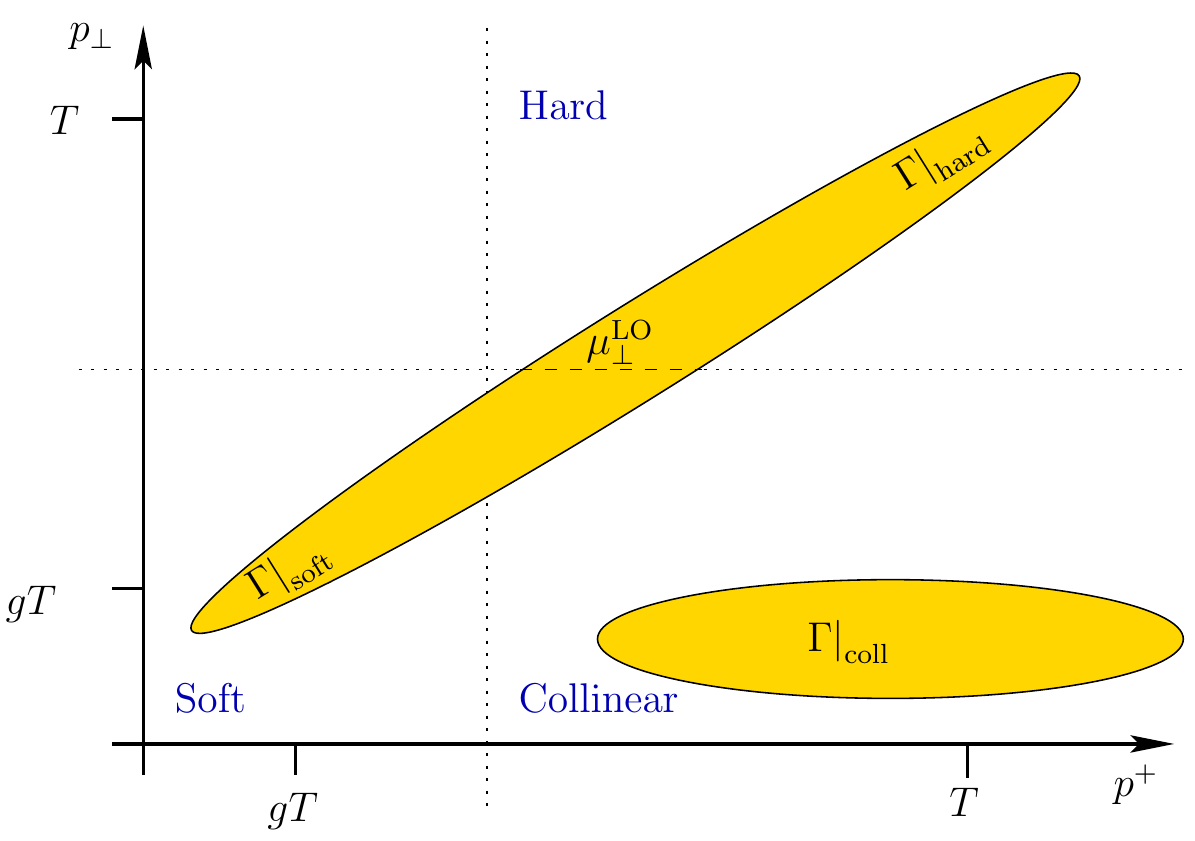}
	\end{center}
	\caption{The momentum regions contributing to the leading-order
	photon rate in a $(p^+,p_\perp)$ plane, where $\mn{P}$ labels the momenta
	indicated in Fig.~\ref{fig_lophoton} (recalling that $k=k^z$, $p^+$ is defined 
	as $p^+\equiv(p^0-p^z)/2$). The $p^-\equiv p^0-p^z$ component follows from momentum conservation.
	``Hard'' denotes here the region where the exchanged momentum is large, and 
	naive perturbation theory, as in Eqs.~\eqref{comptonnaive} and Eq.~\eqref{annihnaive}
	is valid. In the ``soft'' region, all components of $\mn{P}$ are small,
	$\mathcal{O}(gT)$, and HTL resummation becomes necessary. A cutoff $\mu_\perp^\mathrm{LO}$
	separates these first two regions.  In the
	``collinear'' region the light-cone momentum is large, but the transverse momentum
	is soft, so that LPM resummation becomes necessary. As the yellow blobs show, the collinear
	and hard+soft contributions are separated at leading order.
	 Figure taken from \cite{Ghiglieri:2013gia}.}
	\label{fig_lomap}
\end{figure}

We conclude this introduction by  graphically summarizing the momentum
regions contributing to the leading-order photon rate in Fig.~\ref{fig_lomap}.
Mathematically, this corresponds to
\begin{equation}
	\label{fulllophoton}
	\Pi^<_{g^2}(\mn{K})=\Pi^<_{g^2\,\mathrm{naive}}(\mn{K})+
	\Pi^<_{g^2\,\mathrm{soft}}(\mn{K})
	+\Pi^<_{g^2\,\mathrm{coll}}(\mn{K})\,,
\end{equation}
where $\Pi^<_{g^2\,\mathrm{naive}}(\mn{K})$ is given by Eq.~\eqref{totnaive}, 
$\Pi^<_{g^2\,\mathrm{soft}}(\mn{K})$ will be presented in 
Eq.~\eqref{softphotonsumrulefinal} and $\Pi^<_{g^2\,\mathrm{coll}}(\mn{K})$
in Eq.~\eqref{finalphoton}.
\subsection{Soft physics: Hard Thermal Loop resummation}
\label{sub_htl_real}
The emergence of collective effects at frequencies and/or momenta of order $gT$
is a well known fact in plasma physics: indeed, we have just seen how it comes 
about in thermal photon production. In Thermal Field Theory these effects
find a consistent, modern and gauge-invariant definition in the Hard Thermal Loop
(HTL) effective theory. This was originally introduced by Braaten and Pisarski
\cite{Braaten:1989mz,Braaten:1991gm}, by Frenkel and Taylor
\cite{Frenkel:1989br,Frenkel:1991ts}
and by Taylor and Wong \cite{Taylor:1990ia}.
Their connection to a kinetic picture
for the underlying hard modes has been illustrated in great detail in the review
of Blaizot and Iancu \cite{Blaizot:2001nr}. Here we will briefly cover all these aspects,
referring to textbooks such as \cite{Bellac:2011kqa,Kapusta:2006pm}  and to the review
\cite{Blaizot:2001nr} for more detailed expositions on the diagrammatic
derivation/effective action and the kinetic connection, respectively.
Finally, the role of Hard Thermal Loops in imaginary-time calculations will be reviewed later on in
Sec.~\ref{sec:htlnpointeffaction}.

\subsubsection{Heuristic introduction}
\label{sub_heur_htl}

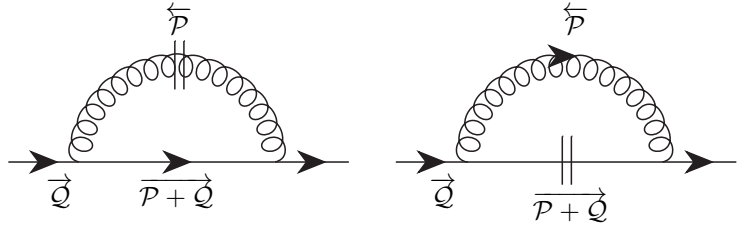
\begin{figure}[t!]
	\begin{center}
		\includegraphics[width=10cm]{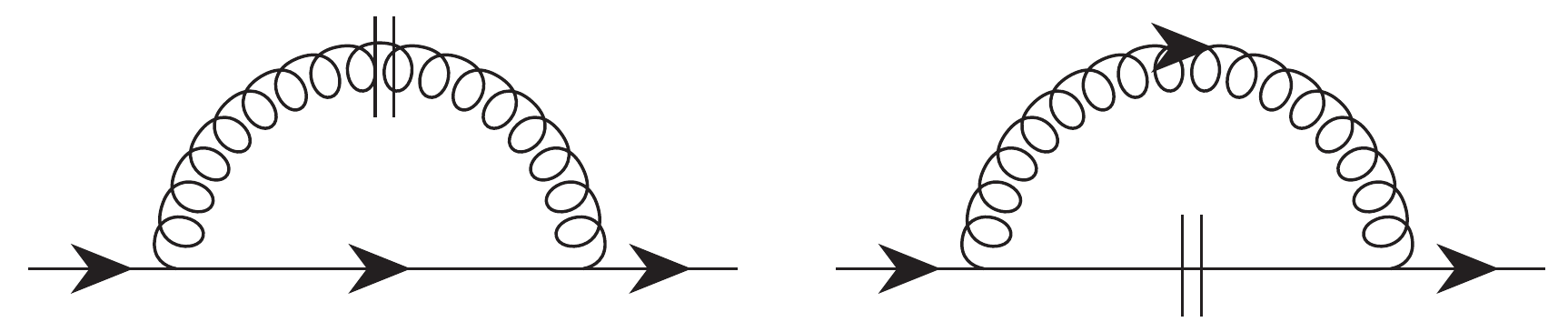}
		\put(-70,60){$\overleftarrow{\mn{P}}$}
		\put(-81,-10){$\overrightarrow{\mn{P+Q}}$}
		\put(-120,-5){$\overrightarrow{\mn{Q}}$}
		\put(-265,-5){$\overrightarrow{\mn{Q}}$}
		\put(-230,-5){$\overrightarrow{\mn{P+Q}}$}
		\put(-220,60){$\overleftarrow{\mn{P}}$}
	\end{center}
	\caption{The two $r/a$ assignments for the retarded fermion self energy. We recall
	that the momenta of fermions need to be aligned with the direction of fermion flow
	(not indicated by the arrows, which instead indicate causation).}
	\label{fig_fermion_self}
\end{figure}

To introduce HTLs and their connection to the kinetic picture in a simple, pedagogic
way, let us start from the diagrammatic derivation of the fermionic HTL in the $r/a$
formalism. This computation will also serve as an example
of a real-time calculation in  said formalism. 

The extension of the discussion of Sec.~\ref{sec_rabasis} to QCD is straightforward. For what concerns the quark-gluon vertex, only $rra$ and $aaa$ assignments are possible, with the latter again suppressed by a factor of $1/4$. Furthermore, the $aaa$ vertex cannot contribute to 
the retarded self energy, since neither an $aa$ propagator nor an $rrr$ vertex are available.
There are then only two assignments of the $r/a$ indices contributing to the retarded 
self energy $\Sigma^R(\mn{P})$, shown in Fig.~\ref{fig_fermion_self}.  This further highlights the advantages of this basis: the retarded
self energy, from which all other self energies can be derived using 
Eq.~\eqref{kmspiaa} and the other relations discussed in Sec.~\ref{sub_self_energies},
is obtained from two assignments only, with a transparent connection to causality and
statistics.

A straightforward application of the Feynman rules of
Sec.~\ref{sec_rabasis} yields
\begin{equation}
-i\Sigma^R(\mn{Q})%&=&
=(-ig)^2C_F \int\frac{d^4\mn{P}}{(2\pi)^4}
\gamma^\mu\big[S^R(\mn{P}+\mn{Q})G_{\mu\nu}^{rr}(\mn P)%\nn\\
	%&&\hspace{3.3cm}
+S_{rr}(\mn{P}+\mn{Q})G_{\mu\nu}^{A}(\mn P)\big]\gamma^\nu,
\label{fermionself}
\end{equation}
where the integration has been kept in exactly 4 dimensions because
we only want to extract the HTL, which is finite. As we
have mentioned before, the Hard Thermal Loop amplitudes are gauge invariant. A complete
field-theoretical proof of this property was given in \cite{Kobes:1990dc}.
We exploit this invariance and continue the computation in Feynman gauge, which
slightly simplifies the intermediate expressions. Using the propagators listed in \ref{app_horror}, we find
\begin{eqnarray}
\Sigma^R(\mn{Q})&=&g^2C_F \int\frac{d^4\mn{P}}{(2\pi)^4}
\gamma^\mu(\slashed{\mn{P}}+\slashed{\mn{Q}})\gamma_\mu \bigg[  \frac{	\nbe(|p^0|)2\pi\delta(\mn{P}^2)}
	{(\mn{P}+\mn{Q})^2- i\epsilon( p^0+q^0)}
\nn\\
	&&\hspace{3.3cm}
	-\frac{\nfd(|p^0+q^0|)
2\pi\delta((\mn{P}+\mn{Q})^2) }{\mn{P}^2+ i\epsilon p^0}\bigg],
\label{fermionselfexp}
\end{eqnarray}
where we have only kept the thermal (statistical) part of the symmetric
propagators, i.e.~$1/2\pm n(|p^0|)\to\pm n(|p^0|)$. 
Since all integrals are finite, we can easily shift
$\mn{P}\to-\mn{P}-\mn{Q}$ on the second line, obtaining
\begin{equation}
\Sigma^R(\mn{Q})=g^2C_F \int\frac{d^4\mn{P}}{(2\pi)^4}\frac{4\pi\delta(\mn{P}^2) }
{(\mn{P}+\mn{Q})^2- i\epsilon( p^0{+}q^0)}
\big[ (\slashed{\mn{P}}+\slashed{\mn{Q}})\,\nbe(|p^0|)
	+\slashed{\mn{P}}\,\nfd(|p^0|)\big].
\label{fermionselfexpsymm}
\end{equation}
Up to now we have not taken any hierarchical approximations: the full thermal
part of the Feynman-gauge quark self energy can be obtained from Eq.~\eqref{fermionselfexpsymm}
by first performing the frequency integration over the $\delta$ function and then
performing the angular integrations. To this end, it is convenient to 
project the Dirac structure on the two vectors $\slashed{\mn{Q}}$ and 
$\slashed{u}$, with $u^\mu=(1,0,0,0)$ the plasma frame.
 The resulting $p$ integration can then only be carried
out numerically. We refer to \cite{Weldon:1982bn} for the expression of the 
integrand. 

On the other hand, since we are interested in extracting the HTL
contribution, we can  now take the assumptions underlying that theory, which
requires the extraction of the leading term for a soft external quark
interacting with a hard loop. We  thus have to expand for
$\mn{Q}\sim gT \ll\mn{P}\sim T$ and take the leading term, leading to
\begin{equation}
\Sigma^R(\mn{Q})=g^2C_F \int\frac{d^4\mn{P}}{(2\pi)^4}2\pi\delta(\mn{P}^2)
\left(\nbe(|p^0|)+\nfd(|p^0|)\right)
  \frac{\slashed{\mn{P}}}
	{\mn{P}\cdot\mn{Q}- i\epsilon p^0}.
\label{fermionselfexphtl}
\end{equation}
Upon defining $v\equiv\mn{P}/p^0=(1,\p/p^0)$ we see that the angular part factors
out, yielding
\begin{equation}
\Sigma^R(\mn{Q})=g^2C_F \int\frac{d^4\mn{P}}{(2\pi)^4}2\pi\delta(\mn{P}^2)
\left(\nbe(|p^0|)+\nfd(|p^0|)\right)
  \frac{\slashed{v}}
	{v\cdot\mn{Q}- i\epsilon }.
\label{fermionselfexphtlfactor}
\end{equation}
Performing here the $p^0$ and $p$ integrations, we obtain
\begin{equation}
	\Sigma^R(\mn{Q})=\frac{\mm}{2} \int\frac{d\Omega_v}{4\pi}
  \frac{\slashed{v}}
	{v\cdot\mn{Q}- i\epsilon },
\label{fermionselfexphtlfinal}
\end{equation}
where $\mm\equiv g^2C_FT^2/4$ is the \emph{asymptotic mass} of the quarks, as we shall
illustrate later on.\footnote{We write the fermionic asymptotic mass
with a lowercase $m_\infty$ and the gluonic one with an uppercase $M_\infty$.}
Here we wish to further elaborate on the structure that has emerged
from our calculation: an angular integration over the eikonal propagator
$\slashed{v}/v\cdot \mn{Q}$ resulting from integrating out the off-shell hard leg of the Hard
Thermal Loop. It is here that the connection to the kinetic picture appears: 
$\slashed{v}/v\cdot \mn{Q}$ is nothing
but the (retarded) propagator of the \emph{induced fermionic source}, in the language
of \cite{Blaizot:2001nr}: the quark-gluon loop in the HTL approximation has reduced
to this structure, which shares the same color-triplet nature of the original quark. This
consideration, combined with gauge invariance, suggests that $\slashed{v}/v\cdot \mn{Q}$ is
just the first term in the (Fourier-transformed)
expansion of $\slashed{v}/v\cdot D$, with $D$ the covariant
derivative, which is indeed borne out by explicit computations of higher-point
functions. More generally, it has been shown \cite{Taylor:1990ia,Braaten:1991gm} that
HTL amplitudes with two external quark lines can be generated by adding  an
extra, effective term to the QCD Lagrangian. This term reads, in Minkowskian signature
\begin{equation}
	\label{deltaLfHTL}
	\delta\mathcal{L}_f=i\frac{\mm}{2}\,\overline\psi \int\frac{d\Omega_v}{4\pi}
	\frac{\slashed{v}}{v\cdot D}\,\psi,
\end{equation}
which generates all fermionic HTLs with two external, soft quark lines and
an arbitrary number of soft external gluons. All these retarded amplitudes are
gauge-invariant and proportional to $g^2T^2$.  % While that is certainly true for the purely gluonic ones
It can be shown that there are no HTLs, i.e.~no amplitudes proportional to $g^2T^2$,
 with more than two external fermion lines,
so Eq.~\eqref{deltaLfHTL}  generates \emph{all} fermionic HTLs.
Furthermore, once the $v\cdot D$ denominator is taken into account, the two-quark
function scales like $gT$ and the $qqg$ amplitude scales like $g$. Hence the
former scales exactly like the denominator of a fermion propagator for soft $\mn{Q}$
and the latter like the bare $qqg$ vertex of QCD. In both cases, this signals a breakdown
of the loop expansion of the bare theory and a need for \emph{HTL resummation},
whose consequences for propagators and vertices we shall explain later.

For HTLs with external gluons only, a similar procedure applies, with
the added complication that the retarded two-point gluon HTL requires the next order
in the expansion for $\mn{Q}\ll \mn{P}$ of the full one-loop
amplitude. This fact, together with extra intricacies
relating to gauge fixing, has brought us to our previous illustration
using the fermionic HTL.
The retarded gluonic HTL turns out to read
\begin{equation}
	\label{gluehtl}
	\Pi^{\mu\nu}_R(\mn{Q})=m_D^2\int\frac{d\Omega_v}{4\pi}\left(\delta^{\mu}_0\delta^{\nu}_0
	+v^\mu v^\nu\frac{q^0}{v\cdot\mn{Q}- i\epsilon }\right),
\end{equation}
where $m_D$ is the leading-order \emph{Debye mass}, which reads $m_D^2=(N_c+T_Fn_f)g^2T^2/3$, 
with  $T_F=1/2$.
Eq.~\eqref{gluehtl} shows a similar structure to Eq.~\eqref{fermionselfexphtlfinal}. The main
differences arise in the different numerator structure for the
eikonal propagator of the gluonic source, and in the presence of the extra
term for temporal gluons. The corresponding Lagrangian term
generating all $n\ge 2$-point gluonic functions reads
\cite{Braaten:1991gm} 
\begin{equation}
	\label{deltaLgHTL}
	\delta\mathcal{L}_g=\frac{m_D^2}{2}\mathrm{Tr}\,\int\frac{d\Omega_v}{4\pi}
	F^{\mu\alpha}\frac{v_\alpha v_\beta}{(v\cdot D)^2}\,F^\beta_{\;\,\mu}.
\end{equation}

Also in the gluonic case, the retarded two-point function,
Eq.~\eqref{gluehtl}, is of order $g^2T^2$, and so are all retarded $n{>}2$ point
functions generated by Eq.~\eqref{deltaLgHTL}. However, in the $ra$ formalism,
different orderings have different power countings. Let us consider the $aa$
two-point function. The KMS condition in Eq.~\eqref{kmspiaa} leads to, in the
boson case
\begin{equation}
	\label{gluehtlaa}
	\Pi^{\mu\nu}_{aa}(\mn{Q})=\frac{T}{q^0}\left(\Pi^{\mu\nu}_R(\mn{Q})-
	\Pi^{\mu\nu}_A(\mn{Q})\right)= i T m_D^2\int\frac{d\Omega_v}{4\pi}
	\,v^\mu v^\nu \,2\pi\delta(v\cdot\mn{Q})\,,
\end{equation}
which can be obtained from the leading-order term in the $\mn{Q}\ll\mn{P}$
expansion, differently to the retarded self energy. We present such a
derivation (for the gauge contribution only) in \ref{app_horror}, showing
explicitly its gauge invariance as well.
We also remark that, due to the factor of $1/g$ from the Bose-enhanced
soft thermal distribution
$\nbe(q^0)\approx T/q^0$, it is of order $gT^2$. Why then consider the retarded HTL at all,
if the $aa$ is larger? The answer lies in the power-counting rules
worked out by Caron-Huot \cite{CaronHuot:2007nw}
for the gluonic HTLs in the $r/a$ formalism. In this two-point example,
the $1/g$-enhanced $aa$ self energy \emph{has to} connect to the other pieces
of the considered amplitude by retarded and advanced propagators, since
there is no $aa$ propagator. These propagators scale like $1/g^2T^2$. The
retarded and advanced self energies, which have an $r$ index, can connect
with an $rr$ propagator. This, in turn, is related to the retarded and
advanced ones by the KMS condition, Eq.~\eqref{kmsrr},
 which enhances it by $1/g$ with respect
to them, making the two cases to be of equivalent order.

Extending these arguments beyond the two-point function, Caron-Huot found that
the gluonic HTL theory in the $r/a$ formalism requires all $n$-point functions
with one or two external $a$ lines, the former case being, as we have mentioned
in Sec.~\ref{sub_self_energies} (see also footnote~\ref{amputated_foot}),  the fully retarded/advanced functions, which
are also those
obtained by analytical continuation from Euclidean functions. Gluonic HTLs with
more than two external $a$ lines do not exist, meaning that gluonic amplitudes with
more than two $a$ external lines represent corrections beyond the HTL theory.

\begin{figure}[t!]
	\renewcommand{\u}{\unitlength}
	\begin{picture}(365,83)(-30,0)
	\put(0,50){\scalebox{-1}[1]{ \includegraphics[width=60\u,height=19\u]{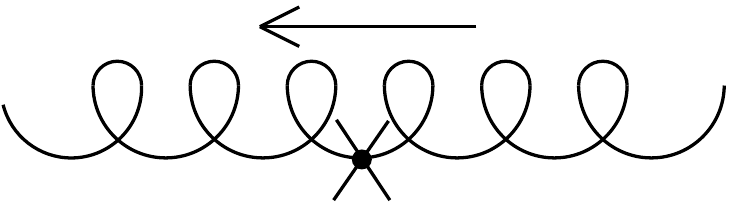}}}
	\put(0,25){\scalebox{-1}[1]{ \includegraphics[width=60\u,height=16\u]{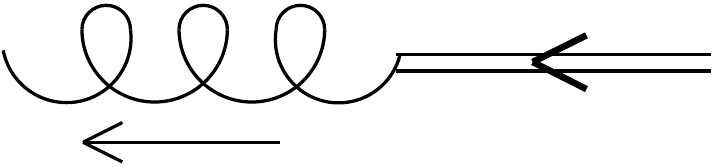}}}
	\put(0,3){\scalebox{-1}[1]{\includegraphics[width=60\u,height=16\u]{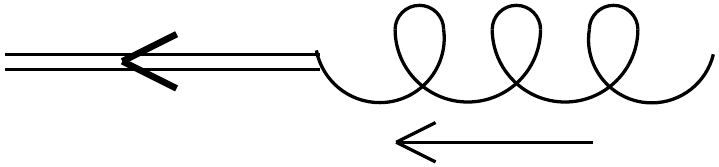}}}
	\put(-30,55){(a)}
	\put(-30,31){(b)}
	\put(-30,8){(c)}
	\put(-8,60){$\mu$}
	\put(62,60){$\nu$}
	\put(-8,13){$\mu$}
	\put(62,35){$\mu$}
	\put(43,1){$\mn{Q}$}
	\put(75,56){$= im_D^2 \delta^\mu_0\delta^\nu_0$}
	\put(75,34){$= iT\,v^\mu$}
	\put(75,12){$= iq^0\,v^\mu$}

	\put(200,55){\scalebox{-1}[1]{\includegraphics[width=40\u,height=22\u]{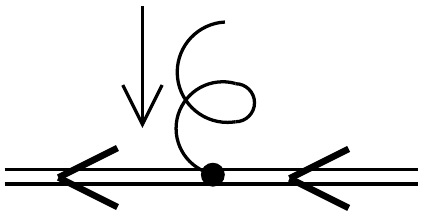}}}
	\put(218,78){$\mu,b$}
	\put(193,58){$c$}
	\put(241,58){$a$}
	\put(252,56){$=-v^\mu f^{abc}$}
	\put(200,29){\scalebox{-1}[1]{\includegraphics[width=40\u,height=8\u] {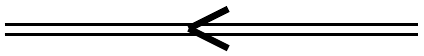}}}
	\put(215,38){$\mn{Q}$}
	\put(252,32){$\displaystyle = \frac{-i}{v\cdot \mn{Q}-i\epsilon}$}
	\put(200,06){\includegraphics[width=40\u,height=10\u]{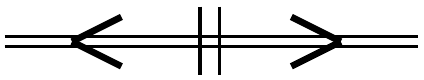}}
	\put(215,16){$\mn{Q}$}
	\put(252,08){$= 2\pi\delta(v\cdot \mn{Q})$}
	\put(170,55){(d)}
	\put(170,31){(e)}
	\put(170,08){(f)}
	%\put(0,90){Each solid line: $\frac{m_D^2}{T} \int \dv$}
	\end{picture} %}
	%\end{center}
	\caption{Effective Feynman rules introduced in \cite{CaronHuot:2007nw}
	 for the gluonic HTL theory in the $r/a$ formalism.
	The arrows follow the graphical notation for $r/a$ diagrams introduced in
	Sec.~\ref{sec_rabasis}. The identity in color space, $\delta^{ab}$, is
	not explicitly shown for the two-point functions. The double line, e.g. in (e),
	is the hard current; a factor $(m_D^2/T)\int\frac{d\Omega_v}{4\pi}$ must
	be assigned to every such disjoint double line 
	appearing in a diagram. The application of these rules
	yields the amputated amplitude $i\Gamma$. %, which for the two-point
	%function equals $-i\Pi$.
	Figure taken and adapted from \cite{CaronHuot:2007nw}.}
	\label{fig_simon_rules}
\end{figure}

Besides these very important power-counting arguments, Caron-Huot also presented
a set of Feynman rules to generate these gluonic HTLs with one or two
$a$ indices. They are presented in Fig.~\ref{fig_simon_rules} and are based on
the propagation of the hard induced current, which is represented as a double
line. These rules thus allow to generate the required HTLs in a simple way,
while at the same time preserving a manifest link with the kinetic picture
through this hard current propagator, which
represents an effective two-particle state. Intuitively, we can go back from these
effective rules to the (sum of) the original one-loop graphs in QCD by
``opening up'' these double lines to recreate the hard loop they describe.

As an example, let us derive the  retarded HTL from these rules. We have
\begin{equation}
i\Gamma^{\mu\nu}=im_D^2 \delta^\mu_0\delta^\nu_0+	
\frac{m_D^2}{T}\int\frac{d\Omega_v}{4\pi} (iT v^\mu)\frac{-i}{v\cdot\mn{Q}-i\epsilon}
(i q^0v^\nu)\,
\end{equation}
where the first term comes from term (a), while the second from the succession of (c), (e) and (b)
which in turn create, propagate and annihilate the hard current. As expected, this agrees with
Eq.~\eqref{gluehtl}. Similarly, the $aa$ self-energy can be obtained by dropping the contribution
from (a) and having a succession of (b), (f) and (b), reproducing Eq.~\eqref{gluehtlaa}.

Finally, we remark that this discussion implies the existence, in the HTL-resummed
theory, of $raa$ and $rraa$ vertices, which, as we have argued in Sec.\ref{sec_rabasis},
do not exist in the bare theory. Furthermore, while the $rra$ vertex
(either bare or resummed) scales like $g^2T$ for soft external gluons, this $raa$
vertex, by the arguments we have just presented, scales like $gT$, that is, is
enhanced, the enhancement being compensated by the need to connect
to the other sections of the graph via one less enhanced $rr$ propagator.

A similar analysis for the HTLs with external quark lines does not exist in
the literature. Clearly, the same power-counting rules do not apply there, because the
fermionic KMS condition $\Sigma^{aa}(\mn{Q})=(1/2-\nfd(q^0))(\Sigma^R(\mn{Q})-
\Sigma^A(\mn{Q}))$
implies that the $aa$ self energy is \emph{suppressed} by a factor of $g$
with respect to the retarded/advanced ones, since $1/2-\nfd(q^0)\approx q^0/(4T)
\sim g$. It would be interesting to work out in detail the rules and
power counting for the fermionic HTLs as well, along the lines
of \cite{CaronHuot:2007nw}. A partial analysis has been presented
in App.~E of \cite{Ghiglieri:2013gia}.

As we have mentioned, the HTL theory resums the leading thermal behaviour for soft
external lines of momenta $\mn{P}_i$. Corrections in $\mn{P}_i/T$ and in $g$ are of course expected;
recently, an EFT approach to systematically tackle the former has been proposed in 
\cite{Manuel:2014dza,Manuel:2016wqs,Carignano:2017ovz,Carignano:2019ofj} under the name of On-Shell Effective Field Theory
(OSEFT) and has been developed to study the subleading contributions to the QED HTLs. The interplay between the 
$\mn{P}_i/T$ and $g$ corrections was analyzed in \cite{Mirza:2013ula}.

\subsubsection{Collective modes}
We now turn to the analysis of the main features of the two-point functions.
In the gluonic case, $\Pi^{\mu\nu}(\mn{Q})$ is still transverse to $\mn{Q}$,
but there exist two separate functions, $\Pi_L$ and $\Pi_T$, which are respectively
longitudinal and transverse with respect to $\bf{q}$. In the standard convention
 $\Pi_L(\mn{Q})=(\mn{Q}^2/q^2)\Pi^{00}(\mn{Q})$ and
$\Pi_T(\mn{Q})=(\delta^{ij}-\hat{q}^i\hat{q}^j)\Pi^{ij}(\mn{Q})/2$.
One can perform the angular integrations of the retarded function
given in Eq.~\eqref{gluehtl} to obtain
\begin{equation}
\Pi_R^{00}(\mn{Q})=m_D^2\left(1-\frac{q^0}{2q}\ln\frac{q^0+q+i\epsilon}{q^0-q+i\epsilon}\right),
\qquad \Pi^R_T(\mn{Q})=\frac{m_D^2}{2}
-\frac{\mn{Q}^2}{2q^2}\Pi_R^{00}(\mn{Q})\,.
\label{htlpi}
\end{equation}
The resummed retarded propagators then follow from Eq.~\eqref{retresum}
and from the bare ones in \ref{app_horror}. In Coulomb gauge we define
$G^{00}_R(\mn{Q})\equiv G^R_L(\mn{Q})$ and
$G^{ij}_R(\mn{Q})\equiv(\delta^{ij}-\hat q^i\hat q^j) G^R_T(\mn{Q})$. They read 
\begin{eqnarray}
\label{htllong}
G^{00}_R(\mn{Q})&=&\frac{i}{\displaystyle
q^2+m_D^2\left(1-\frac{q^0}{2q}\ln\frac{q^0+q+i\epsilon}{q^0-q+i\epsilon}\right)},\\
\nn G^{ij}_R(\mn{Q})&=&%(\delta^{ij}-\hat p^i\hat p^j)G^T_R(\)=
\left.\frac{-i(\delta^{ij}-\hat q^i\hat q^j)}
     {\displaystyle \mn{Q}^2+\frac{m_D^2}2 \left(\frac{q_0^2}{q^2}
       -\left(\frac{q_0^2}{q^2}-1\right)\frac{q^0}{2q}
       \ln\frac{q^0{+}q}{q^0{-}q}\right)}\right\vert_{q^0=q^0+i\epsilon}.\\
&& 	\label{htltrans}
\end{eqnarray}

\begin{figure}[t!]
	\begin{center}
		\includegraphics[width=10cm]{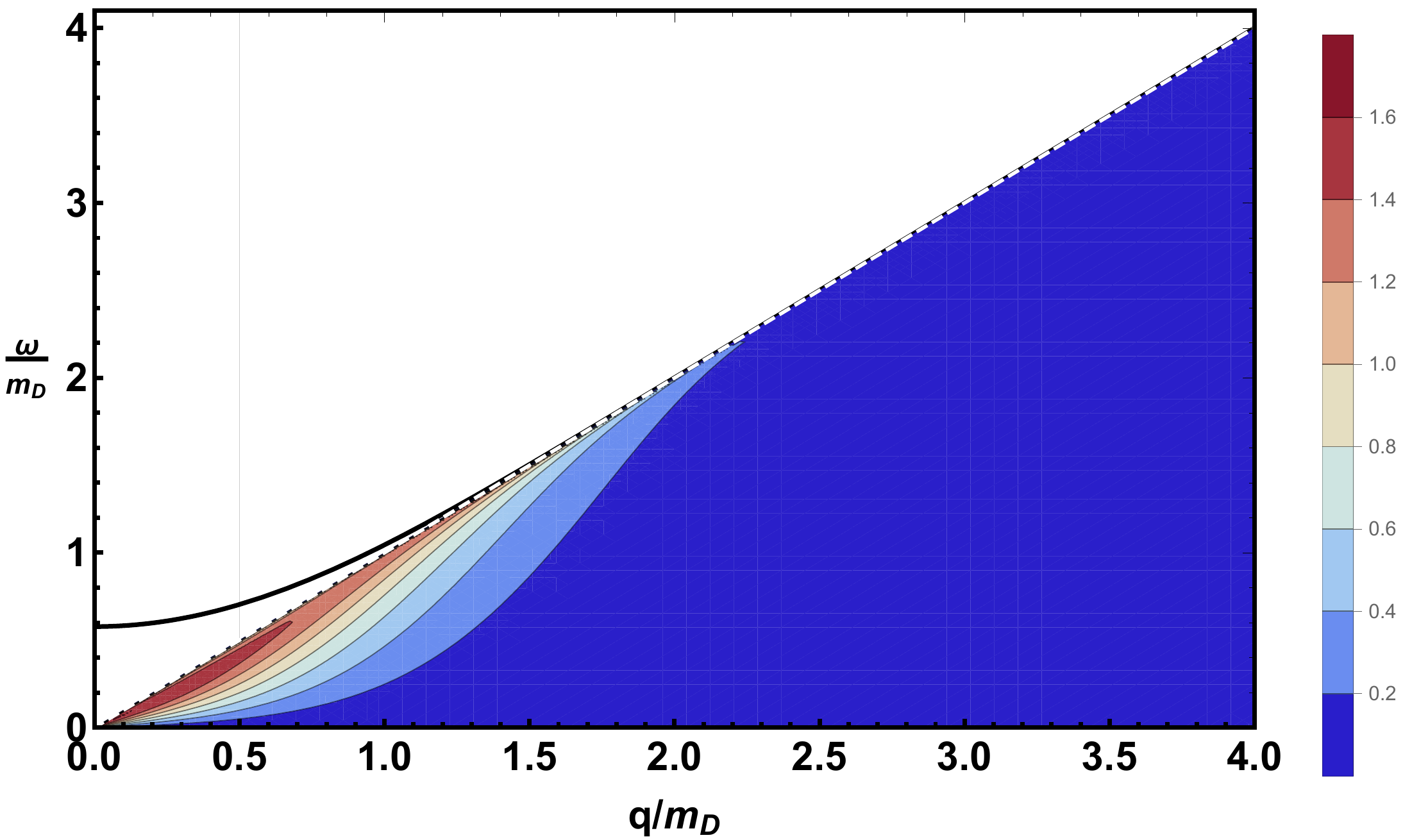}
	\end{center}
	\caption{The longitudinal structure of the
	HTL propagator. The light-cone
	bisector is drawn in dashed white. In the time-like region above we plot the
	dispersion relation (location of the pole),
	and in the space-like Landau cut below we plot the
	contours of the spectral density there in units of $m_D^2$.
	 In Fig.~\ref{fig_sections} we plot
	the longitudinal and transverse spectral functions along
	the vertical line at $q=0.5 m_D$.
	}
	\label{fig_htl_contours_long}
\end{figure}
\begin{figure}[t!]
	\begin{center}
		\includegraphics[width=10cm]{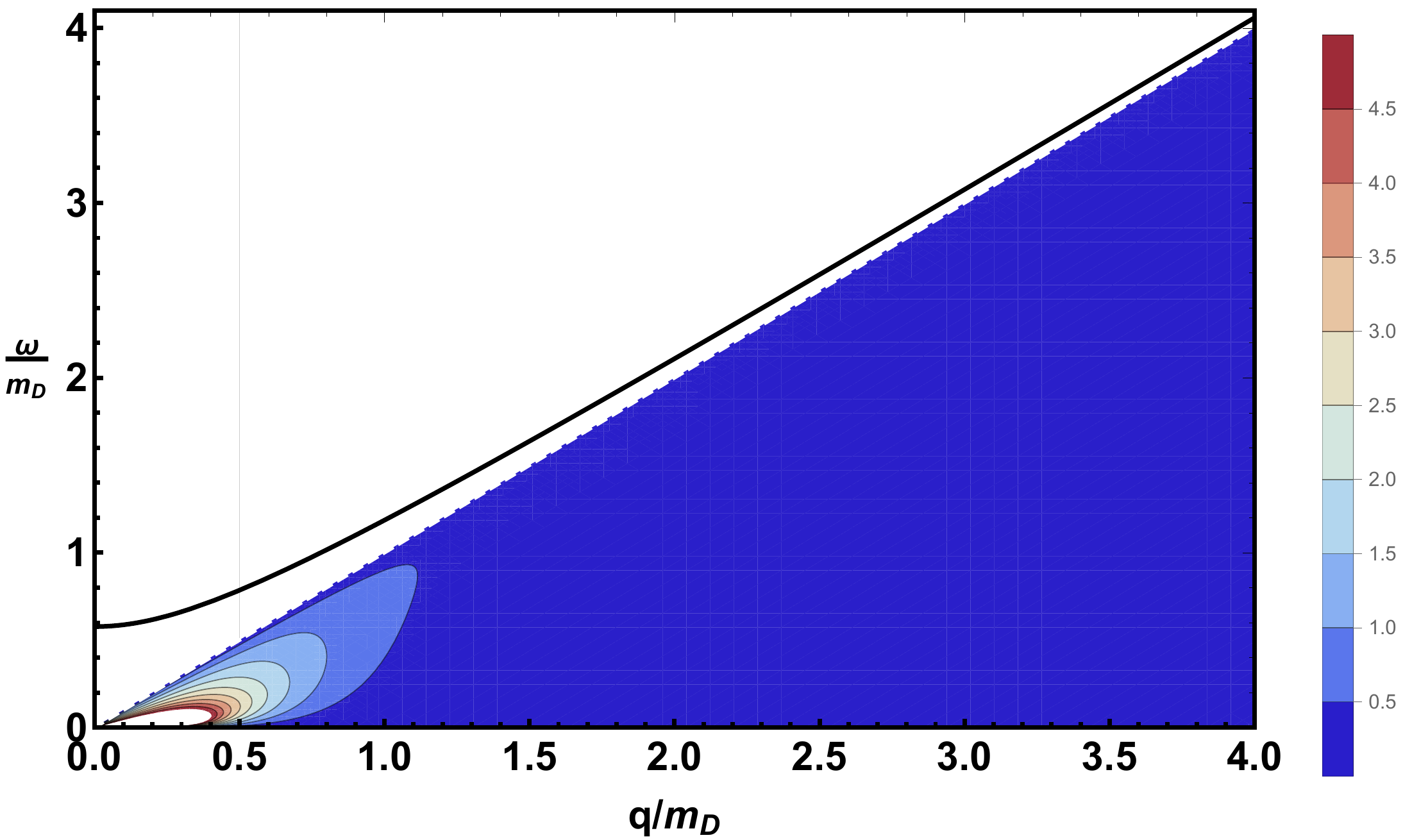}
	\end{center}
	\caption{The transverse structure of the
	HTL propagators. The graphical notation is as in Fig.~\ref{fig_htl_contours_long}.
	The white area in the contour plot represents values above the maximum of the
	color scale.
	}
	\label{fig_htl_contours_trans}
\end{figure}
We can now summarize the main features of these propagators. In the time-like
region both the longitudinal and transverse ones feature \emph{plasmon}
poles: collective plasma oscillations at the scale $gT$. At vanishing three-momentum,
the distinction between longitudinal and transverse modes
vanishes and the plasmon dispersion relation reduces to the well-known
plasma frequency $\omega_L(q=0)=\omega_T(q=0)=\pm m_D/\sqrt{3}$. At asymptotically
large momenta $q\gg m_D$, on the other hand, the longitudinal pole approaches
the light cone, but its residue vanishes exponentially \cite{Pisarski:1989cs}. The
two transverse modes instead survive, with unitary residue and \emph{asymptotic mass}
$M_\infty$, i.e.~$\omega_T(q\gg m_D)=\sqrt{q^2+ M_\infty^2}$, $M_\infty=m_D/\sqrt{2}$ \cite{Kalashnikov:1979cy,Weldon:1982aq}. It is also important to note that, in this
region, the HTL result agrees with the full one-loop result: hard gluons with
$q^0\sim q\sim T$ and $q^0-q\ll T$ do acquire a mass given by $M_\infty$.
At generic momenta $q\sim m_D$,
the poles have to be found numerically: they are shown in Figs.~\ref{fig_htl_contours_long} and 
\ref{fig_htl_contours_trans}
in the time-like region above the red light-cone bisector.

Let us now look at the space-like region: here, the logarithms in Eqs.~\eqref{htllong}
and \eqref{htltrans} clearly acquire an imaginary part, which in turn induces a non-vanishing
spectral function at $\mn{Q}^2>0$. This is called \emph{Landau damping} from its QED
analogue, and corresponds to the scattering of  virtual gluons off the hard
constituents of the plasma with $\mn{P}\sim T$.
The contours of the spectral functions in the \emph{Landau cut} (from the branch cut
of the logarithm) are shown under the red bisector in Figs.~\ref{fig_htl_contours_long} and 
\ref{fig_htl_contours_trans}.

Another feature of the HTL propagator in the space-like region is \emph{Debye screening}:
if we consider the propagators in the static limit $q^0\to 0$, appropriate for studying
time-independent chromoelectric and magnetic fields at large distances $r\sim1/gT$, we find
\begin{equation}
	G^R_L(0,q)=\frac{i}{q^2+m_D^2},\qquad G^R_T(0,q)=\frac{-i}{q^2}.
	\label{debyescreen}
\end{equation}
Hence, static chromoelectric fields are screened: at distances larger
than the Debye radius $r_D=1/m_D$ they vanish exponentially.
Static chromomagnetic fields are not screened in the HTL effective theory.
At even larger distances, the non-perturbative physics arising at
the scale $g^2T$, which will be discussed
in more detail in Sec.~\ref{sec_DR}, will eventually screen these
fields. In the static domain, the Euclidean techniques described later in
this review are applicable. To describe the dynamics of the chromomagnetic modes
at that scale one can use the effective Hamiltonian derived by B\"odeker
\cite{Bodeker:1998hm,Bodeker:2000da}.

\begin{figure}[t!]
	\begin{center}
		\includegraphics[width=6cm]{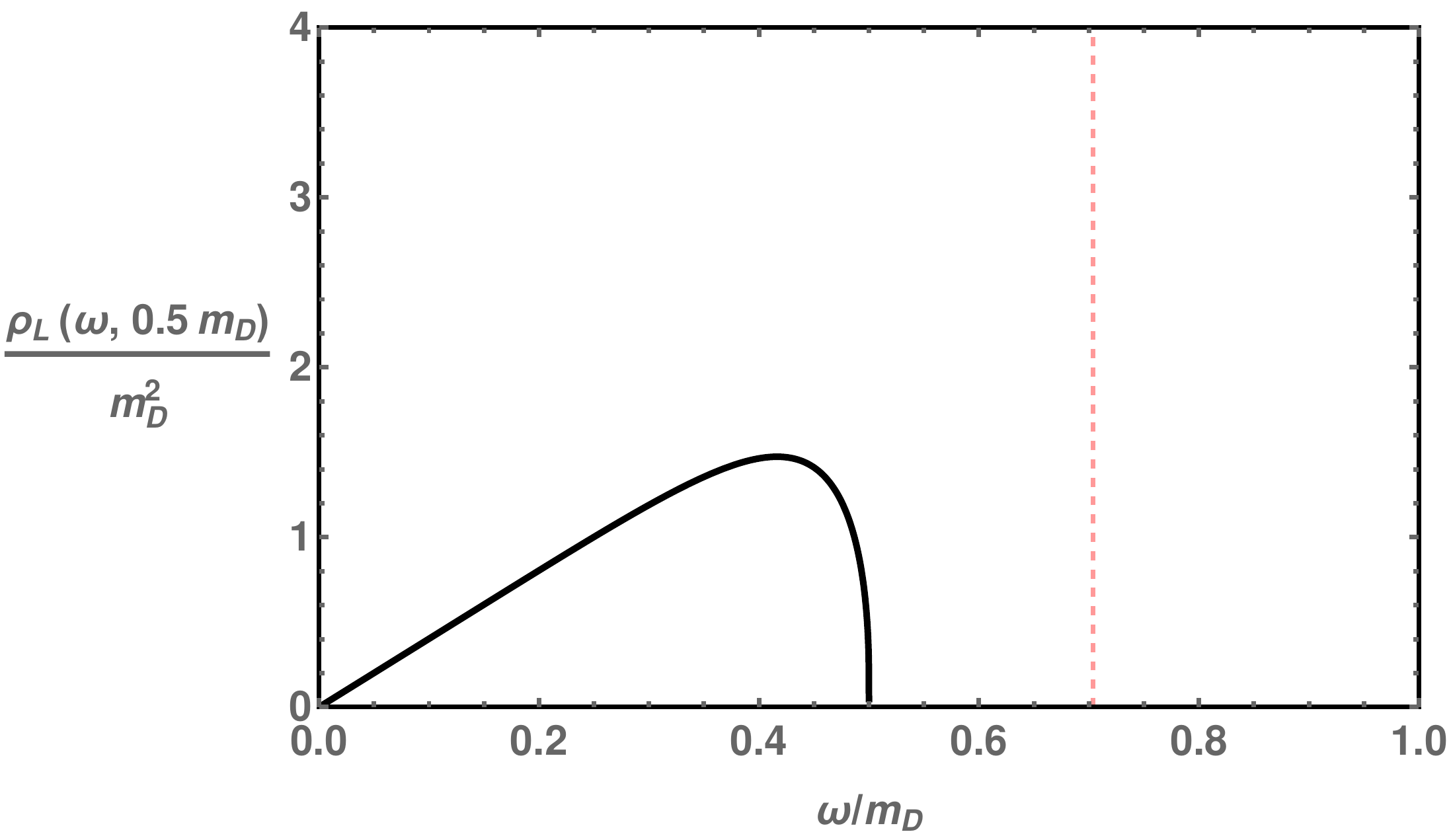}
		\includegraphics[width=6cm]{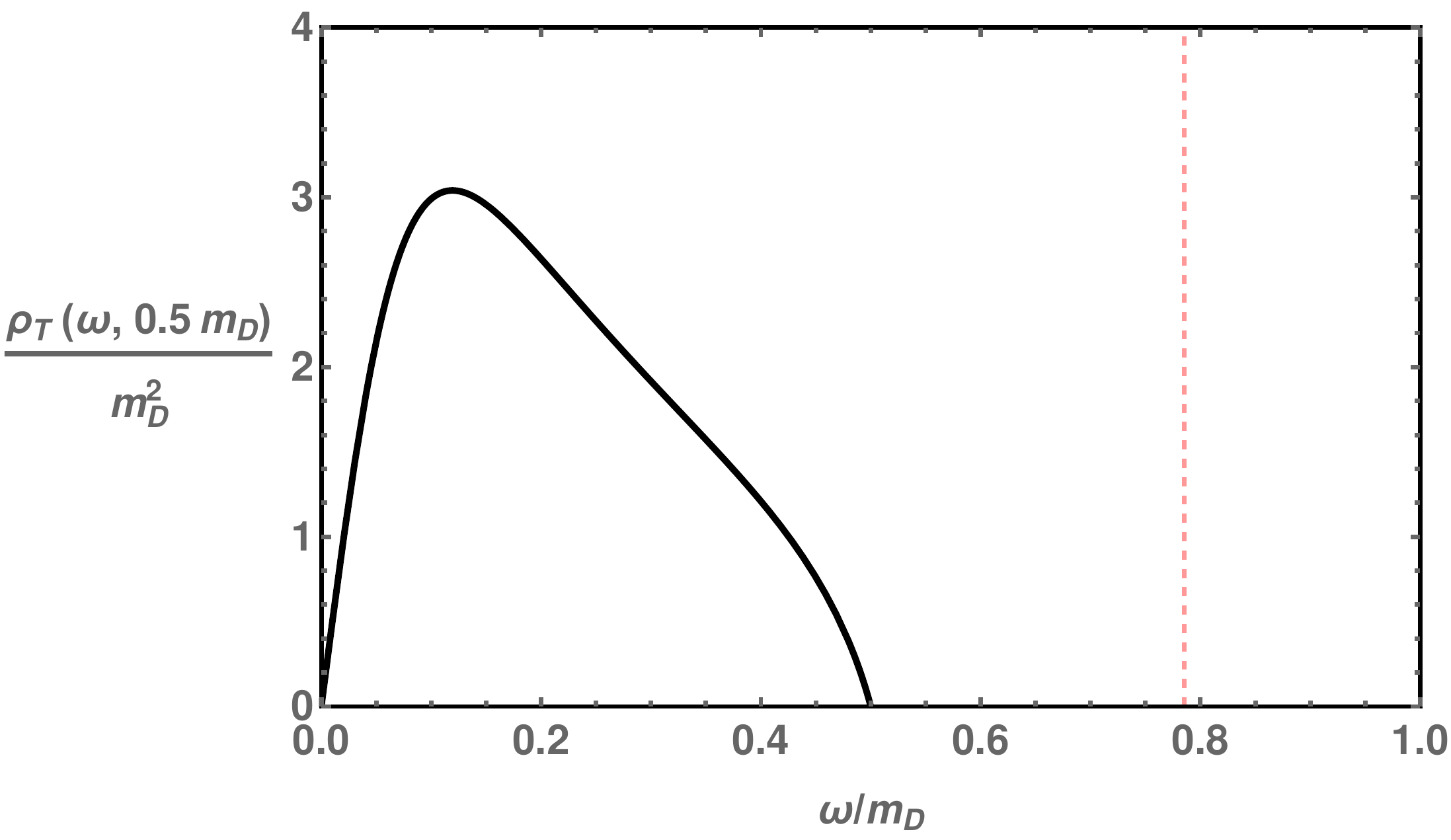}
	\end{center}
	\caption{The longitudinal and transverse HTL
	spectral functions on the left and right respectively. In both plots
	the three-momentum is fixed at $q=0.5 m_D$. We show the space-like
	Landau cut in solid black and the Dirac $\delta$-function at the time-like plasmon
	pole in dashed red.}
	\label{fig_sections}
\end{figure}
In Fig.~\ref{fig_sections} we plot the gluon HTL spectral functions at a fixed value
of momentum, that is, following the vertical lines in Figs.~\ref{fig_htl_contours_long} 
and \ref{fig_htl_contours_trans} . These
plots clearly show the structure of the spectral function in the Landau cut, while the
plasmon pole is a Dirac $\delta$-function. Indeed, in the propagators~\eqref{htllong}
and \eqref{htltrans}, plasmons have zero width. This is just a leading-order effect:
the more precise statement is that the position of the plasmon pole, determined by the real
part of the self energy, is of order $gT$, while the \emph{width} of the plasmon, also
called \emph{gluon damping rate}, is of order $g^2T$. This means that the first arises from a
one-loop diagram in the HTL limit, i.e.~with hard momenta running through it, while
the latter requires soft momenta through the loop and thus a consistent HTL resummation,
including both resummed propagators and vertices.
Indeed, the determination of the gluon damping rate at vanishing momentum within the HTL
theory and the proof of its gauge invariance
represented one of the first successes of the
HTL approach \cite{Braaten:1989kk,Braaten:1990it}, as well as one of the first computational \emph{tours de force} within
the theory. With a similar approach, the $\mathcal{O}(g^2T)$ correction to the plasma
frequency was computed in \cite{Schulz:1993gf}.

\begin{figure}[t!]
	\begin{center}
		\includegraphics[width=9cm]{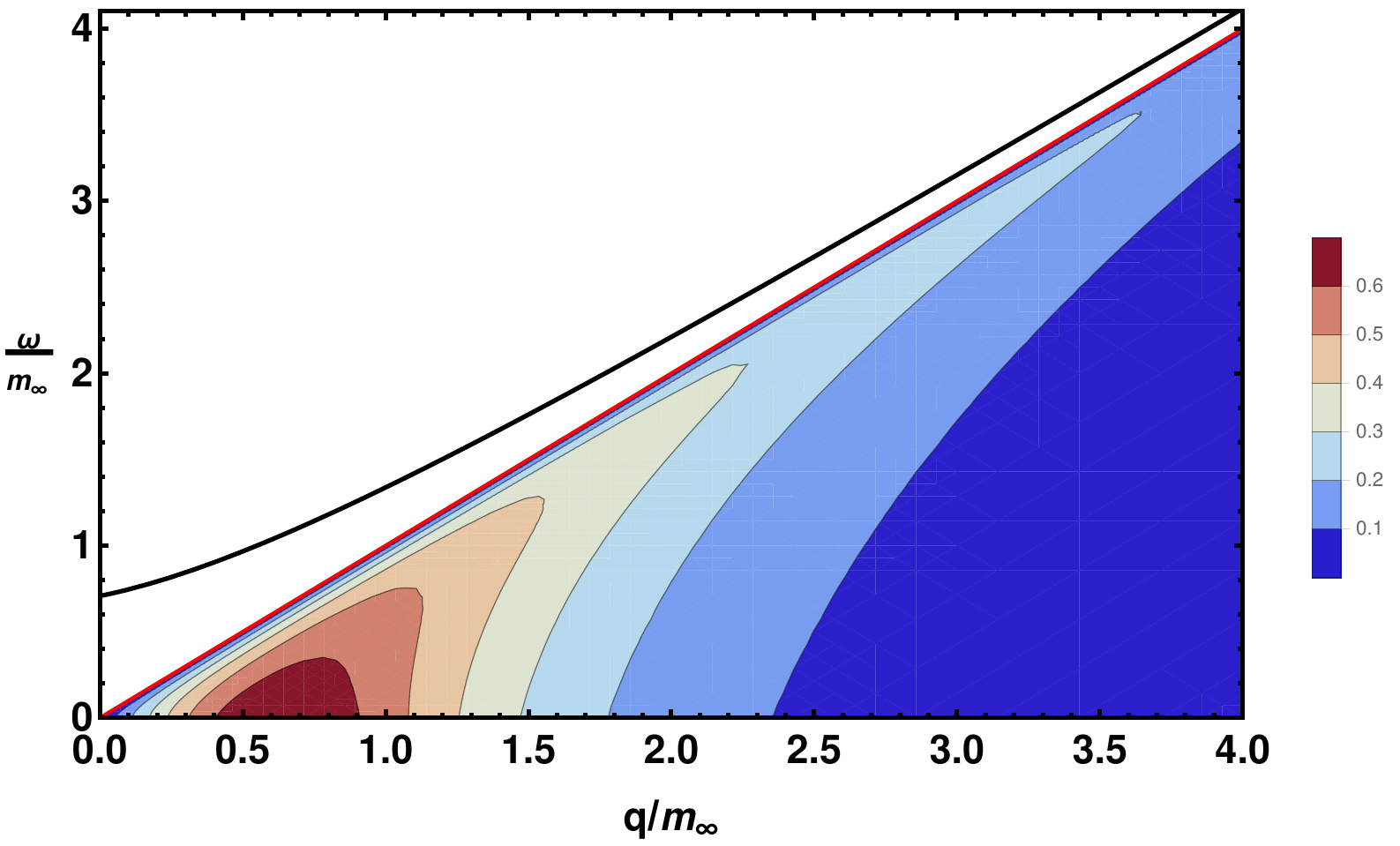}
		\includegraphics[width=9cm]{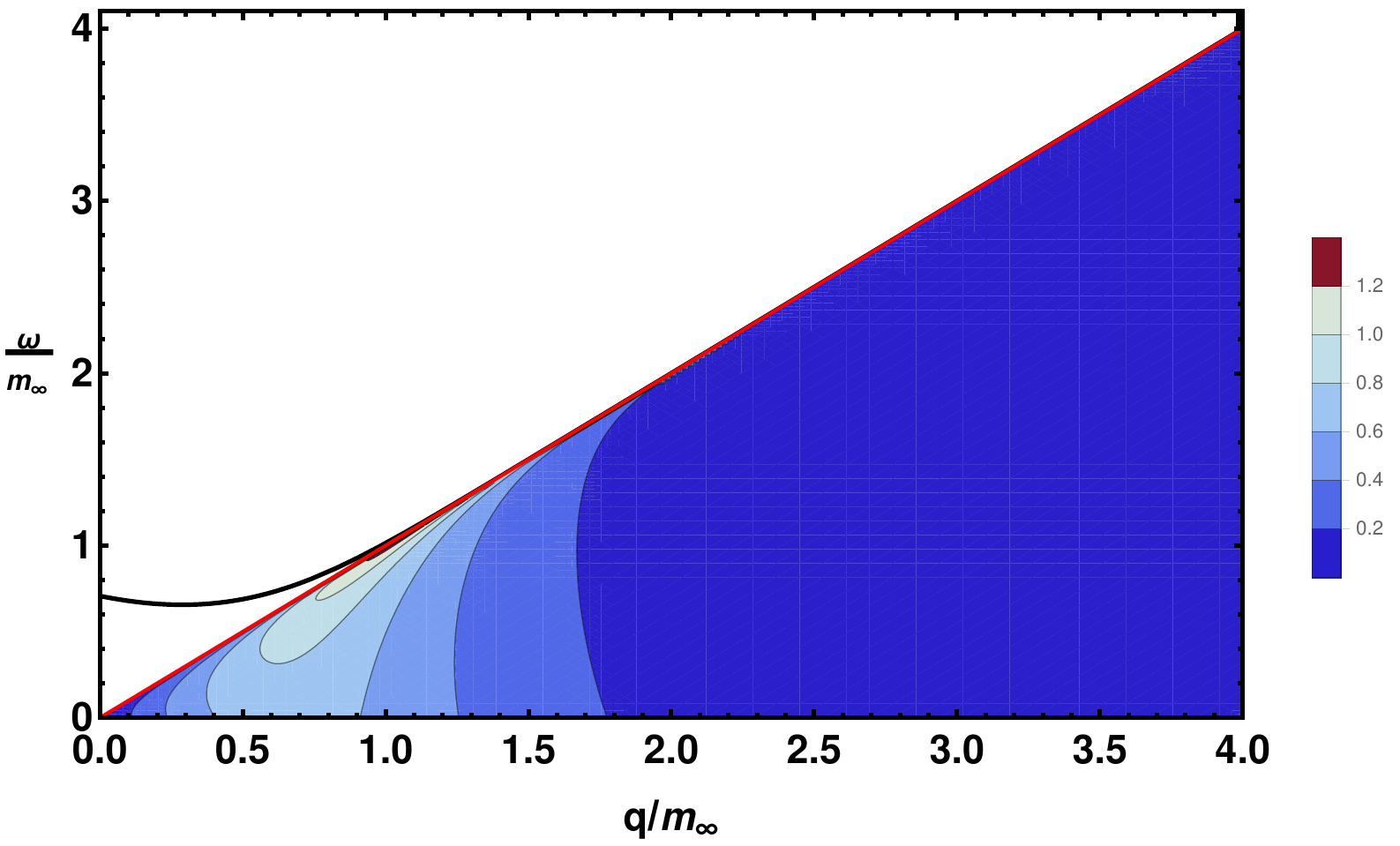}
	\end{center}
	\caption{The positive (top) and negative (bottom) chirality-to-helicity
	fermion modes in the HTL theory. In both figures, we show the position
	of the time-like pole above the light-cone bisector and the contours
	of the spectral function in the Landau cut below the bisector, in units
	of $m_\infty$. Note
	that each spectral function is not odd in $\omega$, as the
	self energies obey $\Sigma^\pm_R(-\omega,q)=-\Sigma^\mp_A(\omega,q)$.}
	\label{fig_fermion_htl}
\end{figure}

For what concerns the discussion of the collective modes of fermions, there exist
many parallels with what we have just illustrated for gluons. While we refer to reviews such
as \cite{Blaizot:2001nr} or textbooks such as \cite{Bellac:2011kqa} for more details, we
give a brief summary of the differences and similarities. Rather than longitudinal and
transverse modes, the retarded self energy $\Sigma^R$ given in
Eq.~\eqref{fermionselfexphtlfinal} can be decomposed in modes with positive
or negative chirality-to-helicity ratios. In detail, one finds that
\begin{equation}
	\label{htlfermiondef}
	S_{R}(\mn{Q})=h^+_\q S^+_{R}(\mn{Q})+h^-_\q S^-_{R}(\mn{Q})\,,
\end{equation}
where $h^\pm_\q\equiv (\gamma^0\mp \gamma^i\hat{q}^i)/2$ and
	\begin{equation}
		\label{htlfermion}
	S^{\pm}_R(\mn{Q})=\frac{i}{q^0\mp (q+\Sigma^\pm_R(q^0/q))}=
	\frac{i}{\displaystyle q^0\mp\left[q+\frac{m_\infty^2}{2q}\left(1
	-\frac{q^0\mp q}{2q}\ln\left(\frac{q^0+q}{q^0-q}\right)\right)\right]},
	%\right\vert_{q^0=q^0+i\epsilon},
\end{equation}
where $q^0$ is understood to be $q^0+i\epsilon$.
At positive (negative)
 frequencies the massless bare theory
only has a positive (negative) chirality-to-helicity mode, with $\omega^+(q)=q$
($\omega^-(q)=-q$) . In
the HTL theory, both modes develop time-like poles. At vanishing momentum
$\omega^+(q=0)=\omega^-(q=0)=\mm/\sqrt{2}$ is the fermionic analogue of the
plasma frequency. At asymptotic momenta, $\omega^+(q\gg \mm)=q+\mm/(2q)$ on the other hand clearly develops
an asymptotic mass $\mm$, while  $\omega^-(q\gg \mm)=q$, with exponentially
vanishing residue \cite{Pisarski:1989wb}. Also in this case, the asymptotic
limit agrees with the full one-loop result for hard fermions.
At intermediate momenta the negative
chirality-to-helicity mode, called the \emph{plasmino}, displays non-monotonic
behavior, as shown in Fig.~\ref{fig_fermion_htl}.
Such a mode can be understood as a collective
excitation where the positive frequency fermion mixes with
the negative frequency anti-fermion. Indeed, for small $q$ the behavior of
$\omega^-(q)$ is  that of a negative energy state: it decreases as $q$ increases.
These time-like modes, whose pole position is of order $gT$, are long-lived: their
width is of order $g^2T$, as in the case of gluonic excitations. The \emph{quark
damping rate} thus requires a similar HTL-resummed calculation, which was presented,
for vanishing momentum, in \cite{Braaten:1992gd} (see also \cite{Nakkagawa:1992ew} for
a discussion of gauge invariance).

In the space-like region Landau damping manifests itself also for soft quarks, corresponding
physically to scatterings of the soft, virtual, space-like quark with
the hard constituents of the medium. The contours of the quark HTL spectral
functions in the Landau cut are shown in Fig.~\ref{fig_fermion_htl}.

\subsubsection{Sum rules}

We now turn to an illustration of \emph{sum rules} that can be obtained from the analytical
properties of the amplitudes, owing to causality. These sum rules also provide insights
into the physical picture behind the HTL amplitudes. We start by illustrating the
classic sum rules, which can be found in textbooks such as \cite{Bellac:2011kqa}. In
the chromoelectric case we have
\begin{align}
I_E&\equiv\frac{1}{d_A}\int d^3\x\, e^{-i \q\cdot\x} \langle E^{i\,a}(t=0,\x)
E^{i\,a}(0,\mathbf{0})\rangle\nn\\
 & =  \int \frac{d\omega}{2\pi} \, \frac T\omega
\big[2\omega^2  \rho_T(\omega,q)+q^2\rho_L(\omega,q)\big]
%&= T \left(2+q^2\frac{m_D^2}{q^2(q^2+m_D^2)}\right)
=T\left(2+\frac{m_D^2}{q^2+m_D^2}\right),
\label{elecsumrule}
\end{align}
with $d_A = N_c^2-1$ standing for the dimension of the adjoint representation of SU($N_c$).
We only consider the field-based definition on the first line at leading-order, so we
omit the Wilson line connecting the two fields and  ensuring gauge invariance.
The result on the second line can be easily obtained
from the analytical properties of the spectral function, which is the difference
of the retarded and advanced propagators. These are in turn  analytic on the
upper and lower half-planes in $\omega$, as dictated
by causality. The retarded (advanced) integration can then be
closed above (below) the real axis without encountering any non-analytic structures
from the propagators themselves. The pole at $\omega=0$ from the Bose--Einstein distribution
contributes to the longitudinal integration only. The longitudinal and
transverse propagators in Eqs.~\eqref{htllong} and
\eqref{htltrans} decay as $1/\omega$ at large $\omega$, so both generate a
contribution when closing the contours away from the real axis. The sum of these
contributions yields Eq.~\eqref{elecsumrule} \cite{Pisarski:1989cs,LeBellac:1994qg,
Vanderheyden:1996bw}.

\begin{figure}[t!]
	\begin{center}
		\includegraphics[width=6.3cm]{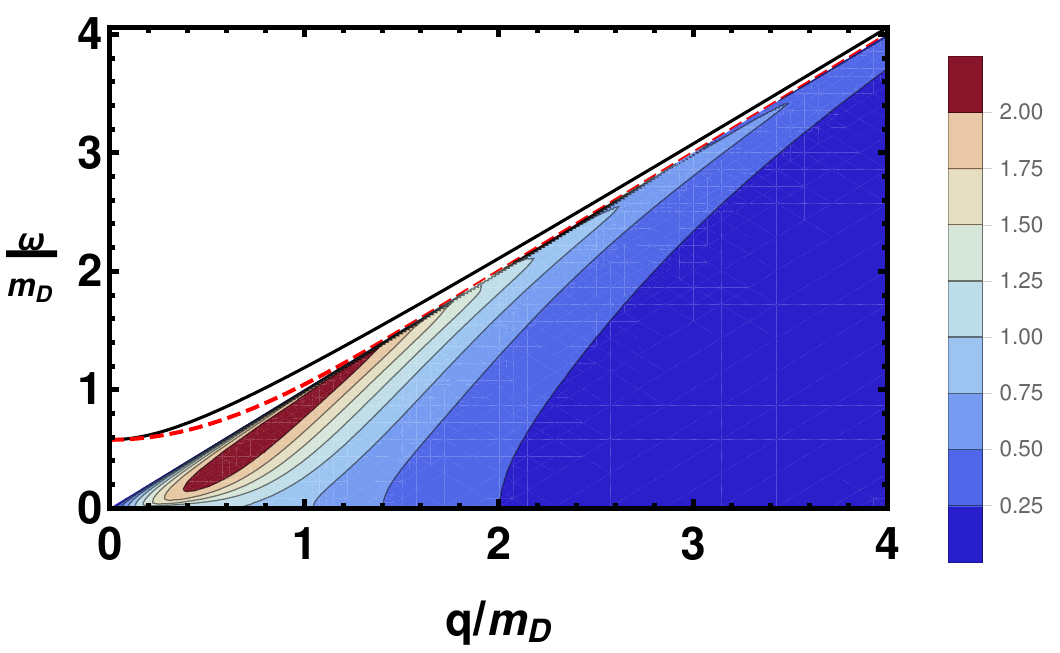}
		\includegraphics[width=5.6cm]{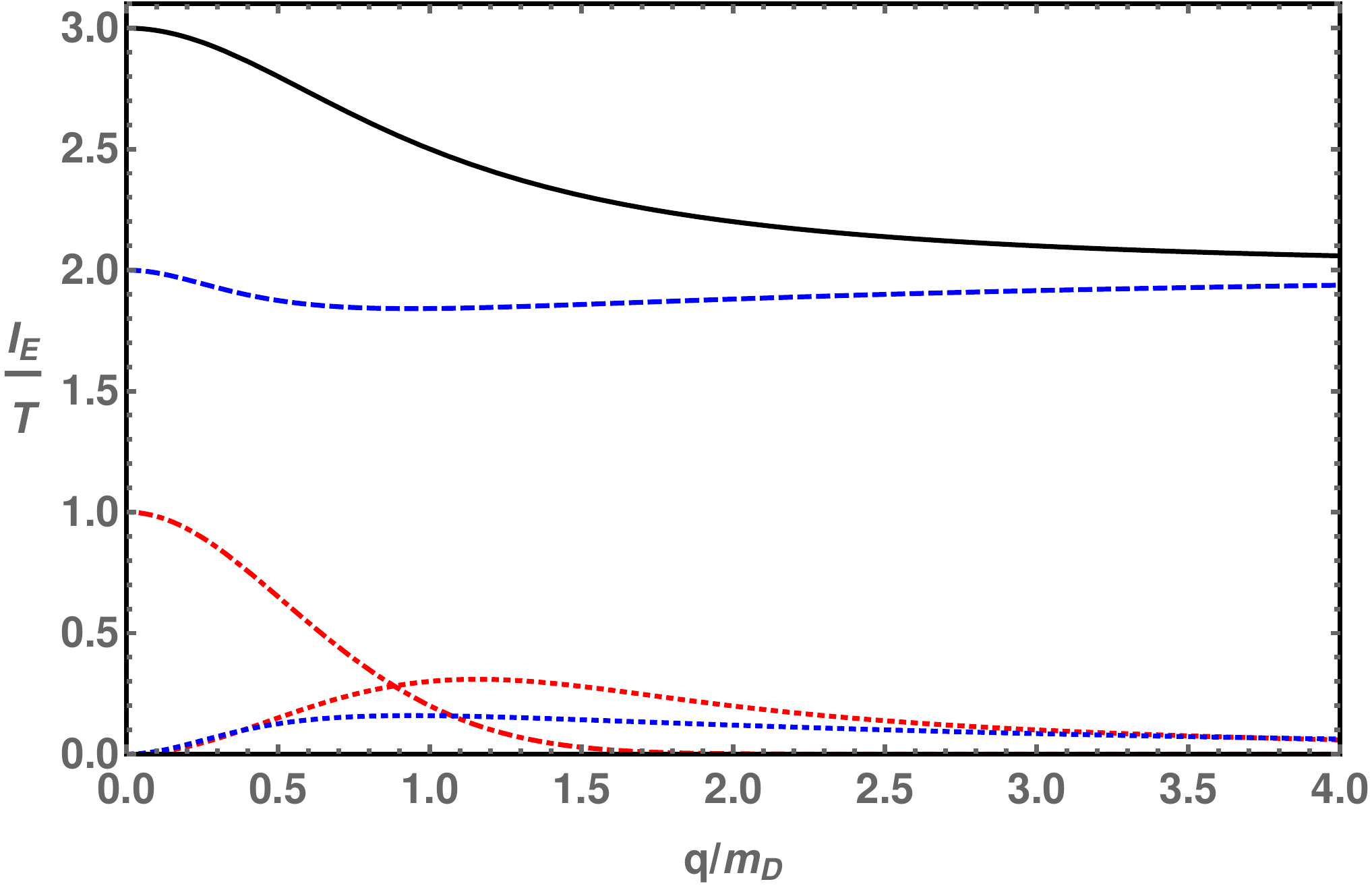}
	\end{center}
	\caption{Contributions to $I_E$ in Eq.~\eqref{elecsumrule}. On the left we plot, for
	$\omega>q$, the position of the transverse (solid black) and longitudinal (dashed red)
	plasmon poles. Under the $\omega=q$ bisector, for $q>\omega$, we plot the contours
	of the Landau cut contribution to Eq.~\eqref{elecsumrule}, divided by $Tm_D^2$. 
	On the right,
	we plot in dashed blue and dot-dashed red the contributions from the transverse and
	longitudinal poles, respectively. The transverse and longitudinal cut contributions
	are drawn in dotted blue and red. The overall total is plotted in solid black.}
	\label{fig_EE}
\end{figure}

\begin{figure}[t!]
	\begin{center}
		\includegraphics[width=6.3cm]{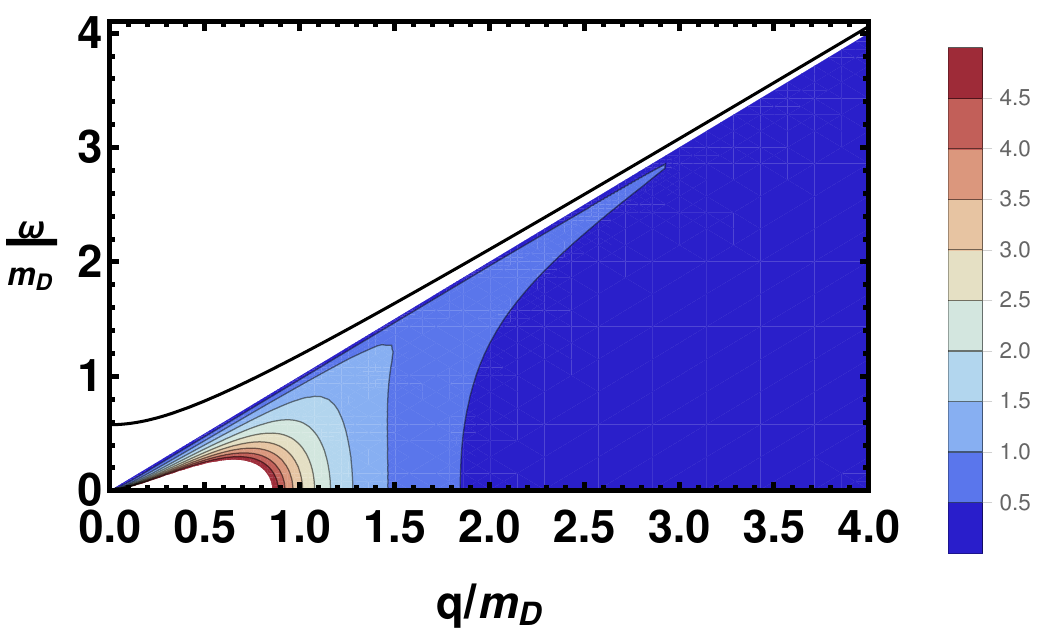}
		\includegraphics[width=5.6cm]{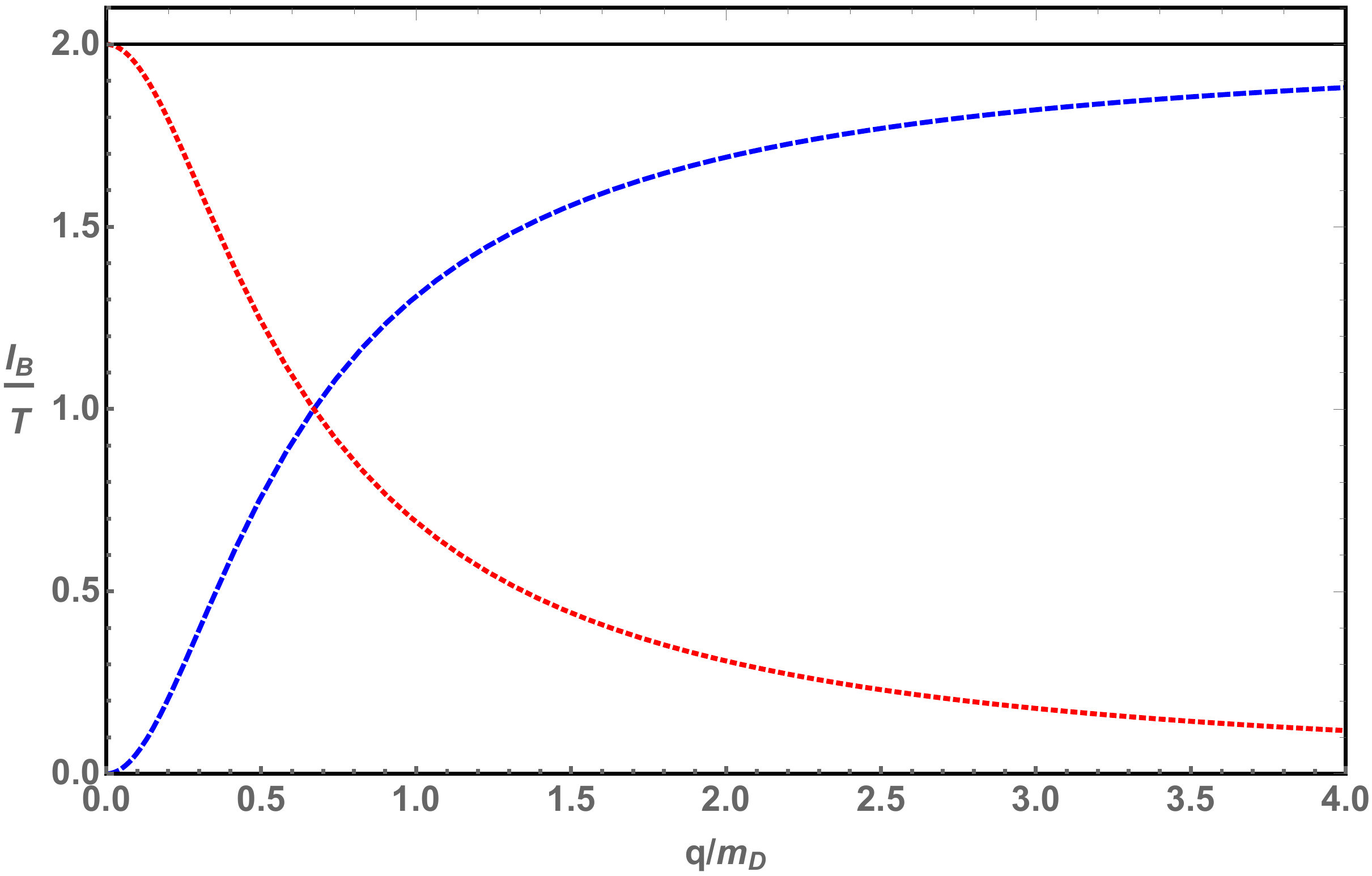}
	\end{center}
	\caption{Contributions to $I_B$ in Eq.~\eqref{magsumrule}.
    As in Fig.~\ref{fig_EE}, on the left we plot, for
   	$\omega>q$, the position of the transverse
   	plasmon pole in solid black and, for $q>\omega$, the contours
	   of the Landau cut contribution to Eq.~\eqref{magsumrule}, divided by $Tm_D^2$.
	   The white area in the contour plot represents values above the maximum 
	   of the color scale.
	On the right, we plot in dashed blue  the contribution from the transverse pole and
	in dotted red the contribution from the transverse cut.
	The overall total is plotted in solid black.}
	\label{fig_BB}
\end{figure}

With the same methods we can also look at the magnetic correlator, which reads
\begin{equation}
I_B\equiv \frac{1}{d_A} \int d^3\x\, e^{-i \q\cdot\x} \langle B^{i\,a}(t=0,\x)
B^{i\,a}(0,\mathbf{0})\rangle
= \int \frac{d \omega}{2\pi}  \frac{T}{\omega}2 q^2\rho_T(\omega,q)=2T\,.
\label{magsumrule}
\end{equation}
As anticipated, the contour sum rules not only lead us to the simple closed form results of
Eqs.~\eqref{elecsumrule} and \eqref{magsumrule}, but they furthermore make
the underlying physical picture more transparent. At large $q$,
we expect to see only the two transverse degrees of freedom, equally distributed
by equipartitioning, which is indeed the case in both the electric and the magnetic
condensates, which both become $2T$.
At vanishing $q$, we on the other hand expect
to see the three degenerate polarizations of plasmons in the electric case, i.e.~chromoelectric fields,
which
is again borne out by Eq.~\eqref{elecsumrule} that reduces to $3T$ at vanishing $q$.
On the other hand, as we will discuss in Sec.~\ref{sub_eucl_feynrules},
 the gauge invariance of MQCD---the effective, static theory of chromomagnetic modes at the scale $g^2T$ which shall
be illustrated later in Sec.~\ref{sec_DR}---prevents from generating a third degree of freedom
for chromomagnetic fields in the IR, which is why
$I_B$ is constant as a function of $q$.
These considerations are reflected not just in the integrated results, but in
the integrands as well: in the limit of small $q$, the integral of $\rho_T$
appearing in $I_E$ is dominated by its pole ($\omega^2\ge q^2$) part,
whereas $I_B$ is dominated by its cut ($q^2>\omega^2$) part.
This reflects the fact that at small $q$, the plasmon contains only electric fields
oscillating with the hard particles, while the magnetic fields are unscreened. In
Figs.~\ref{fig_EE} and \ref{fig_BB} we plot separately the pole and cut contributions
to these integrals.

Similar sum rules, motivated again by causality, can be derived in the fermion
case as well and can be found in textbooks \cite{Bellac:2011kqa}.

As we have remarked, causality is responsible for the sum rules we have just illustrated.
In a way, this is a textbook application of causality, as it relies on the basic property
of analyticity of the retarded propagator on the upper half of the complex $\omega$ plane. However, causality allows for stronger statements, which can be used
to derive sum rules that apply on the light cone.
To this end, let us consider the light-cone causality of retarded propagators,
which implies
\begin{equation}
	\label{causality}
	D^{R}(q^+, q^-, \q_\perp) = \int dx^+ dx^- d^2 x_\perp \; e^{i (q^+ x^- + q^- x^+ - \q_\perp \cdot \x_\perp) }  D^{R}(x^+, x^-, \x_\perp)
\end{equation}
is an analytic function of $q^+\equiv(q^0+q_z)/2$ on the upper half-plane at fixed
$q^-\equiv q^0-q_z$ and $q_{\perp}$. This is  because  the retarded response
function is only non-zero in the forward light
cone $2x^+x^-\ge x_\perp^2$.
Thus the integral in Eq.~\eqref{causality} has support only for $x^- > 0$,
and the Fourier integral provides an analytic continuation in the upper half $q^+$ plane,
due to the decreasing exponential $e^{iq^+ x^-}$ \cite{CaronHuot:2008ni}. In other
words, retarded functions are analytical on the upper half-plane in
any time-like variable. With some caveats, this applies also for soft light-like
variables, up to corrections that are beyond the scope of this review. We refer the interested reader to
\cite{CaronHuot:2008ni} for the original derivation for bosons and for a discussion
of these caveats and to \cite{Ghiglieri:2015zma} for a more pedagogical review.

To see a first application of the above property, let us consider the correlator
\begin{equation}
F_\perp\equiv\frac{1}{d_A}\int_{-\infty}^{+\infty}dx^+\int d^2x_\perp e^{-i \q_\perp\cdot\x_\perp}
\langle F^{-\perp\,a}(x^+,x^-=0,\x_\perp)\,F^{-\,a}_{\;\;\perp}(0,0,{\bf 0})\rangle\,,
\label{defkindasortaqhat}
\end{equation}
where the repeated $\perp$ index implies a summation over the two transverse
directions. As in Eqs.~\eqref{elecsumrule} and
\eqref{magsumrule} before, we have omitted the Wilson lines necessary for gauge invariance
beyond leading order. Physically, $F^{-\perp}$ can be viewed as the transverse
component of the Lorentz force for an eikonal source propagating at $v=c$
in the $\hat{z}$ direction. Indeed, the expression above is related
to the transverse momentum broadening coefficient $\hat{q}$, which we shall further
describe later on in Sec.~\ref{sec_lpm}.

Let us now turn to the evaluation of Eq.~\eqref{defkindasortaqhat}.
As the separation between the two field-strength tensors is, at non-zero $x_\perp$, space-like,
the analytical properties mentioned above apply for the retarded propagator.
Eq.~\eqref{defkindasortaqhat}, however, describes a Wightman correlator, which
in momentum space is related to the retarded one as
\begin{equation}
	F_\perp=\int\frac{dq^+}{(2\pi)}\frac {T}{q^+}q_\perp^2\left[\rho_L(q^+,q^-=0,q_\perp)
	+\frac{q_\perp^2}{q^2}\rho_T(q^+,q^-=0,q_\perp)\right],
\end{equation}
where we have further assumed  $q_\perp\sim gT$ and used the Coulomb gauge propagators
in Eqs.~\eqref{htllong} and \eqref{htltrans}. The retarded (advanced)
propagators entering the spectral functions above are then analytical above (below)
the real $q^+$ axis. The only non-analytical features are a branch cut
all along the real axis and the zero-frequency pole
of the soft limit of the Bose--Einstein distribution. As the spectral functions are actually
$\mathcal{O}(q^+)$ at the origin, this pole can be treated with a principal value
prescription, yielding \cite{Aurenche:2002pd,CaronHuot:2008ni}
\begin{align}
	F_\perp&=i \frac {T}{2}q_\perp^2\left[G_L^R(0,0,q_\perp)+G_L^A(0,0,q_\perp)+
G_T^R(0,0,q_\perp)+G_T^A(0,0,q_\perp)\right],\nn\\
&=T\,q_\perp^2\left[\frac{1}{q_\perp^2}-\frac{1}{q_\perp^2+m_D^2}\right].
\label{finalqhat}
\end{align}
This shows how the soft contribution to this light-cone operator has become much
more straightforward: it has reduced to the Euclidean zero mode, which is three-dimensional
and can be dealt with using EQCD, which is much simpler than the HTL theory, as
Sec.~\ref{sec_DR} will elucidate. Indeed, Eq.~\eqref{finalqhat}
is just the difference between the propagators of the massless, spatial gauge bosons and
the massive $A^0$ scalar arising from dimensional reduction of the temporal gauge field.
In summary, the soft contribution to Euclidean operators,
that is operators whose soft contribution is dominated by the Euclidean zero mode,
is time-independent at spacelike and lightlike separations. We refer to the derivations
in \cite{CaronHuot:2008ni} and  \cite{Ghiglieri:2015zma} for the details of a more formal connection
to the Euclidean formalism and for the reduction to the zero mode at soft momenta. In a nutshell,
\cite{CaronHuot:2008ni} shows how the mapping to EQCD can be seen as arising from a 
\emph{complex boost}. Take the two-point Wightman function of a scalar field
$\phi$ at equal times, $D^>(0,\x)\equiv\langle\phi(0,\x)\phi(0)\rangle$. As Sec.~\ref{sec:IV}
showed, $D^>(t)=D_E(it)$, with $D_E$ the Euclidean correlator. The Matsubara formalism, which shall
be discussed at length in Sec.~\ref{sec:imagtimeform}, then yields
\begin{equation}
	\label{simonstart}
	D^>(0,\x)=T\sum_n\int\frac{d^3p}{(2\pi)^3}e^{i\p\cdot\x} D_E(\omega_n,p),
\end{equation} 
where the sum runs over the Matsubara frequencies $\omega_n=2\pi n T$. But any space-like separated
two-point function is at equal times in a suitable frame, so that $D^>(t,\x)$, with $t/x_z\le 1$,
which would naively read
\begin{equation}
	\label{simonstart2}
	D^>(t,\x)=T\sum_n\int\frac{d^3p}{(2\pi)^3}e^{-i t i \omega_n + i\p\cdot\x} D_E(\omega_n,p),
\end{equation}
can be ``boosted'' to an equal-time form by the change $p_z\to p_z+i \omega_n t/x_z$,
which is allowed by the analyticity arguments mentioned before as long as $t/x_z\le 1$, yielding
\begin{equation}
	\label{simonstart3}
	D^>(t,\x)=T\sum_n\int\frac{d^3p}{(2\pi)^3}e^{i\p\cdot\x} D_E(\omega_n,p_x,p_y,p_z+i\omega_n t/x_z).
\end{equation}
Whenever this sum is dominated by the zero-mode, $\omega_n=0$, as in the case of Eq.~\eqref{finalqhat},
then Eq.~\eqref{simonstart3} shows clearly how the soft contribution becomes time-independent and three-dimensional,
i.e.
\begin{equation}
	\label{simonstart3d}
	D^>(t,\x)_\mathrm{soft}=T\int\frac{d^3p}{(2\pi)^3}e^{i\p\cdot\x} D_E(0,p_x,p_y,p_z).
\end{equation}

However, not all operators are dominated by the zero mode. Let us consider the longitudinal
analogue of Eq.~\eqref{defkindasortaqhat}, i.e.
\begin{equation}
F_L\equiv\frac{1}{d_A}\int_{-\infty}^{+\infty}dx^+\int d^2x_\perp e^{-i \q_\perp\cdot\x_\perp}
\langle F^{-z\,a}(x^+,x^-=0,\x_\perp)\,F^{-z\,a}(0,0,{\bf 0})\rangle\,,
\label{defkindasortaql}
\end{equation}
which is quite clearly related to the longitudinal component of the Lorentz force for the
same eikonal source. At LO for soft momenta this becomes
\begin{equation}
	\label{qlsumrule}
	F_L=T \int\frac{dq^+}{(2\pi)}\,q^+\left[\rho_L(q^+,q^-=0,q_\perp)
	+\frac{q_\perp^2}{q^2}\rho_T(q^+,q^-=0,q_\perp)\right],
\end{equation}
which evidently is not sensitive to the zero mode. In this case a different type
of light-cone sum rule applies, based on the same analyticity properties rooted in causality.
When dealing with the retarded (advanced) contribution to the spectral function we are then
free to deform the integration contour away from the real axis in the upper (lower)
half plane,
on an arc at fixed, large $|q^+|$. On this arc, $|q^+|\gg m_D,q_\perp$ and the structures
in Eq.~\eqref{htllong} and \eqref{htltrans} reduce to their asymptotic, light-like limit,
which can only depend on the asymptotic thermal mass of transverse excitations. Indeed we find
\cite{Peigne:2007sd,Ghiglieri:2015zma,Ghiglieri:2015ala}
\begin{equation}
	F_L=T\left[1-\frac{q_\perp^2}{q_\perp^2+M_\infty^2}\right],
	\label{finalql}
\end{equation}
where the first term in brackets arises from the longitudinal modes and the second from
the transverse ones: as expected, the result depends on $M_\infty$.

For what concerns fermions, which have odd Matsubara frequencies, no equivalent of
Eq.~\eqref{finalqhat} can exist. There exists however an equivalent of
Eqs.~\eqref{defkindasortaql} and \eqref{finalql} \cite{Besak:2012qm,Ghiglieri:2013gia},
which depend on $m_\infty$ and are of relevance for thermal photon production
\cite{Ghiglieri:2013gia} and for right-handed neutrino production in the Early Universe
\cite{Besak:2012qm}. In the former case the sum rule permits a simple,
analytical evaluation of the leading-order contribution to the photon rate from 
soft quark momentum; let us briefly see how it comes about, to tie back
to our discussion at the beginning of Sec.~\ref{sec_soft_collinear}. While
 Eq.~\eqref{zerothphoton} vanishes when both quark propagators are bare,
 which is appropriate when they are both hard, $\mn{P}-\mn{K}\sim T$
 and $\mn{P}\sim T$, this is no longer the case when one of these two
 is soft, $\mn{P}\sim gT$, which, as we have argued, is where a logarithmic IR
 divergence appears in the naive treatment of Eqs.~\eqref{comptonnaive} and
 \eqref{annihnaive}.

\begin{figure}[t!]
	\begin{center}
		\begin{minipage}{0.13\textwidth}
			\hspace{-0.5cm}$\Pi^<_{g^2\;\mathrm{soft}}(\mn{K})=$\\ \vspace{0.5cm}
			\end{minipage}
			\begin{minipage}{0.3\textwidth}
			\includegraphics[width=4cm]{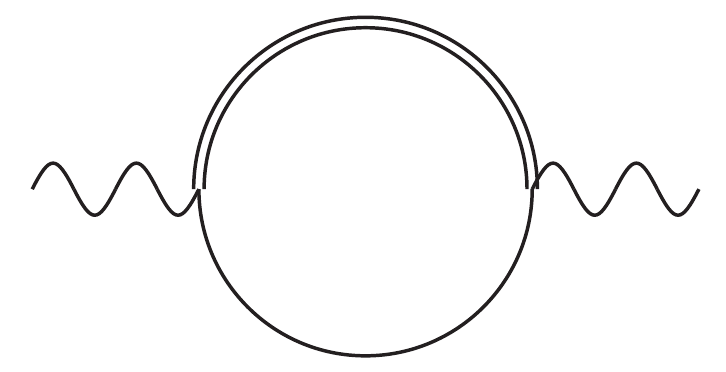}\\\vspace{0.1cm}
			\end{minipage}
			\includegraphics[width=\textwidth]{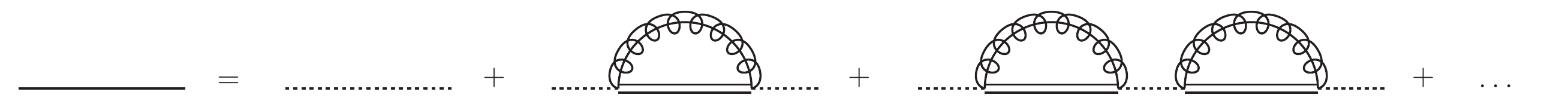}
	\end{center}
	\caption{The diagram contributing to the soft region at leading order, Eq.~\eqref{sofphoton}. 
	The double lines are  hard quarks, such as
	$S(\mn{P}+\mn{K})$ in the diagram above, whereas the dotted single line 
	represents the bare soft propagator $S(\mn{P})$. The single plain line is the HTL-resummed soft 
	quark propagator and curly lines with an extra line
	running through them are hard gluons. Figure taken from~\cite{Ghiglieri:2013gia,
	Ghiglieri:2014kma}.
	}
	\label{fig_soft_photon}
\end{figure}
To properly deal with the $\mn{P}\sim gT$ region we must perform HTL resummation, as shown
diagrammatically in Fig.~\ref{fig_soft_photon}. This gives
\begin{equation}
	\label{sofphoton}
	\Pi^<_{g^2\;\mathrm{soft}}(\mn{K})=-2\sum_i\int \frac{d^4\mn{P}}{(2\pi)^4}\mathrm{Tr}
	\big[(e Q_i \gamma^\mu)S^<_\mathrm{hard}(\mn{P}+\mn{K})(e Q_i \gamma_\mu)
	S^<_\mathrm{soft}(-\mn{P})\big],
\end{equation}
with a factor of 2 accounting for the $\mn{P}-\mn{K}\sim gT$ region. Using the
explicit forms of the propagators (see \ref{app_horror}), we find 
\begin{align}
	\Pi^<_{g^2\;\mathrm{soft}}(\mn{K})=&2 e^2\sum_i Q_i^2 \int \frac{d^4\mn{P}}{(2\pi)^4}
	\mathrm{Tr}
	\big[\gamma^\mu (-\slashed{\mn{P}}-\slashed{\mn{K}}) \gamma_\mu
	(S_R(\mn{P})-S_A(\mn{P}))\big]\nn\\
	\label{sofphoton2}
	&\times (\theta(-p^0-k^0)-\nfd(|k^0+p^0|))2\pi\delta((\mn{P}+\mn{K})^2)(1-\nfd(p^0)),
\end{align}
where we used $S^<(-\mn{P})=-S^>(\mn{P})=-(1-\nfd(p^0))(S_R(\mn{P})-S_A(\mn{P}))$.
We can now make use of the fact that $\mn{K}\gg\mn{P}$, so that
$\delta((\mn{P}+\mn{K})^2)\approx\delta(2kp^-)$, 
with the previously defined light-cone coordinates. Futhermore, for soft
momenta $\nfd(p^0\sim gT)=\frac12$. Hence
\begin{align}
	\Pi^<_{g^2\;\mathrm{soft}}(\mn{K})=&- e^2\sum_i Q_i^2\frac{\nfd(k)}{k} \int \frac{d^4\mn{P}}{(2\pi)^4}
	\mathrm{Tr}
	\big[\slashed{\mn{K}}
	(S_R(\mn{P})-S_A(\mn{P}))\big]%\nn\\
	\label{sofphoton3}
	%&\times 
	2\pi\delta(p^-).%(1-\nfd(p^0)),
\end{align}
Using the explicit form in Eq.~\eqref{htlfermiondef} to take the trace, we find
\begin{align}
	\Pi^<_{g^2\;\mathrm{soft}}(\mn{K})=&2 e^2\sum_i Q_i^2\nfd(k) \int \frac{d^4\mn{P}}{(2\pi)^4}
	\bigg[\bigg(1-\frac{p^z}{p}\bigg)(S^+_R(\mn{P})-S^+_A(\mn{P}))\nn\\
	&\hspace{3.5cm}+
	\bigg(1+\frac{p^z}{p}\bigg)(S^-_R(\mn{P})-S^-_A(\mn{P}))\bigg]%\nn\\
	\label{sofphotontrace}
	%&\times 
	2\pi\delta(p^-),%(1-\nfd(p^0)),
\end{align}
where the $\delta$ function puts the hard quark on shell, resulting in
\begin{align}
	\Pi^<_{g^2\;\mathrm{soft}}(\mn{K})=&2 e^2\sum_i Q_i^2\nfd(k) \int \frac{dp^+d^2p_\perp}
	{(2\pi)^3}
	\bigg[\bigg(1-\frac{p^+}{p}\bigg)\rho^+(p^+,p^-=0,p_\perp)\nn\\
	&\hspace{3.5cm}+
	\bigg(1+\frac{p^+}{p}\bigg)\rho^-(p^+,p^-=0,p_\perp)\bigg].
	\label{sofphotonsumrule}
\end{align}

In the above result, we recognize the fermionic analogue of Eq.~\eqref{qlsumrule}:
the $\rho^\pm\equiv S^\pm_R-S^\pm_A$ fermionic spectral function play the role of those
of longitudinal and transverse gluons. The analyticity in the upper (lower)
half of the complex $p^+$ plane of the retarded (advanced) functions
allow us to deform the integration away from the real axis to the 
arcs at large, complex $p^+$, where the propagators again 
greatly simplify, becoming sensitive only to the shift in the dispersion 
relation at the light cone. Indeed  we find \cite{Ghiglieri:2013gia}
\begin{equation}
	\Pi^<_{g^2\;\mathrm{soft}}(\mn{K})=2 e^2\sum_i Q_i^2\nfd(k) \int \frac{d^2p_\perp}
	{(2\pi)^2}
\bigg[1-\frac{p_\perp^2}{p_\perp^2+m_\infty^2}\bigg],
	\label{softphotonsumrulefinal}
\end{equation}
where at large  positive $p^+$ the constant term in square brackets comes
from the $S^-$ contribution and the other term from the $S^+$ contribution. 

As expected, Eq.~\eqref{softphotonsumrulefinal} is UV log-divergent. The divergence
can be regularized with a cutoff $\mu_\perp^\mathrm{LO}$ on $p_\perp$,
as shown graphically in Fig.~\ref{fig_lomap}; 
when regularizing Eqs.~\eqref{comptonnaive} and \eqref{annihnaive} in
the same scheme (see \cite{Arnold:2001ms} and footnote~7 of \cite{Ghiglieri:2013gia} 
for details) one recovers a finite, cutoff-independent result. We recall
that a complete leading-order photon rate necessitates also the evaluation
of the collinear contribution, which is the subject of the next section.

\subsection{Collinear physics:  LPM resummation}
\label{sec_lpm}
So far we have been talking about the complications that arise from soft kinematics,
where (one of) the particles in the discussion has all four-momentum components
that are soft, i.e.~$\mathcal{O}(g T)$.
As we mentioned in the 
introduction to Sec.~\ref{sec_soft_collinear},
 there is another kinematic region where intricacies arise, as
noticed by Aurenche et al.
\cite{Aurenche:1996sh,Aurenche:1998nw,Aurenche:1999tq,Aurenche:2000gf}
in the context of thermal photon production:
 the collinear region. There the particles are hard with
momenta of order $T$, their virtualities are of order $g^2 T^2$ and their angular
separations are of  order $g$, as we showed in Figs.~\ref{fig_photon_collinear}
and \ref{fig_lomap}.
  Where the physics is sensitive
to this kinematic region, it is necessary to perform a further resummation,
different from HTL, in order to
correctly describe the physics. Such resummation scheme traces back to
the  works of
Landau, Pomeranchuk \cite{Landau:1953um,Landau:1953gr}, and  Migdal \cite{Migdal:1956tc}
(LPM)
in the context of bremsstrahlung in QED, later
generalised to the physics of QCD by Baier, Dokshitzer, Mueller, Peign\'e, and Schiff \cite{Baier:1994bd, Baier:1996kr}  and Zakharov \cite{Zakharov:1996fv, Zakharov:1997uu}. In the context of Thermal Field Theory,
this was introduced by Arnold, Moore, and Yaffe \cite{Arnold:2001ba,Arnold:2002ja},
whose formalism we will follow in this review, in its position-space
formulation. We are in particular
indebted to the heuristic derivation in \cite{CaronHuot:2010bp} and
to the extended and detailed derivation in \cite{Arnold:2015qya}. A useful
\emph{Rosetta stone} between the many different formalisms and associated notations
can be found in App.~A of \cite{Arnold:2008iy}.

\subsubsection{Introduction and physical picture}
\label{subsub_heur_lpm}

The physical origin of the complication is related to the quantum mechanical formation times of scatterings in medium. Consider for concreteness the splitting of a hard parton (the parent) to two softer partons (the children); for simplicity, we start by considering a ``democratic'' splitting where the two offspring particles both carry an $\mn{O}(1)$ fraction of their parent's momentum. 

For massless particles such a splitting is kinematically disallowed in vacuum,\footnote{This is true for QCD, as no channel exists where the parent could spontaneously
decay into the children or vice versa, since gluons are massless in vacuum. This remains
true also when thermal masses are included, as we will do later, since the gluon and quark
asymptotic thermal masses obey $M_\infty^2<4m_\infty^2$.
However, when considering the emission
of virtual photons with virtuality $-\mn{K}^2>4 m_\infty^2$, the direct
(Born) annihilation of a quark-antiquark pair into
a virtual photon becomes possible. As long
as $-\mn{K}^2\lesssim g^2T^2$, the formalism we are describing remains valid with
minor modifications \cite{Aurenche:2002wq,Carrington:2007gt}. The same formalism
has also been applied to the collinear production of right-handed neutrinos
\cite{Anisimov:2010gy,Besak:2012qm}, where parents
and children are three distinct particles (Higgs scalars, left-handed leptons and
right-handed neutrinos), with different Born channels available depending on the
mass hierarchy.} but if the particle is pushed off shell by an interaction, the splitting becomes possible. The splitting process of a massless hard particle with frequency $E$ and virtuality $\mn{Q}^2$ is associated with a quantum mechanical formation time, which is the time it takes to separate the wave functions of the children partons such that they can be identified as two independent on-shell particles. For a hard particle with frequency $E$ and virtuality $\mn{Q}^2$ the formation time is given by $\tau_{\rm form} \sim E/\mn{Q}^2$, i.e.~the closer the particle is to its mass shell, the longer the quantum mechanical formation time of the process. This is so because the closer the particle is to the mass shell, the more collinear the splitting process needs to be to satisfy the kinematical constraints.  

When the hard particle traverses the medium, it undergoes collisions that exchange momentum with it. If the interactions with the medium are independent, the momentum transfer can be described with a (transverse) momentum broadening coefficient $\hat q$ which gives the mean transverse momentum squared acquired by the hard particle per unit time, $\hat  q \sim \langle k_\perp^2\rangle/t$. These interactions do not need to keep the particle on shell, and the hard particle will acquire parametrically the same amount of virtuality, $\mn{Q}^2 \sim \hat q t $. 

During the formation time of a splitting, the hard particle has time to acquire $\mn{Q}^2 \sim \hat q \tau_\text{form}$ of virtuality. Because of the acquired virtuality, the particle will be able to split in a time $\tau_\text{form} \sim E/\mn{Q}^2$. The longer the particle moves in the medium, the farther off shell it goes and the faster it can split. The formation time can be solved self-consistently giving 
\begin{equation}
 \tau_\text{form} \sim \frac{E}{\mn{Q}^2} \sim \frac{E}{\hat q \tau_\text{form}} \sim \sqrt{\frac{E}{\hat q} }.
\end{equation}
This is the quantum mechanical minimum time it takes to form an in-medium splitting of on-shell particles. Of course, in a weakly coupled theory, not everything that can happen happens, and only an $\alpha_s$ fraction of all possible splittings will take place. To this end,  the in-medium splitting rate is parametrically of order $\Gamma \sim \frac{\alpha_s}{ \tau_{\rm form}}\sim \frac{\alpha_s\sqrt{\hat q}}{\sqrt{E}}$.

If there are multiple interactions with the medium during $\tau_{\rm form}$, it is
quantum-mechanically indistinguishable which one of the interactions was responsible for
the splitting, and therefore there will be interference between processes
where the splitting happens, say, at $t=0$ and at $t={\tau_{\rm form}}$.
As the formation time sets an upper bound on the splitting rate and the
resulting rate is suppressed compared to the ``naive'' rate set by
the scattering rate of the medium, this interference is destructive. 
This suppression is the Landau-Pomeranchuk-Migdal (LPM) effect.

Let us now try to be slightly more quantitative in discussing this soft
scattering. In the previous Sec.~\ref{sub_htl_real}, we have
in Eqs.~\eqref{defkindasortaqhat}-\eqref{finalqhat} introduced and computed $F_\perp$, the (soft contribution to)
the correlator of two field strength tensors in the transverse channel.
As we mentioned there, that correlator is related to $\hat{q}$, as proven
formally in \cite{D'Eramo:2010ak,Benzke:2012sz}.
The soft contribution to $\hat{q}$ then reads
\begin{equation}
	\hat{q}_\mathrm{soft}=g^2C_R \int^\mu \frac{d^2q_\perp}{(2\pi)^2}F_\perp=g^2C_R
	T \int^\mu \frac{d^2q_\perp}{(2\pi)^2}\,
	\frac{m_D^2}{q_\perp^2+m_D^2}=\frac{g^2 C_R T m_D^2}{2\pi}\ln\frac{\mu}{m_D},
	\label{loqhat}
\end{equation}
where $gT\ll\mu\ll T$ is a cutoff used to restrict ourselves to the soft region and
$C_R$ is the quadratic Casimir in the representation of the source.
But $\hat{q}$ can also naturally be seen as the second moment of the (differential)
soft scattering rate $d\Gamma/d^2\q_\perp$, oftentimes called $\mathcal{C}(q_\perp)$ in the literature,
\begin{equation}
	\label{dgamma}
	\hat{q}_\mathrm{soft}= \int^\mu \frac{d^2q_\perp}{(2\pi)^2}\,q_\perp^2 \,
	\frac{d\Gamma}{d^2\q_\perp}\,,
	%\;\;\Longrightarrow\;\;
\end{equation}
 which brings us to the identification \cite{Aurenche:2002pd}
\begin{equation}
	\label{locqperp}
	\frac{d\Gamma}{d^2\q_\perp }=
		\frac{g^2 C_R T m_D^2}{q_\perp^2(q_\perp^2+m_D^2)}.
\end{equation}

From the above considerations, we see a power counting emerge: $\hat{q}\sim g^4T^3$,
$\Gamma\sim g^2T$ (recall that $d^2\q_\perp \sim g^2 T^2$ for soft $q_\perp$).%
\footnote{\label{foot_IR}
Eq.~\eqref{locqperp} implies in principle an IR divergent
$\Gamma$, arising from the unscreened magnetostatic gluons. As we have mentioned
in the previous Section, a non-perturbative formulation is required to deal
with that physics. However, as we explained, it enters at the scale $g^2T$ where, as we shall
show, a cancellation will happen, rendering precise knowledge of the form
of the scattering kernel there irrelevant for a leading- or next-to-leading order
discussion of LPM resummation. Hence, Eq.~\eqref{locqperp} will suffice
for our derivation of the leading-order rate for LPM-resummed collinear radiation.}
We remark that these are estimates for an infinite, static and equilibrated medium,
in keeping with the spirit of this review. The formalism we are introducing is also
well suited for media that are inhomogeneous along the longitudinal direction; we refer
to \cite{Arnold:2008iy,CaronHuot:2010bp} for details regarding this issue.
Hence, the power counting above implies that the equilibrium inverse
 soft scattering  rate 
is $\tau_{\rm scat}=1/\Gamma\sim 1/g^2 T$ and
is parametrically independent of  the frequency of the propagating particles.
The formation time of the splitting, $\tau_{\rm form} \sim \sqrt{E/\hat{q}}$,
 increases with the energy of
the parent particle for $E\gtrsim T$.
LPM interference happens when $\tau_{\rm form}\gtrsim \tau_{\rm scat}$
and thus  $E\gtrsim T$. In the following, we will discuss at some length
the derivation of the radiation rate, without assuming any hierarchy between $E$ and $T$.
The formulae we obtain will thus be valid either when $E\gg T$, as is most often
the case when studying jet modification in the quark-gluon plasma
\cite{d'Enterria:2009am,Wiedemann:2009sh,Majumder:2010qh,Mehtar-Tani:2013pia,Roland:2014jsa,Qin:2015srf,Connors:2017ptx},
or when $E\sim T$, as is the case when computing the collinear contribution to the
photon production rate from the Quark-Gluon Plasma (QGP) \cite{Arnold:2001ba} or when solving for transport
coefficients \cite{Arnold:2003zc}, which get a contribution from collinear processes between
the thermal ($\mn{P}\sim T$) constituents of the plasma.

\subsubsection{Heuristic derivation of LPM resummation}

\def\ix{{\rm i}}
\def\xx{{\rm x}}
\def\xbx{{\bar{\rm x}}}
\def\fx{{\rm f}}
\def\Bigdlangle{\Big\langle\!\!\Big\langle}
\def\Bigdrangle{\Big\rangle\!\!\Big\rangle}
\begin{figure}[t!]
	\begin{center}
		\includegraphics[width=11cm]{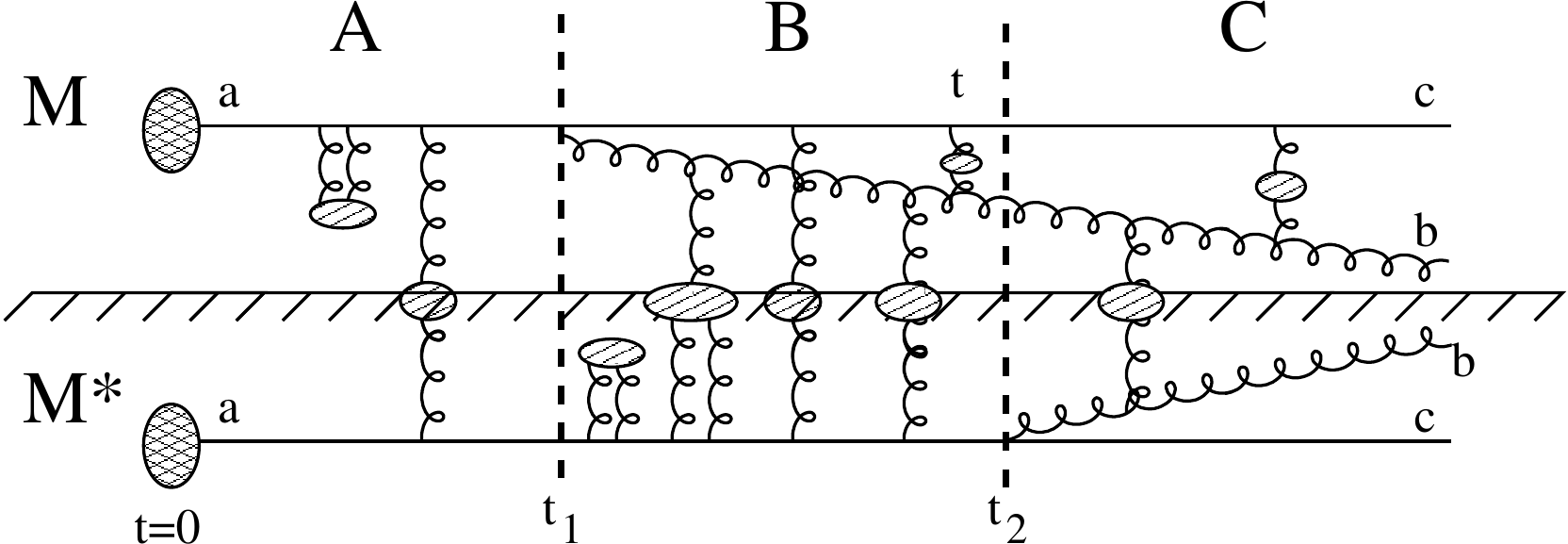}
	\end{center}
	\caption{The three regions discussed in the text in
	the context of gluon emission. The horizontal
	line across the figure represents the cut between the amplitude $\mathcal{M}$
	above and the conjugate amplitude $\mathcal{M}^*$ below. Time flows monotonically
	from left to right, as a consequence of the eikonal approximation.
	The  gluonic ``ladders'' connecting the three collinear particles a,b,c
	through the blobs, as well as the self energies of a,b,c, are intended to be
	HTL $rr$ propagators, with the blobs the HTLs themselves. Intuitively, the assignments
	for these propagators
	have to be $rr$, so that the statistical fluctuations of the medium are sampled.
	An exhaustive diagrammatic power counting and derivation arriving to this conclusion
	can be found in \cite{Arnold:2001ba}.
	Figure taken from \cite{CaronHuot:2010bp}.}
	\label{fig_collinear}
\end{figure}
Having explained the rationale for the LPM resummation as well as its relevance
for the physics of the QGP, we now turn to the derivation of the radiation
rate. A strict diagrammatic derivation of collinear photon radiation from
the QGP was introduced in \cite{Arnold:2001ba}. While undoubtedly exhaustive,
it may leave the underlying physical picture encumbered by the intricacies
of technical detail, which is why we present a slightly
more heuristic derivation following \cite{CaronHuot:2010bp,Arnold:2015qya}.

To move forward, we first examine Fig.~\ref{fig_collinear}. It shows a typical
diagram entering LPM resummation, where the horizontal cut separates
the amplitude $\mathcal{M}$ from the conjugate amplitude $\mathcal{M}^*$
and time flows from left
to right. Under the two assumptions of eikonal propagation
for the hard particles, justified by the separation of scales between
$T$ and $gT$, and of instantaneous soft collisions (compared
to the formation time), justified by the duration of the collision
$\mathcal{O}(1/gT)$ being much smaller than the time between
collisions, $\mathcal{O}(1/g^2T)$, the diagram
is naturally divided in three regions by the two times $t_1$ and $t_2$
($t_\xx$ and $t_\xbx$ in the equations to follow). The key point is that
soft scatterings happening
in region A, that is before any emission has taken place either in $\mathcal{M}$
or in $\mathcal{M}^*$, or in region C, after the emission has taken place in both
amplitudes, do not contribute to LPM interference and thus need not be resummed. Intuitively,
as explained in \cite{CaronHuot:2010bp},
scatterings in region A modify by small amounts the transverse momentum
and energy of the emitter, and can be reabsorbed in a (slight) change of the jet axis, a
freedom which we will also exploit later on. Scatterings in region C, on the other hand,
modify by small amounts the transverse momenta and energy of the emitted particles.
But we are not interested in a rate that is differential in transverse momentum, and the
small energy changes are negligible compared
to the much larger ones deriving from collinear
radiation. Under these approximations, the transverse-momentum
integration is time-independent after $t_2$ ($t_\xbx$) \cite{CaronHuot:2010bp},
hence the irrelevance of region C. However, if one wants to remain
differential in transverse momentum, a more complicated formalism
\cite{Apolinario:2014csa} is necessary to account
for regions A and C as well. 

Based on the above argument, we introduce
the following expression for the differential probability $dI/dx$
to radiate a photon ($\mn{K}$) with momentum fraction $x$ from a quark with
energy $E$, i.e.~$x=k/E$---a result that we will later
generalize to gluon radiation. We follow the notation, derivation 
and exposition of \cite{Arnold:2015qya}, to which we are greatly indebted.
The probability $dI/dx$ then reads
\begin{align}
   &\frac{dI}{dx}
   = 2 \Re \Biggl\{
       \frac{E}{2\pi V_\perp}
       \int_{t_\xx < t_\xbx} dt_\xx \> dt_\xbx
       \sum_{\rm pol.}
       \int_{\p_\fx,\k_\fx}    \int_{\p_\xx,\bar\p_\xbx}
\label{eq:1brem2}
\\
    &   \Bigdlangle
       \biggl(
         %\int_{\p_\xx}
         \langle \p_\fx \k_\fx, t_\xbx | \p_\xx \k_\fx,t_\xx \rangle
         \langle \p_\xx \k_\fx | {-}i \, \delta H | \p_\ix \rangle
       \biggr)%\times
       \biggl(
        % \int_{\bar\p_\xbx}
         \langle \p_\ix,t_\xx| \bar\p_\xbx, t_\xbx \rangle
         \langle  \bar\p_\xbx| i \, \delta H |\p_\fx \k_\fx \rangle
       \biggr)
       \Bigdrangle
   \Biggr\}, \nonumber
\end{align}
where the $\p_i$ stand for the transverse momenta of the quark at
different points in the amplitude and conjugate amplitude, and the $\k_i$ for those of
the radiated photon (we drop the $\perp$ label to avoid overloading our notation).
We assume that the quarks and the radiated photon have most of their momenta
along $z$ ($p_z\sim k_z\sim T$, $\p\sim\k\sim gT$), but we have not  picked a frame where $\k$ is zero: while
it is clearly an obvious choice in the case of photon radiation, it is not
in the case of gluon radiation. We shall comment more extensively on this later.

The hard quarks propagate through the background of (soft) medium gauge fields, with
the double brackets $\langle \langle\ldots\rangle\rangle$ defining a thermal average
over this background. The factors within the first round bracket on the second line describe
the amplitude:
from ket to bra, one has first
the (collinear) splitting matrix element
$ \langle \p_\xx \k_\fx | {-}i \, \delta H | \p_\ix \rangle$, where $\delta H$
is the part of the Hamiltonian containing the vertices for the hard particles.
It is followed by the Propagation Under the Influence of the medium (PUI) 
$\langle \p_\fx \k_\fx, t_\xbx | \p_\xx \k_\fx,t_\xx \rangle$
from $t_\xx$ to $t_\xbx$, with the photon as a spectator, hence $\k$ is unchanged there.
The second round bracket contains similarly the conjugate amplitude.
We take (twice) the real part to combine the interference term we are explicitly
considering
with its complex conjugate, where emission happens first in the conjugate amplitude.
The sum is over the final-state polarization of the quark and photon, and $V_\perp$
represents the  transverse volume, which  will, as expected, drop out from the final
results. Finally, the time integration
is restricted to region B in Fig.~\ref{fig_collinear}.
In equilibrium, the integrand can only depend on
$\Delta t\equiv t_\xbx-t_\xx$: we will make use of this simplification later, but for
now we find the formulation above slightly more instructive.

\def\Hilbert{{\mathbb H}}

Our goal is now to (briefly) show how the two-particle, four-dimensional
 in-medium QFT evolution
of the amplitude can be reduced to a simpler one-particle, two-dimensional
quantum mechanics problem (note that this holds also
in the three-particle case of gluon radiation). Let us rewrite our PUI
time-evolution in the amplitude as
\begin {equation}
  \langle \p_\fx \k_\fx, t_\xbx | \p_\xx \k_\fx,t_\xx \rangle
  =
  \langle \p_\fx \k_\fx| e^{-iH_{(2)} (t_\xbx-t_\xx)} |\p_\xx \k_\fx \rangle ,
\end {equation}
where $H_{(2)}$ is the PUI Hamiltonian for  two
hard particles through the medium: the quark  and
the photon.  All $H_{(i)}$ conserve the number
of hard particles and are thus distinct from $\delta H$.%
\footnote{Formally, we find that the best way to obtain
such a separation would be through Soft Collinear Effective Theory
\cite{Bauer:2000ew,Bauer:2000yr,Bauer:2001ct,Bauer:2001yt,Bauer:2002nz,Beneke:2002ph} (see 
\cite{Becher:2014oda} for a textbook).}
The PUI in the conjugate amplitude is
similarly given by
\begin{equation}
  \langle \p_\ix,t_\xx| \bar\p_\xbx, t_\xbx \rangle
  =
  \langle \p_\ix| e^{+iH_{(1)} (t_\xbx-t_\xx)} |\bar\p_\xbx \rangle .
\end{equation}
%where $H_{(1)}$ is the PUI Hamiltonian for one hard particle.
Since the medium %interactions with the
enters only in $H_{(1)}$ and $H_{(2)}$, the medium average in
Eq.~\eqref{eq:1brem2} can be restricted to
\begin{equation}
   \Bigdlangle
   e^{-iH_{(2)} (t_\xbx-t_\xx)} |\p_\xx \k_\fx \rangle
   \langle \p_\ix| e^{+iH_{(1)}(t_\xbx-t_\xx)}
   \Bigdrangle ,
\label{eq:timeevolve}
\end{equation}
that is, the time evolution of an initial
$|\p_\xx \k_\fx \rangle \langle \p_\ix|$ starting from time $t_\xx$.

The object $|\p_\xx \k_\fx \rangle \langle \p_\ix|$ lives in
\begin{equation}
  \bar \Hilbert_{\rm q} \otimes \Hilbert_{{\rm q},\gamma}
  =
  \bar \Hilbert_{\rm q} \otimes \Hilbert_{\rm q} \otimes \Hilbert_\gamma ,
\label{eq:Hilbert}
\end{equation}
where
$\Hilbert_{\rm q}$ is the Hilbert space of states of
a hard quark, and
$\Hilbert_{{\rm q},\gamma}$ is the Hilbert space of states with a collinear
quark-photon pair.  A central point of Zakharov's pioneering
approach \cite{Zakharov:1996fv,Zakharov:1997uu} is to rethink this
product as  a single Hilbert space of three particles:
one quark, one photon, and one \emph{conjugated quark}.
Correspondingly, we rewrite (\ref{eq:timeevolve}) in the form
of a 3-particle evolution
\begin {equation}
   \Bigdlangle
   e^{-iH_{(\bar 1+2)} (t_\xbx-t_\xx)} |\p_1,\p_2,\p_3 \rangle
   \Bigdrangle ,
\label {eq:timeevolve2}
\end {equation}
where $H_{(\bar 1+2)} = H_{(2)} - H_{(1)}$. The minus sign
stems from the fact that
$H_{(2)}$ acts on the two particles associated with the
amplitude and $H_{(1)}$ on the one associated with the conjugate
amplitude, in a way that is analogous to the
relative minus sign in the Schwinger--Keldysh formalism in Eq.~\eqref{eq:bc}.
For convenience we assign a minus sign to the
momenta in the {\it conjugate}\/ amplitude, so that
(\ref{eq:timeevolve}) may be recast in the
form (\ref{eq:timeevolve2}) with
\begin {equation}
   (\p_1,\p_2,\p_3) = (-\p_\ix,\p_\xx,\k)
   \qquad \mbox{at $t=t_\xx$.}
\end {equation}
With this sign convention for the $\p_i$, momentum conservation
$\p_\ix = \p_\xx+\k_\fx$ implies that
\begin {equation}
   \p_1+\p_2+\p_3 = 0 .
\end {equation}

The medium average in Eq.~\eqref{eq:timeevolve2} only affects the
PUI evolution operator. We can thus
rewrite (\ref{eq:timeevolve2}) as
\begin{equation}
   e^{-i{\cal H} (t_\xbx-t_\xx)}
   |\p_1,\p_2,\p_3 \rangle
   \equiv
   \Bigdlangle
   e^{-iH_{(\bar 1+2)} (t_\xbx-t_\xx)}
   \Bigdrangle
   |\p_1,\p_2,\p_3 \rangle ,
   \label{hmedPUI}
\end{equation}
where ${\cal H}$ is going to be the Hamiltonian of our effective
3-particle quantum mechanics problem.
Note that ${\cal H}$ need not be Hermitian, even though
$H_{(\bar 1+2)}$ is.

\def\b{{\bm b}}

The above effective Hamiltonian takes the
form of a kinetic term and a potential, with medium
effects in both. 
In detail
\begin{equation}
   {\cal H} =
   \frac{\p_1^2+m_{\infty\,1}^2}{2p^+_1} + \frac{\p_2^2+m_{\infty\,2}^2}{2p^+_2}
   + \frac{\p_3^2+m_{\infty\,3}^2}{2p^+_3}
   + V(\b_1,\b_2,\b_3) .
\label{eq:calH2}
\end{equation}
In thermal equilibrium, $\mathcal{H}$ is time independent. We note that
 the large $p^+_i$ components act as  non-relativistic ``masses''
for the two-dimensional, transverse kinetic part above.
We have used a similar minus sign convention for defining
the light-cone momenta $p^+_i$ as we did above for defining
transverse momenta $\p_i$.  In our case here, $p^+_1$ is negative, i.e.
% [[keep the E notation?]]
\begin{equation}
   (p^+_1,p^+_2,p^+_3) = E(-1,1{-}x,x) .
\label{eq:x123}
\end{equation}

The kinetic terms in Eq.~\eqref{eq:calH2} contain the first effect
arising from the medium: the asymptotic masses.
As we have explained in Sec.~\ref{sub_htl_real}, the dispersion
relations of hard, $\mn{P}\sim T$ particles read 
$\varepsilon_{\p_i}=\sqrt{p_i^2+m_{\infty\,i}^2}$.
Including this term in the kinetic Hamiltionian and expanding for large
$p^+_i$, with $\p_i\sim m_{\infty\,i}\sim g T$ and $p^-\sim g^2T^2/p^+$, as required
by $\mn{P}^2\sim g^2T^2$, one finds
\begin{equation}
    H_{(2)}^\mathrm{kin}-H_{(1)}^\mathrm{kin}
    = (\varepsilon_{\p_2}+\varepsilon_{\p_3})-\varepsilon_{\p_1}\approx
\frac{\p_2^2+m_{\infty\,2}^2}{2p^+_2}   + \frac{\p_3^2+m_{\infty\,3}^2}{2p^+_3}
+\frac{\p_1^2+m_{\infty\,1}^2}{2p^+_1}.
\label{eq:calHfree}
\end{equation}
If particle $i$ is a quark, $m_{\infty\,i}=m_\infty$, if a gluon,
$m_{\infty\,i}=M_\infty$, and if a photon, $m_{\infty\,i}=0$. When $E\gg T$,
these masses become negligible.

The second effect of interactions with the medium is to add an imaginary
part ($\mathrm{Re}\,V=0$) to ${\cal H}$, related to the soft scattering
rate $d\Gamma/d^2\q_\perp$.
For the case of a weakly-coupled QCD plasma radiating a photon
this becomes
\begin{equation}
   V(\b_1,\b_2,\b_3) =
      -i g^2C_F \Delta\bar\Gamma(\b_2{-}\b_1) %\left[
        % \, \bar\Gamma(0)
         %- g^2C_F\, \bar\Gamma(\b_2{-}\b_1)
     % \right] ,
\label{eq:Vqed}
\end{equation}
where
\begin{equation}
   \Delta\bar\Gamma(\b)
   \equiv
   \bar\Gamma(0)
   -
   \bar\Gamma(\b)
   = \int d^2q_\perp \> \frac{d\bar\Gamma}{d^2\q_\perp}
          \, (1-e^{i\b\cdot\q_\perp})
   .
\label{eq:DGamma2}
\end{equation}
Here, $g^2C_F$ is the effective coupling of the medium to the hard quark, while $\b_3$ corresponds
to the photon,  Above, $ \bar\Gamma(0)\equiv\Gamma / g^2C_F$
is the coupling- and Casimir-stripped
rate of elastic scattering from the medium and can be obtained from Eq.~\eqref{locqperp}.
Its generalization
$\Gamma(\b)$ is defined as the Fourier transform of the differential
rate of scattering $d\Gamma/d^2\q_\perp$ with respect to
the transverse momentum transfer $\q_\perp$.
The second term in (\ref{eq:DGamma2}) corresponds to
background field
correlations between the amplitude and conjugate amplitude in Fig.~\ref{fig_collinear}.
The first term corresponds to the self energies of charged particle
lines arising from correlations between the amplitude
and itself, or between the conjugate amplitude and itself. These latter
terms
are clearly also those responsible, through their real parts and at hard momentum, for the
asymptotic masses in Eq.~\eqref{eq:calH2}.
The relative sign in (\ref{eq:DGamma2}) arises
because the second term corresponds, in the
language of ${\cal H}$, to the interaction of a quark
and a conjugated quark.  As we remarked in
footnote~\ref{foot_IR}, $\bar\Gamma(0)$ is not IR-finite in
perturbation theory. $\Delta\bar\Gamma(\b_2{-}\b_1)$ in Eq.~\eqref{eq:Vqed},
on the other hand, is, as the leading IR behavior cancels with the subtracted
$\bar\Gamma(\b_2{-}\b_1)$. The contribution of the non-perturbative scale $g^2T$
is then suppressed by a relative factor of $g^2$.

It is also important to remark that
the potential $V$ may  be related
to the value of real-time Wilson loops
which contain two long, parallel, light-like lines separated
by $\b = \b_2-\b_1$ and closed by transverse gauge links, as it
first emerged from the position-space, path-integral approach of
\cite{Zakharov:1996fv,Zakharov:1997uu},
later formalized through SCET in \cite{D'Eramo:2010ak,Benzke:2012sz}.

In the case of gluon radiation, Eq.~\eqref{eq:Vqed} turns into
\begin{align}
   V(\b_1,\b_2,\b_3) &=
      -i g^2 \biggl[
         \tfrac12(C_1{+}C_2{-}C_3) \,
             \Delta\bar\Gamma(\b_2{-}\b_1)\nn
\\
         &\hspace{-0.4cm}+ \tfrac12(C_2{+}C_3{-}C_1) \,
             \Delta\bar\Gamma(\b_3{-}\b_2)
         + \tfrac12(C_3{+}C_1{-}C_2) \,
             \Delta\bar\Gamma(\b_1{-}\b_3)
      \biggr] ,
\label{eq:Vqcd}
\end{align}
where
$C_i$ is the quadratic Casimir for particle $i$. In the $q\to q\gamma$ case
we were considering previously, $C_1=C_2=C_F$ and $C_3=0$, whereby Eq.~\eqref{eq:Vqed}
is recovered.

\def\B{{\bm B}}
\def\grad{{\bm\nabla}}

The next critical step in the reduction to the one-particle quantum-mechanical
problem is the observation that we can choose our $z$ axis to point at a
slightly different direction while maintaining the collinear approximations we
have made. Indeed, in the photon radiation case the most sensibile choice
is, as we mentioned before, to choose the $z$ axis along the photon direction.
In the more general gluon case, this rotation under the collinear approximation
implies that the rate should be invariant for $(\p_i,p_{iz})\to(\p_i+p_{iz}{\bm\xi},p_{iz})$.
Since $p_{iz}\approx p^+_i=x_iE$ and because of the zero sum of the $\p_i$
and $x_i$, this invariance implies the existence of a single independent combination,
\def\P{{\bm P}}
\begin{equation}
   \P \equiv x_2 \p_1 - x_1 \p_2
   = x_3 \p_2 - x_2 \p_3
   = x_1 \p_3 - x_3 \p_1 .
\label{eq:P123}
\end{equation}
Indeed, the 3-particle kinetic energy term in ${\cal H}$, Eq.~\eqref{eq:calH2}, can
be rewritten in terms of $\P$ in the form of a 1-particle kinetic energy:
\begin{equation}
   \frac{\p_1^2+m_{\infty\,1}^2}{2p^+_1} + \frac{\p_2^2+m_{\infty\,2}^2}{2p^+_2} +
   \frac{\p_3^2+m_{\infty\,3}^2}{2p^+_3}
   =
   % -\frac{\P^2-m_{\infty\,1}^2x_2 x_3-m_{\infty\,2}^2x_1 x_3-m_{\infty\,3}^2x_1 x_2}
%    {2x_1 x_2 x_3 E }
 -\frac{\P^2}{2x_1 x_2 x_3 E }+\sum_i\frac{m_{\infty\,i}^2}{2p^+_i}.
\end{equation}
From (\ref{eq:x123}) we recall that $x_1 x_2 x_3$  is negative since
our $x_1$ is negative.
The corresponding position variable is
\begin{equation}
   \B \equiv \frac{\b_1-\b_2}{(x_1+x_2)}
   = \frac{\b_2-\b_3}{(x_2+x_3)}
   = \frac{\b_3-\b_1}{(x_3+x_1)} \,,
\label{eq:B123}
\end{equation}
so that Eq.~\eqref{eq:calH2} becomes in the most general gluon radiation case
\begin{align}
   {\cal H} =
   -\frac{\P^2}{2x_1x_2x_3E}+\sum_i\frac{m_{\infty\,i}^2}{2p^+_i}
      -i g^2 \biggl[
         \tfrac12(C_1{+}C_2{-}C_3) \,
             \Delta\bar\Gamma(-x_3\B)\nn
\\
         + \tfrac12(C_2{+}C_3{-}C_1) \,
             \Delta\bar\Gamma(-x_1\B)
         + \tfrac12(C_3{+}C_1{-}C_2) \,
             \Delta\bar\Gamma(-x_2\B)
      \biggr] .
	  \label{final1part}
\end{align}

\def \als{\alpha_\mathrm{s}}
We now have all the ingredients to get to the final formulae. The time
evolution given by Eq.~\eqref{hmedPUI}, with the PUI Hamiltonian given
above in Eq.~\eqref{final1part}, evolves the momentum states in medium
as a one-particle quantum-mechanics problem. Hence, the radiation probability
can be shown to become
\begin{equation}
   \frac{dI}{dx}
   =
   \frac{\als P_{1\to 2}(x)}{[x(1-x)E]^2}
   \Re \int_{t_\xx < t_\xbx} dt_\xbx \> dt_\xx
   \grad_{\B_\xbx} \cdot \grad_{\B_\xx}
   \langle \B_\xbx,t_\xbx | \B_\xx,t_\xx \rangle
   \Bigr|_{\B_\xbx = \B_\xx = 0} ,
\label{eq:dI123}
\end{equation}
where  $  \langle \B_\xbx,t_\xbx | \B_\xx,t_\xx \rangle$ is the
propagator of the one-particle state described by the PUI
Hamiltonian, Eq.~\eqref{final1part}. In more detail, it is a Green's function
of this Schr\"odinger equation
\begin{equation}
	\label{shroddy}
	i\partial_t \psi(\B,t)=\mathcal{H}\,\psi(\B,t),
\end{equation}
with initial condition
\begin{equation}
	\label{greensfun}
	  \langle \B_\xbx,t_\xx | \B_\xx,t_\xx \rangle=\delta^{2}(\B_\xbx-\B_\xx).
\end{equation}
 The matrix elements of the
splitting Hamiltonian $\delta H$, originally present in Eq.~\eqref{eq:1brem2},
are responsible for the two factors of $\P$ ($\grad_\B$ in position space) and
for  $\als P_{1\to 2}(x)$, where $P_{1\to 2}(x)$ are the standard spin-averaged
DGLAP splitting functions. They read
\begin{equation}
	P_{q\to gq}(x)=C_F\frac{1+(1-x)^2}{x},\quad P_{g\to gg}(x)=C_A\frac{1+x^4+(1-x)^4}{x(1-x)}
	\label{dglap}
\end{equation}
Although we have not written the gluon analogue of Eq.~\eqref{eq:1brem2}, Eq.~\eqref{eq:dI123}
holds both for photon and gluon radiation. In the photon case, $\mathcal{H}$ 
contains Eq.~\eqref{eq:Vqed}, and the DGLAP splitting function
is easily obtained from Eq.~\eqref{dglap},
\begin{equation}
	P_{q\to\gamma q}(x)=\frac{1+(1-x)^2}{x}.
	\label{dglapphoton}
\end{equation}

Finally, in equilibrium one usually works with the differential rate rather than the
probability. The former is obtained by differentiating in time the latter, yielding
\begin{equation}
   \frac{d\Gamma}{dx}
   =
   \frac{\als P_{1\to 2}(x)}{[x(1-x)E]^2}
   \Re \int_0^\infty d(\Delta t) \>
   \grad_{\B_\xbx} \cdot \grad_{\B_\xx}
   \langle \B_\xbx,\Delta t | \B_\xx,0 \rangle
   \Bigr|_{\B_\xbx = \B_\xx = 0}.
\label{eq:drate123}
\end{equation}
To make a further simplification, let us define \cite{Arnold:2008iy}
\begin{equation}
	{\bm f}(\B_\xbx,t)
	\equiv2i\big[\grad_{\B_\xx}\langle \B_\xbx,t| \B_\xx,0 \rangle
	\big]_{B_\xx=0}.
\end{equation}
It also solves Eq.~\eqref{shroddy}, with an analogous initial condition. If we now introduce
the time-integrated
\begin{equation}
	{\bm f}(\B)\equiv\int_0^\infty dt {\bm f}(\B,t),
\end{equation}
where we dropped the $\xbx$ label, 
and integrate both sides of the Schr\"odinger equation, Eq.~\eqref{shroddy}, noting
that  ${\bm f}(\B,t)$ vanishes at large times because of the imaginary part
of $\mathcal{H}$, we obtain
\begin{equation}
	-2\grad_{\B}\delta^2(\B)=\mathcal{H}\,{\bm f}(\B),
	\label{AMYeq}
\end{equation}
and thus, in a time-integrated form,
\begin{equation}
   \frac{d\Gamma}{dx}
   =
   \frac{\als P_{1\to 2}(x)}{[x(1-x)E]^2}
   \Re \bigg[(2i)^{-1}\grad_{\B}\cdot {\bm f}(\B)\bigg]_{B=0},
%    =\frac{\als P_{1\to 2}(x)}{2[x(1-x)E]^2}
%    \Im \bigg[\grad_{\B_\xbx}\cdot {\bm f}(\B_\xbx)\bigg]_{B_\xbx=0},
\label{eq:drateAMY}
\end{equation}
which is the (Fourier transform of) the form originally obtained by Arnold,
Moore and Yaffe \cite{Arnold:2001ba,Arnold:2002ja}. Further details
on the equivalence of the two formulations can be found in \cite{Arnold:2008iy},
whose derivation we have followed for these last steps. We also refer
to \cite{Arnold:2015qya} for details on the handling of the $t\to 0$ divergence
in the integration of Eq.~\eqref{eq:drate123}, which is related to the
(vanishing) vacuum contribution to the splitting rate.

We now have all the ingredients to write the collinear contribution to the leading-order
photon rate. Eq.~\eqref{eq:drateAMY} with the splitting kernel of
Eq.~\eqref{dglapphoton} for $E=p$ and $k=x p$ gives the rate
for a quark $p$ to emit a photon with momentum $k$. We then have to integrate
over the momenta $p$, with the appropriate statistical functions, to find the
photon rate,
\begin{align}
	\frac{d\Gamma_{q\to\gamma q}}{d^3k}=&\frac{1}{(2\pi)^3}
	\int_k^\infty dp\, \nfd(p)(1-\nfd(p-k))\frac{k^2}{ p^3 }
	2N_c\frac{d\Gamma_{q\to\gamma q}}{dx}\nn\\
 &\hspace{-1.7cm} =\frac{\alpha N_c \sum_i Q_i^2}{(2\pi)^32 k} 
 \int_k^\infty dp\, \nfd(p)(1-\nfd(p-k))
   \frac{ p^2+(p{-}k)^2}{ p^2 (p{-}k)^2}
  \Im \big[2\grad_{\B}\cdot {\bm f}(\B)\big]_{B=0},
\end{align}
where on the first line $2 N_c$ accounts for the spin and color multiplicity
of the quark and  $k^2/p^3$  translates  from the 
$\p$-based phase space to the $\k$-based phase space and accounts for
$dx=dk/p$.

The integration above accounts only for the $q\to\gamma q$ process.
At vanishing chemical potentials, the $\bar q\to\gamma \bar q$ process
is identical,\footnote{
	The case of non-zero chemical potentials is considered in \cite{Gervais:2012wd}
	for thermal photon production. For the analogous case 
	of right-handed neutrino production in the electroweak plasma,
	the dependence on the chemical potentials has been derived in
	\cite{Ghiglieri:2017gjz,Ghiglieri:2018wbs}.
} while the $q\bar q\to\gamma $ process can be obtained
from a simple crossing of the above, leading to
\begin{align}
	\frac{d\Gamma_{\mathrm{coll}}}{d^3k}&=
	\frac{\Pi^<_{g^2\,\mathrm{coll}}(\mn{K})}{(2\pi)^32k}
	=\frac{\alpha N_c \sum_i Q_i^2 }{(2\pi)^32 k}
	\int_{-\infty}^{+\infty} dp\, \nfd(p)(1-\nfd(p-k))\nn\\
  &\hspace{4.5cm}\times \frac{ p^2{+}(p{-}k)^2}{ p^2 (p{-}k)^2}
  \Im \big[2\grad_{\B}\cdot {\bm f}(\B)\big]_{B=0},
  \label{finalphoton}
\end{align}
where the integration encompasses the $\bar q\to\gamma \bar q$ process
in the $-\infty<p<0$ range, the $q\bar q\to\gamma $ one in the
$0<p<k$ range, and the  $q\to\gamma q$ one for $p>k$.
This is the standard form 
featured in papers on thermal photon production, see e.g.~Eq.~(2.1)
of \cite{Arnold:2001ms}, where $\Im \big[2\grad_{\B}\cdot {\bm f}(\B)\big]_{B=0}$
is written as its Fourier transform integrated over all momenta, or Eq.~(2) of
\cite{Ghiglieri:2014vua}, which is directly in this form.
 We refer to
\cite{Aurenche:2002wq,Ghisoiu:2014mha,Ghiglieri:2014kma}
for methods for the numerical solution of Eq.~\eqref{AMYeq}. A code is available
in the arXiv package of \cite{Ghiglieri:2014kma}.

%% file: realtimeresults.tex
% !TEX root = review.tex

\section{Applications of the real-time formalism \label{sec:realresults}}

After dedicating Secs.~\ref{sec:realtimept} and \ref{sec_soft_collinear}
to a review of the methods of real-time perturbation theory at finite
temperature and to the resummations that are oftentimes necessary
when tackling calculations in thermal QCD where real times and/or Minkowski 
space play an important role, it is time to show how these methods
are applied to a series of topics in hot QCD, of relevance for heavy ion physics.
We shall keep a focus on the methodological state of the art, rather than
a historical perspective or a phenomenological one, but refer
to reviews covering the latter point whenever possible.

\subsection{Electromagnetic radiation}
\label{sub_res_photon}
As our discussion of soft and collinear dynamics in the previous section
was narrated around the leading-order determination of the thermal photon
production rate, it seems only natural to start our discussion of the 
state of the art of real-time applications with it. As we already remarked there,
the complete determination of the rate proved to be a challenging endeavor, requiring the proper
handling of both soft \cite{Kapusta:1991qp,Baier:1991em} and collinear \cite{Arnold:2001ba}
modes, with the complete LO results published in \cite{Arnold:2001ms}.

The first higher-order correction to this leading-order result has been determined 
in \cite{Ghiglieri:2013gia}. 
Contrary to what happens in ordinary perturbation theory
in vacuum, at finite temperature the loop expansion parameter is not necessarily
$\als$. As we will see in many examples throughout this review, soft gluon loops are penalized
by $g$ only, rather than $g^2$, because of the $1/g$ Bose enhancement we
discussed previously. As the soft modes contribute to the leading-order photon rate, it
is no surprise that the NLO corrections computed in \cite{Ghiglieri:2013gia}
contribute a relative order $g$ correction, i.e.~an order $\alpha g^3$ contribution to the photon rate. 

The computation performed in \cite{Ghiglieri:2013gia} required studying all kinematic
regions where soft gluon loops could be added to the LO graphs. 
In Fig.~\ref{fig_lomap}, these would
be the soft and the collinear regions. Furthermore, the leading-order rate
was obtained by integrating Eqs.~\eqref{comptonnaive}, \eqref{annihnaive} and \eqref{finalphoton}
over the entire phase space. This includes 
small, $\mathcal{O}(g)$ regions of the phase space where the approximations underlying these equations
fail, introducing an $\mathcal{O}(g)$ ambiguity in the LO rate that needs to be handled 
properly at NLO. The region where this happens was termed \emph{semi-collinear} in 
\cite{Ghiglieri:2013gia} and sits in Fig.~\ref{fig_lomap} in the empty area between the
hard, soft and collinear regions. 

The evaluation of all these regions in \cite{Ghiglieri:2013gia} relied heavily on the
causality-based sum rules on the light cone. In the soft region, the
 top diagram in Fig.~\ref{fig_soft_photon} would in principle need to be complemented (and complicated)
 by the addition of an extra soft gluon attaching to the soft quark via HTL-resummed vertices.
 This apparently nightmarish brute-force HTL computation---see the coming Sec.~\ref{sub_res_heavy}
 for an example of comparable intricacy---was however avoided through the extension
 to NLO of the sum rule discussed in the equations leading to Eq.~\eqref{softphotonsumrulefinal},
 yielding a compact, closed-form result for the NLO soft contribution.
 
 %indent fixed
  In the collinear sector,
 the NLO corrections only affect the PUI Hamiltonian discussed in Eq.~\eqref{eq:calH2} in Sec.~\ref{sec_lpm},
 owing to the factorization between soft medium effects and the hard splittings we described there. What is
needed are thus the NLO corrections to the asymptotic mass and to the soft
scattering kernel $\frac{d\Gamma}{d^2\q_\perp}$. Both were computed by Caron-Huot using the mapping to the
three-dimensional Euclidean theory discussed around Eq.~\eqref{finalqhat}; the results can be found in
\cite{CaronHuot:2008uw} and \cite{CaronHuot:2008ni}, respectively, and we will return to the case of of $\frac{d\Gamma}{d^2\q_\perp}$
in Sec.~\ref{sub_res_transport}. The perturbation to Eq.~\eqref{finalphoton}
from these corrections was finally determined in \cite{Ghiglieri:2013gia}. Finally, the semi-collinear region was again proven to factorize into a DGLAP splitting kernel times a soft operator,
which was also determined using the Euclidean mapping discussed above.

\begin{figure}[t!]
	\begin{center}
		\includegraphics[width=8cm]{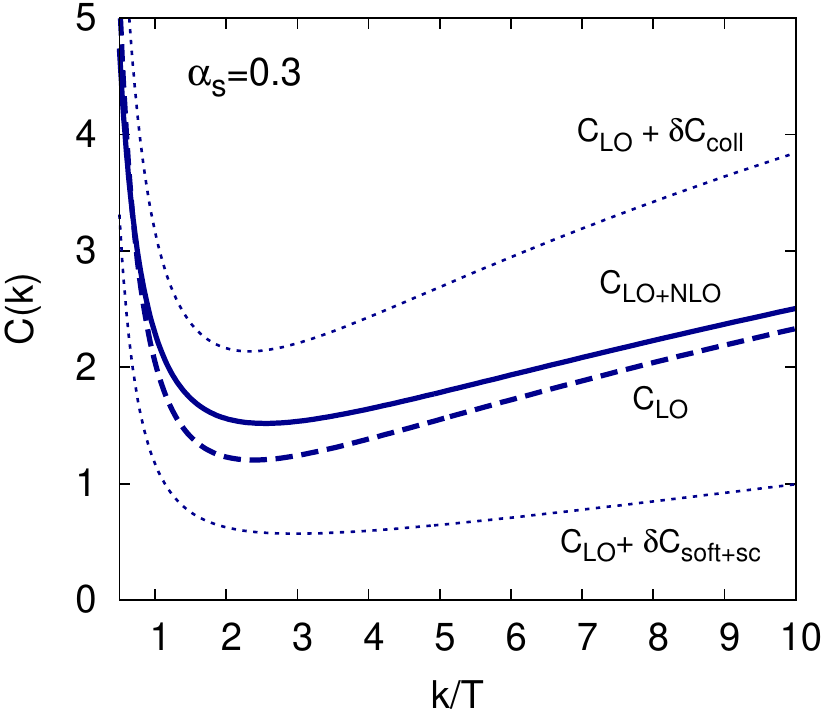}
	\end{center}
	\caption{The thermal photon production rate at LO \cite{Arnold:2001ms} and NLO \cite{Ghiglieri:2013gia}, with
	$C(k)\equiv\Pi^<(\mn{K})/(4 \alpha\sum_i Q_i^2\nfd(k)m_\infty^2)$. $C_\mathrm{LO}$ is the LO rate,
	$\delta C_\mathrm{coll}$ is the collinear correction only, $\delta C_\mathrm{soft+sc}$ is the soft and semi-collinear
	correction only and $C_\mathrm{LO+NLO}$ is the full NLO result. Figure  taken from \cite{Ghiglieri:2013gia}.}
	\label{fig_nlo_photon}
\end{figure}

Assembling together the different  contributions, the results of \cite{Ghiglieri:2013gia} are summarized in Fig.~\ref{fig_nlo_photon},
which shows how the contribution from the NLO collinear modes are large and positive, while those
from the soft and semi-collinear modes are of similar magnitude but opposite sign. They thus largely cancel, leaving only a 20\%
to 30\% increase in the photon rate at $\als=0.3$.

Photons are only a part of the broader concept of \emph{electromagnetic radiation} emitted thermally from a quark-gluon plasma; dileptons,
i.e.~lepton-antilepton pairs, represent the other significant emission. Their production is given by \cite{McLerran:1984ay}
\begin{equation}
	\label{dileptonrate}
	\frac{dN_{l^+l^-}}{d^4\mn{X}d^4\mn{K}}\equiv\frac{d\Gamma_{l^+l^-}}{d^4\mn{K}}=\frac{-2\alpha}{3(2\pi)^4\mn{K}^2}\Pi^{<}(\mn{K})\,,
\end{equation}
where $\Pi^<(\mn{K})$ is to be evaluated for any $-\mn{K}^2>4m_l^2$ and we are assuming to be far from any mass
threshold. The connection to Eq.~\eqref{photonrate}, the photon rate, is straightforward: $\Pi^<(\mn{K})$
tells us about the production of a photon, be it real, $k^0=k$, or virtual, $k^0>k$. In the latter case the prefactor
of $-2\alpha/(3\mn{K}^2)$ describes its propagation and conversion to the $l^+l^-$ pair. 

From a theorist's perspective, the task remains the same: the determination of the Wightman function $\Pi^<$ in thermal 
QCD. However, with respect to the photon case, which only depends on the frequency, the dilepton rate depends on two kinematical
variables, to be picked among the frequency $k^0$, the momentum $k$ and the virtuality $-\mn{K}^2$, which is oftentimes termed
the dilepton mass $M^2=-\mn{K}^2$. Depending on the interplay of these parameters, different scales and techniques play a role
in the determination of $\Pi^<(\mn{K}^2)$.

\begin{figure}[t!]
\begin{center}
	\includegraphics[width=12cm]{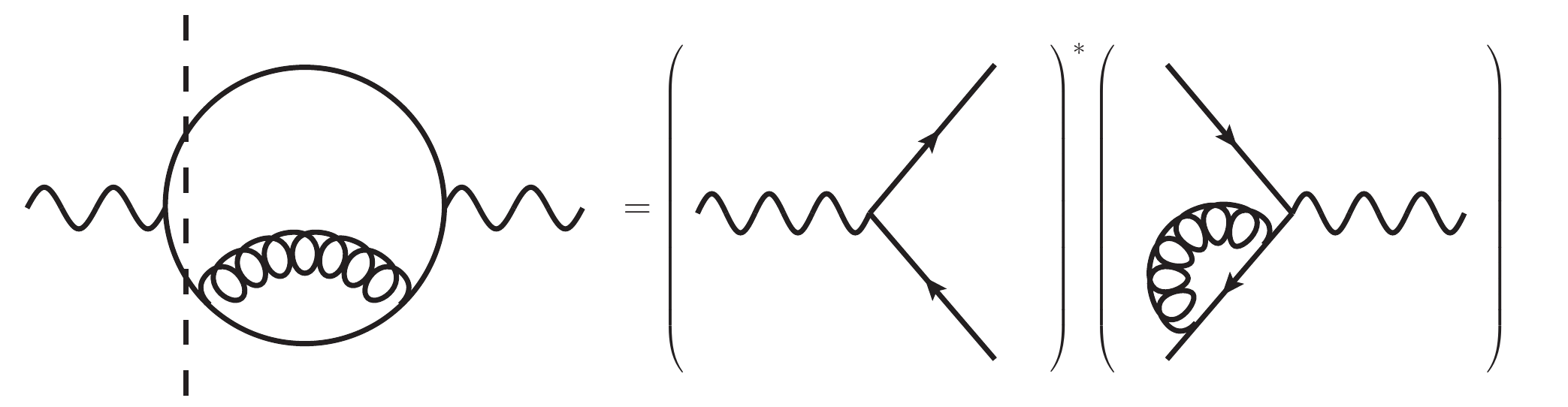}
\end{center}
\caption{One of the \emph{virtual cuts} representing the interference between the Born process and its first
virtual correction. Figure adapted from \cite{Ghiglieri:2015nba}.}
\label{fig_dileptoncut}
\end{figure}

Historically, the first region to be covered was that of vanishing momentum, $k=0$. For $k^0=M\gtrsim T$ the first
orders in perturbation theory are dominated by the hard modes, so that the expansion resembles the familiar
zero-temperature case. The leading order, $\mathcal{O}(\alpha g^0)$, is given by the Born term \cite{McLerran:1984ay}, as shown in Fig.~\ref{fig_dy},
which is non-zero at positive $M$ (we neglect the lepton mass in the following). The NLO corrections
come at order $\alpha g^2$ from the diagrams in Fig.~\ref{fig_lophoton}. In the photon case, only the cut going through the
gluon, shown in Fig.~\ref{fig_lo_cuts}, was non-vanishing. 
In this case, we are kinematically  allowed to avoid cutting the gluon, yielding  cuts  such
as the one shown in Fig.~\ref{fig_dileptoncut}.
As discussed in
the previous section, these cuts describe the interference between the Born process and its first virtual correction. In the literature,
the cuts through the gluon are oftentimes termed \emph{real cuts} or \emph{processes}, 
while the others are the \emph{virtual} ones. At vanishing momentum, these NLO corrections were evaluated in
\cite{Baier:1988xv,Gabellini:1989yk,Altherr:1989fc}. The main challenge lies in the fact that, taken separately, the real 
and virtual processes present soft and collinear divergences when intermediate propagators approach the mass shell. Their
sum is IR safe, so care must be taken in consistently regularizing them (see also \cite{Baier:1989ub} for an early
application of these methods to neutron decays in the Early Universe). 

At zero momentum and small frequency, $k^0\sim gT$, Braaten, Pisarski
 and Yuan \cite{Braaten:1990wp} determined the dilepton rate
through one of the first brute-force HTL-re\-sum\-med computations: they argued that the leading-order contribution
comes from the one-loop graph in Fig.~\ref{fig_dy} where both quarks are soft and the vertices are HTL-re\-sum\-med, too. It was, however, realized later \cite{Aurenche:1996sh,Aurenche:1998nw,Moore:2006qn} that other processes,
not captured by HTL resummation, contribute at LO as well. Finally, the determination of the rate down to $k^0\sim g^4T$
was performed in \cite{Moore:2006qn}, requiring techniques that shall be described later when illustrating
transport coefficients, as the rate at zero momentum and vanishing frequency is related to the electric 
conductivity of the plasma.

At nonzero momenta, the region $M\gtrsim T$ was evaluated at NLO in \cite{Laine:2013vma}. Conceptually one encounters the same issue of intermediate regularization as in the zero-frequency case,
but the technical details are much more intricate. The Euclidean techniques employed 
in \cite{Laine:2013vma}, as well as those based on the Operator Product Expansion for  $M\gg T$,
will be discussed in more detail in Sec.~\ref{sec:beyondqcd}.

If instead $k\sim k^0\sim T$ and the mass is small, $M\sim g T$, the dilepton calculation does not differ
dramatically from the photon one. At leading order, one only needs to modify the collinear
part of the rate---as shown in \cite{Aurenche:2002wq,Carrington:2007gt}---to account for the dependence on the non-vanishing
mass. This small-$M$ rate was extended to NLO, employing the techniques of the NLO photon rate discussed above,
in \cite{Ghiglieri:2014kma}. A procedure to smoothly connect the rate at $M\sim g T$ to the one
in \cite{Laine:2013lka} at $M\sim T$ was on the other hand devised in \cite{Ghisoiu:2014mha,Ghiglieri:2014kma}. It is 
based on the observation that as $M$ grows in the low $M$ calculation, the collinear part of the rate becomes dominated by the Born
term in the collinear limit. One then needs to add to the high-$M$ calculation 
the collinear part of the low-$M$ one, with the first terms of this high-$M$ expansion subtracted off 
to avoid double countings. Through this procedure, one can obtain the electromagnetic rate
at NLO for $k\gtrsim T$ at any $k^0\ge 0$.

\begin{figure}[t!]
	\begin{center}
		\includegraphics[width=8cm]{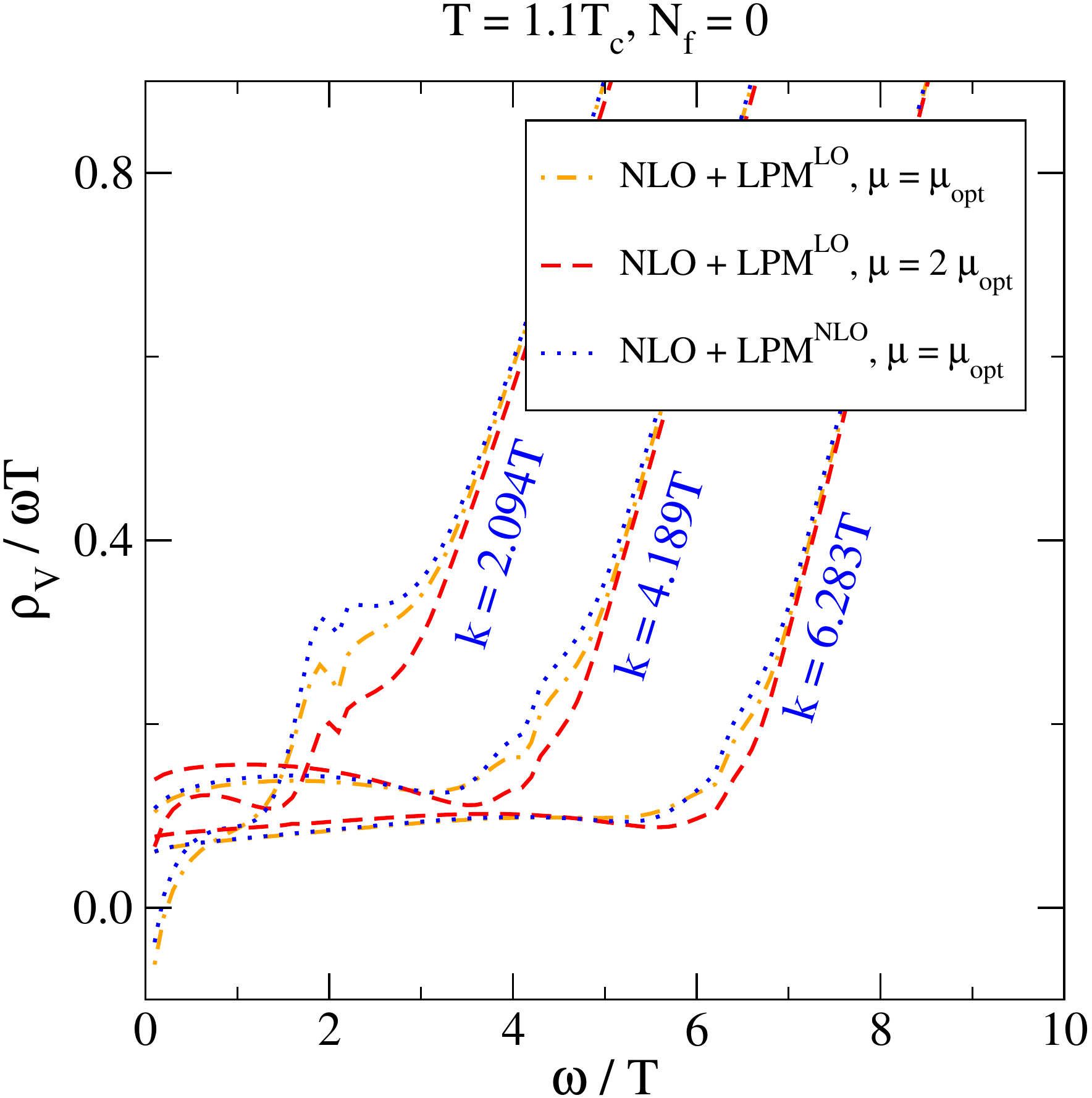}
	\end{center}
	\caption{The electromagnetic spectral function  $\rho_V$ of quenched QCD at leading- and next-to-leading order over the 
	entire kinematical range, as presented in \cite{Jackson:2019yao}. Please note that the definition
	of the spectral function in \cite{Jackson:2019yao} is one half of our definition in
	Eqs.~\eqref{defrho} and \eqref{rhoboson}. The gauge coupling is obtained from 5-loop running,
	with the $\overline{\mathrm{MS}}$ scale set to (multiples of) $\mu_\mathrm{opt}\equiv
	\sqrt{(\pi T)^2+\omega^2-k^2}$. The ``NLO+$\mathrm{LPM}^\mathrm{LO}$'' curves
	do not include the NLO corrections of \cite{Ghiglieri:2014kma} at small $M$, while 
	the ``NLO+$\mathrm{LPM}^\mathrm{NLO}$'' do. Figure taken from \cite{Jackson:2019yao}.}
	\label{fig_greg_mikko}
\end{figure}
In Fig.~\ref{fig_greg_mikko} we show the state of the art of this procedure, as obtained in \cite{Jackson:2019yao}. 
What is plotted here is the electromagnetic
spectral function $\rho$; we recall that the KMS relation, Eq.~\eqref{kmsmombosonless}, relates 
it to $\Pi^<$. In this figure, the spectral function is plotted also for $k^0<k$. While pointless
for the electromagnetic rates, this region is important for comparisons with the lattice data: due to its strictly Euclidean formulation, lattice QCD can only determine
\begin{equation}
	G_E(\tau,k)\equiv\int d^3x e^{-i\k\cdot\x}\left\langle J^i(\tau,\x) J^i(0)-
	J^0(\tau,\x) J^0(0)\right\rangle.
\end{equation}
As we explained in Sec.~\ref{sec:IV} when discussing the analyticity bands of the Wightman
functions, $G^<(t)=G_E(-it)$. Combining this with the KMS relation, Eq.~\eqref{kmsmombosonless},
we find
\begin{equation}
	\label{rhoeuclid}
	G_E(\tau,k)=G^<(i\tau,k)=\int\frac{d\omega}{2\pi}e^{\omega \tau}n_\mathrm{B}(\omega)\rho(\omega,k)\,,
\end{equation}
that is, the Euclidean correlator corresponds to a convolution integral of the spectral function
with a finite-temperature kernel.\footnote{In the literature, the odd nature
of $\rho(\omega)$ is employed to recast the r.h.s of Eq.~\eqref{rhoeuclid} as
\begin{equation}
	\label{rhoeuclidlattice}
	G_E(\tau,k)=\int_0^\infty\frac{d\omega}{2\pi}\frac{\cosh(\omega(\tau-1/(2T)))}{\sinh(\omega/(2T))}\rho(\omega,k)\,.
\end{equation}
We prefer 
the form in Eq.~\eqref{rhoeuclid},
as the connection to the underlying TFT correlators is explicit there.}

As such, the extraction of $\rho$ from a discrete set of datapoints obtained from lattice QCD
is an ill-posed numerical problem, the detailed discussion of which is beyond the scope of this report; instead, we refer to \cite{Meyer:2011gj}
for a review of the topic. On the other hand, once the perturbative spectral function is known
at all values of $\omega$, it becomes possible to compare the perturbative $G_E$ with the lattice-determined
one. Fig.~\ref{fig_greg_mikko} plots then the most advanced determination of the spectral function,
where the methods of \cite{Laine:2013vpa,Laine:2013lka,Laine:2013vma} have been applied 
to determine
the spectral function also in the space-like domain at large $\mn{K}^2$ \cite{Jackson:2019mop}. The aforementioned subtraction
procedure was applied to connect the results to the limit of small $\mn{K}^2$.
As the plot shows, the structure around the light-cone is quite rich, with visible effects from the inclusion
of the NLO small-$\mn{K}^2$ corrections of \cite{Ghiglieri:2014kma} ($\mathrm{LPM}^\mathrm{NLO}$ curves). 
At growing $M$, one sees instead a sharp rise which is largely dominated by vacuum physics (in vacuum 
the spectral function at time-like frequencies grows like $M^2$, as dictated by Lorentz invariance).
The rather low temperature in Fig.~\ref{fig_greg_mikko} was chosen so as to compare with lattice
data \cite{Ghiglieri:2016tvj}. This reference also presents a method to fit the low-frequency
part of the spectral function to the lattice data. A different type of spectral function,
better suited for the extraction of the real photon rate, was proposed in \cite{Brandt:2017vgl,Brandt:2019shg,Ce:2020tmx},
together with lattice data for QCD with two light flavors and a different recipe for the
fitting of the spectral function at the photon point. These first studies seem to suggest
that perturbative and non-perturbative determinations of the photon rate might differ at the $50\%$ level
at these low temperatures, which would be a rather interesting conclusion from the standpoint
of phenomenology, which traditionally uses the perturbative rates (see \cite{Paquet:2017wji}
for a recent overview).

Finally, we conclude our discussion of electromagnetic radiation by noting that, slightly outside the scope
of our review, methods to extend the equilibrium determinations to systems slightly off-equilibrium (of
relevance for the phenomenology of heavy ion collisions) have been introduced in 
\cite{Shen:2013cca,Hauksson:2017udm}. In the latter reference, the derivation of the collinear rate we 
presented in Sec.~\ref{sec_lpm} was generalized to an arbitrary density matrix for the plasma. We 
will return later to similar issues when discussing transport coefficients and thermalization.

\subsection{Transport coefficients}
\label{sub_res_transport}
In the previous discussion of electromagnetic radiation, we have mentioned the emergence of electric conductivity in the zero-frequency limit of the zero-momentum dilepton rate. This is 
an example of a \emph{transport coefficient}: it measures the relaxation of a conserved quantity
(in this case the electromagnetic charge) back to equilibrium following a perturbation. 
In other words, these coefficients describe the small-frequency behaviour 
of long-wavelength excitations of the medium, which are in turn related to
 conserved (or near-conserved) currents, 
such as the energy-momentum tensor, quark number currents and the electromagnetic current. The latter
currents are directly related, as only quarks carry electromagnetic charge in QCD.

Within this picture, transport coefficients are nothing but matching coefficients
of an effective description for these long-wavelength modes; they encode the physics of the UV modes
that have been integrated out to obtain this IR effective theory.
If we consider 
for instance the case of the energy-momentum tensor, the theory in question is
hydrodynamics.
In this theory, whenever there is a perturbation in flow velocity,
 the
stress-energy tensor (which defines the flux of momentum density) departs from its perfect-fluid form.
In the local (Landau--Lifshitz) fluid rest frame at a point $x$,
the stress tensor, to first order in the velocity gradient,
has the form
\begin{equation}
    \langle T_{ij}(x) \rangle
    =
    \delta_{ij} \, \langle {p} \rangle
    - \eta \sigma_{ij}
    - \zeta \, \delta_{ij} \, \nabla^l \, u_l \,,
	\quad \sigma_{ij}\equiv  \nabla_i \, u_j + \nabla_j \, u_i
		- {\textstyle \frac 23} \, \delta_{ij} \, \nabla^l \, u_l\,,
\label{eq:Tij}
\end{equation}
where 
 $p$ is the equilibrium pressure associated with
the energy density $\langle T_{00}(x) \rangle=e$,
and the coefficients $\eta$ and $\zeta$ are known as
the \emph{shear} and \emph{bulk viscosities}, respectively.
The flow velocity $\bm u$ equals the momentum density divided by the
enthalpy density $e+p=sT$ at vanishing chemical potentials. The two viscosities
are the transport coefficients appearing at the first order in
the gradients of $\bm u$; indeed, %the expansion parameter of
hydrodynamics is a gradient expansion in the flow velocity.
Hydrodynamics, and the shear viscosity in particular, play a very important
role in the phenomenology of heavy ion collisions, as they are central to the
description of the bulk properties of the produced particles. We refer
to \cite{Heinz:2013th,Gale:2013da,Jeon:2015dfa} for reviews on the subject.

For what concerns the conserved quark numbers of QCD,
the associated charge densities
$n_i \equiv j^0_i$ and current density ${\bm j}_i$
satisfy a diffusion equation (see e.g. the discussion in \cite{Arnold:2000dr})
\begin{equation}
    \langle {{\bm j}_{i}} \rangle
    = -D_{q_i q_a} \> {\bm \nabla} \langle n_a \rangle \,,
\label{eq:jia}
\end{equation}
in the local rest frame of the medium.
The coefficient  $D_{q_i q_a}$ is called the quark number diffusion constant. In general
it has a matrix structure in flavor, as shown by our notation, though in practice in the
case of QCD the leading-order and next-to-leading order light-quark diffusion matrix
takes the much simpler form $D_q \delta_{i a}$.\footnote{We thank Guy Moore for the observation
that deviations from this simple structure are to be observed starting at relative order $\alpha_s$.}

Quark number  is in turn tied to the electric conductivity, as quarks
are the charge carries: the diffusion matrix for these
charged species determine
the electric conductivity $\sigma$ through an Einstein relation
(see Refs.~\cite{Arnold:2000dr,Arnold:2003zc}).
We thus have
\begin{equation}
    \sigma = \sum_{ij} \, e^2 Q_i Q_j \, D_{q_i q_j} \, \frac {\partial n_i}{\partial \mu_j} \,,
\label{eq:einstein}
\end{equation}
where the sum runs over  flavors 
with $Q_i$ denoting the corresponding electric charges.

From the field-theoretical point of view, transport coefficients 
 are extracted, up to overall prefactors,
 from the small-frequency limit of the spectral function $\rho$ of the spatial part of the corresponding conserved
 current,
\begin{equation}
	\label{kubo}
	\lim_{\omega\to 0^+}\frac{\rho(\omega,0)}{\omega}.
\end{equation}
Note that we have set the spatial momentum to zero before the frequency here. Even though
partly just a convention---the same transport coefficients can for instance be obtained
for $k=\omega\to 0$ with a different numerical prefactor---we may think of this as 
a way of making sure that the system under consideration is ``large'' with respect to the
underlying microscopic dynamics. These relations linking transport coefficients to
the slope at the origin of appropriate spectral functions are termed \emph{Kubo formulae}
\cite{Kubo:1957mj,Hosoya:1983id}.

Determining transport coefficients perturbatively might  seem to require methods
not unlike those discussed in Sec.~\ref{sub_res_photon} for the determination of
the electromagnetic spectral function at non-zero frequencies and possibly momenta.
However, when taking the zero-frequency limit of Eq.~\eqref{kubo}, we experience
yet-another breakdown of the loop expansion. What is happening in this case is that,
as the frequency approaches zero, it becomes comparable with the rates of elementary
processes in QCD. As we have discussed in Sec.~\ref{subsub_heur_lpm}, the rate
of soft scatterings is $\Gamma_\mathrm{soft}\sim g^2T$. When the exchanged momentum is of order $T$
the rate is instead  $\Gamma_\mathrm{hard}\sim g^4T$, which is also the rate for the collinear
splitting processes derived in Sec.~\ref{sec_lpm}, $\Gamma_\mathrm{coll}\sim g^4T$. While soft scatterings are more frequent,
they are less effective in diffusing momentum over large angles, which is of critical importance
for transport coefficients. Similarly to what happens for $\hat{q}$, they receive an extra
$g^2$ suppression and thus contribute at the same order $g^4T$, with a logarithm of the coupling
emerging from the combined hard and soft scatterings, $\Gamma_\mathrm{2\leftrightarrow2}\sim g^4 T \ln(T/m_D)
\sim g^4 T \ln(1/g)$. 

These hard, soft and collinear processes would then need to be resummed, leading to the expectation that, at leading
order, transport coefficients should be inversely proportional to $g^4\ln(1/g)$.
As shown in \cite{Aarts:2002tn,Aarts:2004sd,Aarts:2005vc,Gagnon:2006hi,Gagnon:2007qt},
 setting out to resum all these processes in a diagrammatic way is not very practical
in a gauge theory. It turns out that the best way to determine the QCD transport coefficients
is through a \emph{linearized kinetic theory}. An effective kinetic theory arises by integrating
out the off-shell quantum fluctuations, so as to retain a Boltzmann equation that describes the
evolution of the \emph{single-particle distributions} of long-lived quasiparticles. In other
words, this requires the assumption that the duration of an individual collision is much shorter
than the mean free time between these. In our case, hard (soft) collisions have a duration of $1/T$ ($1/(gT)$), the inverse
of the exchanged momentum, and a mean free time of $1/\Gamma_\mathrm{hard}$ ($1/\Gamma_\mathrm{soft}$), 
so the criterion is satisfied in both cases, though more stringently so in terms of $g$ for soft scatterings.
Collinear processes last for $\tau_\mathrm{form}\sim 1/(g^2T)$ and occur every $1/\Gamma_\mathrm{coll}\sim 1/(g^4T)$,
so they also satisfy this criterion. In other words, weakly-coupled QCD has well-defined quasiparticles.

The leading order effective kinetic theory incorporating rigorously these processes in the 
collision operator was derived in \cite{Arnold:2002zm}. As the transport coefficients describe
the response of the medium to a perturbation, they are obtained by linearizing the kinetic theory, that is,
taking the first-order term away from equilibrium in the specific direction being considered: for
the electric conductivity, that would be a small local charge density gradient.
In a scalar theory \cite{Jeon:1994if} and in an abelian gauge theory
\cite{Aarts:2002tn,Aarts:2004sd,Aarts:2005vc,Gagnon:2006hi,Gagnon:2007qt} it is possible to prove
that a direct diagrammatic evaluation of $\rho(\omega)/\omega$ involves the  derivation of a resummation
scheme that turns out to be equivalent to what is realized by solving the linearized kinetic theory.

Within this framework, complete leading-order results for the shear viscosity, light-quark number diffusion,
and the related electric conductivity were obtained in \cite{Arnold:2003zc} at finite temperature and vanishing 
density. Per our previous discussion, the leading behaviour for these transport coefficients
is $1/(g^4\ln(1/g))$ times the appropriate power of $T$, e.g.~$T^3$ for $\eta$. However,
differently from the case of a scalar theory, where the logarithm is absent and one
needs only determine the number in front of $1/\lambda^2$, in the case of QCD
one has to determine the functional dependence of the transport coefficients on $g$. In 
principle, one could treat $\ln(1/g)$ as a large parameter and perform an expansion in inverse
logs, i.e.~a first ``leading-log'' term, followed by a ``next-to-leading log'' one. Truncating
here, one has e.g.~$\eta_\mathrm{NLL}=T^3 \eta_1/(g^4\ln(\mu_*/m_D))$ \cite{Arnold:2003zc},
with $\eta_1$ the leading-log coefficient, determined in \cite{Arnold:2000dr}, and 
$\mu_*$ the next-to-leading-log one, determined in \cite{Arnold:2003zc}. A comparison
of this approximate form with the full functional dependence on $g$, determined numerically
in \cite{Arnold:2003zc} as well, shows that the NLL approximation diverges from the numerical 
leading order for $m_D/T\gtrsim 1$.

As shown in \cite{Arnold:2003zc}, there is a significant ambiguity in the definition
of leading-order transport coefficients. For instance, one might resum Hard Thermal Loops 
in the $2\leftrightarrow 2$ processes for all values of the exchanged momentum,
or for soft exchanges only. The difference between these prescriptions is parametrically
of higher order, but can be numerically sizeable as soon as the coupling becomes
of order one, $m_D/T\sim 1$, as shown in \cite{Arnold:2003zc}. One would thus conservatively
assign large theory uncertainties to these perturbative estimates in the region 
of phenomenologically interesting couplings.

Next-to-leading order determinations of these transport coefficients are thus necessary
to better ascertain these uncertainties. The ``light-cone'' theoretical developments we mentioned previously,
that is, the mapping to the 3D Euclidean theory for some soft amplitudes \cite{CaronHuot:2008ni}
and the sum-rule mapping to arcs at large $|q^+|$ discussed around Eq.~\eqref{finalql}
have made this possible. Similarly to what happened in the photon production case,
the former mapping can be used to determine gluon-mediated transverse momentum exchange processes at NLO, as well as
the inputs to the PUI Hamiltonian and the semi-collinear splitting rate, which also enters here.
The latter mapping is used to handle soft-fermion exchange processes and gluon-mediated longitudinal
 momentum exchange processes at NLO. With these advancements, a NLO collision operator for the effective
 kinetic theory discussed previously was derived in \cite{Ghiglieri:2015zma,Ghiglieri:2015ala}
under the approximation that at least one of the partons entering the collision has an energy
much larger than the temperature, as is the case when studying the evolution and energy loss
of the leading partons in a jet traversing the QCD medium. We refer to \cite{Ghiglieri:2015zma} in particular
for a more pedagogic review to the application of the aforementioned advancements in the derivation
of the NLO collision operator.

Based on this work, an ``almost NLO'' determination of the shear viscosity and light-quark diffusion
was completed in \cite{Ghiglieri:2018dib}. The word ``almost'' was used because
 light-cone methods typically keep track of the incoming and 
outgoing momentum of a particle in a collision, but lose track of the momentum which it transfers to the
other participants.  This momentum transfer also affects the departure from equilibrium of
the other particle or particles which receive the momentum; an effect which was not accounted for in these
 NLO determinations, hence the ``almost'' NLO. 
The importance of this effect was however computed in the leading-order case and used 
 to make an estimate for this incomplete treatment.  
 The associated errors turned out to be small, much smaller than the 
 difference between LO and NLO.
\begin{figure}[t!]
	\begin{center}
	\includegraphics[width=6cm]{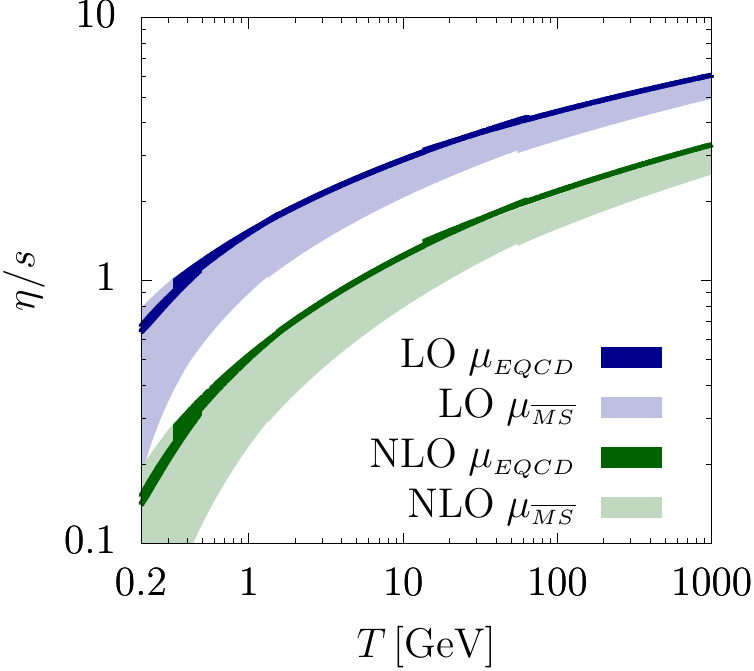}
	\includegraphics[width=6cm]{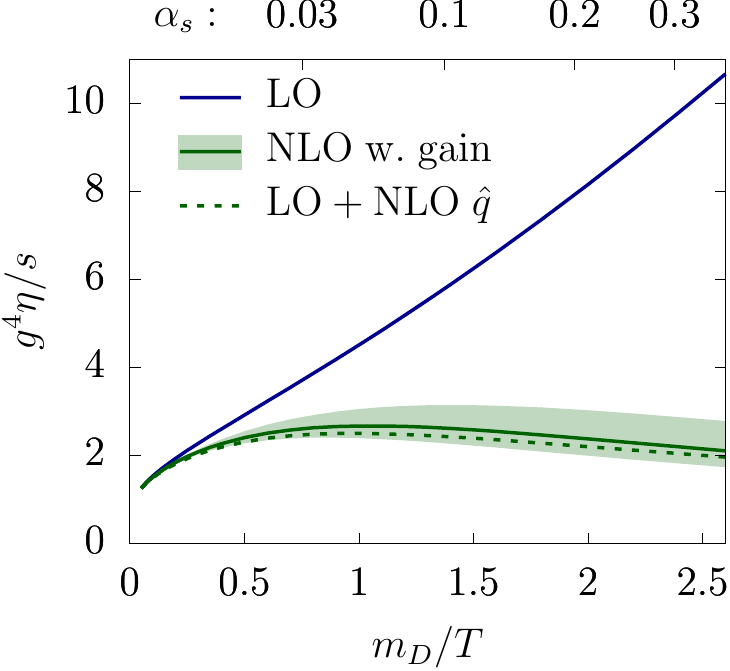}
	\end{center}
	\caption{In both plots, we display the shear viscosity
	over the (Stephan--Boltzmann) entropy density at leading- \cite{Arnold:2003zc} and
	next-to-leading \cite{Ghiglieri:2018dib} order.
	On the left, the horizontal axis is the temperature and the bands come from different
	definitions of the coupling, as explained in the main text. On the right
	we plot the result as a function of coupling, $m_D/T\sim g$ below and $\als$ above. The green band
	corresponds to an estimation of the effect of the terms that were not computed in \cite{Ghiglieri:2018dib},
	while the dashed green line is obtained by adding to the LO collision operator the contribution
	from NLO transverse momentum broadening only. Figures taken from \cite{Ghiglieri:2018dib}.}
	\label{fig_shear}
\end{figure}
In Fig.~\ref{fig_shear}, we display the results of \cite{Arnold:2003zc} and \cite{Ghiglieri:2018dib} for the shear viscosity.
These LO and NLO determinations are still insensitive to genuine vacuum UV divergences and the associated
charge renormalization; in other words, these calculations determine $\eta(g)$. To plot $\eta(T)$ one
needs to fix $g(T)$, with no guidance from the calculation 
on how to perform scale setting. The procedure in \cite{Ghiglieri:2018dib} was to take either a standard $\overline{\mathrm{MS}}$
prescription with the renormalization scale set to multiples of the Matsubara frequency, giving the large bands shown 
in the figure, or to choose instead the effective coupling of EQCD, as computed in \cite{Laine:2005ai} 
and discussed in more detail later on in Sec.~\ref{sec_DR}. This latter coupling has no leading-logarithmic
dependence on the temperature. 

As the plot on the left shows, the ratio between the NLO and LO results varies from $1/2$ at very large temperatures
down to $1/5$ at the QCD transition, where the uncertainty from the coupling becomes large. In this
region, $\eta/s$ is of a size compatible with strong-coupling determinations in holographic theories, to be discussed
later in Sec.~\ref{subsub_beyond_qcd}. The plot on the right shows how the LO and NLO results
start to differ significantly at $m_D/T\gtrsim 0.5$. Also shown is the small uncertainty band
from the estimate of the missing terms and a curve obtained  by adding only the contribution
from transverse momentum exchange, encoded in NLO $\hat{q}$ \cite{CaronHuot:2008ni}, to the LO 
collision operator, showing how it is the dominant NLO contribution. We will return to this
later in this subsection. We refer to \cite{Ghiglieri:2018dib} for the results on light flavor
diffusion, which show a similar pattern to those of $\eta$.

For what concerns the bulk viscosity, its parametric size is much smaller, as the quantity vanishes in a conformal theory: $\zeta$ describes the response to a uniform
compression or rarefaction, which is equivalent to a dilatation. As a conformal theory is invariant
under dilatations, it will not depart from equilibrium following such perturbations. Furthermore,
it can be shown that $\zeta$ depends quadratically on the departure from conformality \cite{Arnold:2006fz}.
If we consider QCD at temperatures where the light quarks can be considered massless and the heavy quarks are absent
from thermal equilibrium, which is indeed the case at the temperatures probed by current heavy-ion collision experiments,
the main contribution to the trace anomaly is thus the $\beta$ function of QCD, $\beta\sim \als^2$. This should
multiply the previous estimates for transport coefficients, so that $\zeta\sim \beta^2/g^4\ln(1/g)\sim\als^2/\ln(1/g)$,
as shown in \cite{Arnold:2006fz}, which then went on to determine its leading-order value by extending
the effective kinetic theory techniques to this case. The contribution from the explicit conformal symmetry breaking
caused by the charm quark mass was computed in \cite{Laine:2014hba}.

We have so far discussed results at vanishing density here. The region where $\mu$ is much larger than $T$
and $m_D$---we anticipate
from Sec.~\ref{sec:imagtimeform} that at finite density and small $T$ $m_D^2\sim g^2\mu^2$---has been explored instead 
in the pioneering study of Heiselberg and Pethick \cite{Heiselberg:1993cr}. In this case there are no collinear splitting
processes to be considered, since the soft scattering rate is not IR enhanced at finite density. 
They found $\eta\sim n \mu\, m_D^{2/3}/(\als^2 T^{5/3})$ for $\mu\gg m_D \gg T$ and $\eta\sim n \mu/(\als^2 T\ln(T/m_D))$ for
$\mu\gg T\gg m_D$.

As we remarked earlier, the large NLO corrections to the shear viscosity
and quark number diffusion coefficient arise mainly from the $\mathcal{O}(g)$ corrections to $\hat{q}$, which we
feel deserve more discussion. They were derived in \cite{CaronHuot:2008ni}, in what was the first application of the mapping to the three-dimensional Euclidean theory we have previously
discussed---a mapping that was presented in that same paper. Through that mapping,
the NLO contribution to $d\Gamma/d^2\q_\perp$ was determined analytically in closed form in a relatively
straightforward computation, to be contrasted with the intricate brute-force HTL-resummed calculation
that would be necessary in the absence of such a mapping (an explicit example of such a computation
will be presented in the coming subsection). 

\begin{figure}[t!]
	\begin{center}
		\includegraphics[width=6cm]{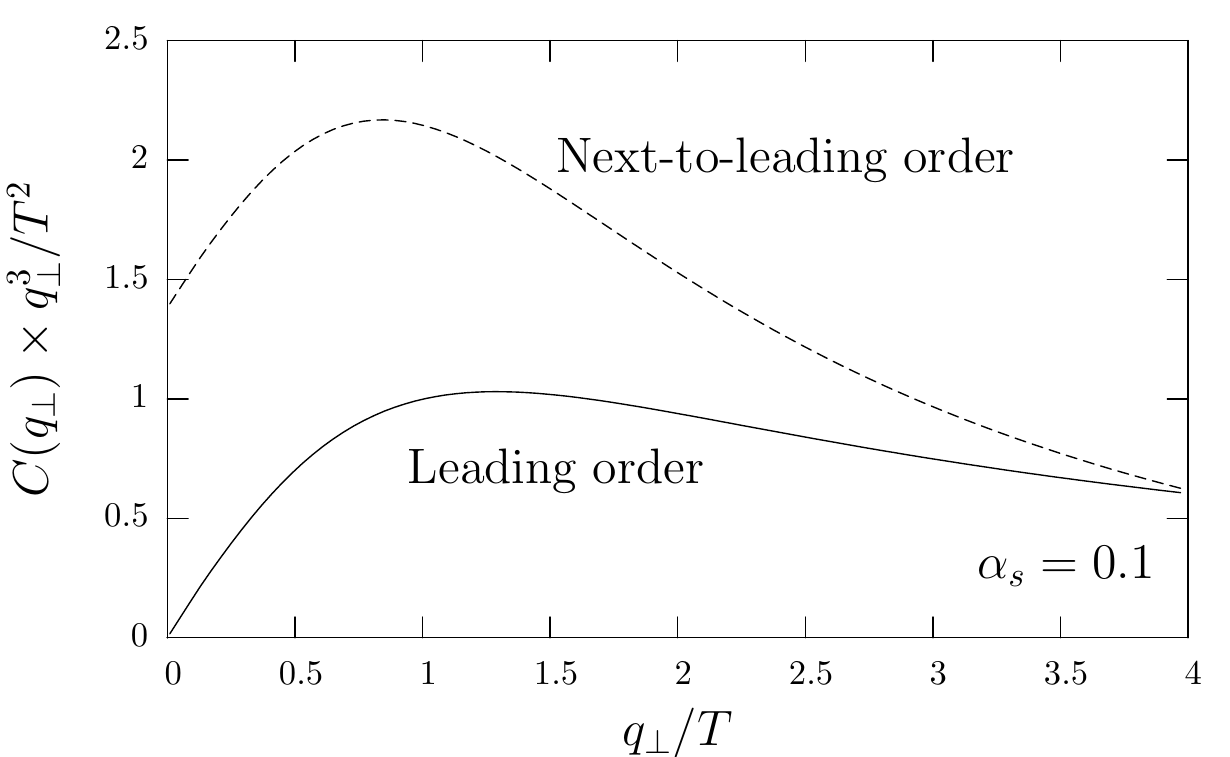}
		\includegraphics[width=6cm]{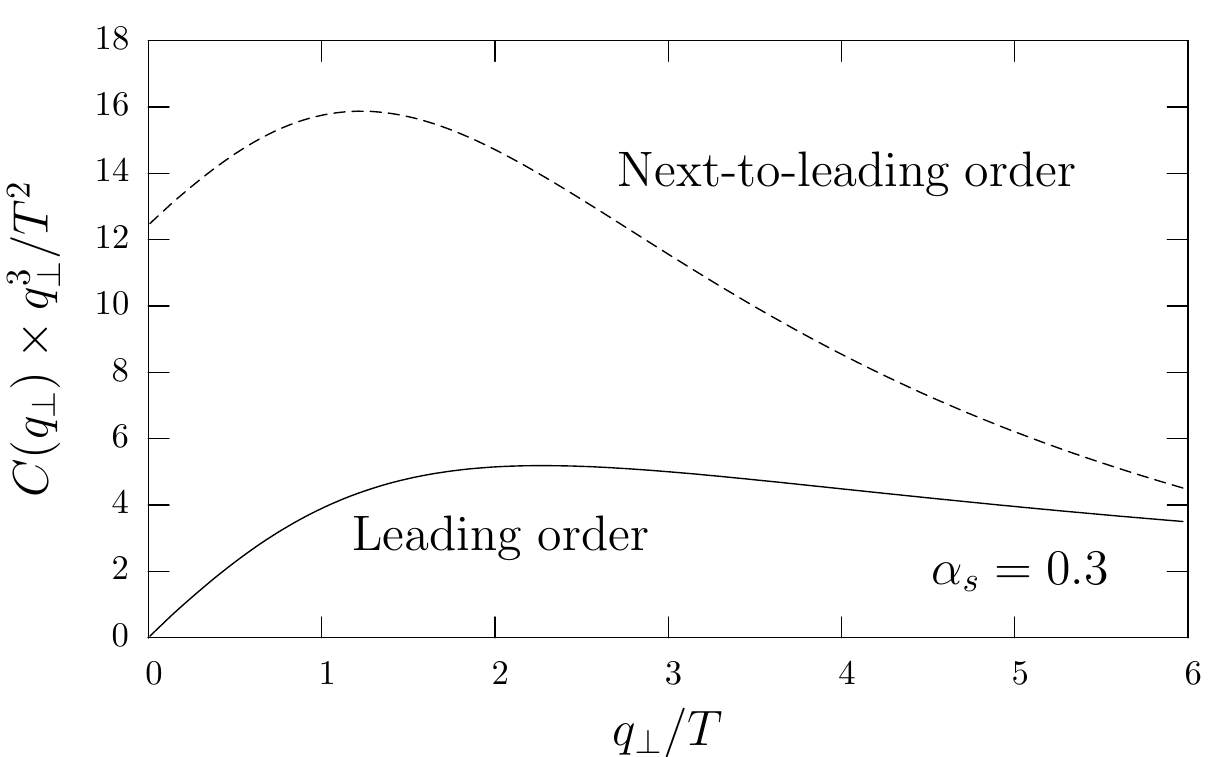}
	\end{center}
	\caption{$d\Gamma/d^2\q_\perp$ (called $\mathcal{C}(q_\perp)$ here and elsewhere in the
	literature) at NLO. The vertical axis is normalized in such a way that the integral of the curves
	shown here is directly proportional to (the soft contribution to) $\hat{q}$. We refer to 
	\cite{Ghiglieri:2018ltw} for a subtlety on the overall normalization of the NLO curves.
	Figure taken from \cite{CaronHuot:2008ni}.}
	\label{fig_qhat}
\end{figure}
In Fig.~\ref{fig_qhat} we show the results of \cite{CaronHuot:2008ni}: one clearly sees
how the NLO curves overtake the LO ones already at $\als=0.1$. It is then not surprising
that $\hat{q}$, directly proportional to the area under these curves, receives a large
$\mathcal{O}(g)$ correction, which in turn drives the corrections to transport coefficients.
We note that the photon production rate is not directly sensitive to $\hat{q}$, which might
explain why in that case the perturbative series seems to converge much better.

Finally, we note that the Euclidean mapping makes the lattice determination of the soft
(and ultrasoft) contributions to $\hat{q}$ possible by simulating the Wilson loop mentioned in 
Sec.~\ref{sec_lpm} within lattice EQCD, without encountering the issues related with analytical 
continuation we have discussed. EQCD determinations of $d\Gamma/d^2\q_\perp$ (in Fourier
space) have 
been presented in \cite{Panero:2013pla} and more recently in \cite{Moore:2019lgw}, with better
control on the UV specifics. See also \cite{Laine:2013lia} for a determination
in classical lattice gauge theory and \cite{Ghiglieri:2015zma} for a review touching
these aspects. These lattice calculations open up the possibility of using
perturbative methods for the modes at the scale $T$ and non-perturbative ones for the softer modes,
though more work is required both on the lattice side and on the matching and factorization
sides.

\subsection{Real-time thermal QCD for heavy flavors and quarkonia}
\label{sub_res_heavy}
Heavy quarks and their bound states have been a key \emph{hard probe} of the hot QCD medium
since the inception of the ultrarelativistic heavy-ion collision program, starting
from the seminal paper of Matsui and Satz \cite{Matsui:1986dk}. This is a vast
topic; to even try to summarise it here would be outside the scope of this report and would
represent a disservice to both the field and the existing reviews, to which we refer the 
interested reader for more details. 
\cite{Rothkopf:2019ipj,Mocsy:2013syh} focus on heavy quarkonia, while for heavy quarks or
a comprehensive perspective on both we refer
to \cite{Andronic:2015wma,Aarts:2016hap}.
That being said, we think there are some aspects that link directly to the methods described in 
Sec.~\ref{sec:realtimept} and the physics reviewed in Sec.~\ref{sec_soft_collinear} that merit
further discussion here.

We start by discussing an observable that is of relevance both for the energy loss of heavy quarks
in the QCD medium \cite{Svetitsky:1987gq,Braaten:1991we,Moore:2004tg} and for the fate of
quarkonia in a non-relativistic Effective Field Theory description \cite{Brambilla:2016wgg,Brambilla:2017zei}.
It is the heavy quark momentum diffusion coefficient $\kappa$, which can be thought of as the non-relativistic
counterpart of $\hat{q}$, i.e.~$\kappa=\langle k^2\rangle/t$, the squared momentum picked up by a non-relativistic heavy quark per
unit time. It was defined in a field-theoretical way in \cite{CasalderreySolana:2006rq} as the insertion
of two chromoelectric fields on a temporal Wilson line, similarly to $\hat{q}$, whose field-theoretical
definition is essentially a boosted version of that for $\kappa$ (see \cite{CaronHuot:2009uh} for the connection to
Euclidean space for $\kappa$). At leading order, $\mathcal{O}(g^4T^3)$, $\kappa$ receives contributions from both hard and soft modes,
similarly to $\hat{q}$. Hence, $\mathcal{O}(g^5T^3)$ corrections from the soft scale are to be expected.
But differently from $\hat{q}$, the $A^0$ fields on the Wilson line and the $E$ fields sit at the same spatial point
and at different times: they are thus time-like separated and there is no hope of using the enormous simplification
introduced by the sum rule leading to Eq.~\eqref{finalqhat}, i.e.~the mapping to the 3D Euclidean theory. The computation
of these $\mathcal{O}(g)$ corrections, as presented in \cite{CaronHuot:2007gq,CaronHuot:2008uh}, 
is thus a daunting brute-force calculation in the Hard Thermal Loop theory.
It required firstly the development of the effective rules described in Sec.~\ref{sub_heur_htl},
secondly their application and the generation of all diagrams, assignments and amplitudes, and finally
the numerical evaluation of 4-dimensional loop integrations over these HTL-resummed amplitudes. 

\begin{figure}[t!]
	\begin{center}
		\includegraphics[width=8cm]{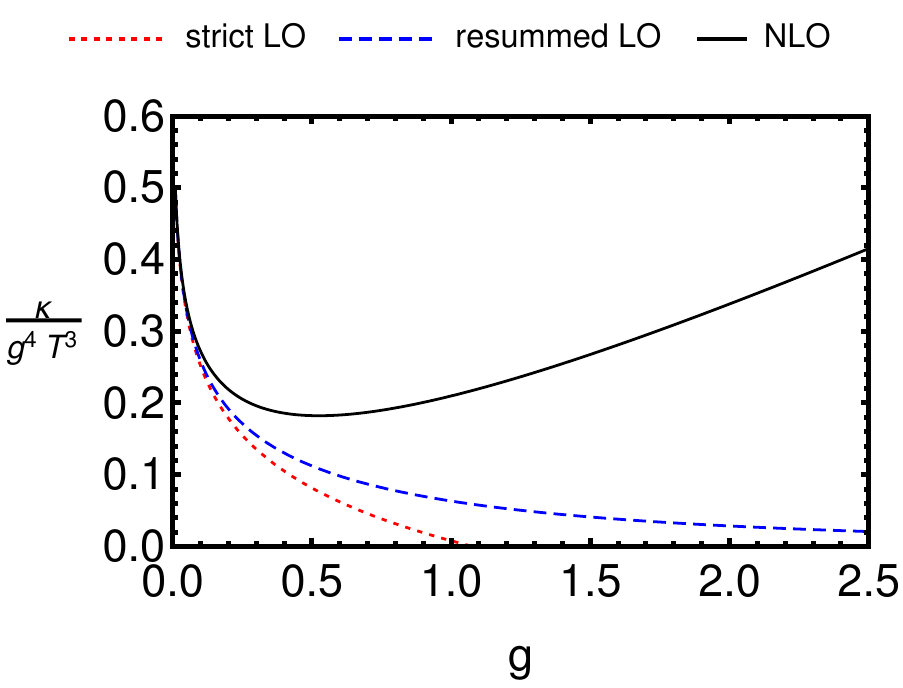}
	\end{center}
	\caption{The heavy quark diffusion coefficient at NLO, as computed in \cite{CaronHuot:2007gq,CaronHuot:2008uh}.
	The ``strict'' and ``resummed'' LO curves differ in that the second resums the Debye mass in the
	propagator for all exchanged momenta. It thus resums a subset of higher order corrections.}
	\label{fig_nlokappa}
\end{figure}
The results of this impressive computational \emph{tour de force} are shown in Fig.~\ref{fig_nlokappa}. The two
different LO definitions differ in how the matching between the soft and hard sectors is performed. Irrespective
of this aspect, whose details are to be found in the original works, the figure shows how
the NLO corrections rapidly ($g\gtrsim 0.5$) overtake the LO results, thus showing again a pattern of bad convergence
similar to what we discussed before in the cases of $\hat{q}$ and transport coefficients. Understanding
precisely the physics responsible for these large corrections in these observables and finding suitable ways
of re-arranging the perturbative expansion remains an important open issue, to 
which we will come back in Sec.~\ref{sec:fut}. Finally, we remark that the Euclidean definition in \cite{CaronHuot:2009uh}
does not allow direct lattice determinations; analytical continuations of the Euclidean results, of the kind discussed in
Sec.~\ref{sub_res_photon}, are necessary, albeit possibly easier due to the
lack of a narrow transport peak \cite{CaronHuot:2009uh}. Results obtained in \cite{Meyer:2010tt,Banerjee:2011ra,Francis:2011gc,Francis:2015daa}---see
also \cite{Brambilla:2019tpt} for an extraction from reconstructed quarkonium spectral functions---show 
a $\kappa$ that is larger than the NLO perturbative results; recent results \cite{Brambilla:2019oaa} point towards
a better agreement at very high temperatures.

For what concerns heavy quark bound states, we wish to discuss an issue where the application of real-time
perturbation theory shows its advantages in comparison with the Euclidean approach: the determination of the
\emph{potential} governing the evolution of the bound state. At $T=0$, this potential can be defined rigorously
in a non-relativistic EFT framework, where one integrates out first the heavy quark mass $m$, obtaining
non-relativistic QCD (NRQCD) \cite{Caswell:1985ui,Bodwin:1994jh}, and then the momentum transfer scale $mv$---with
$v$ the relative velocity---obtaining potential non-relativistic QCD (pNRQCD) \cite{Pineda:1997bj,Brambilla:1999xf}.
In this latter theory, a Schr\"odinger picture appears naturally at the zeroth order of a multipole expansion---pNRQCD
in the weak-coupling limit, $mv\gg \Lambda_\mathrm{QCD}$, is organised as a double expansion in $1/m$ (inherited
from NRQCD) and $r$, the relative coordinate. Within this picture, the potential is just a matching coefficient of the theory.
We refer to
\cite{Brambilla:2004jw} for a review of this approach. 

\begin{figure}[t!]
	\begin{center}
		\includegraphics[width=8cm]{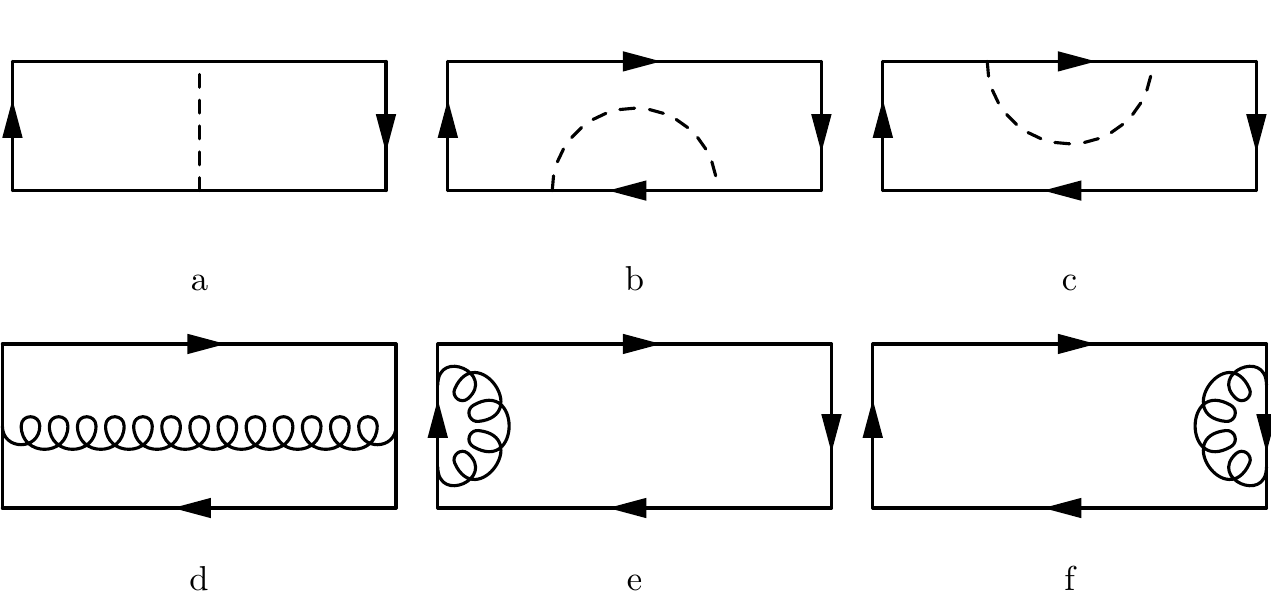}
	\end{center}
	\caption{Leading-order diagrams for the rectangular Wilson loop in a gauge where $G^{0i}=0$. The 
	dashed lines are HTL-resummed temporal gluons and the curly lines spatial gluons. Hence, the horizontal side
	of the Wilson loop is the temporal one, of extent $t$, and the vertical one is the spatial
	one, of extent $r$.}
	\label{fig_wilson}
\end{figure}

Taking this approach to finite temperature requires extra assumptions on the hierarchy between the non-relativistic
scales $m$, $mv$ and $mv^2$ and the thermal scales. Let us look at what happens in the screening regime, i.e.~when
the typical separation is of the order of the electric screening length of the medium, $mv\sim m_D$. Determining
the potential requires the evaluation of a rectangular Wilson loop of time extent $t$ and separation $r\sim 1/m_D$.
In order to single out the potential dynamics at the scale $mv^2$, a $t\to\infty$ limit has to be taken. 
This was first done in \cite{Laine:2006ns}, in the Euclidean formalism. There, one studies a Wilson loop
of Euclidean time extent $\tau<1/T$, evaluating the diagrams in Fig.~\ref{fig_wilson}. As $r\sim 1/m_D$,
HTL resummation is necessary.
Only after the gauge-invariant leading-order amplitude has been evaluated, one can analytically continue
$\tau\to i t$ and take the large time limit. The resulting potential, as found in \cite{Laine:2006ns},
is \emph{complex}, reading
\begin{equation}
	\label{complexpot}
	V(r\sim m_D^{-1})=-C_F\als\left[ \frac{e^{-m_D r}}{ r}+m_D+2i T
	\int_0^\infty \frac{dz\,z}{(z^2+1)^2}\left(1-\frac{\sin( m_D r z)}{m_D r z}\right)\right].
\end{equation}

Before we discuss the significance of the above result, let us show how it is obtained using real-time techniques, as in \cite{Beraudo:2007ky,Brambilla:2008cx}. The major advantage
of this method is that the analytical continuation is performed already at the level 
of the time evolution operator through the introduction of the Schwinger--Keldysh 
contour, as we remarked in Sec.~\ref{sec:IV}. Hence, the infinite-time limit
can be taken early on, making the diagrams on the second line of Fig.~\ref{fig_wilson}
irrelevant in any non-singular gauge. In these gauges, which include the Feynman and 
Coulomb ones, the potential is easily obtained as the Fourier-transform
of the time-ordered temporal propagator at zero frequency, i.e.
\begin{equation}
	\label{complexpot2}
	V(r\sim m_D^{-1})=i g^2 C_F\int\frac{d^3k}{(2\pi)^3}\left(e^{i\k\cdot {\bm r}
	}-1\right)G_{00}^{11}(0,k).
\end{equation}
The time-ordered HTL propagator $G_{00}^{11}(0,k)=G^F_{00}(0,k)$ can be obtained from the retarded HTL propagator,
Eq.~\eqref{htllong}, and from the equation relating it to the time-ordered propagator, Eq.~\eqref{tordom},
i.e.
\begin{equation}
	\label{zerofreqtord}
	G_{00}^{11}(0,k)=\frac{i}{k^2+m_D^2}+\frac{\pi T m_D^2}{k(k^2+m_D^2)^2},
\end{equation}
where the first term comes from the average of the retarded and advanced propagators in Eq.~\eqref{tordom}
and the second from the statistical correlator in Eq.~\eqref{tordom} in the zero-frequency limit. It is 
then easy to see how Eq.~\eqref{complexpot} is obtained. 

This derivation is not only more straightforward, due to the early analytical continuation,
but it also makes the physical picture clearer: the real part of the potential, arising from the
average of the retarded and advanced propagators,  describes screening. The imaginary part on the other hand 
arises from the $rr$ component of $G^{11}$---recall that $G^{11}(\omega)=(G_R(\omega)+G_A(\omega))/2+G_{rr}(\omega)$---and 
is thus Bose-enhanced by $T/m_D$ with respect to the real part. It 
describes  the effect of collisional Landau damping, encoded in the HTL spectral function at space-like momenta; physically,
it describes collisions between the heavy quarks and medium constituents. The relation between these imaginary parts 
and previous approaches based on collisional cross sections integrated over thermal distributions for the incoming
scatterers, e.g.~\cite{Xu:1995eb,Grandchamp:2001pf}, was studied in detail in \cite{Brambilla:2011sg,Brambilla:2013dpa}. We refer to
\cite{Rothkopf:2019ipj} for a review on the implications of the complex potential and for the intricacies
of its non-perturbative determination, due again to the need of analytic continuations of Euclidean lattice data.

Finally, we reiterate that the potential in Eq.~\eqref{complexpot} is valid for $r m_D\sim 1$. The EFT framework
 can be used to derive it systematically in other regimes, such as $r T\ll 1$, which is of relevance for the ground
 states of bottomonium in current heavy-ion collision experiments. In this regime, the potential, as well as the spectrum
 and width for the $\Upsilon(1S)$, were derived in \cite{Brambilla:2010vq}.

\subsection{Applications beyond QCD}
\label{subsub_beyond_qcd}
The methods presented in Sections~\ref{sec:realtimept} and \ref{sec_soft_collinear} have wide applicability, going beyond
the wealth of results in hot and dense QCD we have just reviewed. We dedicate this subsection to a brief overview of 
select results outside of the realm of hot QCD. We start by considering an area in close contact with the latter: 
$\mathcal{N}=4$ supersymmetric Yang--Mills (SYM) theory. Due to the celebrated AdS/CFT correspondence
\cite{Maldacena:1997re,Witten:1998qj,Gubser:1998bc}, the theory's strong coupling regime ($\lambda\equiv g^2 N_c\to\infty$, $N_c\to
\infty$, with $\lambda$ the 't Hooft coupling)
is accessible through computations in 5-dimensional gravity on an AdS background. From this foothold in the 
strong-coupling regime of a non-abelian theory, one can learn useful  lessons about the strong coupling regime of
hot QCD, either through qualitative comparisons with the SYM theory or by studying the gravity duals of theories closer to QCD. For further details, the interested reader is directed to the comprehensive review of Ref.~\cite{CasalderreySolana:2011us}.  

To facilitate the extrapolation of these lessons towards $\mathcal{N}=0$, $N_c=3$ QCD with fundamental Dirac fermions
and with a finite, neither-too-small-nor-too-large coupling,
it is clearly very interesting to investigate the weak-coupling regime of $\mathcal{N}=4$ SYM, so as to have a handle
of how the transition from strong to weak coupling takes place \emph{within the same theory}, and 
to understand the dependence of the results on the type and number of degrees of freedom in the weak-coupling regimes of QCD and SYM, so
as to guide extrapolations at stronger couplings.

To these ends, the methods we reviewed have been applied to the determination of the thermal photon and dilepton rate 
in $\mathcal{N}=4$
SYM in \cite{CaronHuot:2006te}. As this theory does not contain photons, a $U(1)$ subgroup of the $R$ current was gauged,
giving ``electromagnetic'' charge to two of the six adjoint, real scalars and two of the 4 adjoint Weyl fermions of the theory.
The paper presented both weak and strong coupling results, finding the evolution between 
the two regimes to be a rather smooth function of $\lambda$. The dependence on
the details of the theory was also analyzed in detail, finding that 
if the normalization of the $U(1)$ charge is set so that the QCD and SYM dilepton
rates agree in the free limit at large $M$, then the SYM photon rate is much larger than the QCD one at equal $\lambda$,
due to the much larger scattering rates of SYM, coming from the the larger number of matter fields. If instead the theories are compared
at equal $m_D$ or $m_\infty$, the rates become comparable.

\begin{figure}[t!]
	\begin{center}
		\includegraphics[width=8cm]{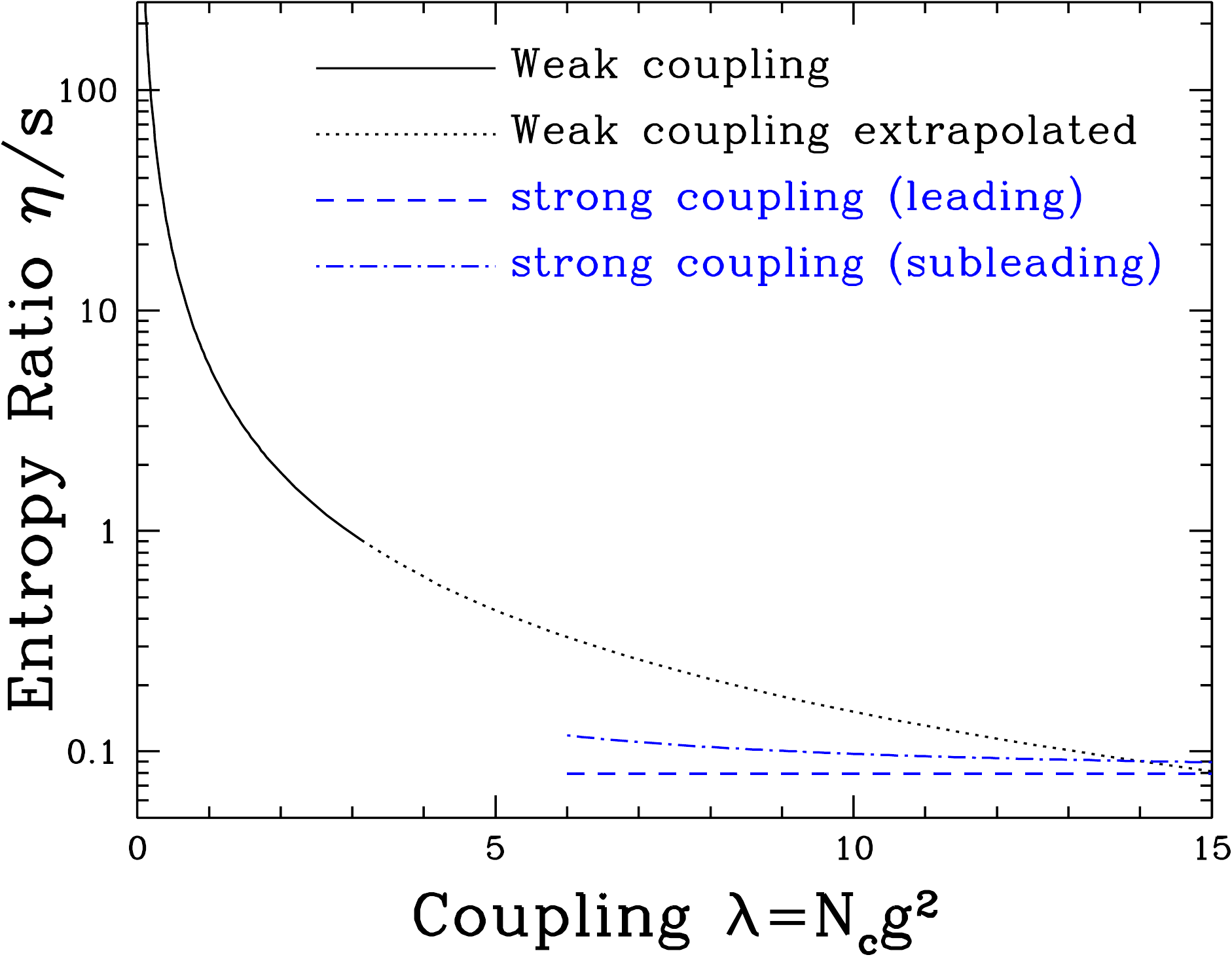}
	\end{center}
	\caption{The $\eta/s$ ratio in $\mathcal{N}=4$ SYM at weak \cite{Huot:2006ys} and strong coupling. In 
	the latter case the leading result is from \cite{Policastro:2001yc,Kovtun:2004de}, while the
	``subleading'' one includes the part of the $\mathcal{O}(1/\lambda^{3/2})$ corrections available
	in 2006. The full $\mathcal{O}(1/\lambda^{3/2})$ corrections can be found in \cite{Buchel:2008ac,Buchel:2008wy}. 
	Figure taken from \cite{Huot:2006ys}.}
	\label{fig_shearsym}
\end{figure}

A similar analysis was performed for the transport coefficients. Ref.~\cite{Huot:2006ys} studied the shear viscosity 
at leading order in $\mathcal{N}=4$ SYM, finding it to be much smaller than in QCD at equal $\lambda$. Again,
this is to be understood as coming from the larger scattering rates in the supersymmetric theory. Comparing at
equal Debye masses, and accounting for the different Casimir factors, the two theories are again in good agreement. For
what concerns the extrapolation between weak and strong coupling, we show in Fig.~\ref{fig_shearsym}
the findings of \cite{Huot:2006ys}, which show how the naive extrapolation of the weak-coupling result
approaches the strong coupling limit already at $\lambda\approx 10$ ($\als\approx 0.3$), where however the
leading-order perturbative curve is out of its region of validity, as denoted graphically by the dotting.
As discussed in Sec.~\ref{sub_res_transport}, NLO results for $\eta/s$ in QCD show large deviations from the LO curve. It 
would be nice to have NLO corrections on the left-hand side of Fig.~\ref{fig_shearsym}, so as to better gauge
the uncertainties in the extrapolation to intermediate couplings. A first step in this direction has been completed in 
\cite{Ghiglieri:2018ltw}, where the soft scattering rate $d\Gamma/d^2\q_\perp$, $\hat{q}$ and the collinear radiation rate
have been determined to NLO in $\mathcal{N}=4$ SYM, showing how the contribution of scalars, absent in QCD, can have large
effects on the collinear rate. Finally, the LO heavy quark diffusion coefficient has been determined in $\mathcal{N}=4$ SYM in \cite{Chesler:2006gr}. 

Another area where the methods we have reviewed have found wide applicability is cosmology. We have already 
hinted about their relevance for the rate of right-handed neutrino production in the early universe. We refer 
to \cite{Biondini:2017rpb,Drewes:2017zyw}  for recent reviews which put the calculations in the physical context of 
this extension of the Standard Model. In general, a useful \emph{Rosetta stone} between
thermal photon production and right-handed neutrino production is the following:
the photon corresponds to the right-handed neutrino, as they are both singlets under the gauge
groups of the plasma. The quarks correspond to the Higgs doublet and left-handed leptons, which
couple to the right-handed neutrino via a Yukawa coupling. Finally, gluons correspond to the
electroweak gauge bosons, which interact with the active leptons and scalars.
 We then wish to highlight some significant applications
of the techniques: as we mentioned, the leading-order collinear production rate was determined in the symmetric phase
in \cite{Anisimov:2010gy,Besak:2012qm}, with the latter paper also obtaining the fermionic sum rule
derived in Eq.~\eqref{softphotonsumrulefinal}, which also enters the production rate
for ultrarelativistic right-handed neutrinos. The collinear rate in the broken phase was 
derived in \cite{Ghiglieri:2016xye}---which also derived the vector boson HTLs of the SM in that phase---at 
zero chemical potential; chemical potentials were considered in both phases in \cite{Ghiglieri:2017gjz,Ghiglieri:2018wbs}. These
derivations relied heavily on the mapping to the Euclidean theory, which was also exploited recently in
\cite{Jackson:2019tnr} to compute a part the active neutrino soft scattering rate $d\Gamma/d^2\q_\perp$ at
NLO in the broken phase. The mapping to the dispersion relation on the arcs introduced around Eq.~\eqref{finalql}
was used to determine another part (a projection on a different Dirac structure) 
to leading order in \cite{Ghiglieri:2018wbs}. For what concerns the dependence on the sterile neutrino mass,
which is the analogue of the dilepton mass in QCD, the procedure of \cite{Ghisoiu:2014mha} to merge the
small and large $M$ results was extended to this model in \cite{Ghisoiu:2014ena}, merging the low-mass
ultrarelativistic results of \cite{Anisimov:2010gy,Besak:2012qm} with the relativistic $M\sim T$ ones
of \cite{Laine:2013lka}, which in turn smoothly extended into the non-relativistic regime $M\gg T$, first studied
in \cite{Salvio:2011sf,Laine:2011pq,Biondini:2013xua}.

The above references are meant to convey the effective two-way
exchange of methods taking place between early-universe cosmology and hot QCD. Other examples of relevant results include
thermal production rates of axions \cite{Graf:2010tv,Graf:2012hb,Salvio:2013iaa}, gravitinos 
\cite{Pradler:2006qh,Pradler:2006hh,Rychkov:2007uq} and gravitational waves \cite{Ghiglieri:2015nfa}.
In all these cases, the collinear, LPM-resummed contribution is absent, as the coupling between
the ``photon'' and the ``quarks'' contains extra derivatives which suppress collinear emission \cite{Salvio:2013iaa}.

%% file: imagtimeformalism.tex
% !TEX root = review.tex

\section{Imaginary-time formalism}
\label{sec:imagtimeform}

We now turn back to the discussion of Sec.~\ref{sec:IV} and the realization after Eq.~(\ref{eq:pif}) that in thermal equilibrium, corresponding to the density matrix 
\begin{equation}
\hat{\rho}_{\rm eq} = \frac{1}{Z}e^{-\beta (\hat{H}-\mu_i \hat{N}_i )},
\end{equation} 
only the Euclidean part of the temporal integrals appearing in different expectation values survives, when no operators with unequal real time arguments are inserted. This leads to important simplifications in the determination of physical quantities, which are efficiently captured in the so-called imaginary-time formalism. The development of this formalism is the topic of this section of our article.

As will become clear in the following, the basics of the imaginary time formalism are considerably more straightforward to absorb than those of its real time counterpart, while the more challenging parts are related to how IR divergences are handled and the convergence of weak coupling expansions improved through resummations or effective theory setups. This fact is reflected in the structure of Secs.~\ref{sec:imagtimeform} and \ref{sec:imagtimeresults}, where particular attention is devoted to the development of effective descriptions for high-temperature and -density QCD matter as well as applications of these techniques to different bulk thermodynamical observables.

\subsection{Introduction}

In the imaginary-time formalism, all Green's functions, and hence the fields themselves, become either periodic or anti-periodic functions of the imaginary time direction as discussed already in Sec.~\ref{sec:IV}.  Looking at one time argument in a generic $n$-point Green's function ${\cal G}$ for a bosonic field, we obtain the relation
\begin{equation}
{\cal G}_{\rm bosonic}(t_i) = {\cal G}_{\rm bosonic}(t_i-i\beta) \hspace{1cm} (\rm bosons) \; ,
\end{equation}
where, for simplicity, we have suppressed all other arguments of ${\cal G}$.  Such a relation can be shown to hold for all field time arguments involved in the definition of the Green's function.  In the case of a fermionic field, due to the anti-commuting nature of the fields one finds instead
\begin{equation}
{\cal G}_{\rm fermionic}(t_i) = -{\cal G}_{\rm fermionic}(t_i - i\beta) \hspace{1cm} (\rm fermions) \; ,
\end{equation}
which implies that a general Green's function for a fermionic field is an anti-periodic function in the imaginary-time direction.  
To proceed, we transform to imaginary time $\tau = -it$.  In this imaginary time $\tau$, all bosonic (fermionic) fields are uniquely defined in the region $0 \leq \tau \leq \beta$ with all other values of $\tau$ obtainable using the periodicity (anti-periodicity) of the fields. This suffices to describe physics in thermal equilibrium, where the system does not depend on the real time $t$, which we are indeed free to set to zero.

Since, in thermal equilibrium, all Green's functions are (anti-)periodic functions of imaginary time, the fields themselves must be (anti-)periodic functions of the imaginary time.  As a result, when one performs a Fourier decomposition of the fields, the Fourier-integral associated with the time direction becomes a discrete Fourier sum.  If the fields are bosonic, then the allowed frequencies conjugate to the imaginary-time direction are $P_0 = \omega_n = 2 \pi n T$ with $n \in \mathbb{Z}$ where $P_0 = -i p_0$ is the zero-component of the Euclidean (imaginary-time) four-momentum, $P = (\omega_n,{\bf p})$.  If the fields are fermionic, then the allowed frequencies are $P_0 = \omega_n = (2n+1) \pi T$ with $n \in \mathbb{Z}$.  The discrete frequencies that result for both bosons and fermions are called {\em Matsubara frequencies}.
The mode expansions then become
\begin{equation}
\phi(\tau,{\bf x}) = \sumint_{\!P} \; \phi(\omega_n,{\bf p}) \, e^{-i(\omega_n \tau - {\bf p}\cdot{\bf x})} \hspace{1cm} {\rm (bosonic \; field)} \, ,
\end{equation}
where
\begin{equation}
\sumint_{\!P} \equiv \, T \! \sum_{P_0 = 2 \pi n T} \, \int \frac{d^3p}{(2\pi)^3} \, ,
\end{equation}
and
\begin{equation}
\phi(\tau,{\bf x}) = \sumint_{\!\{P\}} \; \phi(\omega_n,{\bf p}) \, e^{-i(\omega_n \tau - {\bf p}\cdot{\bf x})} \hspace{1cm} {\rm (fermionic \; field)} \, ,
\end{equation}
where
\begin{equation}
\sumint_{\!\{P\}} \equiv \, T \! \sum_{P_0 = (2 n + 1) \pi T} \, \int \frac{d^3p}{(2\pi)^3} \, .
\end{equation}
If the fields in addition carry a conserved charge $Q$, then in a grand canonical description the Matsubara frequencies are shifted by $-i\mu_Q$, where $\mu_Q$ is the chemical potential associated with the conserved charge.  In QCD, baryon number --- and more generally flavor --- is a conserved quantity, and hence one can introduce quark chemical potentials, $\mu_f$, $f=1,2,...,N_f$, obtaining
\begin{equation}
\sumint_{\!\{P\}} \equiv \, T \! \sum_{P_0 = (2 n + 1) \pi T - i \mu_f} \, \int \frac{d^3p}{(2\pi)^3} \, .
\end{equation}
The quark flavor in question is typically not indicated in this shorthand notation, but must be kept track of carefully in practical calculations.

We note that a more heuristic way to understand the emergence of discrete Matsubara frequencies is to consider the analytic structure of the zero-chemical-potential equilibrium distribution function $n_{\rm eq}(\omega) = (e^{\beta \omega} \pm 1)^{-1}$, where the positive sign gives the Fermi-Dirac and the negative one the Bose-Einstein distribution function.  The distribution functions have singularities when $e^{\beta \omega} = \mp 1$. For bosons, this gives $e^{\beta \omega} = + 1$ which is satisfied by $\omega = \omega_n = 2 \pi n T$ with $n \in \mathbb{Z}$, and for fermions $e^{\beta \omega} = - 1$ which is satisfied by $\omega = \omega_n = (2 n + 1) \pi T$ with $n \in \mathbb{Z}$.  As a result, in integrals involving the equilibrium distribution function times a holomorphic function, one can use Cauchy's theorem to transform the continuous integral into a sum over the corresponding Matsubara frequencies.  If the function itself contains poles, care should be taken when deforming the necessary complex contours; however, the basic idea of deforming complex contours remains in play. These statements can be made precise, and nonzero chemical potentials included, in a straightforward way: for details, the reader is encouraged to consult textbooks, such as \cite{Kapusta:2006pm,Mikko}.

Finally, we close the subsection by mentioning that in the imaginary time formalism the path integrals defining various physical quantities are expressed in terms of the so-called Euclidean action $S_E[\phi] = \int_0^\beta d\tau \int d^3 x \, {\cal L}_{\rm E}[\phi]$, where ${\cal L}_{\rm E}$ in turn reads ${\cal L}_{\rm E}=-{\cal L}_{\rm M}(t\to -i\tau)$.  In this case, the contribution of, say, a bosonic field to the partition function is of the form (see e.g.~\cite{Mikko} for the cases of fermionic and gauge fields)
\begin{equation}
Z = \int_{\phi(\tau=0)}^{\phi(\tau=\beta)} {\cal D}\phi \, e^{- S_E[\phi]} \; . 
\end{equation}
The convergence properties of these types of Euclidean integrals are naturally superior to Minkowskian ones, which makes their evaluation with lattice Monte-Carlo techniques possible, at least in the absence of sizable chemical potentials. This interesting topic, and the problems associated with the infamous Sign Problem of lattice QCD (see e.g. \cite{deForcrand:2010ys}), will, however, not be discussed further in this review.

\subsection{Imaginary-time Feynman rules}\label{sub_eucl_feynrules}

As mentioned above, when working in the imaginary time formalism it is convenient to switch from Minkowski space to the Euclidean one.  As a result, we replace $g_{\mu\nu} \rightarrow \delta_{\mu\nu}$, whereby the anti-commutation relation of the gamma matrices  becomes $\{\gamma_\mu^{\rm E},\gamma_\nu^{\rm E}\} = - 2 \delta_{\mu\nu}$ with $\gamma_0^{\rm E} \equiv i \gamma_0$.  From here on, the label `E' indicating Euclidean gamma matrices will be implicit when working in the imaginary-time formalism.  

With the above definitions, the free quark propagator takes the form
\begin{equation}
S^{ij}_0 = - \delta^{ij} \frac{\slashed{P}-m}{\omega_n^2+{\bf p}^2+m^2} \, ,
\end{equation}
with $i$ and $j$ fundamental representation color indices, while the free gluon propagator in a general covariant gauge reads
\begin{equation}
\left(G_0\right)^{ab}_{\mu\nu} = \frac{\delta^{ab}}{P^2} \left[ \delta_{\mu\nu} - (1-\xi) \frac{P_\mu P_\nu}{P^2} \right] ,
\end{equation}
with $P^2 = \omega_n^2 + {\bf p}^2$, and $a$ and $b$ adjoint color indices.  Finally, we note that, aside from  transforming to imaginary time and using the Euclidean-space gamma matrices, the QCD vertex functions remain the same in the imaginary time formalism (see Appendix B of \cite{Ipp:2003qt} for a comprehensive collection of conventions used in different textbooks and reviews). At the same time, all integrals over internal momentum become either bosonic or fermionic sum-integrals as defined in the previous section.

\subsection*{Example:  One-loop gluon polarization tensor}

\newcommand{\I}{{\cal I}^0}
\newcommand{\It}{\widetilde{\cal I}^0}
\newcommand{\rmi}[1]{{\mbox{\tiny #1}}}

%%%%%%%%%%%%%%%%%%%%%%%%%%%%%%%%%%%%%%%%%%%%%%%%%%%%%%%%%%%%%%%%%%%
\begin{figure}[t]
\begin{center}
\includegraphics[width=0.9 \textwidth]{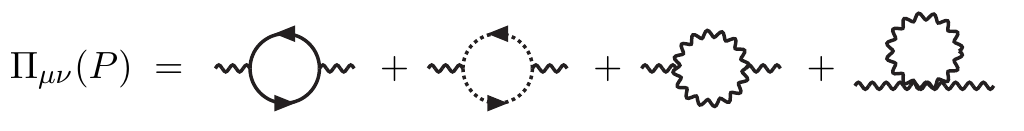}
\end{center}
\caption{The one-loop gluon polarization.  In this figure, wavy lines represent gluons, solid lines with arrows represent quarks and anti-quarks, and dotted lines with arrows represent the ghost field.}
\label{fig:pimunu}
\end{figure}
%%%%%%%%%%%%%%%%%%%%%%%%%%%%%%%%%%%%%%%%%%%%%%%%%%%%%%%%%%%%%%%%%%%

As an example of the application of the imaginary time formalism, consider the behavior of the one-loop correction to the gluon propagator, dubbed the gluon polarization tensor $\Pi_{\mu\nu}(P)$, at finite temperature and zero chemical potential. In the Feynman gauge, $\xi=1$, this quantity is defined by the expression
\begin{equation}
\big(G^{-1}\big)^{ab}_{\mu\nu}(P) = P^2 \delta^{ab}\delta_{\mu\nu} + \Pi_{\mu\nu}^{ab}(P), %\nonumber
\end{equation}
and can be seen to consist of the four one-loop graphs shown in Fig.~\ref{fig:pimunu} and take the algebraic form
\begin{eqnarray}
\Pi_{\mu\nu}^{ab}(P)&=& g^2 \delta^{ab}\Bigg\{C_A\bigg[(D-2)\I_1\delta_{\mu\nu}+2\big(P_{\mu}P_{\nu}-P^2\delta_{\mu\nu}\big)\Pi(P)\nn \\ && \hspace{-1cm} -\frac{D-2}{2}\sumint_{Q}\frac{(2Q-P)_{\mu}(2Q-P)_{\nu}}{Q^2(Q-P)^2}\bigg]\\
&& \hspace{-1cm} -2T_F N_f \bigg[2\It_1\delta_{\mu\nu}+\big(P_{\mu}P_{\nu}-P^2\delta_{\mu\nu}\big)\Pi_\rmi{f}(P)-
\sumint_{\{Q\}} \! \frac{(2Q-P)_{\mu}(2Q-P)_{\nu}}{Q^2(Q-P)^2}\bigg]\Bigg\},\nonumber
\end{eqnarray}
where we have defined
\begin{eqnarray}
\I_1 & \equiv & \sumint_Q \frac{1}{Q^2},\quad \It_1\;\;\equiv\;\; \sumint_{\{Q\}} \frac{1}{Q^2}, \nn \\
\Pi(P)&\equiv&\sumint_Q\frac{1}{Q^2(Q-P)^2}, \quad
\Pi_\rmi{f}(P)\;\;\equiv\;\;\sumint_{\{Q\}}\frac{1}{Q^2(Q-P)^2}. 
\end{eqnarray}
We will now inspect, in detail, how this function behaves in its infrared limit, setting first $p_0=0$ and then letting $p\to 0$. The result of this exercise will be seen to have important implications for the infrared properties of the theory and in particular for the convergence of high-order perturbative calculations.

Setting $P=0$ everywhere, we clearly obtain from the above
\begin{eqnarray}
\Pi_{\mu\nu}^{ab}(p_0=0,p\to 0)&=& g^2 \delta^{ab}\Bigg\{(D-2)C_A\bigg[\I_1\delta_{\mu\nu}-2\sumint_{Q}\frac{Q_\mu Q_{\nu}}{Q^4}\bigg]\nn\\
&-&4T_F N_f\bigg[\It_1\delta_{\mu\nu}-2\,
\sumint_{\{Q\}}\frac{Q_{\mu} Q_{\nu}}{Q^4}\bigg]\Bigg\}, \label{piIR}
\end{eqnarray}
indicating that the most nontrivial object to study is
\begin{equation}
A_{\mu\nu} \equiv \sumint_{Q} \frac{Q_{\mu} Q_{\nu}}{Q^4},
\end{equation}
as well as its fermionic counterpart $\widetilde{A}_{\mu\nu}$. It is useful to note here that, due to rotational and translational invariance, the result for $A_{\mu\nu}$ must be a linear combination of the tensors $\delta_{\mu\nu}$ and $n_{\mu}n_{\nu}$, where $n_{\mu}=\delta_{\mu 0}$ defines the rest frame of the heat bath. This enables us to write 
\begin{equation}
A_{\mu\nu}\equiv A_1\delta_{\mu\nu} +A_2 n_{\mu}n_{\nu},
\end{equation}
from which it is straightforward to obtain
\begin{eqnarray}
A_1 &=& \frac{A_{\mu\mu}-A_{00}}{D-1}, \quad
A_2 \;=\; \frac{-A_{\mu\mu}+D A_{00}}{D-1} \, , \label{eq:A12munu}
\end{eqnarray}
by contracting both sides of Eq.~(\ref{eq:A12munu}) respectively with the two tensors.

Using additionally the fact that $A_{\mu\mu}= \I_1$ and that a differentiation of $\I_1$
 with respect to the temperature produces the relation
\begin{eqnarray}
A_{00} &=& -\frac{1}{2} \I_1,
\end{eqnarray}
we obtain the simple result (valid for $D=4$)
\begin{eqnarray}
A_{\mu\nu}&=&-\frac{1}{2}\I_1 n_\mu n_\nu +\frac{1}{2}\I_1 \delta_{ij},
\end{eqnarray}
where we have for simplicity denoted $\delta_{\mu\nu}-n_\mu n_\nu \equiv \delta_{ij}$. A straightforward generalization of this calculation to the fermionic integral $\widetilde{A}_{\mu\nu}$ finally gives
\begin{eqnarray}
\widetilde{A}_{\mu\nu}&=&-\frac{1}{2}\It_1 n_\mu n_\nu +\frac{1}{2}\It_1 \delta_{ij}.
\end{eqnarray}

Plugging the above results into Eq.~(\ref{piIR}), we get for the IR limit of the gluon polarization tensor
\begin{eqnarray}
\Pi_{\mu\nu}^{ab}(p_0=0,p\to 0)&=& 4g^2 \delta^{ab}n_\mu n_\nu\big(C_A\I_1-2T_F N_f\It_1\big),
\end{eqnarray}
where the two remaining sum-integrals can be straightforwardly computed and $T_F=1/2$ is the Dynkin index of the generators in the fundamental representation. Considering for illustration the fermionic case in detail, we obtain after performing the $3-2\epsilon$ -dimensional momentum integral
\begin{eqnarray}
\widetilde{\mathcal I}_{1}^0
&=& \frac{\Gamma(-1/2+\epsilon) \Lambda^{2\epsilon}T}{(4\pi)^{3/2-\epsilon}} \sum_{k=-\infty}^{\infty}
\frac{\!\!\!\!\!\!\!\!\!\!1}{\Big[\Big((2k+1)\pi T-i\mu\Big)^2\Big]^{-1/2+\epsilon}}  \nonumber \\
&=& \frac{\Gamma(-1/2+\epsilon)T^{2}}{2\sqrt{\pi}}\Bigg(\frac{\Lambda^2}{\pi T^2}\Bigg)^{\epsilon} \text{Re}\Big[\zeta(-1+2\epsilon,1/2-i \bar{\mu})\Big] \nonumber \\
&=&-\frac{T^2}{24}-\frac{\mu^2}{8\pi^2}+{\mathcal O}(\epsilon),
\end{eqnarray}
where we have used the definition of the generalized (Hurwitz) zeta function.

The result obtained has direct physical implications. It means that the zeroth Matsubara mode of the temporal (electrostatic) gluon field $A_0$ obtains a thermal mass at one-loop order and, furthermore, that this leading-order \textit{Debye mass} takes the value
\begin{eqnarray}
m_D^2&=&\frac{g^2}{3N_f}\sum_f\bigg\{(C_A+T_F N_f)T^2+\frac{3\mu_f^2}{\pi^2}T_F N_f\bigg\}.
\end{eqnarray}
The magnetostatic fields ($p_0=0$ component of $A_i$) on the other hand stay unscreened at this order, and in fact only obtain a (non-perturbative) screening mass of order $g^2T$. This fact is related to the gauge transformation properties of the fields: Upon the breaking of Lorentz invariance into mere rotational invariance by the heat bath, four-dimensional gauge invariance is broken to a three-dimensional one. In this process, the electrostatic field becomes an adjoint scalar field and may therefore obtain a nonzero mass, while the magnetostatic fields continue to transform as three-dimensional gauge fields and must therefore remain massless to all orders in perturbation theory. 

Next, we move on to discussing in detail the consequences of the observed energy scales in thermal QCD: the scale $\pi T$, associated with the nonzero Matsubara modes of different fields, as well as $gT$ and $g^2T$, associated with the screening of static gluons, i.e.~their $n=0$ modes.

\subsection{High-temperature limit \label{highTpower}}

As noted above, the imaginary-time formalism is frequently used to study the behavior of bulk thermodynamic quantities at high temperature, meaning in practice the regime where $\pi T\gtrsim \mu$. There, it often turns out that a naive loop expansion of physical quantities is only well-defined for the first few orders of perturbation theory. For example, a closer inspection shows that uncancelled IR divergences enter the expansion of the partition function at three-loop order, and that they can be attributed to long-distance interactions mediated by static gluon fields. A simple way to understand at which perturbative orders terms non-analytic in $\alpha_s$ appear in the weak-coupling expansion of the pressure is to start from the contribution of non-interacting  (but possibly screened) static gluons to the quantity. This takes the schematic form $p_\text{gluons}\sim\int d^3p \,p\, n_B(E_p)$, with $n_B$ denoting the Bose-Einstein distribution function and $E_p$ the dispersion relation of the (electrostatic or magnetostatic) gluons. Inspecting in turn contributions from momenta of orders $\pi T$, $gT$ and $g^2T$, we see the emergence of the following pattern:
\begin{eqnarray}
p_\text{gluons}^{p\sim \pi T}&\sim& T^4n_B(\pi T)\;\sim\; T^4+{\mathcal O}(g^2),\\
p_\text{gluons}^{p\sim g T}&\sim& (gT)^4 n_B(g T)\;\sim\; g^3T^4+{\mathcal O}(g^4),\\
p_\text{gluons}^{p\sim g^2 T}&\sim& (g^2T)^4 n_B(g^2 T)\;\sim\; g^6T^4, \label{eq:magorder}
\end{eqnarray}
where we have taken into account that  $n_B(E)\sim T/E$ if $E\ll T$. It is worth pointing out explicitly that the expansion parameters in the three different terms are of order $g^2n_B(\pi T)\sim g^2$, $g^2n_B(gT)\sim g$, and $g^2n_B(g^2T)\sim 1$, implying in particular that the contribution of magnetostatic gluons to the pressure is fundamentally nonperturbative in nature, which is why we have not included an ${\mathcal O}(g^n)$ term in Eq.~(\ref{eq:magorder}) at all. This complete breakdown of the loop expansion at the scale 
$g^2T$ is called the \emph{Linde problem} \cite{Linde:1980ts}.
In this context, it should be noted that the order at which the nonperturbative contributions make their first appearance in the weak-coupling expansion of a given physical quantity is not universal, but differs from one quantity to the next. An extreme case was presented in Sec.~\ref{sec_lpm}, where we saw how the
scattering rate $\Gamma$  
is affected by ultrasoft contributions already at the leading order, as noted in Footnote~\ref{foot_IR}. We also point to
\cite{York:2008rr}, which shows how a certain second-order transport coefficient, $\lambda_1$, receives a leading-order contribution from
that scale. 
Later in the present section, we will on the other hand observe that for bulk thermodynamic quantities, such as the pressure, the $g^2T$ scale only begins to contribute at the four-loop, or N$^3$LO, order.

In perturbative calculations aimed at reaching high loop orders, resummations of some kind are clearly required to take full account of the contributions of the problematic field modes and thereby cure unphysical IR divergences. In the limit of high temperatures  --- including the case of nonzero density --- there exist several physically motivated schemes for carrying out such resummations, see e.g. \cite{Kajantie:1997tt,Braaten:1995cm,Braaten:1995jr,Andersen:1999fw,Andersen:1999sf,Andersen:1999va,Blaizot:1999ip,Blaizot:1999ap,Blaizot:2000fc,Blaizot:2001vr} and references therein. Most importantly, these include dimensionally reduced effective theories, which take advantage of the scale hierarchies present in the system in the language of effective field theory, and Hard Thermal Loop perturbation theory (HTLpt), which applies the already discussed HTL effective action to the problem. In the following two subsections, we review the associated formalisms and explain, how practical calculations are most efficiently carried out within them.

\subsubsection{Dimensional reduction}
\label{sec_DR}

The method of dimensional reduction is based on the simple observation that in the weak-coupling limit (requiring in practice $T \gg \Lambda_\text{QCD}$), there exists a scale hierarchy between the three energy scales that contribute to bulk thermodynamic observables. Namely, if $g\ll 1$, we clearly have
\begin{eqnarray}
m_\text{mag} \sim g^2 T \ll m_\text{elec} \sim g T \ll m_\text{hard} \sim \pi  T,
\end{eqnarray}
where we denote scales of magnetostatic and electrostatic screening by $m_\text{mag}$ and $m_\text{elec}$, respectively, and the thermal one --- the non-zero Matsubara frequency --- by $m_\text{hard}$. Of the first two scales, the electrostatic screening mass can (to leading order) be obtained from the computation carried out in the previous subsection, i.e.~the IR limit of the one-loop self energy of the $A_0$ field, while the scale of magnetostatic screening appears nonperturbatively. Neglecting the $T=0$ energy scales of different quark masses and the QCD scale $\Lambda_\text{QCD}$, the above three scales are the only ones appearing in the problem, and two of them are furthermore connected with the $n=0$ field modes. It is thus natural to attempt integrating out the largest one, i.e.~$m_\text{hard}$, from the system, amounting to the construction of a three-dimensional effective theory valid for the long-distance static field modes. Such an effective description can be expected to be valid in the limit of high temperatures, the precise meaning of which will be specified later. 

Historically, the construction of dimensionally reduced effective theories for high-temperature QCD dates back to the works of Ginsparg \cite{Ginsparg:1980ef} as well as Appelquist and Pisarski \cite{Appelquist:1981vg} in the early 1980s, but the wider use of the methods began only in mid-1990s, when Kajantie et al.~applied the formalism first to the study of the Electroweak phase transition \cite{Kajantie:1995dw} and later to the context of thermal QCD \cite{Kajantie:1997tt}. Simultaneously to the latter developments, Braaten and Nieto popularized the use of the terms Electrostatic QCD (EQCD) and Magnetostatic QCD (MQCD) to denote the two levels of effective theories obtained by successively integrating out the scales $\pi T$ and $gT$ from full QCD \cite{Braaten:1995jr}, thereby casting the formalism into its modern form. 

As usual in the construction of effective theories, the Lagrangian densities of the two theories can be obtained most straightforwardly  by writing down the most general local Lagrangians respecting all necessary symmetries (most importantly three-dimensional gauge invariance), ordering the operators in terms of their dimensionality, and truncating the result at the desired order. The result of this procedure reads for the case of EQCD \cite{Braaten:1995cm,Braaten:1995jr} 
\begin{eqnarray}
{\mathcal L}_\text{E}& = & \frac{1}{2} \text{Tr} F_{ij}^2 + \text{Tr} [D_i,A_0]^2 + m_\text{E}^2 \text{Tr} A_0^2 
+\lambda_\text{E}^{(1)} (\text{Tr} A_0^2)^2
 +\lambda_\text{E}^{(2)} \text{Tr} A_0^4 \nonumber \\
 &+& i\lambda_\text{E}^{(3)} \text{Tr} A_0^3 + \cdots , 
\label{leqcd}
\end{eqnarray}
where the fields $A_i\equiv A_i^a T^a$, $A_0\equiv A_0^a T^a$ now live in three dimensions, we have denoted 
\begin{eqnarray}
F_{ij}^a&=&\partial_i A_j^a - \partial_j A_i^a + g_\text{E} f^{abc} A_i^b A_j^c,\\
D_i&=&\partial_i-ig_\text{E}A_i, 
\end{eqnarray}
and operators of dimensionality higher than 4 have been suppressed. Further integrating out the temporal gauge field, we similarly obtain the effective theory MQCD
\begin{eqnarray}
{\mathcal L}_\text{M}& = & \frac{1}{2} \text{Tr} F_{ij}^2 + \cdots,
\label{lmqcd}
\end{eqnarray}
where this time 
\begin{eqnarray}
F_{ij}^a&=&\partial_i A_j^a - \partial_j A_i^a + g_\text{M} f^{abc} A_i^b A_j^c.
\end{eqnarray}

At leading order in weak coupling, the degrees of freedom in the above effective theories correspond to the $n=0$ Matsubara modes of the four-dimensional $A_i$ and $A_0$ fields, of which the first transforms as a three-dimensional gauge field and the latter as a scalar in the adjoint representation of SU($N$). It is worth noting that, for MQCD, the only dimensionful scale appearing in the theory is $g_\text{M}^2 = g^2T + {\mathcal O}(g^3)$. This implies that, barring effects from higher-order operators, the MQCD contribution to any physical quantity must necessarily be of the form of some dimensionless number times an appropriate power of this scale, determinable through the dimension of the quantity in question.

Returning momentarily to the symmetries of the original theory, it is interesting to note that the term cubic in $A_0$ in Eq.~(\ref{leqcd}) has a coefficient proportional to the sum of the quark chemical potentials \cite{Hart:2000ha},
\begin{eqnarray}
\lambda_\text{E}^{(3)} = \frac{i g^3}{3\pi^2} \sum_f\mu_f + {\mathcal O}(g^5).
\end{eqnarray}
This reflects the charge conjugation invariance of the original theory, which is broken by a nonzero quark number density. Similarly, one may note that the discrete Z($N$) center symmetry of four-dimensional pure Yang-Mills theory is broken in EQCD even in the limit $N_f=0$, although there has been claimed to be some evidence of its partial dynamical restoration \cite{Kajantie:1998yc}. This is due to the fact that in any perturbative calculation (such as the derivation of the EQCD Lagrangian) one needs to arbitrarily pick one of the $N$ equivalent deconfined vacua of the original theory as the expansion point. The situation can be remedied by generalizing the $A_0$ field of EQCD into a Wilson line type variable, as has been proposed in \cite{Vuorinen:2006nz,deForcrand:2008aw}; as the focus of the present review is in perturbative calculations, this subtle issue will, however, not be discussed any further here.

The parameters of the effective theories can be determined by matching a set of physical quantities, in practice various Green's functions, in EQCD and MQCD to the full theory. This is done by requiring that the effective theories reproduce the long-distance physics of the original one, with ``long distances'' referring to $x\gtrsim 1/(gT)$ and $x\gtrsim 1/(g^2 T)$ for EQCD and MQCD, respectively. An important simplification in these calculations comes from the fact that they can be performed within a \textit{strict loop expansion} in the full theory, i.e.~without invoking any kind of a resummation and using dimensional regularization to regulate both IR and UV divergences. 

\begin{figure}[t]
\begin{center}
\includegraphics[width=1.0 \textwidth]{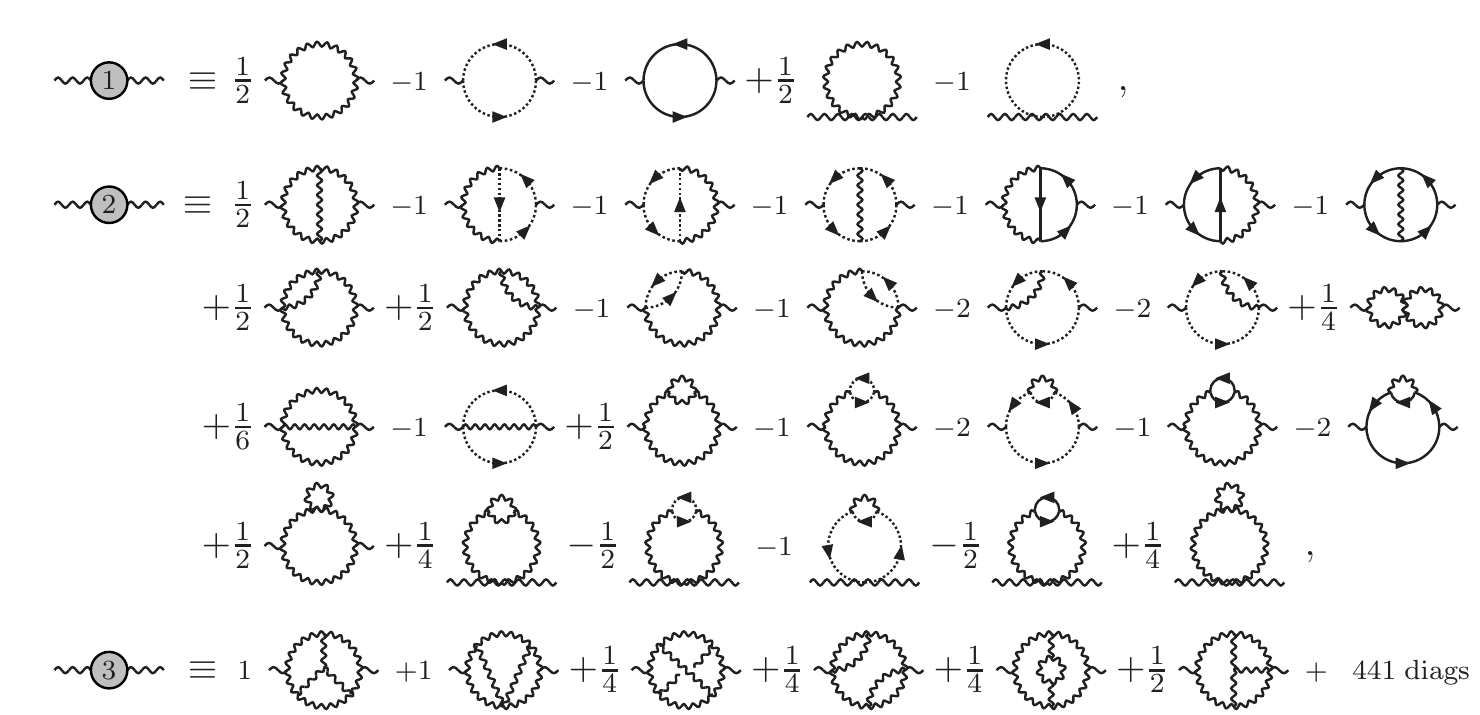}
\end{center}
\caption{The one-, two-, and three-loop graphs contributing to the EQCD parameters $g_\text{E}^2$ and $m_\text{E}$. The figure is taken from \cite{Moeller:2012da}.}
\label{fig:selfen}
\end{figure}

By now, the EQCD and MQCD parameters, i.e.~the operator coefficients visible in the above Lagrangians, have been determined to a high order in perturbation theory, with the current results reaching up to:
\begin{itemize}
\item ${\mathcal O}(g^6)$ for $g_\text{E}^2$ \cite{Laine:2005ai}
\item ${\mathcal O}(g^6)$ for $m_\text{E}^2$ \cite{Ghisoiu:2015uza}
\item ${\mathcal O}(g^6)$ for $\lambda_\text{E}^{(1)}$ and $\lambda_\text{E}^{(2)}$ \cite{Kajantie:1997tt}
\item ${\mathcal O}(g^3)$ for $\lambda_\text{E}^{(3)}$ \cite{Hart:2000ha}
\item ${\mathcal O}(g^6)$ for $g_\text{M}^2$  \cite{Laine:2005ai}
\end{itemize}
In addition, some impressive progress has been achieved in recent years in attempts to proceed to even higher orders; see e.g.~\cite{Laine:2018lgj,Laine:2019uua} and references therein for a summary. The success of these very demanding calculations relies in a large part on the development of integration by parts and tensor reduction techniques at finite temperature, pioneered by York Schr\"oder and collaborators (cf.~e.g.~\cite{Nishimura:2012ee,Ghisoiu:2012yk}). See also fig.~\ref{fig:selfen} for a list of the full theory gluon self energy graphs that contribute to the determination of $g_\text{E}^2$ and $m_\text{E}$ at three-loop order \cite{Moeller:2012da}. 

%%%%%%%%%%%%%%%%%%%%%%%%%%%%%%%%%%%%%%%%%%%%%%%%%%%%%%%%%%%%%%%%%%%
%%%%%%%%%%%%%%%%%%%%%%%%%%%%%%%%%%%%%%%%%%%%%%%%%%%%%%%%%%%%%%%%%%%

\begin{figure}[t]
\begin{center}
\includegraphics[width=1.0 \textwidth]{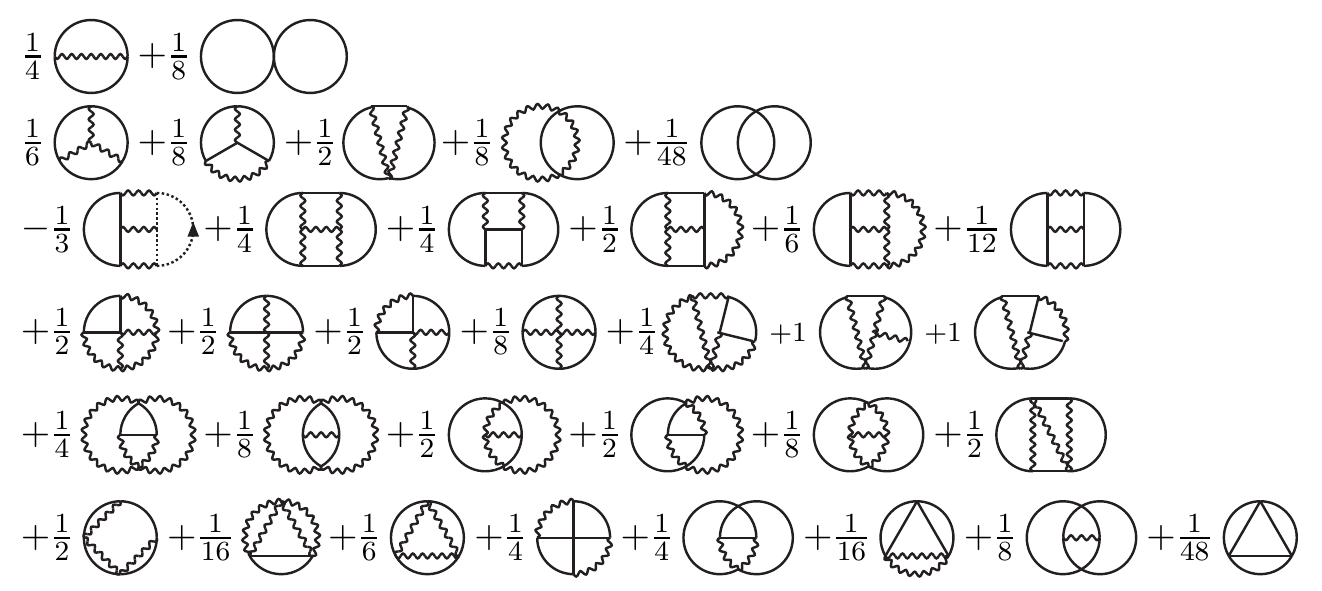}
\end{center}
\caption{The two-, three-, and four-loop two-particle irreducible vacuum graphs contributing to the partition function of EQCD. The figure is taken from \cite{Kajantie:2003ax}.}
\label{fig:EQCDpres}
\end{figure}

Out of all applications of the dimensional reduction machinery in the context of QCD, arguably the most important one concerns the determination of the Equation of State (EoS) of a hot QGP, or the computation of the weak coupling expansion for the pressure of QCD. As explained in some length in \cite{Braaten:1995jr}, the partition function of the full theory can be written as a sum of three parts, 
\begin{eqnarray}
p_\text{QCD} = p_\text{E} + p_\text{M} + p_\text{G},
\end{eqnarray}
where each of the terms on the right hand side has a distinct physical meaning as the contribution of a specific energy (or length) scale to the pressure:
\begin{itemize}
 \item $p_\text{E}$ stands for the contribution of the hard energy scale $\pi T$, and is obtained via a strict loop expansion of the full theory pressure, which means it has the form of an expansion in powers of $g^2$ (up to arbitrarily high orders in principle). This function has been determined to full three-loop order both at $\mu=0$ \cite{Kajantie:2002wa} and at nonzero density \cite{Vuorinen:2003fs}, in addition to which the leading large-$N_f$ term of the four-loop contribution has been computed in \cite{Gynther:2009qf} at vanishing chemical potentials.
 \item $p_\text{M}$ stands for the pressure of EQCD and can be evaluated in a weak coupling expansion within this three-dimensional theory, with an expansion parameter of order $g$. This function has been determined to order $g^6$ in \cite{Kajantie:2003ax} (cf.~also \cite{Vuorinen:2003fs} for the contribution of the operator cubic in $A_0$) in an impressive calculation that included the evaluation of all graphs displayed in fig.~\ref{fig:EQCDpres}. The effective theory has in addition been subjected to non-perturbative lattice studies; see e.g.~\cite{Hietanen:2008xb}.
 \item $p_G$ stands for the nonperturbatively determinable pressure of MQCD. It can be expressed in the form of a dimensionless number times $T(g_M^2T)^3$, with no further perturbative corrections emerging (except for corrections to the parameter $g_M$, determinable within EQCD). The determination of the dimensionless coefficient was completed using a combination of three-dimensional lattice simulations and stochastic perturbation theory  \cite{Hietanen:2004ew,DiRenzo:2006nh}, which is needed to convert the lattice results to continuum regularization. 
\end{itemize}
As expected, the sum of the three contributions is completely IR finite and yields a well-defined result for the full theory pressure accurate in principle to the full $g^6$ order, although some of the hard contributions are still lacking at the moment. It is worth mentioning already at this point that the convergence properties and renormalization scale dependence of this expression can be dramatically improved by not expanding the effective theory contributions $p_\text{M}$ and $p_\text{G}$ in powers of the full theory gauge coupling $g$ \cite{Blaizot:2003iq,Laine:2006cp} --- an issue we shall return in some length in the following section. In this context, we also point out in passing that ref.~\cite{Laine:2006cp} presents a low-loop-order generalization of the determination of basic thermodynamic observables to the case of nonzero quark masses (see also ref.~\cite{Laine:2019uua} for the latest developments on this front).

%%%%%%%%%%%%%%%%%%%%%%%%%%%%%%%%%%%%%%%%%%%%%%%%%%%%%%%%%%%%%%%%%%%
%%%%%%%%%%%%%%%%%%%%%%%%%%%%%%%%%%%%%%%%%%%%%%%%%%%%%%%%%%%%%%%%%%%

\subsubsection{Hard Thermal Loops}
\label{sec:htlnpointeffaction}

An alternative method for performing a high-temperature resummation is based on the Hard Thermal Loop (HTL) effective theory, discussed in some length already in Sec.~\ref{sec_soft_collinear} of this review. As discussed there, the HTL description contains resummed gluon and quark propagators, in addition to which resummed vertices which are necessary in order to maintain gauge invariance.  As it turns out, all of these can be collected into a compact HTL effective action which is manifestly gauge invariant.  Before presenting this effective action, we first discuss explicit expressions for the low order $n$-point functions.  The Feynman rules are presented in Minkowski space, after which we provide a set of simple rules that can be used to obtain the Euclidean imaginary-time expressions.  Additionally, for generality we present expressions valid in $d+1$ space-time dimensions since these are necessary when performing calculations in dimensional regularization. 

Finally, we note that many parallels exist between this section and its real-time counterpart \ref{sub_heur_htl}, including several essentially duplicate relations. We have, however, found it worthwhile to keep both the real- and imaginary-time sections of the review self-consistent for the benefit of readers wishing to concentrate on only one of the two parts.

\paragraph{Minkowski-space HTL gluon propagator}

Based on the results presented in Sec.~4, one finds that the HTL inverse gluon propagator in a general covariant gauge can be expressed in the form
\begin{eqnarray}
\Delta^{-1}(\mn{P})^{\mu\nu}&=&
\Delta^{-1}_{\infty}(\mn{P})^{\mu\nu}-\frac{1}{\xi}
\mn{P}^{\mu}\mn{P}^{\nu} \, ,
\label{Delinv:cov}
\end{eqnarray}
where $\xi$ is the gauge-fixing parameter and
\begin{eqnarray}
\Delta^{-1}_{\infty}(\mn{P})^{\mu\nu}&\equiv&
-\mn{P}^2 g^{\mu \nu} + \mn{P}^\mu \mn{P}^\nu + \Pi^{\mu\nu}(\mn{P}) \, ,
\label{delta-inv:inf0}
\end{eqnarray}
with $\Pi^{\mu\nu}$ being the HTL resummed gluon polarization tensor.  The HTL gluon polarization tensor reads in turn
\begin{eqnarray}
\label{a1}
\Pi^{\mu\nu}(\mn{P})=m_D^2\left[\mn{N}^{\mu} \mn{N}^{\nu}-
{\cal T}^{\mu\nu}(\mn{P},-\mn{P})
\right]\;,
\end{eqnarray}
where $\mn{N}^\mu$ is again the heat-bath four-velocity which satisfies  $\mn{N}\cdot\mn{N}=-1$ and is given by $\mn{N}^\mu = (1,{\bf 0})$ in the local rest frame. The tensor ${\cal T}^{\mu\nu}(\mn{P},\mn{Q})$, which is defined only for momenta that satisfy $\mn{P}+\mn{Q}=0$, is
\begin{eqnarray}
{\cal T}^{\mu\nu}(\mn{P},-\mn{P})=
\left \langle \mn{Y}^{\mu} \mn{Y}^{\nu}{\mn{P}\!\cdot\!\mn{N}\over \mn{P}\!\cdot\!\mn{Y}}
\right\rangle_{\bf\hat{y}} \;,
\label{T2-def}
\end{eqnarray}
where the angular brackets indicate averaging over the $d$ spatial directions of the light-like vector $\mn{Y}=(1,\hat{\bf y})$, with $\hat{\bf y}$ denoting a unit three-vector.
The tensor ${\cal T}^{\mu\nu}$ is symmetric in $\mu$ and $\nu$
and satisfies the identity $\mn{P}_{\mu}{\cal T}^{\mu\nu}(\mn{P},-\mn{P})=(\mn{P}\cdot\!\mn{N})\mn{N}^{\nu}$.
The polarization tensor $\Pi^{\mu\nu}$ is therefore also
symmetric in $\mu$ and $\nu$, is transverse $\mn{P}_{\mu}\Pi^{\mu\nu}(\mn{P})=0$, and satisfies $g_{\mu\nu}\Pi^{\mu\nu}(\mn{P})=m_D^2$.

Just as its full theory counterpart, the HTL gluon polarization tensor can be expressed in terms of two scalar functions,
the transverse and longitudinal polarization functions $\Pi_T$ and $\Pi_L$ (note a slight difference in notation compared to the rest of the review here; elsewhere the definition of $\Pi_L$ includes a factor of $\mn{P}^2/\mathbf{p}^2$)
\begin{eqnarray}
\label{pit2}
\Pi_T(\mn{P})&=&{1\over d-1}\left(
\delta^{ij}-\hat{p}^i\hat{p}^j
\right)\Pi^{ij}(\mn{P})\;, \\
\label{pil2}
\Pi_L(\mn{P})&=&\Pi^{00}(\mn{P})\;,
\end{eqnarray}
where ${\bf \hat p}$ is the unit vector
in the direction of ${\bf p}$.
In terms of these functions, the polarization tensor reads	
\begin{eqnarray}
\label{pi-def}
\Pi^{\mu\nu}(\mn{P}) \;=\;  \Pi_T(\mn{P}) T_p^{\mu\nu}
+ {1\over n_p^2} \Pi_L(\mn{P}) L_p^{\mu\nu}\;,
\end{eqnarray}
where the transverse and longitudinal projectors $T_p$ and $L_p$ are
\begin{eqnarray}
T_p^{\mu\nu}&=&g^{\mu\nu} - {\mn{P}^{\mu}\mn{P}^{\nu} \over \mn{P}^2}
-{\mn{N}_p^{\mu}\mn{N}_p^{\nu}\over \mn{N}_p^2}\;,\\
L_p^{\mu\nu}&=&{\mn{N}_p^{\mu}\mn{N}_p^{\nu} \over \mn{N}_p^2}\;.
\end{eqnarray}
The four-vector $\mn{N}_p^{\mu}$ is defined via $\mn{N}_p^{\mu} \;=\; \mn{N}^{\mu} - ({\mn{N} \cdot \mn{P}}) \, \mn{P}^{\mu}/\mn{P}^2$ and satisfies $\mn{P}\!\cdot\!\mn{N}_p=0$ and $\mn{N}^2_p = -1 - (\mn{N} \cdot \mn{P})^2/\mn{P}^2$. In the local rest frame of the heat bath, one has $\mn{N}_p^2 = - {\bf p}^2/\mn{P}^2$.  Note that the identity $\mn{P}_{\mu}\Pi^{\mu\nu}(\mn{P})=0$ reduces to $(d-1)\Pi_T(\mn{P})+\Pi_L(\mn{P})/\mn{N}^2_p = m_D^2$ which implies that there is only one independent polarization function.

Interestingly, we can express both gluon polarization functions in terms of the function ${\cal T}^{00}$ defined in Eq.~(\ref{T2-def}):
\begin{eqnarray}
\Pi_T(\mn{P})&=& {m_D^2 \over (d-1) \mn{N}_p^2}
\left[ {\cal T}^{00}(\mn{P},-\mn{P}) - 1 - \mn{N}_p^2  \right] ,
\label{PiT-T}
\\
\Pi_L(\mn{P})&=& m_D^2
\left[ 1-{\cal T}^{00}(\mn{P},-\mn{P}) \right] ,
\label{PiT-L}
\end{eqnarray}
%In the definition of ${\cal T}^{\mu \nu}(p,-p)$,
%the angular brackets indicate the angular average over
%the unit vector $\hat{\bf y}$.
For consistency of higher order radiative
corrections, it is essential to take the angular average in the definition of ${\cal T}^{\mu \nu}(\mn{P},-\mn{P})$  in $d=3-2\epsilon$
dimensions and analytically continue to $d=3$ only after all poles in
$\epsilon$ have been cancelled.
Expressing the angular average as an integral over the cosine of an angle,
the expression for the $00$ component of the tensor becomes
\begin{eqnarray}
{\cal T}^{00}(\mn{P},-\mn{P}) &=& {w(\epsilon)\over2}
\int_{-1}^1dc\;(1-c^2)^{-\epsilon}{p^0\over p^0-|{\bf p}|c} \;,
\label{T00-int}
\end{eqnarray}
where the weight function reads $w(\epsilon)= \Gamma({3\over2}-\epsilon)/(\Gamma({3\over2}) \Gamma(1-\epsilon))$.

The integral in Eq.~(\ref{T00-int}) must be defined so that it remains analytic
as $p^0\to\infty$.
It then has a branch cut running from $p^0=-|{\bf p}|$ to $p^0=+|{\bf p}|$, and if we take the limit $\epsilon\rightarrow 0$, the result reduces to
\begin{equation}
{\cal T}^{00}(\mn{P},-\mn{P}) =
{p^0 \over 2|{\bf p}|}
                \ln {p^0 +|{\bf p}| \over p^0-|{\bf p}|}\;,
\end{equation}
i.e.~the function appearing in the usual $d=3$ HTL polarization functions.  Working in $d=3$ and in the rest frame of the heat bath, we hereby obtain
\begin{eqnarray}
\Pi_T(\mn{P})&=& {m_D^2 \over 2 }
\left[ \frac{p_0^2}{{\bf p}^2} - {p^0 \over 2|{\bf p}|} \frac{p_0^2 - {\bf p}^2}{{\bf p}^2}\ln {p^0 +|{\bf p}| \over p^0-|{\bf p}|} \right] ,
\label{PiT-T3d}
\\
\Pi_L(\mn{P})&=& m_D^2
\left[ 1 - {p^0 \over 2|{\bf p}|}
                \ln {p^0 +|{\bf p}| \over p^0-|{\bf p}|} \right] ,
\label{PiT-L3d}
\end{eqnarray}
which in the static limit ($p^0 \rightarrow 0$) produce $\lim_{p^0 \rightarrow 0} \Pi_T = 0$ and $\lim_{p^0 \rightarrow 0} \Pi_L = -m_D^2$.  As discussed in the previous subsection, the vanishing of the static limit of the transverse polarization function means that chromomagnetic fields are not screened, while the finiteness of the static $\Pi_L$ corresponds to the Debye screening of the chromoelectric interaction.

Returning to the HTL gluon propagator, Eq.~(\ref{delta-inv:inf0}) can also be written as
\begin{eqnarray}
\Delta^{-1}_{\infty}(\mn{P})^{\mu\nu}&=&
- {1 \over \Delta_T(\mn{P})}       T_p^{\mu\nu}
+ {1 \over \mn{N}_p^2 \Delta_L(\mn{P})} L_p^{\mu\nu}\;,
\label{delta-inv:inf}
\end{eqnarray}
where $\Delta_T$ and $\Delta_L$ are the transverse and longitudinal
propagators:
\begin{eqnarray}
\Delta_T(\mn{P})&=&{1 \over \mn{P}^2+\Pi_T(\mn{P})}\;,
\label{Delta-T:M}
\\
\Delta_L(\mn{P})&=&{1 \over - \mn{N}_p^2 \mn{P}^2+\Pi_L(\mn{P})}\;.
\label{Delta-L:M}
\end{eqnarray}
Note that for $d=3$ and in the heat bath rest frame the second relation reduces to $\Delta_L^{-1} = {\bf p}^2+\Pi_L$, which, furthermore, becomes in the static limit $\lim_{p^0 \rightarrow 0} \Delta_L^{-1} = {\bf p}^2+m_D^2$. 

Finally, we mention that the general covariant gauge HTL gluon propagator can be obtained by inverting Eq.~(\ref{Delinv:cov}) to obtain
\begin{equation}
\Delta^{\mu\nu}(\mn{P}) = -\Delta_T(\mn{P})T_p^{\mu\nu}
+\Delta_L(\mn{P})\mn{N}_p^{\mu}\mn{N}_p^{\nu}
- \xi {\mn{P}^{\mu}\mn{P}^{\nu} \over \mn{P}^4}
\end{equation}

\paragraph{Minkowski-space HTL quark propagator}

One can also extract the HTL resummed quark propagator using a similar procedure as we outlined for the gluon propagator.  The result reads
\begin{equation}
\label{qprop}
S(\mn{P})={1\over/\!\!\!p+\Sigma(\mn{P})}\;,
\end{equation}
where the quark self energy is given by
\begin{equation}
\label{selfq}
\Sigma(\mn{P})=m_q^2\,/\!\!\!\!{\cal T}(\mn{P})
\;.
\end{equation}
Here, we have defined
\begin{equation}
\label{deftf}
{\cal T}^{\mu}(\mn{P}) = -\left\langle{\mn{Y}^{\mu}\over \mn{P} \cdot \mn{Y}} \right\rangle_{\hat{\bf y}} \, ,
\end{equation}
while for $d=3$ one furthermore obtains
\begin{equation}
m_q^2 = \frac{ C_F}{8} g^2 T^2 \, .
\end{equation}

Expressing the angular average as an integral over the cosine of an angle,
the expression for ${\cal T}^{\mu}(p)$ reads
\begin{eqnarray} 
\label{def-tf}
{\cal T}^{\mu}(\mn{P})={w(\epsilon)\over2}
\int_{-1}^1dc\;(1-c^2)^{-\epsilon}{\mn{Y}^{\mu}\over p^0-|{\bf p}|c}\;.
\end{eqnarray}
As before, the integral in Eq.~(\ref{def-tf}) must be defined so that it is analytic 
as $p^0\to\infty$.
It then has a branch cut running from $p^0=-|{\bf p}|$ to $p^0=+|{\bf p}|$.  
For $d=3$ and in the heat bath rest frame, the fermion self energy reduces to
\begin{eqnarray}\nonumber
\Sigma(\mn{P})&=&
{m_q^2\over 2|{\bf p}|}\gamma^0\ln{p^0+|{\bf p}|\over p^0-|{\bf p|}}
\\&&
+{m_q^2\over |{\bf p}|}{\boldsymbol{\gamma}}\cdot \hat{\bf p}
\left(1-{p^0\over 2|{\bf p}|}\ln{p^0+|{\bf p}|\over p^0-|{\bf p|}}\right) .
\end{eqnarray}

\paragraph{Three-gluon vertex}

The three-gluon vertex
for gluons with outgoing four-momenta $\mn{P}$, $\mn{Q}$, and $\mn{R}$,
Lorentz indices $\mu$, $\nu$, and $\lambda$,
and color indices $a$, $b$, and $c$ reads
\begin{equation}
i\Gamma_{abc}^{\mu\nu\lambda}(\mn{P},\mn{Q},\mn{R})=-gf_{abc}
\Gamma^{\mu\nu\lambda}(\mn{P},\mn{Q},\mn{R})\;,
\end{equation}
where the three-gluon vertex tensor is
\begin{equation}
\Gamma^{\mu\nu\lambda}(\mn{P},\mn{Q},\mn{R})=
g^{\mu\nu}(\mn{P}-\mn{Q})^{\lambda}+
g^{\nu\lambda}(\mn{Q}-\mn{R})^{\mu}+
g^{\lambda\mu}(\mn{R}-\mn{P})^{\nu}
-m_D^2{\cal T}^{\mu\nu\lambda}(\mn{P},\mn{Q},\mn{R})\;.
\label{Gam3}
\end{equation}
The tensor ${\cal T}^{\mu\nu\lambda}$ in the HTL correction term
is defined only for $\mn{P}+\mn{Q}+\mn{R}=0$:
\begin{eqnarray}
{\cal T}^{\mu\nu\lambda}(\mn{P},\mn{Q},\mn{R}) \;=\;
  \Bigg\langle \mn{Y}^{\mu} \mn{Y}^{\nu} \mn{Y}^{\lambda}
\left( {\mn{P}\!\cdot\!\mn{N}\over \mn{P}\!\cdot\!\mn{Y}\;\mn{Q}\!\cdot\!\mn{Y}}
	- {\mn{R}\!\cdot\!\mn{N}\over\!\mn{R}\cdot\!\mn{Y}\;\mn{Q}\!\cdot\!\mn{Y}} \right)
	\Bigg\rangle\;.
\label{T3-def}
\end{eqnarray}
This tensor is totally symmetric in its three indices and traceless in any
pair of indices: $g_{\mu\nu}{\cal T}^{\mu\nu\lambda}=0$.
It is odd (even) under odd (even) permutations of the momenta $\mn{P}$, $\mn{Q}$, and
$\mn{R}$, and it satisfies the identity
\begin{eqnarray}
q_{\mu}{\cal T}^{\mu\nu\lambda}(\mn{P},\mn{Q},\mn{R}) \;=\;
{\cal T}^{\nu\lambda}(\mn{P}+\mn{Q},\mn{R})-
{\cal T}^{\nu\lambda}(\mn{P},\mn{R}+\mn{Q})\;.
\label{ward-t3}
\end{eqnarray}
The three-gluon vertex tensor therefore also obeys the Ward-Takahashi identity
\begin{eqnarray}
p_{\mu}\Gamma^{\mu\nu\lambda}(\mn{P},\mn{Q},\mn{R}) \;=\;
\Delta_{\infty}^{-1}(\mn{Q})^{\nu\lambda}-\Delta_{\infty}^{-1}(\mn{R})^{\nu\lambda}\;.
\label{ward-3}
\end{eqnarray}

\paragraph{Four-gluon vertex}

The four-gluon vertex
for gluons with outgoing momenta $\mn{P}$, $\mn{Q}$, $\mn{R}$, and $\mn{S}$,
Lorentz indices $\mu$, $\nu$, $\lambda$, and $\sigma$,
and color indices $a$, $b$, $c$, and $d$ reads
\begin{eqnarray}
i\Gamma^{\mu\nu\lambda\sigma}_{abcd}(\mn{P},\mn{Q},\mn{R},\mn{S}) &=&
- ig^2\Bigg\{ f_{abx}f_{xcd} \left(g^{\mu\lambda}g^{\nu\sigma}
				-g^{\mu\sigma}g^{\nu\lambda}\right)
\nonumber
\\
&&
\hspace{-.67cm}
+2m_D^2\mbox{tr}\left[T^a\left(T^bT^cT^d+T^dT^cT^b
\right)\right]{\cal T}^{\mu\nu\lambda\sigma}(\mn{P},\mn{Q},\mn{R},\mn{S})
\Bigg\}
\nonumber
\\
&&+ \; 2 \; \mbox{cyclic permutations}\;,
\end{eqnarray}
where the cyclic permutations are of
$(\mn{Q},\nu,b)$, $(\mn{R},\lambda,c)$, and $(\mn{S},\sigma,d)$.
The matrices $T^a$ are the generators of the fundamental representation
of the $SU(3)$ group with the standard normalization
${\rm tr}(T^a T^b) = {1 \over 2} \delta^{ab}$.
The tensor ${\cal T}^{\mu\nu\lambda\sigma}$
in the HTL correction term is defined only for $\mn{P}+\mn{Q}+\mn{R}+\mn{S}=0$, and reads
\begin{eqnarray}
{\cal T}^{\mu\nu\lambda\sigma}(\mn{P},\mn{Q},\mn{R},\mn{S}) &=&
\Bigg\langle y^{\mu} y^{\nu} y^{\lambda} y^{\sigma}
\left( {\mn{P}\!\cdot\!n \over \mn{P}\!\cdot\!y \; \mn{Q}\!\cdot\!y \; (\mn{Q}+\mn{R})\!\cdot\!y}
\right.
\nonumber
\\
&&
\left. \hspace{-1cm}
+{(\mn{P}+\mn{Q})\!\cdot\!n\over \mn{Q}\!\cdot\!y\;\mn{R}\!\cdot\!y\;(\mn{R}+\mn{S})\!\cdot\!y}
+{(\mn{P}+\mn{Q}+\mn{R})\!\cdot\!n\over \mn{R}\!\cdot\!y\;\mn{S}\!\cdot\!y\;(\mn{S}+\mn{P})\!\cdot\!y}\right)
\Bigg\rangle . \hspace{7mm}
\label{T4-def}
\end{eqnarray}
This tensor is totally symmetric in its four indices and traceless in any
pair of indices: $g_{\mu\nu}{\cal T}^{\mu\nu\lambda\sigma}=0$.
It is even under cyclic or anti-cyclic
permutations of the momenta $\mn{P}$, $\mn{Q}$, $\mn{R}$, and $\mn{S}$, and
satisfies the identity
\begin{equation}
\mn{Q}_{\mu}{\cal T}^{\mu\nu\lambda\sigma}(\mn{P},\mn{Q},\mn{R},\mn{S})=
{\cal T}^{\nu\lambda\sigma}(\mn{P}+\mn{Q},\mn{R},\mn{S})
-{\cal T}^{\nu\lambda\sigma}(\mn{P},\mn{R}+\mn{Q},\mn{S}) \, .
\label{ward-t4}
\end{equation}

When the color indices are traced in pairs, the four-gluon vertex becomes
much simpler
\begin{equation}
\delta^{ab} \delta^{cd} i \Gamma_{abcd}^{\mu\nu\lambda\sigma}(\mn{P},\mn{Q},\mn{R},\mn{S})
= -i g^2 N_c (N_c^2-1) \Gamma^{\mu\nu,\lambda\sigma}(\mn{P},\mn{Q},\mn{R},\mn{S}) \;,
\end{equation}
where the color-traced four-gluon vertex tensor is
\begin{equation}
\Gamma^{\mu\nu,\lambda\sigma}(\mn{P},\mn{Q},\mn{R},\mn{S})=
2g^{\mu\nu}g^{\lambda\sigma}
-g^{\mu\lambda}g^{\nu\sigma}
-g^{\mu\sigma}g^{\nu\lambda}
-m_D^2{\cal T}^{\mu\nu\lambda\sigma}(\mn{P},\mn{S},\mn{Q},\mn{R})\;.
\label{Gam4}
\end{equation}
The tensor (\ref{Gam4}) is symmetric
under the interchange of $\mu$ and $\nu$,
under the interchange of $\lambda$ and $\sigma$,
and under the interchange of $(\mu,\nu)$ and $(\lambda,\sigma)$.
It is also symmetric under the interchange of $\mn{P}$ and $\mn{Q}$,
under the interchange of $\mn{R}$ and $\mn{S}$,
and under the interchange of $(\mn{P},\mn{Q})$ and $(\mn{R},\mn{S})$.
Finally, it satisfies the Ward-Takahashi identity
\begin{equation}
\mn{P}_{\mu}\Gamma^{\mu\nu,\lambda\sigma}(\mn{P},\mn{Q},\mn{R},\mn{S})
=\Gamma^{\nu\lambda\sigma}(\mn{Q},\mn{R}+\mn{P},\mn{S})
-\Gamma^{\nu\lambda\sigma}(\mn{Q},\mn{R},\mn{S}+\mn{P})\;.
\label{ward-4}
\end{equation}

\paragraph{Quark-gluon three-vertex}

The HTL resummed quark-gluon vertex with outgoing gluon momentum $\mn{P}$, 
incoming quark momentum $\mn{Q}$, and outgoing quark momentum $\mn{R}$, 
Lorentz index $\mu$, and color index $a$ reads
\begin{equation}
\label{3qgv}
\Gamma^{\mu}_a(\mn{P},\mn{Q},\mn{R})
=gt_a\left(\gamma^{\mu}-m_q^2\tilde{{\cal T}}^{\mu}(\mn{P},\mn{Q},\mn{R})\right) .
\end{equation}
The tensor in the HTL correction term is only defined for $\mn{P}-\mn{Q}+\mn{R}=0$ and is given by
\begin{equation}
\tilde{{\cal T}}^{\mu}(\mn{P},\mn{Q},\mn{R})
=-\left\langle
\mn{Y}^{\mu}\left({\mn{Y}\!\!\!/\over \mn{Q}\!\cdot\!\mn{Y}\;\;\mn{R}\!\cdot\!\mn{Y}}\right)
\right\rangle_{\hat{\bf \mn{Y}}}\;.
\label{T3q-def}
\end{equation}
This tensor is even under the permutation of $\mn{Q}$ and $\mn{R}$.
It satisfies the identity
\begin{equation}
\mn{P}_{\mu}\tilde{\cal T}^{\mu}(\mn{P},\mn{Q},\mn{R})=
\tilde{\cal T}^{\mu}(\mn{R})-\tilde{\cal T}^{\mu}(\mn{Q})\;,
\end{equation}
and the quark-gluon vertex therefore satisfies
the Ward-Takahashi identity
\begin{equation}
\mn{P}_{\mu}\Gamma^{\mu}(\mn{P},\mn{Q},\mn{R})=S^{-1}(\mn{Q})-S^{-1}(\mn{R})\;.
\label{qward1}
\end{equation}

\paragraph{Quark-gluon four-vertex}

We define the quark-gluon four-point vertex with outgoing gluon 
momenta $\mn{P}$ and $\mn{Q}$, incoming fermion momentum $\mn{R}$, and outgoing
fermion momentum $\mn{S}$.  Generally this vertex has both adjoint and
fundamental indices; however, for our presentation we will only  need the quark-gluon four-point vertex traced over the adjoint 
color indices,
\begin{eqnarray}\nonumber
\delta^{ab} \Gamma^{\mu\nu}_{abij}(\mn{P},\mn{Q},\mn{R},\mn{S}) &=& 
    - g^2 m_q^2 C_F \delta_{ij} \tilde{\cal T}^{\mu\nu}(\mn{P},\mn{Q},\mn{R},\mn{S})  \\
&    \equiv &g^2 C_F \delta_{ij} \Gamma^{\mu\nu}  \, .
\label{4qgv}
\end{eqnarray}
The tensor $\tilde{\cal T}^{\mu\nu}$ is only defined for $\mn{P}+\mn{Q}-\mn{R}+\mn{S}=0$
\begin{equation}
\tilde{{\cal T}}^{\mu\nu}(\mn{P},\mn{Q},\mn{R},\mn{S})
= \left\langle
\mn{Y}^{\mu}\mn{Y}^{\nu}\left({1\over \mn{R}\!\cdot\!\mn{Y}}+{1\over \mn{S}\!\cdot\!\mn{Y}}\right)
{\mn{Y}\!\!\!/\over[(\mn{R}-\mn{P})\!\cdot\!\mn{Y}]\;[(\mn{S}+\mn{P})\!\cdot\!\mn{Y}]}
\right\rangle\;.
\label{T4q-def}
\end{equation}
It is is traceless and symmetric in $\mu$ and $\nu$, and  satisfies the Ward-Takahashi identity
\begin{eqnarray}
\mn{P}_{\mu}\Gamma^{\mu\nu}(\mn{P},\mn{Q},\mn{R},\mn{S})=\Gamma^{\nu}(\mn{Q},\mn{R}-\mn{P},\mn{S})-\Gamma^{\nu}(\mn{Q},\mn{R},\mn{S}+\mn{P})\;.
\label{qward2}
\end{eqnarray}

\paragraph{Hard thermal loop effective Lagrangian}

The HTL effective Lagrangian can be written compactly as \cite{Braaten:1991gm}
\begin{equation}
\label{L-HTL}
{\cal L} = {\cal L}_{\rm QCD} + {\cal L}_{\rm HTL} \, ,
\end{equation}
where ${\cal L}_{\rm QCD}$ is the usual vacuum QCD Lagrangian.
The HTL contribution to the effective Lagrangian reads
\begin{equation}
{\cal L}_{\rm HTL} = - {1\over2} m_D^2 {\rm Tr}
\left(G_{\mu\alpha}\left\langle {\mn{Y}^{\alpha}\mn{Y}^{\beta}\over(\mn{Y}\cdot D)^2}
	\right\rangle_{\!\!\mn{Y}}G^{\mu}_{\;\;\beta}\right)
	+ i m_q^2 \bar{\psi}\gamma^\mu \left\langle {\mn{Y}_\mu\over \mn{Y}\cdot D}
	\right\rangle_{\!\!y}\psi
	\, ,  \label{htl_lag}
\end{equation}
where $G^{\mu\nu}$ is the gluon field strength tensor (denoted $F^{\mu\nu}$ elsewhere), $D$ stands for the covariant derivative in the appropriate representation, $\mn{Y}^\mu=(1,{\hat {\bf y}})$ is a light-like vector, and 
$\langle \cdots \rangle$ is the already familiar average over
all possible directions of ${\hat {\bf y}}$.
The HTL effective action is gauge invariant and can generate all 
HTL $n$-point functions~\cite{Braaten:1991gm}, which satisfy the necessary 
Ward-Takahashi identities by construction.  This includes all of the $n$-point functions we have listed thus far.  For example, when the HTL contribution to the effective Lagrangian is expanded in powers of the quark and gluon fields, there will be a term of the form
$$
\int_y \int_z \bar{\psi}(x) \: \Gamma^\mu (x,y,z) \: \psi (y) \: A_\mu (z) \;,
$$
where $\Gamma^\mu (x,y,z)$ is the quark-gluon vertex function. 
To obtain this vertex function, we only need to expand the HTL effective Lagrangian to
leading order in the gluon field strength 
\begin{eqnarray}
{\cal L}^{(\bar\psi A \psi)}_{\rm HTL}(x) &=& i m_q^2
\; \bar{\psi}(x)  \left\langle {\slashed{\mn{Y}} \over \mn{Y}\cdot D} \right\rangle_{\!\!\mn{Y}}
\psi (x)  \nonumber \\
&=& i m_q^2
\; \bar{\psi}(x) \gamma^\mu  \left\langle {\slashed{\mn{Y}}  \over \mn{Y}\cdot \partial}
\sum_{n=0}^{\infty}\bigg( {i \,g \, \mn{Y}\cdot A(x) \over \mn{Y} \cdot \partial} \bigg)^n \right\rangle_{\!\!y}
\psi (x) \, .
\end{eqnarray}
After a Fourier transformation, the ${\cal O}(g^3)$ contribution gives
\begin{equation} 
\label{vertexdef}
\Gamma^\mu_a (\mn{P},\mn{Q},\mn{R}) = i g t^a \: (2\pi)^4 \delta^{(4)}(\mn{P}+\mn{Q}+\mn{R}) \, \Gamma^\mu(\mn{P},\mn{Q},\mn{R}) \, ,
\end{equation}
with
\begin{equation} 
\label{vertex}
\Gamma^\mu(\mn{P},\mn{Q},\mn{R}) = m_q^2 \left\langle
\mn{Y}^{\mu}\left({\mn{Y}\!\!\!/\over \mn{Q}\!\cdot\!\mn{Y}\;\;\mn{R}\!\cdot\!\mn{Y}}\right)
\right\rangle_{\hat{\bf y}} ,
\end{equation}
where $\mn{Q}$ and $\mn{R}$ are the incoming and outgoing quark momenta and $\mn{P}$ is the outgoing gluon momentum.   This corresponds precisely to the HTL correction to the bare QCD vertex presented earlier.

\paragraph{Euclidean space HTL effective Lagrangian and vertex functions}

The HTL effective Lagrangian and vertex functions listed above were specified for Minkowski space.  As mentioned earlier, in the imaginary-time formalism one has discrete imaginary energies, i.e.~the Matsubara frequencies $p_0 = i 2 \pi n T$.  Continuing to use a capital letter for Euclidean momenta, e.g. $P=(P_0,{\bf p})$, the inner product of two Euclidean vectors reads $P \cdot Q = P_0 Q_0 + {\bf p} \cdot {\bf q}$, while the vector that specifies the thermal rest frame remains $n = (1,{\bf 0})$.  The Feynman rules for Minkowski space given in the prior subsections can then be easily adapted to Euclidean space.  The Euclidean tensor corresponding to a given Feynman rule is obtained from the corresponding Minkowski tensor with all indices raised by replacing each Minkowski energy $p^0$ by $iP_0$ and multiplying for every $0$ index by $-i$.  This prescription transforms $\mn{P}=(p^0,{\bf p})$ into $P=(P_0,{\bf p})$,  $g^{\mu \nu}$ into $\delta^{\mu \nu}$, and $\mn{P}\!\cdot\!\mn{Q}$ into $P\!\cdot\!Q$. 

\paragraph{Hard Thermal Loop perturbation theory}

A widely used method for computing QCD thermodynamics which solves the related IR problems via a reorganization of finite temperature perturbation theory and the HTL formalism is called Hard Thermal Loop perturbation theory (HTLpt) \cite{Andersen:1999fw,Andersen:1999sf,Andersen:1999va,Andersen:2002ey,Andersen:2003zk,Andersen:2009tw,Andersen:2009tc,Andersen:2010ct,Andersen:2010wu,Andersen:2011sf,Andersen:2011ug,Haque:2012my,Mogliacci:2013mca,Haque:2013qta,Haque:2013sja,Haque:2014rua,Andersen:2015eoa}.  The HTLpt framework allows for a systematic analytic reorganization of perturbative series based on the HTL effective
Lagrangian.  Additionally, it is manifestly gauge invariant and applicable to calculating both static and dynamical quantities.  The HTLpt approach is an extension of the simpler screened perturbation theory which has been applied to scalar field theories \cite{Karsch:1997gj,Chiku:1998kd,Andersen:2000yj,Andersen:2001ez,Andersen:2008bz}.

In HTLpt the Lagrangian density is written in the form
\begin{eqnarray}
{\mathcal L}= \Big[ {\mathcal L}_{\rm QCD}
+ (1- \delta) {\mathcal L}_{\rm HTL} \Big]_{g \to \sqrt{\delta} g}
+ \Delta{\mathcal L}_{\rm HTL} \; ,
\label{L-HTLQCD}
\end{eqnarray}
where ${\mathcal L}_{\rm HTL}$ is the HTL contribution to the HTL effective Lagrangian given in Eq.~(\ref{htl_lag}) and $\Delta{\mathcal L}_{\rm HTL}$ collects any additional counterterms necessary for renormalization.  The first term is the usual QCD Lagrangian
\begin{eqnarray}
{\mathcal L}_{\rm QCD}=-{1\over2}{\rm Tr}\left(G_{\mu\nu}G^{\mu\nu}\right)
+{\mathcal L}_{\rm gf}+{\mathcal L}_{\rm ghost}+\Delta{\mathcal L}_{\rm QCD},
\label{L-QCD}
\end{eqnarray}
where $G_{\mu\nu}=\partial_{\mu}A_{\nu}-\partial_{\nu}A_{\mu} -ig[A_{\mu},A_{\nu}]$ is the gluon field strength
and $A_{\mu}$ is the gluon field expressed as a $N_c\times N_c$ matrix in the $SU(N_c)$ algebra.  The ghost term ${\mathcal L}_{\rm ghost}$ depends on the choice of
the gauge-fixing term ${\mathcal L}_{\rm gf}$.  The final term, $\Delta{\mathcal L}_{\rm QCD}$, collects all vacuum counterterms necessary for renormalization at $T=0$.

\begin{figure}[t]
\begin{center}
\includegraphics[width=\textwidth]{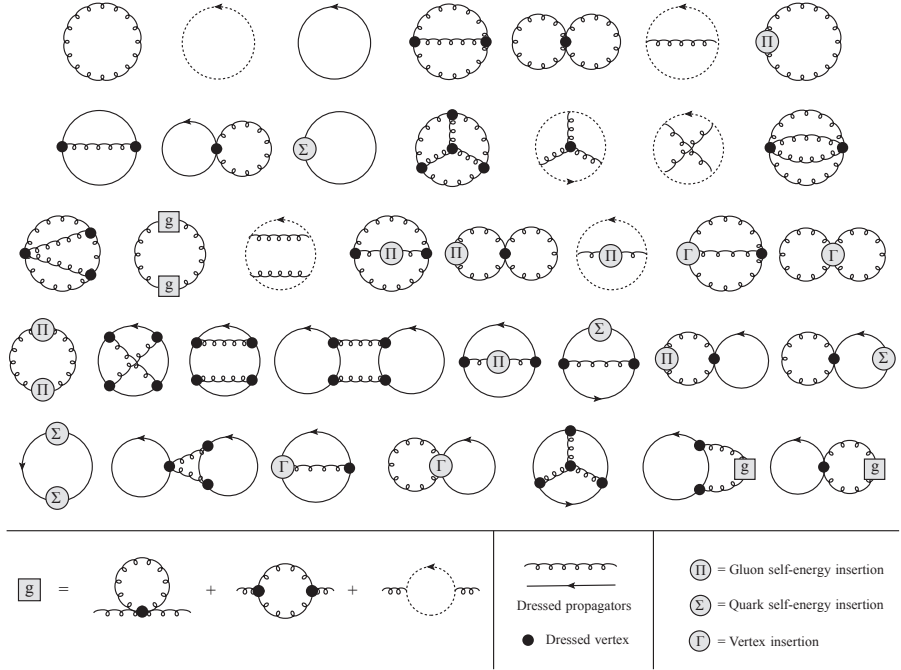}
\end{center}
\vspace{-6mm}
\caption{Diagrams that contribute to the NNLO thermodynamic potential using the HTLpt reorganization.}
\label{fig:htlptnnlo}
\end{figure}

The coefficient $\delta$ appearing in Eq.~(\ref{L-HTLQCD}) serves as the expansion parameter in HTLpt.  If $\delta=1$, then there is no modification of the vacuum QCD Lagrangian.  To proceed we Taylor expand the generating functional around $\delta=0$.  To order $\delta^0$ one has freely propagating HTL quasiparticles and higher orders in $\delta$ include higher and higher order quasiparticle interactions.  If we were able to expand the result to all orders in $\delta$ there would be no dependence on the mass parameters $m_D$ and $m_q$ appearing in Eq.~(\ref{htl_lag}); however, at any finite order of expansion, one needs a prescription for choosing the mass parameters.  In higher order calculations, one usually fixes the parameters $m_D$ and $m_q$ by employing a variational prescription which requires that the first derivative of the pressure with respect to both $m_D$ and $m_q$ vanishes such that the free energy is minimized; however, at high loop orders the variational prescription has been shown to break down in the sense that the resulting solutions are no longer real valued.  In practice, the solution has been to use the highest-order perturbative expressions for the mass scales available from EQCD \cite{Haque:2013sja,Haque:2014rua}.  Finally, we note that, in practice, the integrals resulting from the diagrams shown in Fig.~\ref{fig:htlptnnlo} are expanded in a power series in $m_D$ and $m_q$ in order to evaluate them.  Terms which would naively contribute to order $g^5$ if $m_D \sim m_q \sim g$ are kept in the final result.

%%%%%%%%%%%%%%%%%%%%%%%%%%%%%%%%%%%%%%%%%%%%%%%%%%%%%%%%%%%%%%%%%%%
%%%%%%%%%%%%%%%%%%%%%%%%%%%%%%%%%%%%%%%%%%%%%%%%%%%%%%%%%%%%%%%%%%%

Equation \eqref{L-HTLQCD} provides a systematic way to resum HTLs in a gauge-invariant manner.  At each order in the HTLpt $\delta$-expansion, results are infinite-order power series in the strong coupling constant.  One can Taylor expand the results obtained at each order in $\delta$ to make contact with naive perturbation theory calculations.  At order $\delta^0$ (LO), the resulting series only reproduces the ${\cal O}(g^3)$ term correctly, but the results are automatically free from electric-scale infrared divergences.  At order $\delta^1$ (NLO), the Taylor-expanded result correctly reproduces the order ${\cal O}(g^2)$ and ${\cal O}(g^3)$ contributions and, at order $\delta^2$, one reproduces all known perturbative coefficients through ${\cal O}(g^5)$.  As with the LO HTLpt calculation, both the NLO and NNLO calculations are automatically free of electric-scale infrared divergences.

\subsection{Low temperatures and high densities \label{lowTsec}}

In this section, we have so far implicitly assumed the temperature in the system to be high enough, so that it resides in the deconfined phase. In this case, the phase is called quark-gluon plasma, which is relevant for the description of the early universe and ultrarelativistic heavy ion collisions. At high enough baryon densities, deconfined matter, however, exists all the way to the zero-temperature limit, and may in particular be realized in the cores of the most massive neutron stars \cite{Lattimer:2004pg,Annala:2019puf}. Recalling the severe limitations of lattice QCD in the description of physics at nonzero density, one clearly needs to develop machinery for perturbative thermal field theory calculations also at small or vanishing temperatures.

%%%%%%%%%%%%%%%%%%%%%%%%%%%%%%%%%%%%%%%%%%%%%%%%%%%%%%%%%%%%%%%%%%%
\begin{figure}[t]
\begin{center}
\includegraphics[width=1.0 \textwidth]{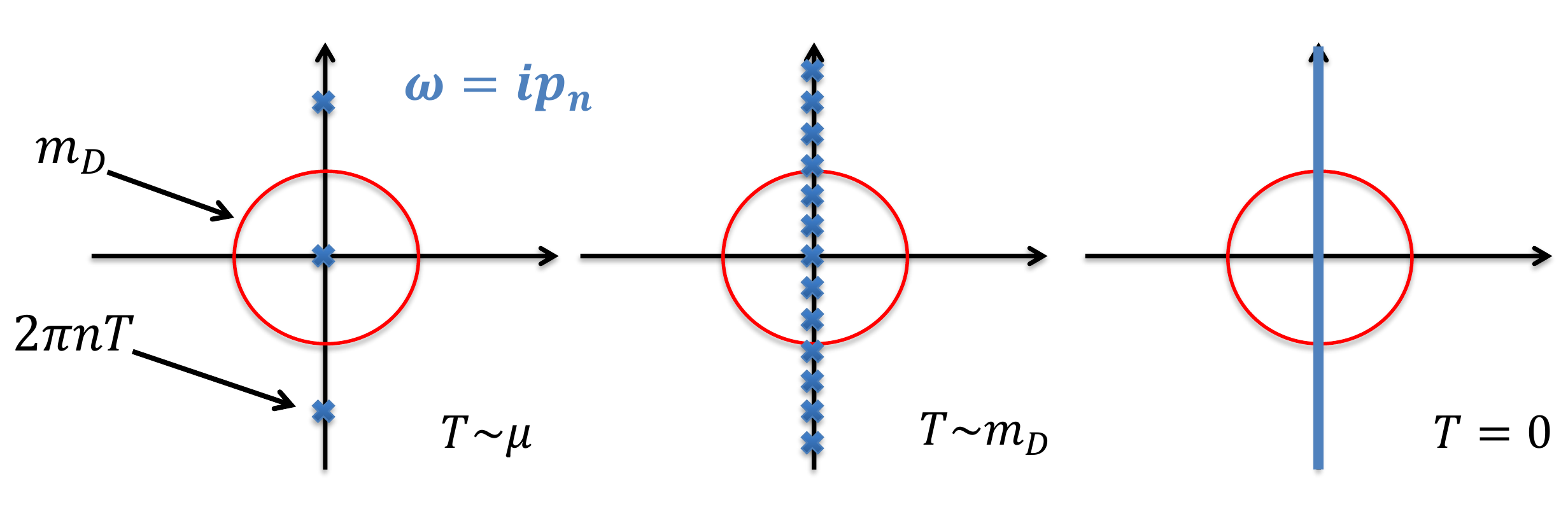}
\end{center}
\caption{An illustration of how bosonic Matsubara modes $i2\pi nT$ become more densely spaced on the complex frequency plane as the temperature is decreased towards and below the scale of the low-temperature Debye screening mass $m_D\sim g\mu$. The frequencies located inside the circle of radius $m_D$ should be treated in a nonperturbative way, while their number diverges the strict $T\to 0$ limit.}
\label{fig:smallT}
\end{figure}
%%%%%%%%%%%%%%%%%%%%%%%%%%%%%%%%%%%%%%%%%%%%%%%%%%%%%%%%%%%%%%%%%%%

There are two main differences between perturbative computations carried out at high and low temperatures. The first one is largely technical and has to do with the fact that the evaluation of multi-loop Feynman integrals at $T=0$ and finite chemical potentials is most easily carried out with methods that resemble zero-temperature pQCD techniques more than those used at finite temperature. The second difference, on the other hand, has to do with the properties of the IR sector of the theory, and is summarized in Fig.~\ref{fig:smallT}. Whereas at high temperatures, it suffices to single out the bosonic zero Matsubara modes and develop a dimensionally reduced effective theory framework for them, at small or zero temperature there are vastly more three-dimensional modes in need of a nonperturbative treatment. In the strict $T\to 0$ limit, the IR-sensitive sector of the theory in fact becomes four-dimensional, which is reflected in the fact that there are an infinite number of Matsubara modes that fit inside the red circle of radius $m_D$ in Fig.~\ref{fig:smallT} (right). On the other hand, the famous Linde problem, related to the nonperturbative contributions from the scale $g^2T$ at high temperatures, is however absent at $T=0$, so that e.g.~the weak-coupling expansion of the pressure is in principle well-defined to arbitrary orders in the coupling $g$. 

Below, we first cover the strict $T=0$ limit, touching both the techniques used in recent multi-loop calculations and the description of the IR sector in this particular case. After this, we proceed to the limit of small but nonzero temperatures, explaining how one can very efficiently combine the HTLpt and EQCD frameworks to provide a result for the QCD pressure that is valid to order $g^5$ at all values of $T/\mu$.

\subsubsection{The strict zero-temperature limit \label{zeroTsec}}

With neutron star matter applications in mind, it is a meaningful starting point to first set the temperature strictly to zero: for all neutron stars older than a few seconds, the temperature scale is vastly smaller than the baryon chemical potential due to cooling via neutrino emission. At very high density, the ground state of QCD is known to be a Color-Flavor-Locked (CFL) color superconductor, as has been shown through a consistent weak-coupling calculation \cite{Son:1998uk}. In our presentation, we will, however, not concentrate on the physics of quark pairing, but simply refer the interested reader to the review article \cite{Alford:2007xm}. Apart from simplicity, the reason for this is that from the viewpoint of most bulk thermodynamic quantities evaluated at perturbatively large densities, pairing is of subleading importance: it contributes to the energy density or pressure of the system at the parametric order $\Delta^2 \mu_B^2$, where the non-perturbative parameter $\Delta$ stands for the superconducting gap, while the much larger non-superconducting contribution is proportional to $\mu_B^4$. Note, however, that for some other Euclidean quantities, such as specific heats, the situation may be different.

With the above considerations in mind, we are led to inspect the thermodynamics of QCD in its deconfined but unpaired phase at $T=0$ and nonzero quark chemical potentials. In this case, the sum-integrations reduce to ordinary four-dimensional integrals, $\int \frac{d^4\! P}{(2\pi)^4},$ with the quark chemical potentials present in the fermionic propagators through the shift $P_0\to P_0-i\mu_q$. Otherwise the Feynman rules stay unaltered, i.e.~they are simply the $T=0$ limits of the finite-temperature Feynman rules of the imaginary time formalism.

The methods used in the evaluation of Feynman graphs at zero temperature and nonzero chemical potentials differ qualitatively from those typically encountered in thermal field theory. Assuming we have first `scalarized' the diagrams, i.e.~taken care of the Lorentz and color algebra, the next task becomes to perform the integrals over the 0-components of all (both fermionic and bosonic) momenta. Here, so-called ``cutting rules'' have turned out to be a very efficient book-keeping tool \cite{Ghisoiu:2016swa}:  for each $n$ between 0 and the number of loops in the graph, we 
\begin{enumerate}
\item Remove $n$ internal scalarized fermionic propagators from the graph in question, 
\item Place the corresponding momenta on shell, i.e.~set $P_0\to i E(p)$ with $E(p)\equiv \sqrt{p^2+m^2}$, and 
\item Integrate the thus generated amplitude with respect to the three-momenta $p$ with the measure
\begin{equation}
-\int\frac{{\rm d}^3p}{(2\pi)^3}\frac{\theta(\mu-E({p}))}{2E({p})}
\end{equation}
while setting $\mu=0$ inside the amplitude. 
\end{enumerate}
Finally, we sum over all the terms generated, both at every fixed value of $n$ and over the index $n$ itself. In this context, it should be noted that in \cite{Ghisoiu:2016swa} the chemical potential was assumed to appear in the fermionic momenta in the form $P_0+i\mu$, differing from our convention by the sign of the imaginary part. It is easy to verify that this does not affect the evaluation of scalarized vacuum-type Feynman integrals, but in the case of external fermion lines or an odd number of zero components of momenta appearing in the numerator of the integrand, an extra minus sign may appear. The easiest course of action then is to explicitly redefine all integration momenta via $P_0\to -P_0$ prior to the application of the cutting rules.

%%%%%%%%%%%%%%%%%%%%%%%%%%%%%%%%%%%%%%%%%%%%%%%%%%%%%%%%%%%%%%%%%%%
\begin{figure}[t]
\begin{center}
\includegraphics[width=0.9 \textwidth]{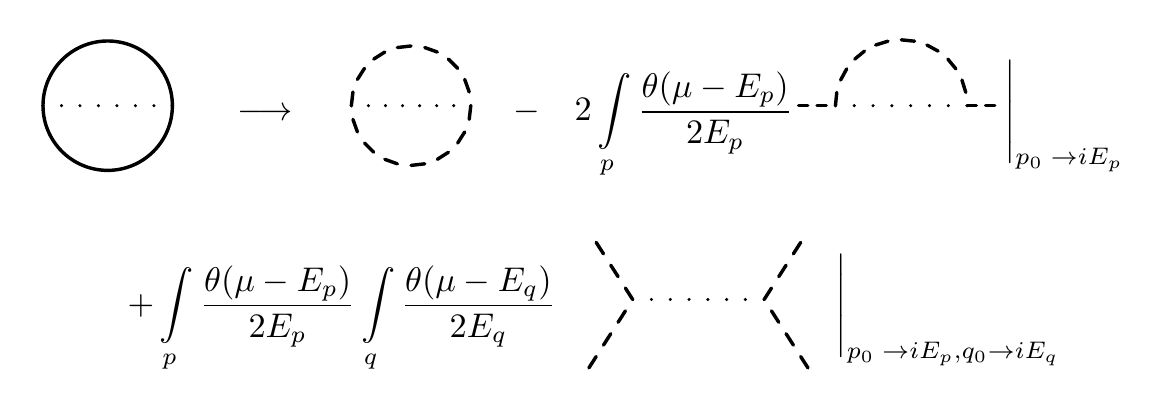}
\end{center}
\caption{An illustration of the cutting procedure amenable to zero-temperature, finite-$\mu$ Feynman graphs, described in the main text. The solid lines stand for scalarized fermionic propagators, and the dashed lines for massless bosonic ones. On the right hand side of the relation, all $\mu$-dependence resides inside the $\theta$-functions, i.e.~$\mu$ has been set to zero in the amplitudes, where the momentum $p$ flows along the upper and $q$ along the lower cut lines.}
\label{fig:cutting}
\end{figure}
%%%%%%%%%%%%%%%%%%%%%%%%%%%%%%%%%%%%%%%%%%%%%%%%%%%%%%%%%%%%%%%%%%%

Following the above procedure, we see that an $N$-loop vacuum graph, relevant for the determination of the EoS, gets reduced into $N+1$ distinct parts: an $N$-loop vacuum diagram with $\mu=0$, a sum of on-shell two-point amplitudes integrated with the above measure, a sum of on-shell 4-point amplitudes with two associated integrations, etc. --- all the way to $2N$-point amplitudes integrated over $N$ 3-momenta. For an explicit illustration of this procedure in the case of a simple two-loop graph, see fig.~\ref{fig:cutting} and Ref.~\cite{Ghisoiu:2016swa}. As explained in detail in this reference, a crucial simplification in computations utilizing the cutting rules originates from the fact that the required amplitudes can be evaluated at vanishing chemical potential, implying that one may take full advantage of zero-temperature QFT results and tools, such as integration-by-part relations (see e.g.~the Mathematica package FIRE \cite{Smirnov:2008iw} and references \cite{Bonciani:2003hc,Schroder:2005va}).

Just as at high temperatures, an issue complicating perturbative calculations at $T=0$ is the emergence of IR divergences in a strict loop expansion. As alluded to above, the main difference is that unlike at high $T$, we can no longer single out just one (static) Matsubara mode as the IR sensitive one, but all modes satisfying $P^2 = P_0^2 + p^2 \lesssim g^2\mu^2$ need to be treated in a nonperturbative way. The higher dimensionality of the soft sector in principle somewhat alleviates the IR problems, and e.g.~the leading  non-analytic contribution to the $T=0$ partition function is of order $g^4\ln g$ instead of the odd power $g^3$ encountered at high temperatures. Unfortunately, this does not imply that the implementation of a nonperturbative treatment for the soft sector would be simpler than at high $T$; on the contrary, in the absence of a dimensionally reduced effective theory, the traditional approach in zero-temperature calculations has been to resort to technically complicated explicit resummations of full theory diagrams (see e.g.~\cite{Ipp:2006ij}). In recent years, such tour de force calculations have, however, been significantly streamlined in computations such as those presented in \cite{Kurkela:2016was,Gorda:2018gpy}, whose logic we shall follow below. 

There are a few key insights that greatly simplify the determination of thermodynamic observables at high density and zero (or small) temperature. First, the IR problems discussed above all originate from gluon fields, so to remove them it suffices to resum gluon propagators and vertices in diagrammatic expansions. If one adds and subtracts from this ``resummed'' pressure the same quantity evaluated in a naive loop expansion,
\begin{eqnarray}
p_\text{QCD}^\text{resummed} &=& (p_\text{QCD}^\text{resummed} - p_\text{QCD}^\text{naive}) + p_\text{QCD}^\text{naive},
\label{presorg}
\end{eqnarray}
we observe that the quantity inside parentheses contains all the terms in the weak-coupling expansion of the pressure non-analytic in $\alpha_s$. Note that here the term ``resummation'' should be understood as summing together unspecified (infinite) classes of full-theory diagrams.

Our second observation is the following: whenever a gluonic momentum is hard, i.e.~of order $\mu$, in a resummed Feynman diagram, the corresponding propagator may be expanded in powers of the self energy, because no IR problems can by definition occur at large momenta. This implies on one hand that at least one gluonic line in the resummed diagrams must be soft, or of order $g\mu$, in order for there to be a nonzero contribution from the two terms inside the parentheses in Eq.~(\ref{presorg}), and on the other hand that whenever a gluonic line remains resummed, the self energy appearing in the said propagator may be replaced by the corresponding HTL version. A simple corollary of this is that the terms inside the parentheses start contributing to the pressure only at order $g^4\ln g \mu^4$: their leading effect comes in the form of the well-known one-loop ``ring sum'', i.e.~the pressure of non-interacting but HTL-dressed gluons,
\begin{eqnarray}
\!\!\!\! p_\text{HTL}^\text{ring}&=&-\frac{(d-1)d_A}{2}\int_P\ln\bigg[1+\frac{\Pi_T(P)}{P^2}\bigg] - \frac{d_A}{2}\int_P\ln\bigg[1+\frac{\Pi_L(P)}{P^2}\bigg]\, ,\; \label{eq:HTLring1}
\end{eqnarray}
which for $p\sim g\mu$ clearly produces a result of parametric order $g^4\mu^4$. Due to the fact that this is a one-loop integral, only one logarithm can arise from the integration. It is interesting to compare these observations to the emergence of soft contributions both in the case of thermal photon production, discussed in Sec.~\ref{sec_soft_collinear}, and bulk thermodynamic quantities at high temperature, cf.~Sec.~\ref{highTpower}. Clearly, the order at which IR sensitive field modes produce the first non-analytic term in the weak-coupling expansion of a given quantity depends sensitively on both the nature of the observable in question and the values of $T$ and $\mu$ in the system under inspection.

Finally, an important simplification occurs if we are only after the coefficients of the logarithms arising from the above resummed diagrams. The appearance of these logs can namely be traced back to so-called ``semisoft'' momenta, satisfying $g\mu\ll P\ll \mu$ \cite{Gorda:2018gpy}, which allow for particularly useful approximations.\footnote{This statement originates from the simple observation that logarithms of $\alpha_s$ necessarily originate from integrals where a logarithmic IR divergence is cured by dynamics at the scale $g\mu$, i.e.~from contributions to the pressure proportional to $\int_{g\mu}^\mu \frac{dp}{p}\sim \ln g$.} In this kinematic regime, we may continue to use the HTL limit for the self energies but simultaneously expand the propagators in powers of the HTL self energy. Finally, as noted in \cite{Gorda:2018gpy}, at least the leading logarithms $g^4\ln g$ and  $g^6\ln^2g$ can be obtained utilizing one further simplification, namely replacing the HTL self energies by their on-shell limits, whereby they reduce to the simple forms $\Pi_T(iP_0=p,p)= M_\infty^2$, $\Pi_L(iP_0=p,p)= 0$. In the case of Eq.~(\ref{eq:HTLring1}) above, this makes the determination of the coefficient of the ${\mathcal O}(g^4\ln g)$ term downright trivial. First, expanding the logarithms, replacing the self energies by their on-shell limits, and noting that logarithmically divergent massive integrals are the only ones capable of producing logs, one obtains
\begin{eqnarray}
p_\text{HTL}^\text{ring}&=&-\frac{(d-1)d_A}{2}\int_P\bigg(\frac{M_\infty^2}{P^2}-\frac{M_\infty^4}{2(P^2)^2}+\cdots\bigg) \nonumber\\ 
&=&\frac{(d-1)d_A M_\infty^4}{4}\int_P\frac{1}{(P^2)^2}+{\mathcal O}(g^4)\, .
\end{eqnarray}
Next, we may concentrate on the part of the $P$-integral running between $\Lambda_1\sim g\mu$ and $\Lambda_2\sim \mu$, which produces upon setting $d=3$
\begin{eqnarray}
p_\text{HTL}^\text{ring}&=&\frac{d_A M_\infty^4}{2}\int_P\frac{1}{(P^2)^2}+{\mathcal O}(g^4) \;=\; 
\frac{d_A M_\infty^4}{(4\pi)^2}\int_{\Lambda_1}^{\Lambda_2}\frac{dP}{P}+{\mathcal O}(g^4)\nonumber\\
&=&-\frac{d_A M_\infty^4}{(4\pi)^2}\ln g +{\mathcal O}(g^4)\, .
\end{eqnarray}
This can easily be verified to coincide with the known ${\mathcal O}(g^4\ln  g)$ term in the weak-coupling expansion of the pressure, originally derived in a considerably more cumbersome fashion \cite{Freedman:1976ub}.

The current state-of-the-art pressure calculation of order $g^6\ln^2 g$, performed in \cite{Gorda:2018gpy}, utilizes the two-loop pressure of the HTL effective theory, derived in \cite{Andersen:2002ey}, and the above observation of all logarithms originating from the semisoft momentum scale. Even the next order  $g^6\ln  g$ in the weak-coupling expansion can be obtained with closely related methods, and it is only at the full order $g^6$ that one needs to e.g.~perform the daunting task of evaluating all full theory four-loop vacuum diagrams (albeit with no resummations). Completing this order in the expansion will be a task qualitatively harder than figuring out the coefficients of the logarithms discussed above.

\subsubsection{Small but nonzero temperatures
\label{smallbutnonzeroTsec}}

Although the history of perturbative computations in both the limits of high temperatures and  $T=0$ is extensive, the case of small but nonzero temperatures received far less attention until the early 2000's. At that point, it was first discovered that both in QED and QCD, the low-temperature specific heats display an anomalous ``non-Fermi-liquid'' behavior in the $T=0$ limit \cite{Ipp:2003cj,Gerhold:2004tb}, shortly after which a proof-of-principle calculation was completed for the QCD pressure that covered all values of the temperature \cite{Ipp:2006ij}. All these computations, however, utilized machinery that is somewhat outdated by modern standards, and we shall therefore not discuss them further here.

The state-of-the-art framework designed for evaluating bulk thermodynamic quantities in QCD at arbitrary values of $T/\mu$ was introduced in \cite{Kurkela:2016was} and largely follows the ideas laid out in the previous subsection. The key observation made in this work --- which actually predates the $T=0$ calculation of \cite{Gorda:2018gpy} by more than two years --- was to note that at low orders in perturbation theory one can replace the two terms inside the parentheses in Eq.~(\ref{presorg}) by their counterparts determined in a (yet unspecified) effective description for the IR sensitive degrees of freedom of QCD. This leads to the result
\begin{eqnarray}
p_\text{QCD}^\text{resummed} &=& p_\text{IR}^\text{resummed} - p_\text{IR}^\text{naive} + p_\text{QCD}^\text{naive}\, , \label{eq:pIRres}
\end{eqnarray}
where ``IR'' refers to the soft effective theory. The justification for this is simple: those parts of $p_\text{QCD}^\text{resummed} - p_\text{QCD}^\text{naive}$ that are not correctly reproduced by the effective IR theory exactly cancel in the difference of the two terms. Similarly, when using Eq.~(\ref{eq:pIRres}) one needs not worry excessively about the identity of the field modes for which the IR theory is built, as long as all modes in need of resummation are included. Should some modes of the IR theory be hard and contribute to the pressure only perturbatively, the subtraction of the ``naive'' term will make sure that they are not double counted in the final result.

To optimally exploit the above insights in the determination of the QCD pressure at small but nonzero temperatures, one may use a mixture of the EQCD and HTL effective descriptions for the different Matsubara modes. Indeed, in the treatment of \cite{Kurkela:2016was}, the bosonic zero mode sector of the theory was described via EQCD, leaving the HTL effective theory to resum all other soft contributions that contribute in particular at small temperatures. This lead one to the decomposition, valid to order $g^5$, 
\begin{eqnarray}
\!\!\!\! p_\rmi{QCD} &=& p_\rmi{QCD}^\rmi{naive} + \underbrace{p_\rmi{DR}^\rmi{res} - p_\rmi{DR}^\rmi{naive}}_{p_\rmi{DR}^\rmi{corr}} + \underbrace{p_\rmi{HTL}^\rmi{res} -  p_\rmi{HTL}^\rmi{naive}}_{p_\rmi{HTL}^\rmi{corr}}, \label{resgen}
\end{eqnarray}
where $p_\rmi{DR}$ stands for the EQCD pressure and $p_\rmi{HTL}$ for the HTL ring sum, with the zero Matsubara mode contribution excluded. For further details of this result, we refer the interested reader to the original reference \cite{Kurkela:2016was}.

%% file: imagtimeresults.tex
% !TEX root = review.tex

\section{Applications of the imaginary time formalism}
\label{sec:imagtimeresults}

The most important applications of the imaginary time formalism in thermal QCD concern the determination of various bulk thermodynamic quantities, typically performed in the grand canonical ensemble. The most fundamental of these quantities is the grand potential itself, giving the pressure as a function of temperature and quark chemical potentials, from which several other quantities can be derived using simple thermodynamic relations. Noteworthy examples are e.g.~the trace anomaly $\epsilon-3p$ that measures the deviation of the system from the conformal limit, as well as quark number susceptibilities that probe the effects of finite density while being measurable using nonperturbative lattice simulations. The Equation of State, or the functional relationship between the pressure and energy density, can also be determined from the grand potential as soon as the values of the quark chemical potentials are fixed through e.g.~requirements of charge neutrality and beta equilibrium.

As discussed in some length above, the history of thermal perturbation theory is plagued by problems related to the contributions of infrared sensitive soft field modes to physical quantities, ultimately leading to the breakdown of naive, and sometimes even resummed, weak coupling expansions \cite{Linde:1980ts}. Already at relatively low loop orders, these issues lead to a poor convergence of perturbative results when presented in terms of (generalized) power series in the gauge coupling of the full theory, $g$. Until roughly the turn of the millennium, these problems were thought to completely invalidate the use of perturbation theory in thermal QCD, but several advances since then have improved the situation considerably.

In this section, we demonstrate that systematic efforts to build an effective description for the IR degrees of freedom in thermal QCD have lead to a qualitative improvement in the status of weak coupling calculations. At high temperatures, two popular frameworks have been introduced for this purpose: Dimensional Reduction, building on the effective theory EQCD \cite{Kajantie:1997tt,Braaten:1995cm,Braaten:1995jr}, and Hard Thermal Loop perturbation theory, or HTLpt \cite{Andersen:1999fw}, which 
were the subject of Sections~\ref{sec_DR} and \ref{sec:htlnpointeffaction} respectively. Both of these setups offer systematically improvable schemes for resumming weak coupling expansions of bulk thermodynamic quantities, which will be seen to significantly improve the convergence of perturbative expansions and extend the applicability of the weak coupling method to moderately low energy densities. It should be stressed, though, that perturbative weak coupling calculations always miss some nonperturbative contributions that become increasingly important at low energies. For this reason, their use is restricted to the description of the deconfined phase of QCD, and in particular cannot be used to directly probe the phase structure of the theory, such as the existence of a possible tricritical point.

When discussing bulk thermodynamic quantities, it is important to distinguish between the regions of small (or vanishing) and sizable baryon densities. In the former, lattice QCD remains applicable, and as a first-principles nonperturbative method provides reliable results for quantities such as the EoS, trace anomaly, and quark number susceptibilities. At the same time, it serves as an efficient test bed for the predictions of perturbation theory, whose true value becomes apparent in particular at larger baryon densities. Nonzero quark chemical potentials namely provide no obstacle for weak coupling calculations, whereas they are known to invalidate lattice simulations due to the infamous Sign Problem \cite{deForcrand:2010ys}. At the moment, the results of lattice QCD can be extended at most to chemical potentials $\mu_B\approx \pi T$, where the most useful tool has turned out to be the so-called Taylor expansion method (see e.g.~\cite{Karsch:2010hm,Endrodi:2011gv} and references therein). At higher densities, perturbation theory on the other hand remains the only first principles computational method available.

In the following, we shall explore in detail the predictions of modern thermal perturbation theory for the most important bulk thermodynamic quantities describing deconfined QCD matter, comparing the DR and HTLpt results to each other and to those of lattice QCD whenever the latter are available. The results are divided into two subsections: First, we cover the EoS and trace anomaly at zero density and then explore the effects of small but nonzero density by considering quark number susceptibilities as well as the pressure and trace anomaly at moderate values of $\mu_B$. After these topics, we move on to the thermodynamics of cold and dense quark matter, where the DR and HTLpt resummations are no longer valid in their standard form, but one needs to find fundamentally new ways to deal with IR physics. Finally, to close our discussion, in Sec.~\ref{sec:beyondqcd} we  briefly comment on observables beyond bulk thermodynamic quantities, such as different Euclidean correlators, that have been determined within the imaginary time formalism. In this section, we also extend our discussion to theories other than QCD, covering similar calculations in the weakly interacting part of the Standard Model, various Beyond the Standard Model theories, as well as e.g.~${\mathcal N}=4$ Super Yang-Mills theory.

In all results discussed in this section, we consider the case of three colors, $N_c=3$, and three \textit{massless} dynamical quark flavors, i.e.~set $N_f=3$. This choice is natural considering that in most physical applications the up and down quark masses are clearly negligible and even the strange quark mass of ${\mathcal O}$(100 MeV) can be considered small, while the three heavier quarks have not been excited. The order of the running gauge coupling is always chosen to be consistent with the loop order of the perturbative result. This means that in the DR result, containing all perturbative contributions up to and including order $g^5$,\footnote{The justification for not including the partially known ${\mathcal O}(g^6)$ term in the DR pressure will be presented shortly.} we employ the two-loop running coupling, while in the three-loop HTLpt result we use the one-loop $\alpha_s$. In both cases, the QCD scale $\lambdamsbar$ is fixed such that $\alpha_s=0.326$ at the energy scale of 1.5 GeV \cite{Bazavov:2012ka}. For the two-loop running, this results in $\lambdamsbar=283$ MeV, and for one-loop running in $\lambdamsbar=176$ MeV.  In all cases, the renormalization scale $\bar{\Lambda}$ is varied by a factor of 2 around a midpoint value, which is chosen as $2\pi T$ at zero density and $2\mu_B/3$ at zero temperature; between these two extremes, it is natural to choose the parameter to be the root sum square of the $\mu=0$ and $T=0$ values. 

Finally, we note that while we aim to provide a rather comprehensive look at topical applications of the imaginary time formalism, the list of references we provide is by no means exhaustive. For one thing, we concentrate on perturbative field theory, and thus only refer to lattice works when comparing to specific nonperturbative results, and secondly, we give emphasis on recent state-of-the-art works, and only cite the most important historical references. For a more comprehensive list of references, we refer the reader to ref.~\cite{Brambilla:2014jmp}.

\subsection{Bulk thermodynamics at vanishing density}

We begin from the bulk thermodynamic properties of hot QGP at vanishing quark chemical potentials, i.e.~at zero baryon (and isospin) density. On the HTLpt side, the state-of-the-art three-loop pressure and trace anomaly can be found from ref.~\cite{Haque:2014rua}, which followed a series of earlier works, including most importantly \cite{Andersen:2002ey,Andersen:2003zk,Andersen:2010ct,Andersen:2011sf}. The DR results we use are on the other hand based on the ${\mathcal O}(g^5)$ work of \cite{Zhai:1995ac}
(see also the earlier works of \cite{Shuryak:1977ut,Kapusta:1979fh,Toimela:1982hv,Arnold:1994ps,Arnold:1994eb}),
but a crucial extra ingredient is the resummation proposed in \cite{Blaizot:2003iq,Laine:2006cp}. This resummation amounts to presenting the result as a function of the EQCD parameters, and not expanding it in powers of the full-theory coupling, which has been seen to significantly improve its convergence properties. 

In connection with the QCD pressure, we note that the $g^6\ln\,g$ term in the weak coupling expansion of the quantity has been determined in \cite{Kajantie:2002wa}, and even certain parts of the full four-loop result of ${\mathcal O}(g^6)$ are known by now \cite{Kajantie:2003ax,DiRenzo:2006nh,Gynther:2009qf}. We have, however, decided to not use these terms in our results, owing to the ambiguity related to choosing the ``constant inside the log'' within the $g^6\, \ln \, g$
term that has a sizable impact on the result. It has been demonstrated in \cite{Laine:2006cp} that fitting this single parameter to lattice results at low temperatures leads to excellent agreement with lattice data over a wide temperature range, and to this end, the DR results we display may be rightfully considered to not represent the current state of the art. The upshot of our convention is, however, that no optimization of the result is required --- or even possible --- and that no complications arise when proceeding to nonzero density or quark number susceptibilities. In this respect, our results differ from those presented in \cite{Andersen:2012wr}, and are in fact presented here for the first time.

%%%%%%%%%%%%%%%%%%%%%%%%%%%%%%%%%%%%%%%%%%%%%%%%%%%%
\begin{figure}[t]
\begin{center}
\includegraphics[width=0.48\linewidth]{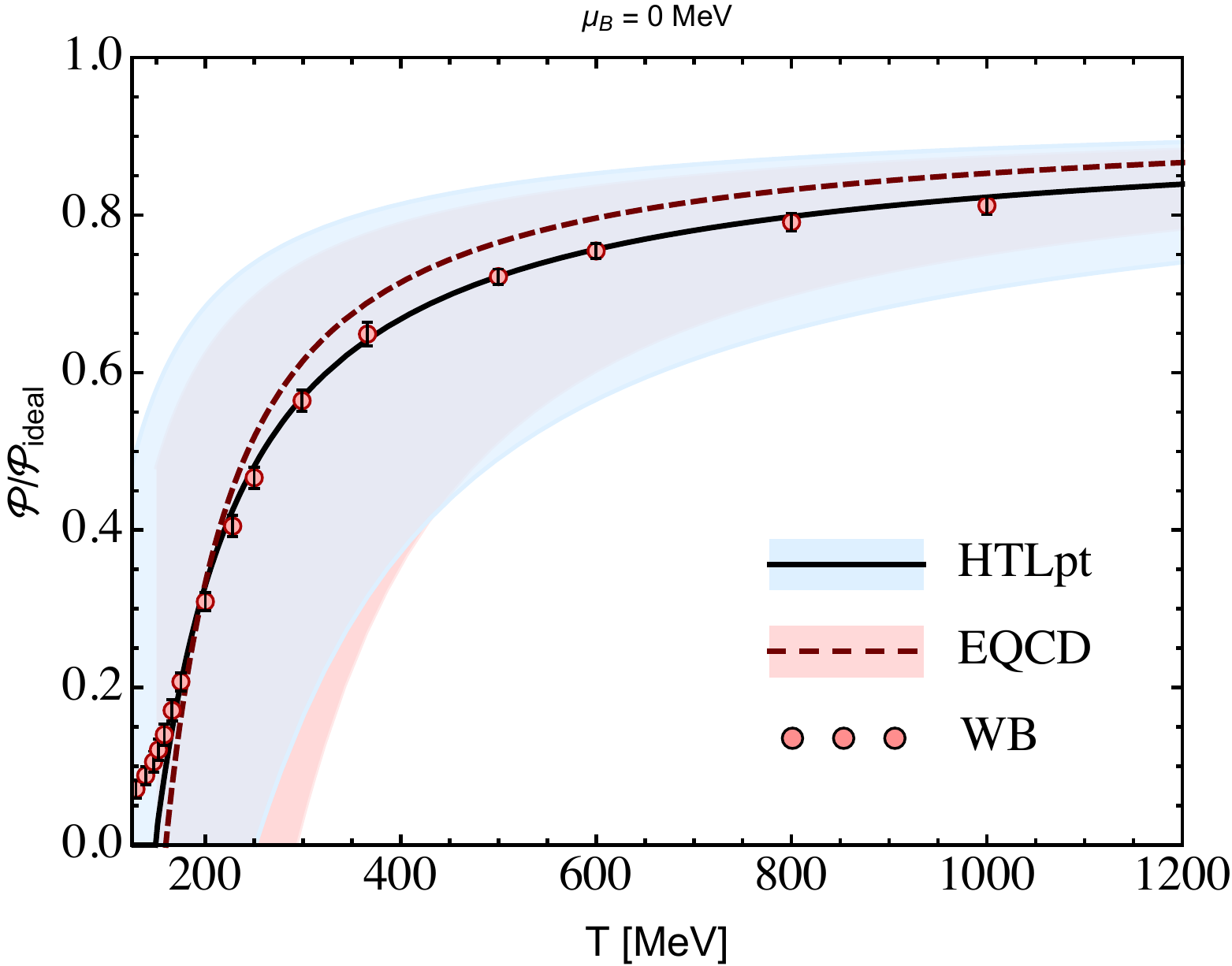}$\;\;\;$
\includegraphics[width=0.48\linewidth]{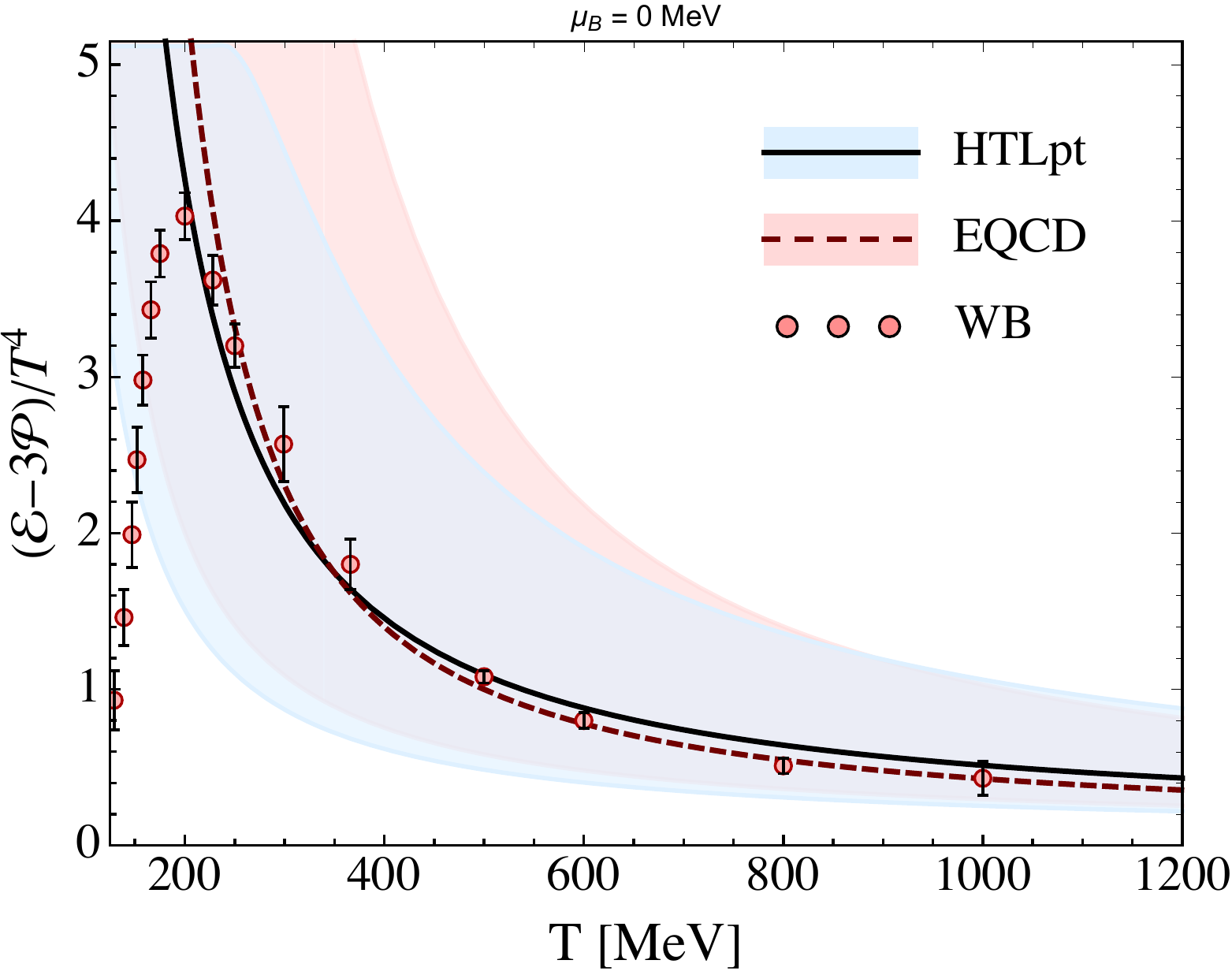}
\end{center}
\caption{The QCD pressure and trace anomaly at $\mu_B=0$.  In both panels we compare the perturbative results with lattice data from the  Wuppertal-Budapest (WB) collaboration~\cite{Borsanyi:2010cj}.}
\label{fig:thermo0}
\end{figure}
%%%%%%%%%%%%%%%%%%%%%%%%%%%%%%%%%%%%%%%%%%%%%%%%%%%%

In fig.~\ref{fig:thermo0}, we display the two most important quantities characterizing the bulk thermodynamic properties of zero-density QGP: the pressure and trace anomaly as functions of temperature. We observe that the HTLpt and DR predictions are in remarkably good agreement with each other, and furthermore that they correctly capture the behavior of the lattice results of \cite{Borsanyi:2010cj} down to temperatures of the order of 200 MeV. The midpoint values of the renormalization scale even turn out to reside extremely close to the datapoints for a wide temperature range, but this is likely a fortuitous coincidence and should not be given too much weight.

\subsection{Probing nonzero densities}

%%%%%%%%%%%%%%%%%%%%%%%%%%%%%%%%%%%%%%%%%%%%%%%%%%%%
\begin{figure}[t]
\begin{center}
\includegraphics[width=0.48\linewidth]{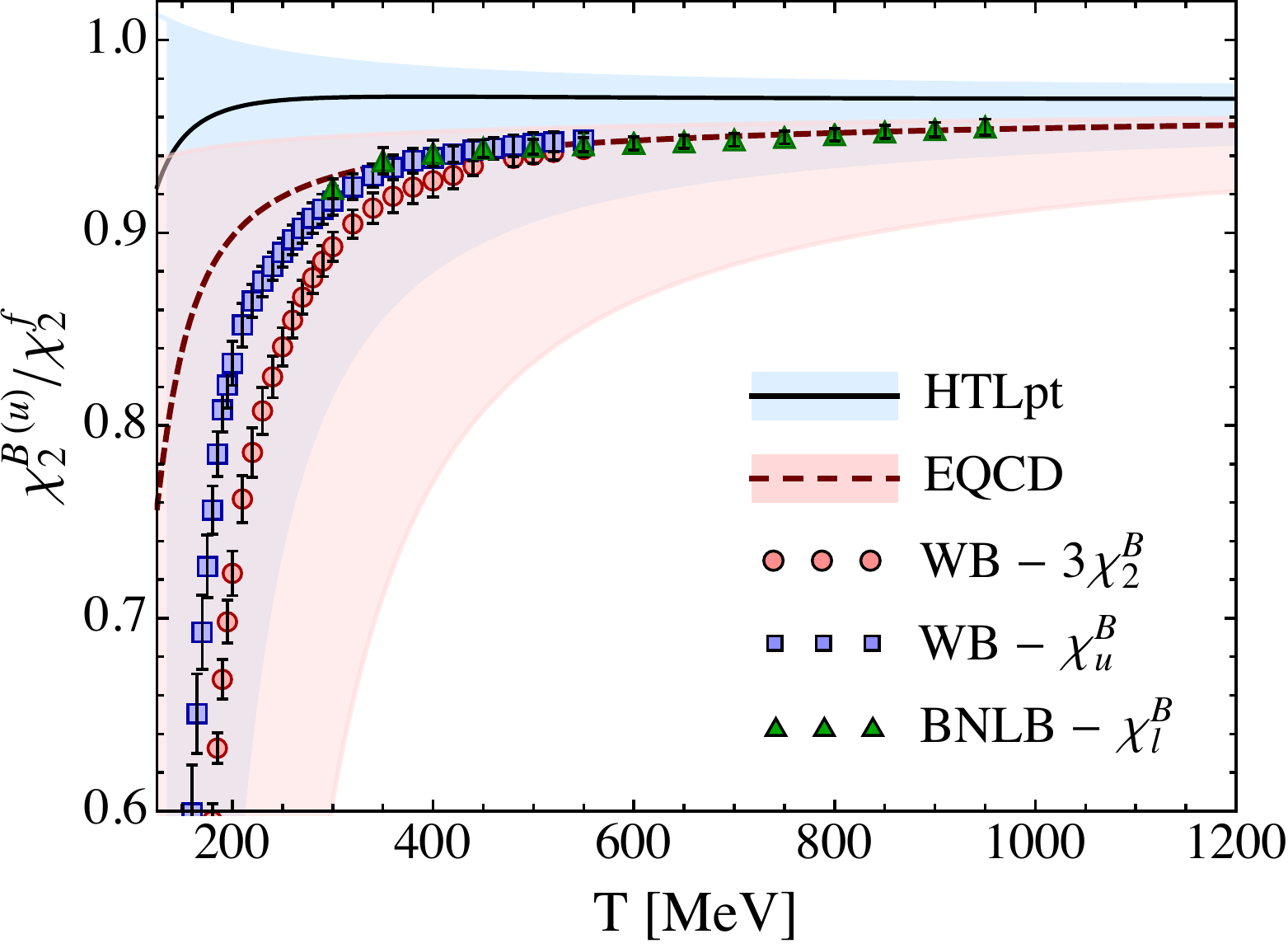}$\;\;\;$
\includegraphics[width=0.48\linewidth]{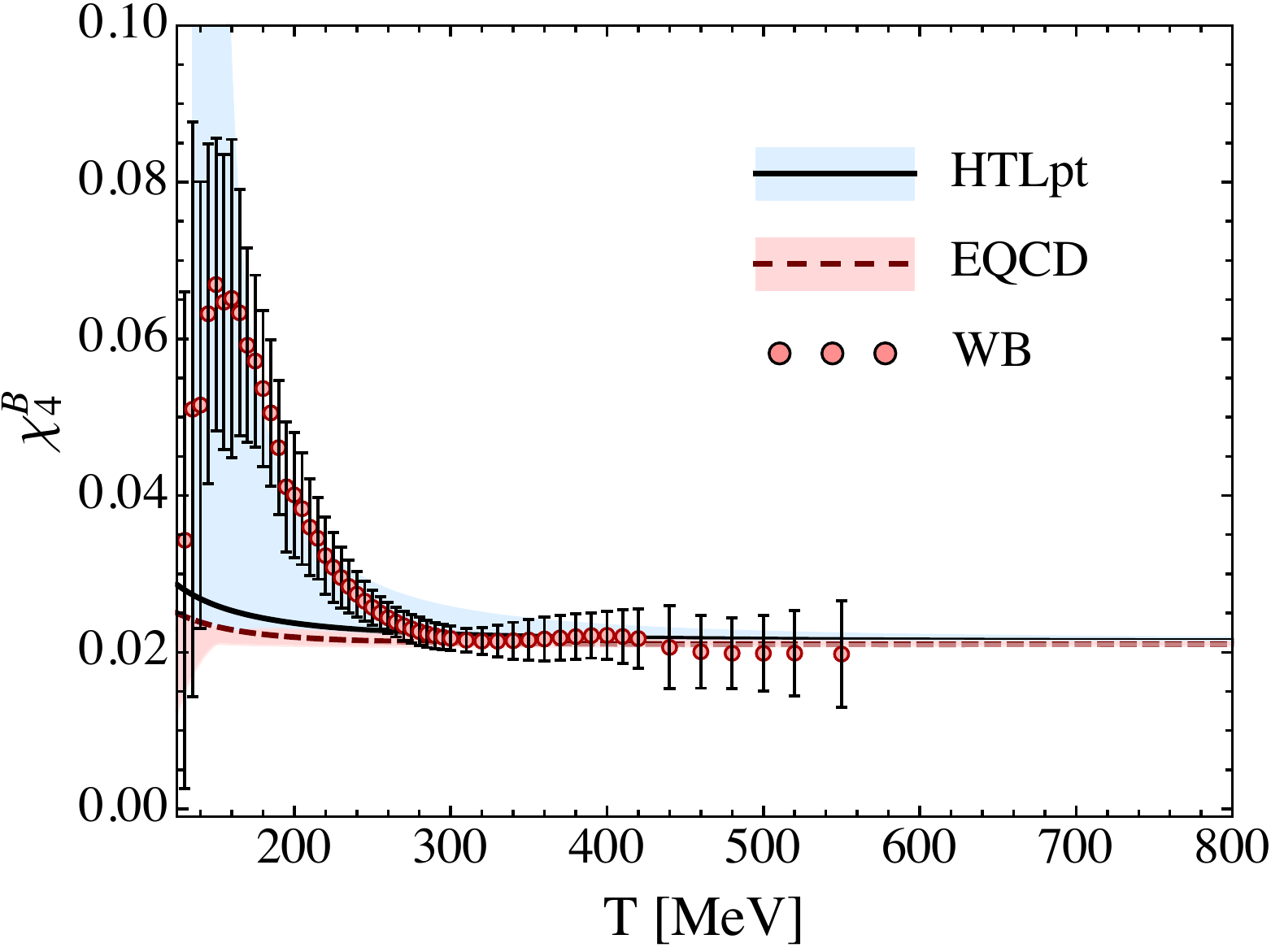}
\end{center}
\caption{Left: the 2nd order light quark (and baryon) number susceptibilities. Right: The 4th baryon number susceptibility.  In both panels we compare with lattice data from the  Wuppertal-Budapest (WB) \cite{Borsanyi:2012rr,Borsanyi:2013hza} and BNLB  collaborations \cite{Bazavov:2013uja}.}
\label{fig:suscept1}
\end{figure}
%%%%%%%%%%%%%%%%%%%%%%%%%%%%%%%%%%%%%%%%%%%%%%%%%%%%

Next, we move on to quantities that probe the finite-density part of the QCD phase diagram, yet are directly measurable on the lattice, i.e.~various susceptibilities. These quantities are defined as derivatives of the pressure with respect to different chemical potentials dual to conserved charges. A commonly studied subclass are the diagonal and off-diagonal quark number susceptibilities (QNSs)
\begin{eqnarray}
\chi_{ijk}\left(T\right)&\equiv&\frac{\partial^{i+j+k}\; p\left(T,\mu_u,\mu_d,\mu_s\right)}{\partial\mu_u^i\, \partial\mu_d^j \, \partial\mu_s^k}\bigg|_{\mu=0} \, ,
\end{eqnarray}
where the indices $u,\,d,\,s$ refer to the three lightest quark flavors. Alternatively, we may consider derivatives with respect to chemical potentials corresponding to the baryon number $B$, electric charge $Q$, and strangeness $S$, related to the quark chemical potentials via
\begin{eqnarray}
\mu_u&=&\frac{1}{3}\mu_B+\frac{2}{3}\mu_Q,\\
\mu_d&=&\frac{1}{3}\mu_B-\frac{1}{3}\mu_Q,\\
\mu_s&=&\frac{1}{3}\mu_B-\frac{1}{3}\mu_Q-\mu_S.
\end{eqnarray}
From these results, it is trivial to derive linear relations between susceptibilities in the $\{u,d,s\}$ and $\{B,Q,S\}$ bases.

\begin{figure}[t]
\begin{center}
\includegraphics[width=0.48\linewidth]{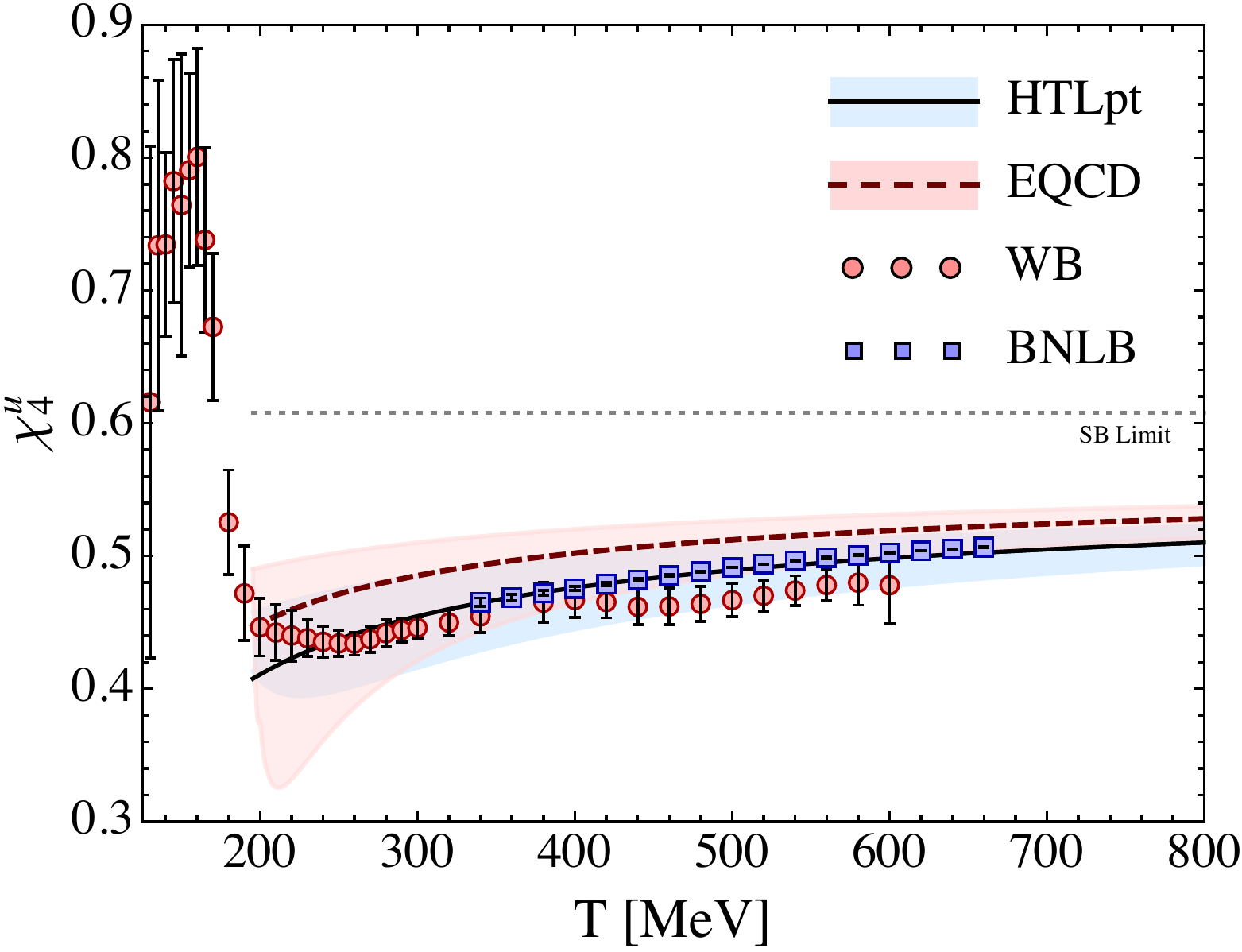}$\;\;\;$
\includegraphics[width=0.48\linewidth]{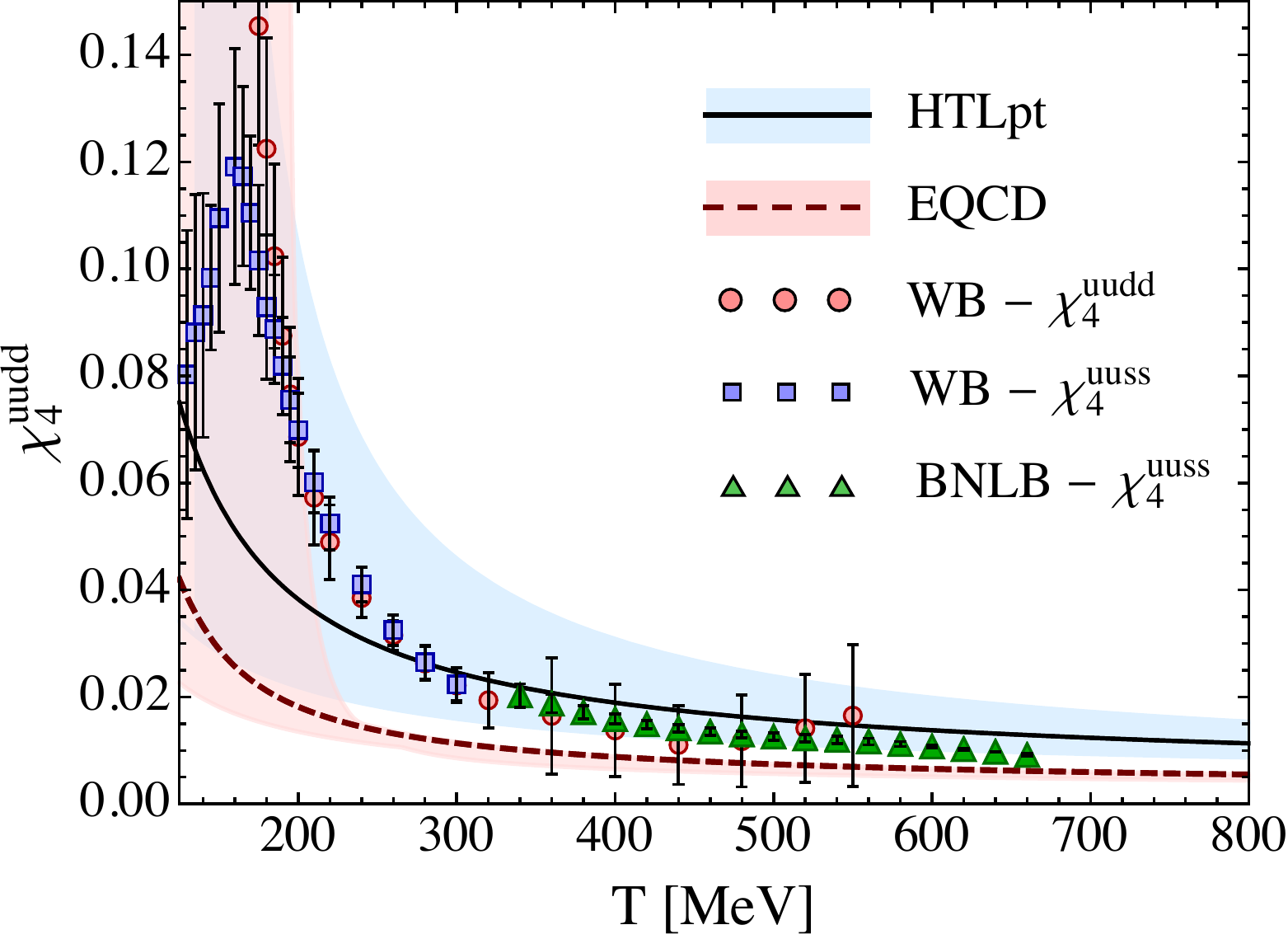}
\end{center}
\caption{Left: The 4th order diagonal light QNS. Right: The 4th order off-diagonal light QNS.  In both panels we compare with lattice data from the  Wuppertal-Budapest (WB) \cite{Borsanyi:2012rr,Borsanyi:2013hza} and BNLB  collaborations \cite{Ding:2015fca}.}
\label{fig:suscept3}
\end{figure}

Different susceptibilities have been considered up to the full two- and three-loop orders within the HTLpt framework in \cite{Andersen:2012wr,Haque:2013qta,Haque:2013sja}, respectively, and up to ${\mathcal O}(g^6\ln\,g)$ using the DR resummation \cite{Vuorinen:2002ue,Andersen:2012wr} (see also refs.~\cite{Blaizot:2001vr,Blaizot:2002xz} for related work). In fig.~\ref{fig:suscept1} (left), we first look at the second order diagonal QNS $\chi_2 \equiv \chi_{uu}$, which coincides with the corresponding baryon number susceptibility up to a rescaling. We observe a good agreement of both the HTLpt and DR bands with lattice data, although the midpoint of the DR one happens to lie somewhat closer to the lattice results. This is, however, clearly coincidental, as for the fourth order baryon and quark number susceptibilities, shown in figs.~\ref{fig:suscept1} (right) and \ref{fig:suscept3} (left), the situation is much less clear. In all three cases, we conclude that lattice data are well described by our perturbative predictions from temperatures of ca.~300 MeV onwards, with the latter missing only the dramatic rise of the lattice results for the fourth order susceptibilities at low $T$.

Finally, in fig.~\ref{fig:suscept3} (right) we display the fourth order off-diagonal susceptibility $\chi_4^{uudd}$, which has the interesting feature that its weak coupling expansion begins only at order $g^3$. In this case, we see that it is the HTLpt result that appears to provide a better description of the lattice data, with the possible exception of the very highest temperatures. This time, the increasing trend of the lattice data at decreasing temperature is at least partially reflected in the perturbative results. The lattice data used in our four figures are from \cite{Borsanyi:2012rr,Borsanyi:2013hza,Bazavov:2013uja,Ding:2015fca}.

%%%%%%%%%%%%%%%%%%%%%%%%%%%%%%%%%%%%%%%%%%%%%%%%%%%%
%%%%%%%%%%%%%%%%%%%%%%%%%%%%%%%%%%%%%%%%%%%%%%%%%%%%

\begin{figure}[t]
\begin{center}
\includegraphics[width=0.48\linewidth]{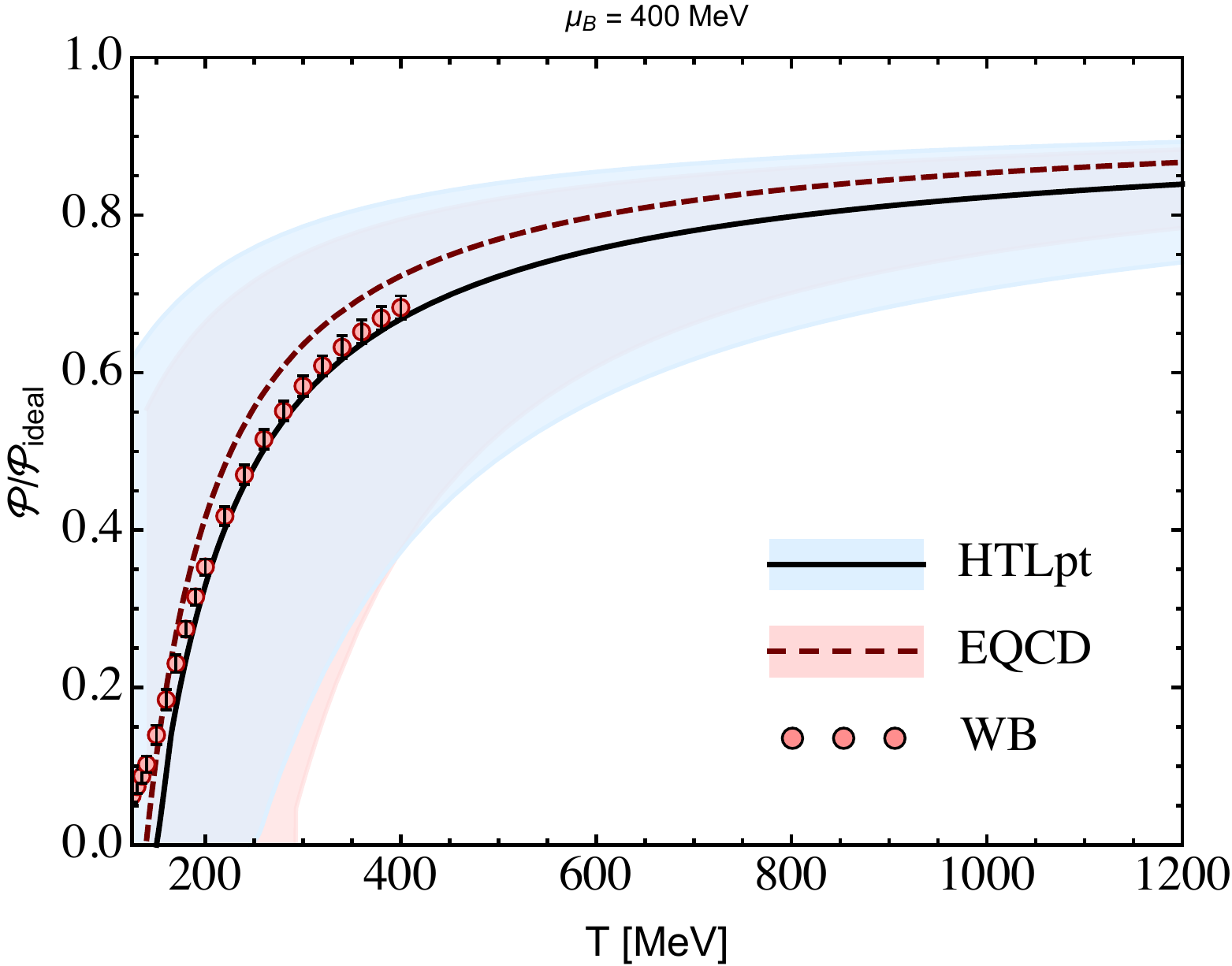}$\;\;\;$
\includegraphics[width=0.48\linewidth]{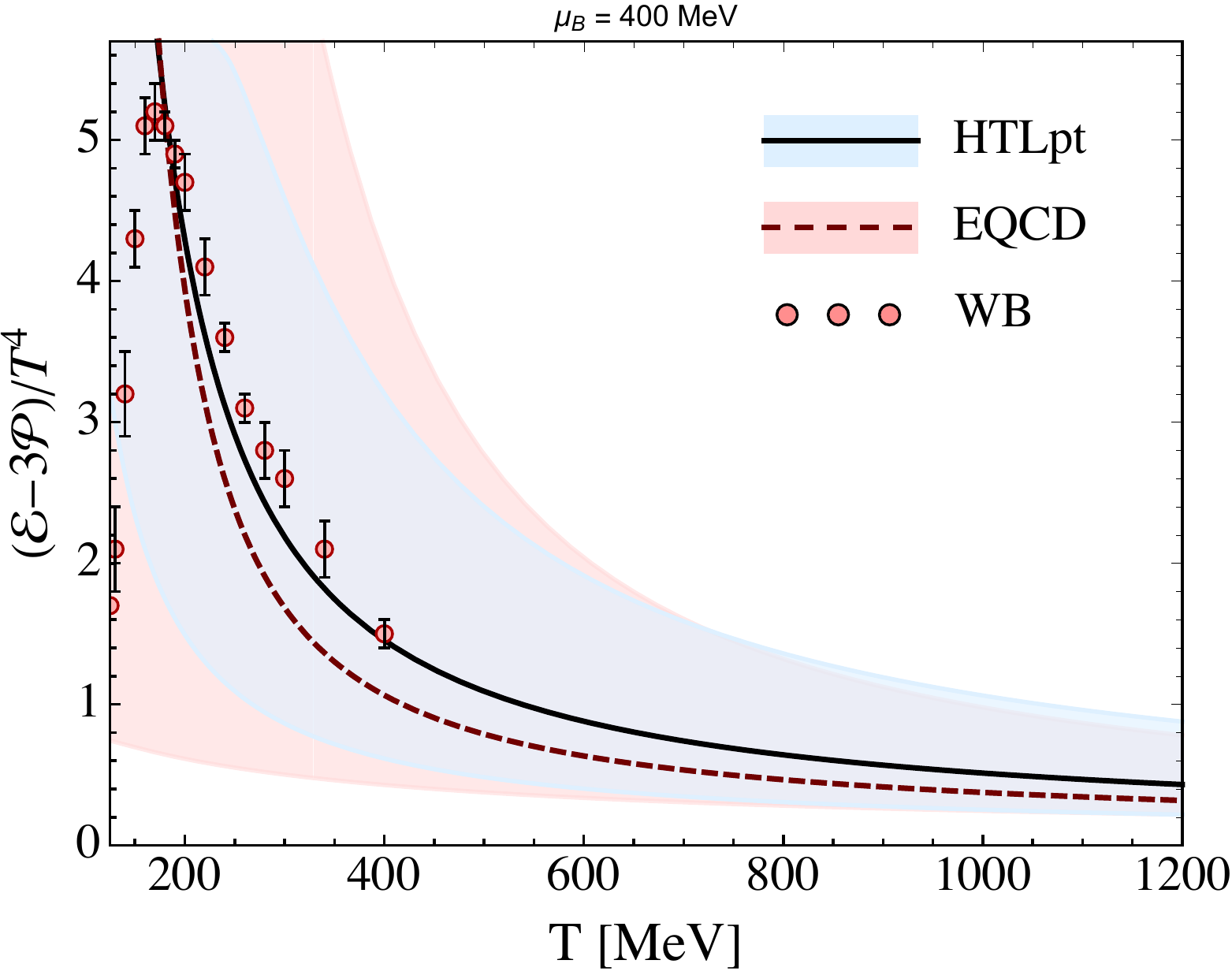}
\end{center}
\caption{The pressure and trace anomaly in three-flavor QCD at $\mu_B=400$ MeV.  In both panels, we compare our results with lattice data from the Wuppertal-Budapest (WB) collaboration~\cite{Borsanyi:2012cr}.}
\label{fig:thermomu}
\end{figure}

While the susceptibilities are typically determined at zero chemical potentials and are therefore computable on the lattice, they also allow predicting the behavior of the Equation of State at small but nonvanishing densities via a Taylor expansion of the quantity in powers of $\mu/T$. This facilitates a comparison of our analytic results for the finite-density EoS with lattice data at small and moderate values of $\mu_B$, which we choose to fix to 400 MeV in fig.~\ref{fig:thermomu}. The qualitative features seen in these plots for the pressure and trace anomaly remain similar to the case of vanishing density, and the perturbative regime again appears to begin at temperatures around 200 MeV. It should, however, be noted that despite these successes the details of the QCD phase diagram, such as the existence and location of a possible tricritical point, are outside the scope of such perturbative studies.

\subsection{Cold and cool quark matter}

The value of perturbative methods in the description of bulk thermodynamic quantities becomes most pronounced in the limit where (some) chemical potentials become larger than roughly $\pi T$. In this region, the cornerstone method of modern lattice QCD at nonzero density, Taylor expansions,  run into serious problems and can no longer be used to reliably estimate the behavior of thermodynamic quantities. This is a severe restriction in particular for the study of cold quark matter, relevant for the physics of neutron star cores, and implies that the only first principles quantum field theory method available to tackle the problem is perturbative QCD.

%%%%%%%%%%%%%%%%%%%%%%%%%%%%%%%%%%%%%%%%%%%%%%%%%%%%
%%%%%%%%%%%%%%%%%%%%%%%%%%%%%%%%%%%%%%%%%%%%%%%%%%%%

Recalling that thermal perturbation theory does not suffer from a Sign Problem of any kind, it should not come as a surprise that the perturbative results discussed in the above subsections remain valid all the way to very small temperatures. It is in fact only in the limit where the electric screening scale $m_D$, proportional to $g\mu$ at small $T$, becomes of the same order as the temperature that the treatment of infrared physics via the dimensionally reduced theory EQCD becomes problematic \cite{Ipp:2006ij}. This issue is related to the fact that while at high and moderate temperatures the IR sensitive field modes are all static, i.e.~three-dimensional, this is no longer the case at zero or very small $T$, where the discrete Matsubara frequencies merge into a continuous momentum variable $p_0$, as explained in fig.~\ref{fig:smallT} of the previous section. As briefly discussed there, the problem of constructing an optimal effective description for the soft modes at any temperature was resolved in \cite{Kurkela:2016was}, where a novel resummation scheme was introduced by combining an EQCD treatment for the static sector of the theory with an HTL resummation of the nonstatic modes. The new scheme was used in \cite{Kurkela:2016was} to derive an ${\mathcal O}(g^5)$ result for the QCD pressure, valid at \textit{all} ratios of $\mu$ and $T$.

%%%%%%%%%%%%%%%%%%%%%%%%%%%%%%%%%%%%%%%%%%%%%%%%%%%%
\begin{figure}[t]
\begin{center}
\includegraphics[width=0.47\linewidth]{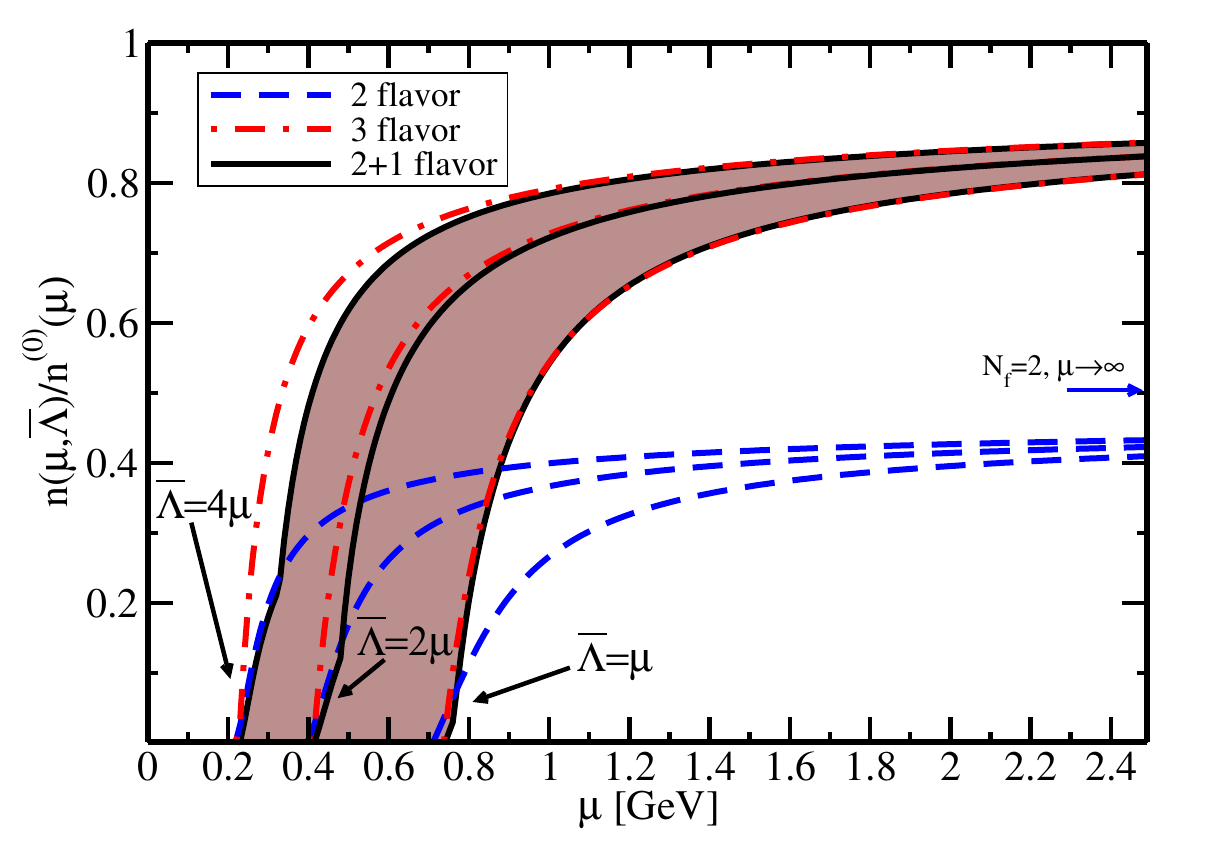}$\,$
\includegraphics[width=0.51\linewidth]{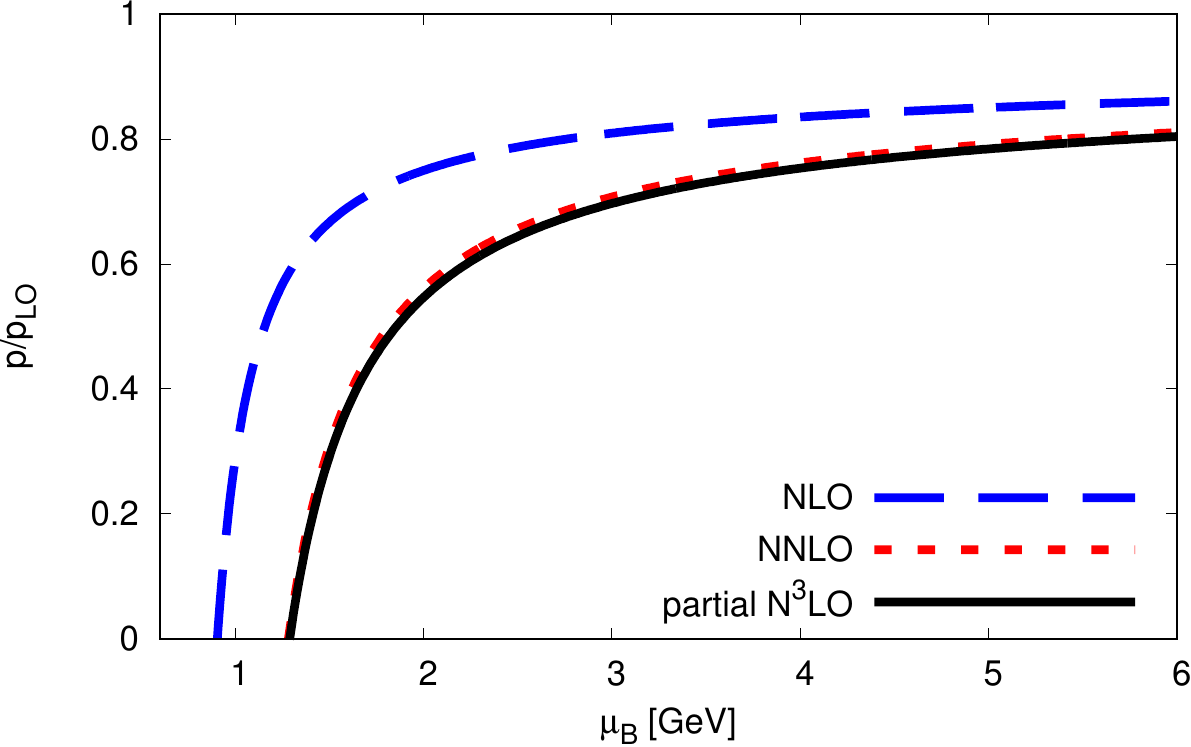}
\end{center}
\caption{Left: the baryon number density at $T=0$ as a function of quark chemical potential, with the limits of beta equilibrium and charge neutrality implemented. The figure has been taken from \cite{Kurkela:2009gj}, i.e.~a three-loop calculation featuring nonzero strange quark mass. Right: The effect of the leading four-loop logarithm of order $g^6\ln^2 g$ on the $T=0$ pressure of massless three-flavor QCD, obtained by fixing the renormalization scale to its mid value (cf.~the left figure) and displaying the pressure at three successive orders: NLO or ${\mathcal O}(g^2)$, NNLO or ${\mathcal O}(g^4)$, and  partial NNNLO or ${\mathcal O}(g^6\ln^2 g)$.}
\label{fig:thermoT0}
\end{figure}
%%%%%%%%%%%%%%%%%%%%%%%%%%%%%%%%%%%%%%%%%%%%%%%%%%%%

Before moving on to practical results, let us briefly discuss one subtlety inherent in all applications of thermal perturbation theory to the context of neutron stars, where --- unlike in applications motivated by heavy-ion physics --- weak interactions can typically \emph{not} be ignored. This implies that different quark numbers are no longer conserved quantities, but nontrivial relations exist between their respective chemical potentials. In quiescent neutron stars, matter is typically taken to be locally charge neutral and in chemical (beta) equilibrium. The former of these requirements can be represented as the simple condition 
\begin{equation}
\frac{2}{3}n_u -  \frac{1}{3}n_d -\frac{1}{3}n_s -n_e  =0, \label{neutrality}
\end{equation}
where the four functions stand for the number densities of the up, down and strange quarks as well as electrons. Chemical equilibrium on the other hand relies on the processes
\begin{eqnarray}
d\rightarrow u+e+\bar{\nu}_e,& & u+e\rightarrow d+\nu_e\, , \\ \nonumber
s\rightarrow u+e+\bar{\nu}_e,& & u+e\rightarrow s+\nu_e\, , \\ \nonumber
s+u &\leftrightarrow& d+u,\nonumber
\end{eqnarray}
which imply the conditions
\begin{equation}
\mu_s=\mu_d\equiv\mu\,,\qquad \mu_u=\mu-\mu_e\, 
\end{equation}
that remain valid assuming neutrinos escape the system quickly and need not be taken into account. These altogether three conditions for the four variables $\mu_u$, $\mu_d$, $\mu_s$, and $\mu_e$ suffice to reduce the number of free parameters to one, which is typically taken to be the down quark chemical potential, also dubbed simply quark chemical potential $\mu$. This means that to obtain the needed EoS from a perturbative calculation, one needs to first evaluate various derivatives of the pressure with respect to the quark chemical potentials, and then (numerically) solve Eq.~(\ref{neutrality}) above.

In the limit of exactly zero temperature, the weak coupling expansion of the QCD EoS has been worked out up to and including the full three-loop order, i.e.~${\mathcal O}(g^4)$ in the gauge coupling, as well as the leading logarithmic term at four loops, of ${\mathcal O}(g^6\ln^2 g)$ \cite{Gorda:2018gpy}. The three-loop result was first derived already in the late 1970s in a calculation that relied heavily on (rather inaccurate) numerics and was performed in the on-shell scheme with zero quark masses \cite{Freedman:1976ub}. The result was subsequently converted to an analytic form in the $\msbar$ scheme nearly 30 years later \cite{Vuorinen:2003fs}, followed by the inclusion of quark masses in \cite{Kurkela:2009gj} (see also ref.~\cite{Kneur:2019tao}). These results are depicted in fig.~\ref{fig:thermoT0} (left), but unfortunately cannot be directly compared to any other first principles calculation due to the Sign Problem of lattice QCD discussed above. Nevertheless, these results have found very important uses in the phenomenology of neutron stars; see e.g.~refs.~\cite{Weissenborn:2011qu,Fraga:2013qra,Kurkela:2014vha,Kojo:2014rca,Heinimann:2016zbx,Gorda:2016uag,Annala:2017llu,Annala:2019puf} and references therein. In the right panel of  fig.~\ref{fig:thermoT0}, we finally display for comparison a figure taken from \cite{Gorda:2018gpy} that displays the numerical effect of the leading four-loop logarithm on the QCD pressure; as can be seen from the plot, this new term plays a minuscule role at practically all densities.

%%%%%%%%%%%%%%%%%%%%%%%%%%%%%%%%%%%%%%%%%%%%%%%%%%%%
\begin{figure}[t]
\begin{center}
\includegraphics[width=\linewidth]{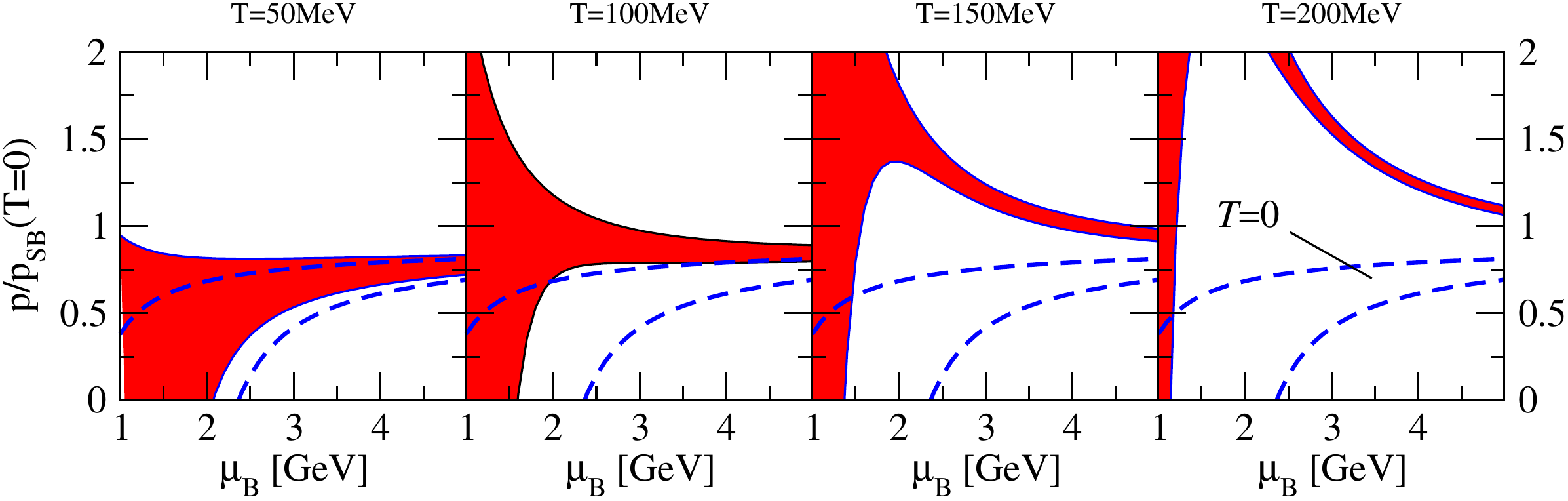}
\end{center}
\caption{The QCD pressure at small fixed values of $T$. The figure is taken from \cite{Kurkela:2016was}.}
\label{fig:thermomu2}
\end{figure}
%%%%%%%%%%%%%%%%%%%%%%%%%%%%%%%%%%%%%%%%%%%%%%%%%%%%

At small but nonzero temperatures, the physically most interesting question is related to the manner, in which the bulk thermodynamic properties of the system transition from a high-temperature behavior towards their $T=0$ limits. Within the past two decades, this problem has been addressed in many different ways, ranging from Hard Dense Loop computations of the low-temperature specific heat \cite{Ipp:2003cj,Schafer:2004zf,Gerhold:2004tb,Gerhold:2005uu}, revealing so-called non-Fermi-Liquid behavior, to explicit resummations of the pressure within full QCD \cite{Ipp:2006ij}. As proposed in these references, and later confirmed in \cite{Kurkela:2016was}, the relevant parameter characterizing the transition between the two regimes is $T/m_D\sim T/(g\mu_B)$, while the leading corrections to the zero-temperature pressure come in the form of logarithms and non-integer powers of this  parameter. In fig.~\ref{fig:thermomu2}, we display the behavior of the EoS as a function of the baryochemical potential for four fixed values of the temperature \cite{Kurkela:2016was}. These results are of direct relevance e.g.~to simulations of neutron-star mergers.

\subsection{Beyond bulk thermodynamics and QCD \label{sec:beyondqcd}}

In addition to the bulk thermodynamic quantities discussed above, the imaginary time
formalism has been applied to the determination of a wide range of complementary
observables in QCD and other quantum field theories. Many of these quantities are
derivable from Euclidean two-point functions, with examples ranging from different
screening masses to spectral functions, available through an analytic continuation
of the correlator to  Minkowskian signature. Below, we briefly review
some example computations in the context of thermal QCD, and thereafter list a number of
key references concerning similar exercises in other relevant QFTs. The list of
references provided is, however, by no means exhaustive, and interested readers are
referred to more extensive reviews and textbooks in the field, including
e.g.~\cite{Brambilla:2014jmp,Mikko,Kapusta:2006pm,Kraemmer:2003gd}.

Examples of purely Euclidean calculations in QCD include the determination of
various screening masses \cite{Laine:2003bd,Laine:2009dh} --- also in relation 
to real-time rates \cite{Brandt:2014uda}--- as well as spatial
and imaginary time correlators
%Jacopo: added three references by the Munich group, moved Burnier:2010rp below
\cite{Burnier:2009bk,Brambilla:2010xn,Laine:2010tc,
Laine:2010fe,Burnier:2012ze,Berwein:2012mw,Burnier:2013vsa,Berwein:2015ayt,Berwein:2017thy}.
Besides their intrinsic physical value
%JACOPO: added physics motivations. Mike, can you add more reviews?
e.g.~for the study of thermal modifications of heavy quark-antiquark bound states in the QCD medium
(see \cite{Mocsy:2013syh,Brambilla:2014jmp} for reviews), such results can be used to test holographic models of QCD as well as to verify lattice results (see e.g.~\cite{Bazavov:2016uvm,Bazavov:2018wmo}) and known sum rules (see e.g.~\cite{Kajantie:2013gab} and references therein) or
to extract $\als$ from lattice data \cite{Bazavov:2019qoo}.

On the Minkowskian side, spectral functions can be determined as imaginary
parts of retarded Green's functions that in turn are obtained either directly
in the real time formalism, cf.~Sec.~\ref{sec:realtimept}, or via a Euclidean
correlator by means of an analytic continuation. As explained in some detail e.g.~in
\cite{Laine:2013vma} (see also \cite{Ghiglieri:2018wbs} for the case of finite
chemical potentials), the latter procedure amounts to the replacement of the
discrete Matsubara frequency $p_n$ of the external momentum of the correlator by the
combination $-i(\omega+i\epsilon)$, where $\omega$ is a continuous
real-valued frequency. This procedure yields the spectral function $\rho$ in the form
\begin{eqnarray}
 \rho(\omega,\mathbf{p})&=&2{\rm Im}\big[\Pi^E(P)\big]_{P\to(-i[\omega+i\epsilon],\mathbf{p})}\,,
\end{eqnarray}
where $\Pi^E$ stands for the Euclidean self energy. If the imaginary part is not taken, this replacement simply maps the Euclidean Green's
functions to retarded correlators, up to convention-dependent factors of $i$ and 2.

Spectral functions corresponding to many different operators have been
determined using the imaginary time formalism, with the physical quantities of
interest being often various production or decay rates. In the realm of QCD,
examples include e.g.~heavy quark observables
\cite{Burnier:2008ia,Burnier:2010rp}, dilepton production rates
\cite{Laine:2013vma},
as well as correlators of the bulk and shear components of the energy momentum
tensor \cite{Laine:2011xm,Zhu:2012be,Vuorinen:2015wla}. In these studies,
one is typically particularly interested in the IR ($\omega\to 0$) structure of
the spectral functions, which, as we have seen in Sec.~\ref{sub_res_transport},
 is related to different transport coefficients via
the Kubo formulae. As we saw, one runs into a technical problem in this region: obtaining
information about the IR limit usually requires extremely complicated
resummations in lieu of standard loop expansions. To this end, in most
existing calculations the frequency is assumed to be of the order of the hard
scale in the problem (typically $T$) or at least $\omega\sim gT$, for which a
simple HTL resummation typically suffices.
In the opposite extreme, $\omega\gg T$, it was shown in ref.~\cite{CaronHuot:2009ns}  that
standard Operator Product Expansion techniques, which are defined in the
Euclidean regime $Q^2\gg T^2$ (or equivalently the  deep space-like regime
$\mathcal{Q}^2\gg T^2$), become applicable also in the
deep time-like regime $-\mathcal{Q}^2\gg T^2$. From a clever application of
analytical continuation and cutting rules it then becomes possible to extract
the $\omega\gg T$ asymptotics of spectral functions. In the case of QCD, the
results of \cite{CaronHuot:2009ns} have been used to study the thermal width
of the Higgs boson in \cite{Ghiglieri:2019lzz}.

Concerning somewhat more formal developments in high-temperature effective theories, 
there have been extensive efforts to go beyond the HTL effective Lagrangian 
in the description of soft excitations in QED and QCD as already discussed 
in Sec.~\ref{sec_soft_collinear}, see~e.g.~\cite{Mirza:2013ula,Manuel:2016wqs,Carignano:2017ovz,Carignano:2019ofj}. 
Within the dimensional reduction approach, high-order results for the weak-coupling expansion of the pressure have on 
the other hand been derived in theories somewhat simpler than QCD, such as massless scalar $\phi^4$ theory 
\cite{Gynther:2007bw,Andersen:2009ct}. Finally, the solvability of QCD in the limit of a large number 
of quark flavors has lead to a series of interesting works on the large-$N_f$ thermodynamics of the 
theory \cite{Moore:2002md,Ipp:2003jy,Ipp:2003yz,Blaizot:2005fd,Blaizot:2005wr}, and similar methods 
have recently been applied to the study of lower-dimensional exactly solvable QFTs by Romatschke and 
collaborators \cite{Romatschke:2019rjk,Romatschke:2019ybu,Romatschke:2019qbx}.

Finally, we note that the methods of imaginary-time perturbation theory have been frequently applied
to the study of the Electroweak sector of the Standard Model as well as to
different Beyond the Standard Model (BSM) theories. Bulk thermodynamic quantities
in the Standard Model, in particular the EoS, have been determined to the full
three-loop order in \cite{Gynther:2005dj,Gynther:2005av,Laine:2015kra}, while in
${\mathcal N}=4$ Super Yang-Mills theory, the entropy density has been determined
in an HTLpt-related approach \cite{Blaizot:2006tk},
EQCD frameworks \cite{Nieto:1999kc}, and using explicit resummations \cite{Kim:1999sg,VazquezMozo:1999ic}. In the latter theory, thermal correlators for heavy fundamental particles have also been considered in \cite{Chesler:2009yg}.

Dimensionally reduced effective theories akin to EQCD have in addition been derived not only for the weakly interacting part of the Standard Model \cite{Kajantie:1995dw} (where they in fact preceded the application of the same methods to QCD), but also in many BSM models
\cite{Brauner:2016fla,Andersen:2017ika,Niemi:2018asa}, with motivation
stemming from a desire to nonperturbatively study the Electroweak phase transition
using three-dimensional lattice simulations. Finally, the
Euclidean methods developed for determining spectral functions in
QCD (see also \cite{Laine:2013vpa}) have found applications in the evaluation
of the production rate of right-handed neutrinos in the early universe as well as
in determining the sterile neutrino dark matter spectrum
\cite{Asaka:2006rw,Asaka:2006nq,Laine:2011pq,Laine:2013lka,Ghiglieri:2018wbs}.

%% file: futuredirections.tex
% !TEX root = review.tex

\section{Conclusions and future directions \label{sec:fut}}

In a quantum field theory characterized by sizable  couplings in most phenomenologically interesting settings, the machinery of perturbative field theory is often considered a last resort --- to be used in situations, where no other first principles method is available. This is to some extent the case also in thermal QCD, where nonperturbative lattice simulations are the commonly accepted method of choice whenever applicable. In practice, the availability of this method is, however, restricted to the limits of thermal equilibrium, Euclidean quantities, and small baryon densities, leaving considerable room for applications of thermal perturbation theory, the principal theme of the present review article. Thankfully, somewhat contrary to the traditional common lore in the field, a creative application of perturbative methods and results has oftentimes led to advances also outside the realm of very small couplings and correspondingly ultrahigh energies.

The goals of our review have been twofold. First, we wanted to provide a pedagogical, yet sufficiently detailed introduction to the methods used in modern perturbative calculations, so that the interested reader may use the review for self-study, naturally complemented by more extensive textbooks, such as \cite{Kapusta:1989tk,Kapusta:2006pm,Bellac:2011kqa,Mikko}. In this respect, we note that our sections \ref{sec:realtimept}--\ref{sec_soft_collinear} on the real-time formalism offer the first detailed and self-consistent account of the modern real-time methods, marking the first extensive treatise on the subject since the classic textbook of Le Bellac from 1996 \cite{Bellac:2011kqa}. Second, we have sought to review recent research in the field in a way that gives a fair overview of the most important modern applications of perturbative QCD in the contexts of deconfined hot quark-gluon plasma and dense quark matter. Here, we have tried to highlight problems that have actively preoccupied the community in recent years, summarizing the current state of affairs in the form of a collection of up-to-date results. The choice of results covered has been guided by a conscious choice to stick to first-principles perturbative calculations, thereby entirely excluding many important complementary approaches, such as lattice QCD, holographic and Functional Renormalization Group (FRG) techniques. For a comprehensive account of recent developments in these closely related fields, we refer the interested reader instead to Ref.~\cite{Brambilla:2014jmp}.

At the moment of writing the review, it is safe to say that some of the topics we have discussed represent nearly closed chapters of research in the sense that a consensus has been reached in the field and the most important questions satisfactorily answered. A prominent example of this is the bulk thermodynamics of QCD at high temperatures and small or vanishing density, which was for a long time the most pressing question in lattice and pQCD calculations at high $T$ (for just a few highlight papers, see \cite{Borsanyi:2010cj,Borsanyi:2012rr,Bellwied:2015lba,Haque:2013sja,Mogliacci:2013mca}).  At the same time, new challenges have arisen or become highlighted by recent experimental advances, including perhaps most importantly the study of transport in deconfined QCD matter in and out of equilibrium and the desire to quantitatively address the bulk thermodynamic properties of dense Quark Matter (QM) possibly present inside neutron stars. To close the discussion, we will next briefly comment on our view of the prospects of significant future progress on these two topics.

How QCD behaves when pushed away from thermal equilibrium is a frontier with many currently unanswered questions. While there has been a lot of progress in recent years in following the time evolution of simple out-of-equilibrium settings \cite{Kurkela:2011ti,Kurkela:2014tea,Kurkela:2015qoa,Kurkela:2018wud,Kurkela:2018xxd}, several open questions exist related to the dynamics of, in particular, anisotropic systems. In systems that exhibit momentum space anisotropies, one encounters the rich physics of non-abelian plasma-instabilities whose complex dynamics in the non-linear regime remains poorly understood
\cite{Mrowczynski:1993qm,Mrowczynski:1996vh,Romatschke:2003ms,Arnold:2003rq,Mrowczynski:2004kv,Romatschke:2004jh,Arnold:2004ti,Rebhan:2004ur,Rebhan:2005re,Arnold:2005qs,Arnold:2005ef,Arnold:2005vb,Romatschke:2006wg,Rebhan:2008uj,Rebhan:2009ku,Ipp:2010uy,Kurkela:2011ti,Attems:2012js} (see Ref.~\cite{Mrowczynski:2016etf} for a recent review). In the context of real-time quantities, such as transport coefficients, the non-equilibrium photon production rate, and the heavy-quark potential out of equilibrium, we note that as one pushes the calculations to higher orders, one eventually runs into fundamental issues related to the existence of these instabilities.  For example, already at leading order in $g$ one faces issues related to presence of unstable modes in the calculation of the imaginary part of the heavy-quark potential in an anisotropic quark-gluon plasma \cite{Burnier:2009yu,Dumitru:2009fy,Nopoush:2017zbu}.  In this case, the unstable modes result in pinch singularities, which cause the imaginary part of the heavy-quark potential to diverge \cite{Nopoush:2017zbu}. A similar issue has been recently highlighted in the computation of photon production from a viscous (anisotropic) QGP \cite{Hauksson:2017udm}, where it was shown that holding the momentum anisotropy fixed while taking $g \rightarrow 0$ results in a divergent result.  In fact, most non-equilibrium observables will be affected by this issue at some order in $g$. It seems that to go forward in a systematic manner will require the development of new methods for treating plasma instabilities and their effects on non-equilibrium transport and interactions. Promising advancements have recently been presented in \cite{Hauksson:2020etn}.

As we have seen in our discussions of real-time observables and bulk thermodynamics alike, a rather generic unsolved issue of crucial importance is the need for a better understanding of the dynamics of soft field modes. As we have seen e.g.~in the discussion of transport coefficients in Sec.~\ref{sec:realresults} and of the EoS of dense quark matter in Sec.~\ref{sec:imagtimeresults}, such modes are typically responsible for the poor convergence properties of perturbative expansions. In both cases, one hopes to be able to systematically identify and understand the physics responsible for the large corrections affecting convergence, so that the corresponding perturbative expansions might be rearranged in ways that lead to dramatically improved convergence properties, much like what has happened in the case of bulk thermodynamic quantities at high temperatures during the past 20 years, cf.~Secs.~\ref{sec:imagtimeform} and \ref{sec:imagtimeresults}.

Finally, in the past five years or so there has been considerable progress in the perturbative study of the bulk thermodynamic properties of unpaired cold QM \cite{Kurkela:2016was,Gorda:2018gpy}, and it is even not out of the question that the full order $g^6$ pressure of $T=0$ QM will be completed before its high-temperature counterpart. As noted above, recent progress in our quantitative understanding of the IR sector of cold and dense QCD has in addition opened up the possibility to dramatically improve the convergence properties of the $T=0$ expansion using a rearrangement of the weak coupling series. Should such a line of work prove fruitful and the region of applicability of the pQCD EoS become successfully continued towards lower densities, our understanding of the thermodynamic properties of NS matter will likely be dramatically improved. At the same time, new and very interesting challenges will, however, inevitably present themselves. So far, all applications of the pQCD EoS to NS physics (see e.g.~\cite{Annala:2019puf} and references therein) have namely applied the perturbative result at such high densities that it has been possible to argue that quark pairing is parametrically negligible. Extending these results to lower densities, the physics of pairing may, however, begin to play a significantly more important role. Taking a fresh look at the first-principles machinery, with which the color-superconducting phases of QCD are tackled, may therefore become necessary in the not-so-distant future.

%% file: appendix-gauge.tex
% !TEX root = review.tex

\section{Real-time Feynman rules}
\label{app_horror}

With the conventions listed in Secs.~\ref{sec:IV} and \ref{sec:realtimept},
the fermion propagators in the $r/a$ basis read
\begin{equation}
S_{R,A}(\mn{P})=\frac{i\slashed{\mn{P}}}{\mn{P}^2\mp i\epsilon p^0},\qquad
S_{rr}(\mn{P})=-\slashed{\mn{P}}\left(\frac12-\nfd(|p^0|)\right)
2\pi\delta(\mn{P}^2).
	\label{rafermion}
\end{equation}
We remind of our nonstandard convention for the Dirac matrices, as noted in 
footnote~\ref{foot_metric}.

In the case of gluons, we list both Coulomb and Feynman gauge results. In the former case,
the bare propagators read
\begin{eqnarray}
	G^{00}_{R,A}(\mn{Q})&=&\frac{i}{q^2},\quad
	G^{00}_{rr}(\mn{Q})=0,\quad 
	G^{ij}_{R,A}(\mn{Q})=\left(\delta^{ij}-\hat{q}^i\hat{q}^j\right)
	\frac{-i}{\mn{Q}^2\mp i\epsilon q^0},\nn\\
	G^{ij}_{rr}(\mn{Q})&=&\left(\delta^{ij}-\hat{q}^i\hat{q}^j\right)
	\left(\frac12+\nbe(|q^0|)\right)
2\pi\delta(\mn{Q}^2),
	\label{ragluoncg}
\end{eqnarray}
while in Feynman gauge they instead take the forms
\begin{equation}
	G^{\mu\nu}_{R,A}(\mn{Q})=
	\frac{-ig^{\mu\nu}}{\mn{Q}^2\mp i\epsilon q^0},\quad
	G^{\mu\nu}_{rr}(\mn{Q})=g^{\mu\nu}
	\left(\frac12+\nbe(|q^0|)\right)2\pi\delta(\mn{Q}^2).
	\label{ragluonfg}
\end{equation}
In this case one also needs  to include ghosts in loops, which propagate with
\begin{equation}
	\tilde G_{R,A}(\mn{Q})=
	\frac{-i}{\mn{Q}^2\mp i\epsilon q^0},\quad
	\tilde G_{rr}(\mn{Q})=
	\left(\frac12+\nbe(|q^0|)\right)2\pi\delta(\mn{Q}^2),
	\label{raghostfg}
\end{equation}
where $\tilde G$ labels the ghost propagator. We remark that it is also possible
 to suppress the thermal part of the ghost
propagator ($\nbe(|q^0|)\to 0$ in Eq.~\eqref{raghostfg}) by only including the
thermal part for the two transverse, physical degrees of freedom \cite{Landshoff:1992ne,Landshoff:1993ag}.

As an example, let us derive the gluonic contribution to $\Pi^{00}_{aa}$ in the Hard Thermal Loop
approximation. In Coulomb gauge, we have, including a symmetry factor
of $1/2$ and neglecting the contribution from the $aaa$ vertex, purely of vacuum nature,  \cite{CaronHuot:2007nw}
\begin{equation}
	i\Pi^{00}_{aa}(\mn{P})=-\frac12g^2C_A\int\frac{d^4\mn{Q}}{(2\pi)^4}4q_0^2G_{rr}^{ij}(\mn{Q})G_{rr}^{ij}(\mn{P}+\mn{Q})
\end{equation}
Upon dropping the vacuum part from this, we find
\begin{equation}
	i\Pi^{00}_{aa}(\mn{P})=-4g^2C_A\int\frac{d^4\mn{Q}}{(2\pi)^4}q_0^2\nbe(\vert q^0\vert)(1+\nbe(\vert q^0\vert))
	(2\pi)^2\delta(\mn{Q}^2)\delta((\mn{P}+\mn{Q})^2),
\end{equation}
where we are consistently taking the HTL approximation $\mn{Q}\gg\mn{P}$. The energy integration can be performed
as in Sec.~\ref{sub_heur_htl}, leading to
\begin{equation}
	i\Pi^{00}_{aa}(\mn{P})=2g^2C_A\int\frac{d^3q}{(2\pi)^3}\nbe (q)(1+\nbe( q))
	2\pi\delta(v\cdot\mn{P})=\frac{g^2C_AT^3}{3}\int \frac{d\Omega_v}{4\pi} 2\pi\delta(v\cdot\mn{P}).
\end{equation}
This agrees with the gluonic part of Eq.~\eqref{gluehtlaa}.

In Feynman gauge, the same procedure, i.e.~the HTL approximation for the thermal part only, leads to
\begin{equation}
	i\Pi^{00}_{aa}(\mn{P})=-\frac12g^2C_A\int\frac{d^4\mn{Q}}{(2\pi)^4}10q_0^2\nbe(\vert q^0\vert)(1+\nbe(\vert q^0\vert))
	(2\pi)^2\delta(\mn{Q}^2)\delta((\mn{P}+\mn{Q})^2).
\end{equation}
The ghost contribution on the other hand has opposite sign and reads
\begin{equation}
	i\Pi^{00}_{aa}(\mn{P})=g^2C_A\int\frac{d^4\mn{Q}}{(2\pi)^4}q_0^2\nbe(\vert q^0\vert)(1+\nbe(\vert q^0\vert))
	(2\pi)^2\delta(\mn{Q}^2)\delta((\mn{P}+\mn{Q})^2),
\end{equation}
so that when the two are summed, agreement with Coulomb gauge is restored.

%% file: review.bbl
\begin{thebibliography}{100}
\expandafter\ifx\csname url\endcsname\relax
  \def\url#1{\texttt{#1}}\fi
\expandafter\ifx\csname urlprefix\endcsname\relax\def\urlprefix{URL }\fi
\expandafter\ifx\csname href\endcsname\relax
  \def\href#1#2{#2} \def\path#1{#1}\fi

\bibitem{Kapusta:1989tk}
J.~I. Kapusta, {Finite Temperature Field Theory}, Vol. 360 of Cambridge
  Monographs on Mathematical Physics, Cambridge University Press, Cambridge,
  1989.

\bibitem{Kapusta:2006pm}
J.~I. Kapusta, C.~Gale, {Finite-temperature field theory: Principles and
  applications}, Cambridge University Press, 2011.

\bibitem{Bellac:2011kqa}
M.~Le~Bellac, {Thermal Field Theory}, Cambridge Monographs on Mathematical
  Physics, Cambridge University Press, 2011.
\newblock \href {https://doi.org/10.1017/CBO9780511721700}
  {\path{doi:10.1017/CBO9780511721700}}.

\bibitem{Mikko}
M.~Laine, A.~Vuorinen, {Basics of Thermal Field Theory}, Lect. Notes Phys. 925
  (2016) pp.1--281.
\newblock \href {https://doi.org/10.1007/978-3-319-31933-9}
  {\path{doi:10.1007/978-3-319-31933-9}}.

\bibitem{Blaizot:2001nr}
J.-P. Blaizot, E.~Iancu, {The Quark gluon plasma: Collective dynamics and hard
  thermal loops}, Phys. Rept. 359 (2002) 355--528.
\newblock \href {http://arxiv.org/abs/hep-ph/0101103}
  {\path{arXiv:hep-ph/0101103}}, \href
  {https://doi.org/10.1016/S0370-1573(01)00061-8}
  {\path{doi:10.1016/S0370-1573(01)00061-8}}.

\bibitem{Kraemmer:2003gd}
U.~Kraemmer, A.~Rebhan, {Advances in perturbative thermal field theory}, Rept.
  Prog. Phys. 67 (2004) 351.
\newblock \href {http://arxiv.org/abs/hep-ph/0310337}
  {\path{arXiv:hep-ph/0310337}}, \href
  {https://doi.org/10.1088/0034-4885/67/3/R05}
  {\path{doi:10.1088/0034-4885/67/3/R05}}.

\bibitem{Peskin}
M.~E. Peskin, D.~V. Schroeder, An Introduction to quantum field theory,
  Addison-Wesley, 1995.

\bibitem{Schwinger:1960qe}
J.~S. Schwinger, {Brownian motion of a quantum oscillator}, J.Math.Phys. 2
  (1961) 407--432.

\bibitem{Keldysh:1964ud}
L.~Keldysh, {Diagram technique for nonequilibrium processes}, Zh.Eksp.Teor.Fiz.
  47 (1964) 1515--1527.

\bibitem{Kubo:1957mj}
R.~Kubo, {Statistical mechanical theory of irreversible processes. 1. General
  theory and simple applications in magnetic and conduction problems}, J. Phys.
  Soc. Jap. 12 (1957) 570--586.
\newblock \href {https://doi.org/10.1143/JPSJ.12.570}
  {\path{doi:10.1143/JPSJ.12.570}}.

\bibitem{Martin:1959jp}
P.~C. Martin, J.~S. Schwinger, {Theory of many particle systems. 1.}, Phys.Rev.
  115 (1959) 1342--1373.
\newblock \href {https://doi.org/10.1103/PhysRev.115.1342}
  {\path{doi:10.1103/PhysRev.115.1342}}.

\bibitem{Chou:1984es}
K.-c. Chou, Z.-b. Su, B.-l. Hao, L.~Yu, {Equilibrium and Nonequilibrium
  Formalisms Made Unified}, Phys. Rept. 118 (1985) 1--131.
\newblock \href {https://doi.org/10.1016/0370-1573(85)90136-X}
  {\path{doi:10.1016/0370-1573(85)90136-X}}.

\bibitem{CaronHuot:2007nw}
S.~Caron-Huot, {Hard thermal loops in the real-time formalism}, JHEP 0904
  (2009) 004.
\newblock \href {http://arxiv.org/abs/0710.5726} {\path{arXiv:0710.5726}},
  \href {https://doi.org/10.1088/1126-6708/2009/04/004}
  {\path{doi:10.1088/1126-6708/2009/04/004}}.

\bibitem{Evans:1991ky}
T.~S. Evans, {N point finite temperature expectation values at real times},
  Nucl. Phys. B374 (1992) 340--370.
\newblock \href {https://doi.org/10.1016/0550-3213(92)90357-H}
  {\path{doi:10.1016/0550-3213(92)90357-H}}.

\bibitem{Kobes:1985kc}
R.~L. Kobes, G.~W. Semenoff, {Discontinuities of Green Functions in Field
  Theory at Finite Temperature and Density}, Nucl. Phys. B260 (1985) 714--746.
\newblock \href {https://doi.org/10.1016/0550-3213(85)90056-2}
  {\path{doi:10.1016/0550-3213(85)90056-2}}.

\bibitem{Kobes:1986za}
R.~L. Kobes, G.~W. Semenoff, {Discontinuities of Green Functions in Field
  Theory at Finite Temperature and Density. 2}, Nucl. Phys. B272 (1986)
  329--364.
\newblock \href {https://doi.org/10.1016/0550-3213(86)90006-4}
  {\path{doi:10.1016/0550-3213(86)90006-4}}.

\bibitem{Guerin:1993ik}
F.~Guerin, {Retarded - advanced N point Green functions in thermal field
  theories}, Nucl.Phys. B432 (1994) 281--314.
\newblock \href {http://arxiv.org/abs/hep-ph/9306210}
  {\path{arXiv:hep-ph/9306210}}, \href
  {https://doi.org/10.1016/0550-3213(94)90603-3}
  {\path{doi:10.1016/0550-3213(94)90603-3}}.

\bibitem{Gelis:1997zv}
F.~Gelis, {Cutting rules in the real time formalisms at finite temperature},
  Nucl.Phys. B508 (1997) 483--505.
\newblock \href {http://arxiv.org/abs/hep-ph/9701410}
  {\path{arXiv:hep-ph/9701410}}, \href
  {https://doi.org/10.1016/S0550-3213(97)00511-7}
  {\path{doi:10.1016/S0550-3213(97)00511-7}}.

\bibitem{Caron-Huot:2007zhp}
S.~Caron-Huot, {Heavy quark energy losses in the quark-gluon plasma : beyond
  leading order}, Master's thesis, McGill U. (2007).

\bibitem{Mueller:2002gd}
A.~H. Mueller, D.~T. Son, {On the Equivalence between the Boltzmann equation
  and classical field theory at large occupation numbers}, Phys. Lett. B582
  (2004) 279--287.
\newblock \href {http://arxiv.org/abs/hep-ph/0212198}
  {\path{arXiv:hep-ph/0212198}}, \href
  {https://doi.org/10.1016/j.physletb.2003.12.047}
  {\path{doi:10.1016/j.physletb.2003.12.047}}.

\bibitem{Bodeker:1995pp}
D.~{B\"odeker}, L.~D. McLerran, A.~V. Smilga, {Really computing nonperturbative
  real time correlation functions}, Phys. Rev. D52 (1995) 4675--4690.
\newblock \href {http://arxiv.org/abs/hep-th/9504123}
  {\path{arXiv:hep-th/9504123}}, \href
  {https://doi.org/10.1103/PhysRevD.52.4675}
  {\path{doi:10.1103/PhysRevD.52.4675}}.

\bibitem{Bodeker:1996wb}
D.~{B\"odeker}, {Classical real time correlation functions and quantum
  corrections at finite temperature}, Nucl. Phys. B486 (1997) 500--514.
\newblock \href {http://arxiv.org/abs/hep-th/9609170}
  {\path{arXiv:hep-th/9609170}}, \href
  {https://doi.org/10.1016/S0550-3213(96)00688-8}
  {\path{doi:10.1016/S0550-3213(96)00688-8}}.

\bibitem{Greiner:1996dx}
C.~Greiner, B.~Muller, {Classical fields near thermal equilibrium}, Phys. Rev.
  D55 (1997) 1026--1046.
\newblock \href {http://arxiv.org/abs/hep-th/9605048}
  {\path{arXiv:hep-th/9605048}}, \href
  {https://doi.org/10.1103/PhysRevD.55.1026}
  {\path{doi:10.1103/PhysRevD.55.1026}}.

\bibitem{Aarts:1996qi}
G.~Aarts, J.~Smit, {Finiteness of hot classical scalar field theory and the
  plasmon damping rate}, Phys. Lett. B393 (1997) 395--402.
\newblock \href {http://arxiv.org/abs/hep-ph/9610415}
  {\path{arXiv:hep-ph/9610415}}, \href
  {https://doi.org/10.1016/S0370-2693(96)01624-3}
  {\path{doi:10.1016/S0370-2693(96)01624-3}}.

\bibitem{Aarts:1997kp}
G.~Aarts, J.~Smit, {Classical approximation for time dependent quantum field
  theory: Diagrammatic analysis for hot scalar fields}, Nucl. Phys. B511 (1998)
  451--478.
\newblock \href {http://arxiv.org/abs/hep-ph/9707342}
  {\path{arXiv:hep-ph/9707342}}, \href
  {https://doi.org/10.1016/S0550-3213(97)00723-2}
  {\path{doi:10.1016/S0550-3213(97)00723-2}}.

\bibitem{Mathieu:2014aba}
V.~Mathieu, A.~H. Mueller, D.~N. Triantafyllopoulos, {The Boltzmann Equation in
  Classical Yang-Mills Theory}, Eur. Phys. J. C74 (2014) 2873.
\newblock \href {http://arxiv.org/abs/1403.1184} {\path{arXiv:1403.1184}},
  \href {https://doi.org/10.1140/epjc/s10052-014-2873-8}
  {\path{doi:10.1140/epjc/s10052-014-2873-8}}.

\bibitem{Jeon:2013zga}
S.~Jeon, {Color Glass Condensate in Schwinger-Keldysh QCD}, Annals Phys. 340
  (2014) 119--170.
\newblock \href {http://arxiv.org/abs/1308.0263} {\path{arXiv:1308.0263}},
  \href {https://doi.org/10.1016/j.aop.2013.09.019}
  {\path{doi:10.1016/j.aop.2013.09.019}}.

\bibitem{Aarts:2001yx}
G.~Aarts, {Spectral function at high temperature in the classical
  approximation}, Phys. Lett. B518 (2001) 315--322.
\newblock \href {http://arxiv.org/abs/hep-ph/0108125}
  {\path{arXiv:hep-ph/0108125}}, \href
  {https://doi.org/10.1016/S0370-2693(01)01081-4}
  {\path{doi:10.1016/S0370-2693(01)01081-4}}.

\bibitem{Laine:2009dd}
M.~Laine, G.~D. Moore, O.~Philipsen, M.~Tassler, {Heavy Quark Thermalization in
  Classical Lattice Gauge Theory: Lessons for Strongly-Coupled QCD}, JHEP 05
  (2009) 014.
\newblock \href {http://arxiv.org/abs/0902.2856} {\path{arXiv:0902.2856}},
  \href {https://doi.org/10.1088/1126-6708/2009/05/014}
  {\path{doi:10.1088/1126-6708/2009/05/014}}.

\bibitem{Moore:2010jd}
G.~D. Moore, M.~Tassler, {The Sphaleron Rate in SU(N) Gauge Theory}, JHEP 02
  (2011) 105.
\newblock \href {http://arxiv.org/abs/1011.1167} {\path{arXiv:1011.1167}},
  \href {https://doi.org/10.1007/JHEP02(2011)105}
  {\path{doi:10.1007/JHEP02(2011)105}}.

\bibitem{Boguslavski:2018beu}
K.~Boguslavski, A.~Kurkela, T.~Lappi, J.~Peuron, {Spectral function for
  overoccupied gluodynamics from real-time lattice simulations}, Phys. Rev.
  D98~(1) (2018) 014006.
\newblock \href {http://arxiv.org/abs/1804.01966} {\path{arXiv:1804.01966}},
  \href {https://doi.org/10.1103/PhysRevD.98.014006}
  {\path{doi:10.1103/PhysRevD.98.014006}}.

\bibitem{Ghiglieri:2014kma}
J.~Ghiglieri, G.~D. Moore, {Low Mass Thermal Dilepton Production at NLO in a
  Weakly Coupled Quark-Gluon Plasma}, JHEP 12 (2014) 029.
\newblock \href {http://arxiv.org/abs/1410.4203} {\path{arXiv:1410.4203}},
  \href {https://doi.org/10.1007/JHEP12(2014)029}
  {\path{doi:10.1007/JHEP12(2014)029}}.

\bibitem{McLerran:1984ay}
L.~D. McLerran, T.~Toimela, {Photon and Dilepton Emission from the Quark -
  Gluon Plasma: Some General Considerations}, Phys. Rev. D31 (1985) 545.
\newblock \href {https://doi.org/10.1103/PhysRevD.31.545}
  {\path{doi:10.1103/PhysRevD.31.545}}.

\bibitem{Ghiglieri:2013gia}
J.~Ghiglieri, J.~Hong, A.~Kurkela, E.~Lu, G.~D. Moore, D.~Teaney,
  {Next-to-leading order thermal photon production in a weakly coupled
  quark-gluon plasma}, JHEP 1305 (2013) 010.
\newblock \href {http://arxiv.org/abs/1302.5970} {\path{arXiv:1302.5970}},
  \href {https://doi.org/10.1007/JHEP05(2013)010}
  {\path{doi:10.1007/JHEP05(2013)010}}.

\bibitem{Burgess:2007zi}
C.~P. Burgess, G.~D. Moore, {The standard model: A primer}, Cambridge
  University Press, 2006.

\bibitem{Kapusta:1991qp}
J.~I. Kapusta, P.~Lichard, D.~Seibert, {High-energy photons from quark - gluon
  plasma versus hot hadronic gas}, Phys.Rev. D44 (1991) 2774--2788.
\newblock \href {https://doi.org/10.1103/PhysRevD.47.4171,
  10.1103/PhysRevD.44.2774} {\path{doi:10.1103/PhysRevD.47.4171,
  10.1103/PhysRevD.44.2774}}.

\bibitem{Baier:1991em}
R.~Baier, H.~Nakkagawa, A.~Niegawa, K.~Redlich, {Production rate of hard
  thermal photons and screening of quark mass singularity}, Z.Phys. C53 (1992)
  433--438.
\newblock \href {https://doi.org/10.1007/BF01625902}
  {\path{doi:10.1007/BF01625902}}.

\bibitem{Arnold:2001ms}
P.~B. Arnold, G.~D. Moore, L.~G. Yaffe, {Photon emission from quark gluon
  plasma: Complete leading order results}, JHEP 0112 (2001) 009.
\newblock \href {http://arxiv.org/abs/hep-ph/0111107}
  {\path{arXiv:hep-ph/0111107}}.

\bibitem{Aurenche:1998nw}
P.~Aurenche, F.~Gelis, R.~Kobes, H.~Zaraket, {Bremsstrahlung and photon
  production in thermal QCD}, Phys.Rev. D58 (1998) 085003.
\newblock \href {http://arxiv.org/abs/hep-ph/9804224}
  {\path{arXiv:hep-ph/9804224}}, \href
  {https://doi.org/10.1103/PhysRevD.58.085003}
  {\path{doi:10.1103/PhysRevD.58.085003}}.

\bibitem{Braaten:1989mz}
E.~Braaten, R.~D. Pisarski, {Soft Amplitudes in Hot Gauge Theories: A General
  Analysis}, Nucl.Phys. B337 (1990) 569.
\newblock \href {https://doi.org/10.1016/0550-3213(90)90508-B}
  {\path{doi:10.1016/0550-3213(90)90508-B}}.

\bibitem{Braaten:1991gm}
E.~Braaten, R.~D. Pisarski, {Simple effective Lagrangian for hard thermal
  loops}, Phys. Rev. D45 (1992) 1827--1830.
\newblock \href {https://doi.org/10.1103/PhysRevD.45.R1827}
  {\path{doi:10.1103/PhysRevD.45.R1827}}.

\bibitem{Frenkel:1989br}
J.~Frenkel, J.~Taylor, {High Temperature Limit of Thermal QCD}, Nucl.Phys. B334
  (1990) 199.
\newblock \href {https://doi.org/10.1016/0550-3213(90)90661-V}
  {\path{doi:10.1016/0550-3213(90)90661-V}}.

\bibitem{Frenkel:1991ts}
J.~Frenkel, J.~C. Taylor, {Hard thermal QCD, forward scattering and effective
  actions}, Nucl. Phys. B374 (1992) 156--168.
\newblock \href {https://doi.org/10.1016/0550-3213(92)90480-Y}
  {\path{doi:10.1016/0550-3213(92)90480-Y}}.

\bibitem{Taylor:1990ia}
J.~C. Taylor, S.~M.~H. Wong, {The Effective Action of Hard Thermal Loops in
  {QCD}}, Nucl. Phys. B346 (1990) 115--128.
\newblock \href {https://doi.org/10.1016/0550-3213(90)90240-E}
  {\path{doi:10.1016/0550-3213(90)90240-E}}.

\bibitem{Kobes:1990dc}
R.~Kobes, G.~Kunstatter, A.~Rebhan, {Gauge dependence identities and their
  application at finite temperature}, Nucl. Phys. B355 (1991) 1--37.
\newblock \href {https://doi.org/10.1016/0550-3213(91)90300-M}
  {\path{doi:10.1016/0550-3213(91)90300-M}}.

\bibitem{Weldon:1982bn}
H.~A. Weldon, {Effective Fermion Masses of Order gT in High Temperature Gauge
  Theories with Exact Chiral Invariance}, Phys. Rev. D26 (1982) 2789.
\newblock \href {https://doi.org/10.1103/PhysRevD.26.2789}
  {\path{doi:10.1103/PhysRevD.26.2789}}.

\bibitem{Manuel:2014dza}
C.~Manuel, J.~M. Torres-Rincon, {Chiral transport equation from the quantum
  Dirac Hamiltonian and the on-shell effective field theory}, Phys. Rev.
  D90~(7) (2014) 076007.
\newblock \href {http://arxiv.org/abs/1404.6409} {\path{arXiv:1404.6409}},
  \href {https://doi.org/10.1103/PhysRevD.90.076007}
  {\path{doi:10.1103/PhysRevD.90.076007}}.

\bibitem{Manuel:2016wqs}
C.~Manuel, J.~Soto, S.~Stetina, {On-shell effective field theory: A systematic
  tool to compute power corrections to the hard thermal loops}, Phys. Rev.
  D94~(2) (2016) 025017, [Erratum: Phys. Rev.D96,no.12,129901(2017)].
\newblock \href {http://arxiv.org/abs/1603.05514} {\path{arXiv:1603.05514}},
  \href {https://doi.org/10.1103/PhysRevD.94.025017,
  10.1103/PhysRevD.96.129901} {\path{doi:10.1103/PhysRevD.94.025017,
  10.1103/PhysRevD.96.129901}}.

\bibitem{Carignano:2017ovz}
S.~Carignano, C.~Manuel, J.~Soto, {Power corrections to the HTL effective
  Lagrangian of QED}, Phys. Lett. B780 (2018) 308--312.
\newblock \href {http://arxiv.org/abs/1712.07949} {\path{arXiv:1712.07949}},
  \href {https://doi.org/10.1016/j.physletb.2018.03.012}
  {\path{doi:10.1016/j.physletb.2018.03.012}}.

\bibitem{Carignano:2019ofj}
S.~Carignano, M.~E. Carrington, J.~Soto, {The HTL Lagrangian at NLO: the photon
  case}\href {http://arxiv.org/abs/1909.10545} {\path{arXiv:1909.10545}}.

\bibitem{Mirza:2013ula}
A.~Mirza, M.~E. Carrington, {Thermal field theory at next-to-leading order in
  the hard thermal loop expansion}, Phys. Rev. D87~(6) (2013) 065008.
\newblock \href {http://arxiv.org/abs/1302.3796} {\path{arXiv:1302.3796}},
  \href {https://doi.org/10.1103/PhysRevD.87.065008}
  {\path{doi:10.1103/PhysRevD.87.065008}}.

\bibitem{Pisarski:1989cs}
R.~D. Pisarski, {Renormalized Gauge Propagator in Hot Gauge Theories}, Physica
  A158 (1989) 146--157.

\bibitem{Kalashnikov:1979cy}
O.~K. Kalashnikov, V.~V. Klimov, {Polarization Tensor in QCD for Finite
  Temperature and Density}, Sov. J. Nucl. Phys. 31 (1980) 699, [Yad.
  Fiz.31,1357(1980)].

\bibitem{Weldon:1982aq}
H.~A. Weldon, {Covariant Calculations at Finite Temperature: The Relativistic
  Plasma}, Phys. Rev. D26 (1982) 1394.
\newblock \href {https://doi.org/10.1103/PhysRevD.26.1394}
  {\path{doi:10.1103/PhysRevD.26.1394}}.

\bibitem{Bodeker:1998hm}
D.~{B\"odeker}, {On the effective dynamics of soft nonAbelian gauge fields at
  finite temperature}, Phys.Lett. B426 (1998) 351--360.
\newblock \href {http://arxiv.org/abs/hep-ph/9801430}
  {\path{arXiv:hep-ph/9801430}}, \href
  {https://doi.org/10.1016/S0370-2693(98)00279-2}
  {\path{doi:10.1016/S0370-2693(98)00279-2}}.

\bibitem{Bodeker:2000da}
D.~{B\"odeker}, {A local Langevin equation for slow long-distance modes of hot
  non-Abelian gauge fields}, Phys. Lett. B516 (2001) 175--182.
\newblock \href {http://arxiv.org/abs/hep-ph/0012304}
  {\path{arXiv:hep-ph/0012304}}, \href
  {https://doi.org/10.1016/S0370-2693(01)00911-X}
  {\path{doi:10.1016/S0370-2693(01)00911-X}}.

\bibitem{Braaten:1989kk}
E.~Braaten, R.~D. Pisarski, {Resummation and Gauge Invariance of the Gluon
  Damping Rate in Hot QCD}, Phys. Rev. Lett. 64 (1990) 1338.
\newblock \href {https://doi.org/10.1103/PhysRevLett.64.1338}
  {\path{doi:10.1103/PhysRevLett.64.1338}}.

\bibitem{Braaten:1990it}
E.~Braaten, R.~D. Pisarski, {Calculation of the gluon damping rate in hot QCD},
  Phys. Rev. D42 (1990) 2156--2160.
\newblock \href {https://doi.org/10.1103/PhysRevD.42.2156}
  {\path{doi:10.1103/PhysRevD.42.2156}}.

\bibitem{Schulz:1993gf}
H.~Schulz, {Gluon plasma frequency: The Next-to-leading order term}, Nucl.
  Phys. B413 (1994) 353--395.
\newblock \href {http://arxiv.org/abs/hep-ph/9306298}
  {\path{arXiv:hep-ph/9306298}}, \href
  {https://doi.org/10.1016/0550-3213(94)90624-6}
  {\path{doi:10.1016/0550-3213(94)90624-6}}.

\bibitem{Pisarski:1989wb}
R.~Pisarski, {Renormalized Fermion Propagator in Hot Gauge Theories},
  Nucl.Phys. A498 (1989) 423C--428C.
\newblock \href {https://doi.org/10.1016/0375-9474(89)90620-9}
  {\path{doi:10.1016/0375-9474(89)90620-9}}.

\bibitem{Braaten:1992gd}
E.~Braaten, R.~D. Pisarski, {Calculation of the quark damping rate in hot QCD},
  Phys. Rev. D46 (1992) 1829--1834.
\newblock \href {https://doi.org/10.1103/PhysRevD.46.1829}
  {\path{doi:10.1103/PhysRevD.46.1829}}.

\bibitem{Nakkagawa:1992ew}
H.~Nakkagawa, A.~Niegawa, B.~Pire, {Resolution of the gauge dependence problem
  of the fermion damping rate in hot gauge theories}, Phys. Lett. B294 (1992)
  396--402.
\newblock \href {https://doi.org/10.1016/0370-2693(92)91540-P}
  {\path{doi:10.1016/0370-2693(92)91540-P}}.

\bibitem{LeBellac:1994qg}
M.~Le~Bellac, P.~Reynaud, {Resummation and infrared safe processes in hot QCD},
  Nucl. Phys. B416 (1994) 801--823.
\newblock \href {https://doi.org/10.1016/0550-3213(94)90556-8}
  {\path{doi:10.1016/0550-3213(94)90556-8}}.

\bibitem{Vanderheyden:1996bw}
B.~Vanderheyden, J.-Y. Ollitrault, {Damping rates of hard momentum particles in
  a cold ultrarelativistic plasma}, Phys. Rev. D56 (1997) 5108--5122.
\newblock \href {http://arxiv.org/abs/hep-ph/9611415}
  {\path{arXiv:hep-ph/9611415}}, \href
  {https://doi.org/10.1103/PhysRevD.56.5108}
  {\path{doi:10.1103/PhysRevD.56.5108}}.

\bibitem{CaronHuot:2008ni}
S.~Caron-Huot, {O(g) plasma effects in jet quenching}, Phys.Rev. D79 (2009)
  065039.
\newblock \href {http://arxiv.org/abs/0811.1603} {\path{arXiv:0811.1603}},
  \href {https://doi.org/10.1103/PhysRevD.79.065039}
  {\path{doi:10.1103/PhysRevD.79.065039}}.

\bibitem{Ghiglieri:2015zma}
J.~Ghiglieri, D.~Teaney, {Parton energy loss and momentum broadening at NLO in
  high temperature QCD plasmas}, Int. J. Mod. Phys. E24~(11) (2015) 1530013,
  also appears in QGP5, ed. X.N. Wang, World Scientific (2015).
\newblock \href {http://arxiv.org/abs/1502.03730} {\path{arXiv:1502.03730}},
  \href {https://doi.org/10.1142/S0218301315300131, 10.1142/9789814663717_0006}
  {\path{doi:10.1142/S0218301315300131, 10.1142/9789814663717_0006}}.

\bibitem{Aurenche:2002pd}
P.~Aurenche, F.~Gelis, H.~Zaraket, {A Simple sum rule for the thermal gluon
  spectral function and applications}, JHEP 0205 (2002) 043.
\newblock \href {http://arxiv.org/abs/hep-ph/0204146}
  {\path{arXiv:hep-ph/0204146}}.

\bibitem{Peigne:2007sd}
S.~Peigne, A.~Peshier, {Collisional Energy Loss of a Fast Muon in a Hot QED
  Plasma}, Phys.Rev. D77 (2008) 014015.
\newblock \href {http://arxiv.org/abs/0710.1266} {\path{arXiv:0710.1266}},
  \href {https://doi.org/10.1103/PhysRevD.77.014015}
  {\path{doi:10.1103/PhysRevD.77.014015}}.

\bibitem{Ghiglieri:2015ala}
J.~Ghiglieri, G.~D. Moore, D.~Teaney, {Jet-Medium Interactions at NLO in a
  Weakly-Coupled Quark-Gluon Plasma}, JHEP 03 (2016) 095.
\newblock \href {http://arxiv.org/abs/1509.07773} {\path{arXiv:1509.07773}},
  \href {https://doi.org/10.1007/JHEP03(2016)095}
  {\path{doi:10.1007/JHEP03(2016)095}}.

\bibitem{Besak:2012qm}
D.~Besak, D.~{B\"odeker}, {Thermal production of ultrarelativistic right-handed
  neutrinos: Complete leading-order results}, JCAP 1203 (2012) 029.
\newblock \href {http://arxiv.org/abs/1202.1288} {\path{arXiv:1202.1288}},
  \href {https://doi.org/10.1088/1475-7516/2012/03/029}
  {\path{doi:10.1088/1475-7516/2012/03/029}}.

\bibitem{Aurenche:1996sh}
P.~Aurenche, F.~Gelis, R.~Kobes, E.~Petitgirard, {Breakdown of the hard thermal
  loop expansion near the light cone}, Z.Phys. C75 (1997) 315--332.
\newblock \href {http://arxiv.org/abs/hep-ph/9609256}
  {\path{arXiv:hep-ph/9609256}}, \href {https://doi.org/10.1007/s002880050475}
  {\path{doi:10.1007/s002880050475}}.

\bibitem{Aurenche:1999tq}
P.~Aurenche, F.~Gelis, H.~Zaraket, {KLN theorem, magnetic mass, and thermal
  photon production}, Phys.Rev. D61 (2000) 116001.
\newblock \href {http://arxiv.org/abs/hep-ph/9911367}
  {\path{arXiv:hep-ph/9911367}}, \href
  {https://doi.org/10.1103/PhysRevD.61.116001}
  {\path{doi:10.1103/PhysRevD.61.116001}}.

\bibitem{Aurenche:2000gf}
P.~Aurenche, F.~Gelis, H.~Zaraket, {Landau-Pomeranchuk-Migdal effect in thermal
  field theory}, Phys.Rev. D62 (2000) 096012.
\newblock \href {http://arxiv.org/abs/hep-ph/0003326}
  {\path{arXiv:hep-ph/0003326}}, \href
  {https://doi.org/10.1103/PhysRevD.62.096012}
  {\path{doi:10.1103/PhysRevD.62.096012}}.

\bibitem{Landau:1953um}
L.~Landau, I.~Pomeranchuk, {Limits of applicability of the theory of
  bremsstrahlung electrons and pair production at high-energies},
  Dokl.Akad.Nauk Ser.Fiz. 92 (1953) 535--536.

\bibitem{Landau:1953gr}
L.~Landau, I.~Pomeranchuk, {Electron cascade process at very high-energies},
  Dokl.Akad.Nauk Ser.Fiz. 92 (1953) 735--738.

\bibitem{Migdal:1956tc}
A.~B. Migdal, {Bremsstrahlung and pair production in condensed media at
  high-energies}, Phys.Rev. 103 (1956) 1811--1820.
\newblock \href {https://doi.org/10.1103/PhysRev.103.1811}
  {\path{doi:10.1103/PhysRev.103.1811}}.

\bibitem{Baier:1994bd}
R.~Baier, Y.~L. Dokshitzer, S.~Peigne, D.~Schiff, {Induced gluon radiation in a
  QCD medium}, Phys.Lett. B345 (1995) 277--286.
\newblock \href {http://arxiv.org/abs/hep-ph/9411409}
  {\path{arXiv:hep-ph/9411409}}, \href
  {https://doi.org/10.1016/0370-2693(94)01617-L}
  {\path{doi:10.1016/0370-2693(94)01617-L}}.

\bibitem{Baier:1996kr}
R.~Baier, Y.~L. Dokshitzer, A.~H. Mueller, S.~Peigne, D.~Schiff, {Radiative
  energy loss of high-energy quarks and gluons in a finite volume quark - gluon
  plasma}, Nucl.Phys. B483 (1997) 291--320.
\newblock \href {http://arxiv.org/abs/hep-ph/9607355}
  {\path{arXiv:hep-ph/9607355}}, \href
  {https://doi.org/10.1016/S0550-3213(96)00553-6}
  {\path{doi:10.1016/S0550-3213(96)00553-6}}.

\bibitem{Zakharov:1996fv}
B.~Zakharov, {Fully quantum treatment of the Landau-Pomeranchuk-Migdal effect
  in QED and QCD}, JETP Lett. 63 (1996) 952--957.
\newblock \href {http://arxiv.org/abs/hep-ph/9607440}
  {\path{arXiv:hep-ph/9607440}}, \href {https://doi.org/10.1134/1.567126}
  {\path{doi:10.1134/1.567126}}.

\bibitem{Zakharov:1997uu}
B.~Zakharov, {Radiative energy loss of high-energy quarks in finite size
  nuclear matter and quark - gluon plasma}, JETP Lett. 65 (1997) 615--620.
\newblock \href {http://arxiv.org/abs/hep-ph/9704255}
  {\path{arXiv:hep-ph/9704255}}, \href {https://doi.org/10.1134/1.567389}
  {\path{doi:10.1134/1.567389}}.

\bibitem{Arnold:2001ba}
P.~B. Arnold, G.~D. Moore, L.~G. Yaffe, {Photon emission from ultrarelativistic
  plasmas}, JHEP 0111 (2001) 057.
\newblock \href {http://arxiv.org/abs/hep-ph/0109064}
  {\path{arXiv:hep-ph/0109064}}.

\bibitem{Arnold:2002ja}
P.~B. Arnold, G.~D. Moore, L.~G. Yaffe, {Photon and gluon emission in
  relativistic plasmas}, JHEP 0206 (2002) 030.
\newblock \href {http://arxiv.org/abs/hep-ph/0204343}
  {\path{arXiv:hep-ph/0204343}}.

\bibitem{CaronHuot:2010bp}
S.~Caron-Huot, C.~Gale, {Finite-size effects on the radiative energy loss of a
  fast parton in hot and dense strongly interacting matter}, Phys. Rev. C82
  (2010) 064902.
\newblock \href {http://arxiv.org/abs/1006.2379} {\path{arXiv:1006.2379}},
  \href {https://doi.org/10.1103/PhysRevC.82.064902}
  {\path{doi:10.1103/PhysRevC.82.064902}}.

\bibitem{Arnold:2015qya}
P.~Arnold, S.~Iqbal, {The LPM effect in sequential bremsstrahlung}, JHEP 04
  (2015) 070, [Erratum: JHEP09,072(2016)].
\newblock \href {http://arxiv.org/abs/1501.04964} {\path{arXiv:1501.04964}},
  \href {https://doi.org/10.1007/JHEP09(2016)072, 10.1007/JHEP04(2015)070}
  {\path{doi:10.1007/JHEP09(2016)072, 10.1007/JHEP04(2015)070}}.

\bibitem{Arnold:2008iy}
P.~B. Arnold, {Simple Formula for High-Energy Gluon Bremsstrahlung in a Finite,
  Expanding Medium}, Phys.Rev. D79 (2009) 065025.
\newblock \href {http://arxiv.org/abs/0808.2767} {\path{arXiv:0808.2767}},
  \href {https://doi.org/10.1103/PhysRevD.79.065025}
  {\path{doi:10.1103/PhysRevD.79.065025}}.

\bibitem{Aurenche:2002wq}
P.~Aurenche, F.~Gelis, G.~Moore, H.~Zaraket, {Landau-Pomeranchuk-Migdal
  resummation for dilepton production}, JHEP 0212 (2002) 006.
\newblock \href {http://arxiv.org/abs/hep-ph/0211036}
  {\path{arXiv:hep-ph/0211036}}.

\bibitem{Carrington:2007gt}
M.~E. Carrington, A.~Gynther, P.~Aurenche, {Energetic di-leptons from the Quark
  Gluon Plasma}, Phys. Rev. D77 (2008) 045035.
\newblock \href {http://arxiv.org/abs/0711.3943} {\path{arXiv:0711.3943}},
  \href {https://doi.org/10.1103/PhysRevD.77.045035}
  {\path{doi:10.1103/PhysRevD.77.045035}}.

\bibitem{Anisimov:2010gy}
A.~Anisimov, D.~Besak, D.~{B\"odeker}, {Thermal production of relativistic
  Majorana neutrinos: Strong enhancement by multiple soft scattering}, JCAP
  1103 (2011) 042.
\newblock \href {http://arxiv.org/abs/1012.3784} {\path{arXiv:1012.3784}},
  \href {https://doi.org/10.1088/1475-7516/2011/03/042}
  {\path{doi:10.1088/1475-7516/2011/03/042}}.

\bibitem{D'Eramo:2010ak}
F.~D'Eramo, H.~Liu, K.~Rajagopal, {Transverse Momentum Broadening and the Jet
  Quenching Parameter, Redux}, Phys.Rev. D84 (2011) 065015.
\newblock \href {http://arxiv.org/abs/1006.1367} {\path{arXiv:1006.1367}},
  \href {https://doi.org/10.1103/PhysRevD.84.065015}
  {\path{doi:10.1103/PhysRevD.84.065015}}.

\bibitem{Benzke:2012sz}
M.~Benzke, N.~Brambilla, M.~A. Escobedo, A.~Vairo, {Gauge invariant definition
  of the jet quenching parameter}, JHEP 1302 (2013) 129.
\newblock \href {http://arxiv.org/abs/1208.4253} {\path{arXiv:1208.4253}},
  \href {https://doi.org/10.1007/JHEP02(2013)129}
  {\path{doi:10.1007/JHEP02(2013)129}}.

\bibitem{d'Enterria:2009am}
D.~d'Enterria, {Jet quenching}, Springer Verlag, Landholt-Boernstein Vol.
  1-23A.
\newblock \href {http://arxiv.org/abs/0902.2011} {\path{arXiv:0902.2011}}.

\bibitem{Wiedemann:2009sh}
U.~A. Wiedemann, {Jet Quenching in Heavy Ion Collisions} (2010)
  521--562[Landolt-Bornstein23,521(2010)].
\newblock \href {http://arxiv.org/abs/0908.2306} {\path{arXiv:0908.2306}},
  \href {https://doi.org/10.1007/978-3-642-01539-7_17}
  {\path{doi:10.1007/978-3-642-01539-7_17}}.

\bibitem{Majumder:2010qh}
A.~Majumder, M.~Van~Leeuwen, {The Theory and Phenomenology of Perturbative QCD
  Based Jet Quenching}, Prog.Part.Nucl.Phys. A66 (2011) 41--92.
\newblock \href {http://arxiv.org/abs/1002.2206} {\path{arXiv:1002.2206}},
  \href {https://doi.org/10.1016/j.ppnp.2010.09.001}
  {\path{doi:10.1016/j.ppnp.2010.09.001}}.

\bibitem{Mehtar-Tani:2013pia}
Y.~Mehtar-Tani, J.~G. Milhano, K.~Tywoniuk, {Jet physics in heavy-ion
  collisions}, Int.J.Mod.Phys. A28 (2013) 1340013.
\newblock \href {http://arxiv.org/abs/1302.2579} {\path{arXiv:1302.2579}},
  \href {https://doi.org/10.1142/S0217751X13400137}
  {\path{doi:10.1142/S0217751X13400137}}.

\bibitem{Roland:2014jsa}
G.~Roland, K.~Safarik, P.~Steinberg, {Heavy-ion collisions at the LHC},
  Prog.Part.Nucl.Phys. 77 (2014) 70--127.
\newblock \href {https://doi.org/10.1016/j.ppnp.2014.05.001}
  {\path{doi:10.1016/j.ppnp.2014.05.001}}.

\bibitem{Qin:2015srf}
G.-Y. Qin, X.-N. Wang, {Jet quenching in high-energy heavy-ion collisions},
  Int. J. Mod. Phys. E24~(11) (2015) 1530014, [,309(2016)].
\newblock \href {http://arxiv.org/abs/1511.00790} {\path{arXiv:1511.00790}},
  \href {https://doi.org/10.1142/S0218301315300143, 10.1142/9789814663717_0007}
  {\path{doi:10.1142/S0218301315300143, 10.1142/9789814663717_0007}}.

\bibitem{Connors:2017ptx}
M.~Connors, C.~Nattrass, R.~Reed, S.~Salur, {Jet measurements in heavy ion
  physics}, Rev. Mod. Phys. 90 (2018) 025005.
\newblock \href {http://arxiv.org/abs/1705.01974} {\path{arXiv:1705.01974}},
  \href {https://doi.org/10.1103/RevModPhys.90.025005}
  {\path{doi:10.1103/RevModPhys.90.025005}}.

\bibitem{Arnold:2003zc}
P.~B. Arnold, G.~D. Moore, L.~G. Yaffe, {Transport coefficients in high
  temperature gauge theories. 2. Beyond leading log}, JHEP 0305 (2003) 051.
\newblock \href {http://arxiv.org/abs/hep-ph/0302165}
  {\path{arXiv:hep-ph/0302165}}, \href
  {https://doi.org/10.1088/1126-6708/2003/05/051}
  {\path{doi:10.1088/1126-6708/2003/05/051}}.

\bibitem{Apolinario:2014csa}
L.~Apolin\'ario, N.~Armesto, J.~G. Milhano, C.~A. Salgado, {Medium-induced
  gluon radiation and colour decoherence beyond the soft approximation}, JHEP
  02 (2015) 119.
\newblock \href {http://arxiv.org/abs/1407.0599} {\path{arXiv:1407.0599}},
  \href {https://doi.org/10.1007/JHEP02(2015)119}
  {\path{doi:10.1007/JHEP02(2015)119}}.

\bibitem{Bauer:2000ew}
C.~W. Bauer, S.~Fleming, M.~E. Luke, {Summing Sudakov logarithms in B $\to$ X(s
  gamma) in effective field theory}, Phys.Rev. D63 (2000) 014006.
\newblock \href {http://arxiv.org/abs/hep-ph/0005275}
  {\path{arXiv:hep-ph/0005275}}, \href
  {https://doi.org/10.1103/PhysRevD.63.014006}
  {\path{doi:10.1103/PhysRevD.63.014006}}.

\bibitem{Bauer:2000yr}
C.~W. Bauer, S.~Fleming, D.~Pirjol, I.~W. Stewart, {An Effective field theory
  for collinear and soft gluons: Heavy to light decays}, Phys.Rev. D63 (2001)
  114020.
\newblock \href {http://arxiv.org/abs/hep-ph/0011336}
  {\path{arXiv:hep-ph/0011336}}, \href
  {https://doi.org/10.1103/PhysRevD.63.114020}
  {\path{doi:10.1103/PhysRevD.63.114020}}.

\bibitem{Bauer:2001ct}
C.~W. Bauer, I.~W. Stewart, {Invariant operators in collinear effective
  theory}, Phys.Lett. B516 (2001) 134--142.
\newblock \href {http://arxiv.org/abs/hep-ph/0107001}
  {\path{arXiv:hep-ph/0107001}}, \href
  {https://doi.org/10.1016/S0370-2693(01)00902-9}
  {\path{doi:10.1016/S0370-2693(01)00902-9}}.

\bibitem{Bauer:2001yt}
C.~W. Bauer, D.~Pirjol, I.~W. Stewart, {Soft collinear factorization in
  effective field theory}, Phys.Rev. D65 (2002) 054022.
\newblock \href {http://arxiv.org/abs/hep-ph/0109045}
  {\path{arXiv:hep-ph/0109045}}, \href
  {https://doi.org/10.1103/PhysRevD.65.054022}
  {\path{doi:10.1103/PhysRevD.65.054022}}.

\bibitem{Bauer:2002nz}
C.~W. Bauer, S.~Fleming, D.~Pirjol, I.~Z. Rothstein, I.~W. Stewart, {Hard
  scattering factorization from effective field theory}, Phys.Rev. D66 (2002)
  014017.
\newblock \href {http://arxiv.org/abs/hep-ph/0202088}
  {\path{arXiv:hep-ph/0202088}}, \href
  {https://doi.org/10.1103/PhysRevD.66.014017}
  {\path{doi:10.1103/PhysRevD.66.014017}}.

\bibitem{Beneke:2002ph}
M.~Beneke, A.~Chapovsky, M.~Diehl, T.~Feldmann, {Soft collinear effective
  theory and heavy to light currents beyond leading power}, Nucl.Phys. B643
  (2002) 431--476.
\newblock \href {http://arxiv.org/abs/hep-ph/0206152}
  {\path{arXiv:hep-ph/0206152}}, \href
  {https://doi.org/10.1016/S0550-3213(02)00687-9}
  {\path{doi:10.1016/S0550-3213(02)00687-9}}.

\bibitem{Becher:2014oda}
T.~Becher, A.~Broggio, A.~Ferroglia, {Introduction to Soft-Collinear Effective
  Theory}, Lect. Notes Phys. 896 (2015) pp.1--206.
\newblock \href {http://arxiv.org/abs/1410.1892} {\path{arXiv:1410.1892}},
  \href {https://doi.org/10.1007/978-3-319-14848-9}
  {\path{doi:10.1007/978-3-319-14848-9}}.

\bibitem{Gervais:2012wd}
H.~Gervais, S.~Jeon, {Photon Production from a Quark-Gluon-Plasma at Finite
  Baryon Chemical Potential}, Phys.Rev. C86 (2012) 034904.
\newblock \href {http://arxiv.org/abs/1206.6086} {\path{arXiv:1206.6086}},
  \href {https://doi.org/10.1103/PhysRevC.86.034904}
  {\path{doi:10.1103/PhysRevC.86.034904}}.

\bibitem{Ghiglieri:2017gjz}
J.~Ghiglieri, M.~Laine, {GeV-scale hot sterile neutrino oscillations: a
  derivation of evolution equations}, JHEP 05 (2017) 132.
\newblock \href {http://arxiv.org/abs/1703.06087} {\path{arXiv:1703.06087}},
  \href {https://doi.org/10.1007/JHEP05(2017)132}
  {\path{doi:10.1007/JHEP05(2017)132}}.

\bibitem{Ghiglieri:2018wbs}
J.~Ghiglieri, M.~Laine, {Precision study of GeV-scale resonant leptogenesis},
  JHEP 02 (2019) 014.
\newblock \href {http://arxiv.org/abs/1811.01971} {\path{arXiv:1811.01971}},
  \href {https://doi.org/10.1007/JHEP02(2019)014}
  {\path{doi:10.1007/JHEP02(2019)014}}.

\bibitem{Ghiglieri:2014vua}
J.~Ghiglieri, {Next-to-leading order thermal photon production in a
  weakly-coupled plasma}, Nucl. Phys. A932 (2014) 326--333.
\newblock \href {http://arxiv.org/abs/1404.0626} {\path{arXiv:1404.0626}},
  \href {https://doi.org/10.1016/j.nuclphysa.2014.09.048}
  {\path{doi:10.1016/j.nuclphysa.2014.09.048}}.

\bibitem{Ghisoiu:2014mha}
I.~Ghisoiu, M.~Laine, {Interpolation of hard and soft dilepton rates}, JHEP 10
  (2014) 083.
\newblock \href {http://arxiv.org/abs/1407.7955} {\path{arXiv:1407.7955}},
  \href {https://doi.org/10.1007/JHEP10(2014)083}
  {\path{doi:10.1007/JHEP10(2014)083}}.

\bibitem{CaronHuot:2008uw}
S.~Caron-Huot, {On supersymmetry at finite temperature}, Phys.Rev. D79 (2009)
  125002.
\newblock \href {http://arxiv.org/abs/0808.0155} {\path{arXiv:0808.0155}},
  \href {https://doi.org/10.1103/PhysRevD.79.125002}
  {\path{doi:10.1103/PhysRevD.79.125002}}.

\bibitem{Ghiglieri:2015nba}
J.~Ghiglieri, {The thermal dilepton rate at NLO at small and large invariant
  mass}, Nucl. Part. Phys. Proc. 276-278 (2016) 305--308.
\newblock \href {http://arxiv.org/abs/1510.00525} {\path{arXiv:1510.00525}},
  \href {https://doi.org/10.1016/j.nuclphysbps.2016.05.070}
  {\path{doi:10.1016/j.nuclphysbps.2016.05.070}}.

\bibitem{Baier:1988xv}
R.~Baier, B.~Pire, D.~Schiff, {Dilepton production at finite temperature:
  Perturbative treatment at order $\alpha_s$}, Phys. Rev. D38 (1988) 2814.
\newblock \href {https://doi.org/10.1103/PhysRevD.38.2814}
  {\path{doi:10.1103/PhysRevD.38.2814}}.

\bibitem{Gabellini:1989yk}
Y.~Gabellini, T.~Grandou, D.~Poizat, {Electron - Positron Annihilation in
  Thermal {QCD}}, Annals Phys. 202 (1990) 436--466.
\newblock \href {https://doi.org/10.1016/0003-4916(90)90231-C}
  {\path{doi:10.1016/0003-4916(90)90231-C}}.

\bibitem{Altherr:1989fc}
T.~Altherr, P.~Aurenche, {About Fermion Selfenergy Corrections in Perturbative
  Theory at Finite Temperature}, Phys. Rev. D40 (1989) 4171.
\newblock \href {https://doi.org/10.1103/PhysRevD.40.4171}
  {\path{doi:10.1103/PhysRevD.40.4171}}.

\bibitem{Baier:1989ub}
R.~Baier, E.~Pilon, B.~Pire, D.~Schiff, {Finite Temperature Radiative
  Corrections to Early Universe Neutron - Proton Ratio: Cancellation of
  Infrared and Mass Singularities}, Nucl. Phys. B336 (1990) 157--183.
\newblock \href {https://doi.org/10.1016/0550-3213(90)90347-G}
  {\path{doi:10.1016/0550-3213(90)90347-G}}.

\bibitem{Braaten:1990wp}
E.~Braaten, R.~D. Pisarski, T.-C. Yuan, {Production of Soft Dileptons in the
  Quark - Gluon Plasma}, Phys.Rev.Lett. 64 (1990) 2242.
\newblock \href {https://doi.org/10.1103/PhysRevLett.64.2242}
  {\path{doi:10.1103/PhysRevLett.64.2242}}.

\bibitem{Moore:2006qn}
G.~D. Moore, J.-M. Robert, {Dileptons, spectral weights, and conductivity in
  the quark-gluon plasma}\href {http://arxiv.org/abs/hep-ph/0607172}
  {\path{arXiv:hep-ph/0607172}}.

\bibitem{Laine:2013vma}
M.~Laine, {NLO thermal dilepton rate at non-zero momentum}, JHEP 1311 (2013)
  120.
\newblock \href {http://arxiv.org/abs/1310.0164} {\path{arXiv:1310.0164}},
  \href {https://doi.org/10.1007/JHEP11(2013)120}
  {\path{doi:10.1007/JHEP11(2013)120}}.

\bibitem{Laine:2013lka}
M.~Laine, {Thermal right-handed neutrino production rate in the relativistic
  regime}, JHEP 1308 (2013) 138.
\newblock \href {http://arxiv.org/abs/1307.4909} {\path{arXiv:1307.4909}},
  \href {https://doi.org/10.1007/JHEP08(2013)138}
  {\path{doi:10.1007/JHEP08(2013)138}}.

\bibitem{Jackson:2019yao}
G.~Jackson, M.~Laine, {Testing thermal photon and dilepton rates}, JHEP 11
  (2019) 144.
\newblock \href {http://arxiv.org/abs/1910.09567} {\path{arXiv:1910.09567}},
  \href {https://doi.org/10.1007/JHEP11(2019)144}
  {\path{doi:10.1007/JHEP11(2019)144}}.

\bibitem{Meyer:2011gj}
H.~B. Meyer, {Transport Properties of the Quark-Gluon Plasma: A Lattice QCD
  Perspective}, Eur. Phys. J. A47 (2011) 86.
\newblock \href {http://arxiv.org/abs/1104.3708} {\path{arXiv:1104.3708}},
  \href {https://doi.org/10.1140/epja/i2011-11086-3}
  {\path{doi:10.1140/epja/i2011-11086-3}}.

\bibitem{Laine:2013vpa}
M.~Laine, {Thermal 2-loop master spectral function at finite momentum}, JHEP
  1305 (2013) 083.
\newblock \href {http://arxiv.org/abs/1304.0202} {\path{arXiv:1304.0202}},
  \href {https://doi.org/10.1007/JHEP05(2013)083}
  {\path{doi:10.1007/JHEP05(2013)083}}.

\bibitem{Jackson:2019mop}
G.~Jackson, {Two-loop thermal spectral functions with general kinematics},
  Phys. Rev. D100~(11) (2019) 116019.
\newblock \href {http://arxiv.org/abs/1910.07552} {\path{arXiv:1910.07552}},
  \href {https://doi.org/10.1103/PhysRevD.100.116019}
  {\path{doi:10.1103/PhysRevD.100.116019}}.

\bibitem{Ghiglieri:2016tvj}
J.~Ghiglieri, O.~Kaczmarek, M.~Laine, F.~Meyer, {Lattice constraints on the
  thermal photon rate}, Phys. Rev. D94~(1) (2016) 016005.
\newblock \href {http://arxiv.org/abs/1604.07544} {\path{arXiv:1604.07544}},
  \href {https://doi.org/10.1103/PhysRevD.94.016005}
  {\path{doi:10.1103/PhysRevD.94.016005}}.

\bibitem{Brandt:2017vgl}
B.~B. Brandt, A.~Francis, T.~Harris, H.~B. Meyer, A.~Steinberg, {An estimate
  for the thermal photon rate from lattice QCD}, EPJ Web Conf. 175 (2018)
  07044.
\newblock \href {http://arxiv.org/abs/1710.07050} {\path{arXiv:1710.07050}},
  \href {https://doi.org/10.1051/epjconf/201817507044}
  {\path{doi:10.1051/epjconf/201817507044}}.

\bibitem{Brandt:2019shg}
B.~B. Brandt, M.~C\`e, A.~Francis, T.~Harris, H.~B. Meyer, A.~Steinberg,
  A.~Toniato, {Lattice QCD estimate of the quark-gluon plasma photon emission
  rate}, in: {37th International Symposium on Lattice Field Theory (Lattice
  2019) Wuhan, Hubei, China, June 16-22, 2019}, 2019.
\newblock \href {http://arxiv.org/abs/1912.00292} {\path{arXiv:1912.00292}}.

\bibitem{Ce:2020tmx}
M.~C\`e, T.~Harris, H.~B. Meyer, A.~Steinberg, A.~Toniato, {The rate of photon
  production in the quark-gluon plasma from lattice QCD}\href
  {http://arxiv.org/abs/2001.03368} {\path{arXiv:2001.03368}}.

\bibitem{Paquet:2017wji}
J.-F. Paquet, {Probing the space-time evolution of heavy ion collisions with
  photons and dileptons}, Nucl. Phys. A967 (2017) 184--191.
\newblock \href {http://arxiv.org/abs/1704.07842} {\path{arXiv:1704.07842}},
  \href {https://doi.org/10.1016/j.nuclphysa.2017.06.003}
  {\path{doi:10.1016/j.nuclphysa.2017.06.003}}.

\bibitem{Shen:2013cca}
C.~Shen, U.~W. Heinz, J.-F. Paquet, I.~Kozlov, C.~Gale, {Anisotropic flow of
  thermal photons as a quark-gluon plasma viscometer}, Phys. Rev. C91~(2)
  (2015) 024908.
\newblock \href {http://arxiv.org/abs/1308.2111} {\path{arXiv:1308.2111}},
  \href {https://doi.org/10.1103/PhysRevC.91.024908}
  {\path{doi:10.1103/PhysRevC.91.024908}}.

\bibitem{Hauksson:2017udm}
S.~Hauksson, S.~Jeon, C.~Gale, {Photon emission from quark-gluon plasma out of
  equilibrium}, Phys. Rev. C97~(1) (2018) 014901.
\newblock \href {http://arxiv.org/abs/1709.03598} {\path{arXiv:1709.03598}},
  \href {https://doi.org/10.1103/PhysRevC.97.014901}
  {\path{doi:10.1103/PhysRevC.97.014901}}.

\bibitem{Heinz:2013th}
U.~Heinz, R.~Snellings, {Collective flow and viscosity in relativistic
  heavy-ion collisions}, Ann. Rev. Nucl. Part. Sci. 63 (2013) 123--151.
\newblock \href {http://arxiv.org/abs/1301.2826} {\path{arXiv:1301.2826}},
  \href {https://doi.org/10.1146/annurev-nucl-102212-170540}
  {\path{doi:10.1146/annurev-nucl-102212-170540}}.

\bibitem{Gale:2013da}
C.~Gale, S.~Jeon, B.~Schenke, {Hydrodynamic Modeling of Heavy-Ion Collisions},
  Int. J. Mod. Phys. A28 (2013) 1340011.
\newblock \href {http://arxiv.org/abs/1301.5893} {\path{arXiv:1301.5893}},
  \href {https://doi.org/10.1142/S0217751X13400113}
  {\path{doi:10.1142/S0217751X13400113}}.

\bibitem{Jeon:2015dfa}
S.~Jeon, U.~Heinz, {Introduction to Hydrodynamics}, Int. J. Mod. Phys. E24~(10)
  (2015) 1530010.
\newblock \href {http://arxiv.org/abs/1503.03931} {\path{arXiv:1503.03931}},
  \href {https://doi.org/10.1142/S0218301315300106}
  {\path{doi:10.1142/S0218301315300106}}.

\bibitem{Arnold:2000dr}
P.~B. Arnold, G.~D. Moore, L.~G. Yaffe, {Transport coefficients in high
  temperature gauge theories. 1. Leading log results}, JHEP 11 (2000) 001.
\newblock \href {http://arxiv.org/abs/hep-ph/0010177}
  {\path{arXiv:hep-ph/0010177}}, \href
  {https://doi.org/10.1088/1126-6708/2000/11/001}
  {\path{doi:10.1088/1126-6708/2000/11/001}}.

\bibitem{Hosoya:1983id}
A.~Hosoya, M.-a. Sakagami, M.~Takao, {Nonequilibrium Thermodynamics in Field
  Theory: Transport Coefficients}, Annals Phys. 154 (1984) 229.
\newblock \href {https://doi.org/10.1016/0003-4916(84)90144-1}
  {\path{doi:10.1016/0003-4916(84)90144-1}}.

\bibitem{Aarts:2002tn}
G.~Aarts, J.~M. Martinez~Resco, {Ward identity and electrical conductivity in
  hot QED}, JHEP 11 (2002) 022.
\newblock \href {http://arxiv.org/abs/hep-ph/0209048}
  {\path{arXiv:hep-ph/0209048}}, \href
  {https://doi.org/10.1088/1126-6708/2002/11/022}
  {\path{doi:10.1088/1126-6708/2002/11/022}}.

\bibitem{Aarts:2004sd}
G.~Aarts, J.~M. Martinez~Resco, {Shear viscosity in the O(N) model}, JHEP 02
  (2004) 061.
\newblock \href {http://arxiv.org/abs/hep-ph/0402192}
  {\path{arXiv:hep-ph/0402192}}, \href
  {https://doi.org/10.1088/1126-6708/2004/02/061}
  {\path{doi:10.1088/1126-6708/2004/02/061}}.

\bibitem{Aarts:2005vc}
G.~Aarts, J.~M. Martinez~Resco, {Transport coefficients in large N(f) gauge
  theories with massive fermions}, JHEP 03 (2005) 074.
\newblock \href {http://arxiv.org/abs/hep-ph/0503161}
  {\path{arXiv:hep-ph/0503161}}, \href
  {https://doi.org/10.1088/1126-6708/2005/03/074}
  {\path{doi:10.1088/1126-6708/2005/03/074}}.

\bibitem{Gagnon:2006hi}
J.-S. Gagnon, S.~Jeon, {Leading order calculation of electric conductivity in
  hot quantum electrodynamics from diagrammatic methods}, Phys. Rev. D75 (2007)
  025014, [Erratum: Phys. Rev.D76,089902(2007)].
\newblock \href {http://arxiv.org/abs/hep-ph/0610235}
  {\path{arXiv:hep-ph/0610235}}, \href
  {https://doi.org/10.1103/PhysRevD.75.025014, 10.1103/PhysRevD.76.089902}
  {\path{doi:10.1103/PhysRevD.75.025014, 10.1103/PhysRevD.76.089902}}.

\bibitem{Gagnon:2007qt}
J.-S. Gagnon, S.~Jeon, {Leading Order Calculation of Shear Viscosity in Hot
  Quantum Electrodynamics from Diagrammatic Methods}, Phys. Rev. D76 (2007)
  105019.
\newblock \href {http://arxiv.org/abs/0708.1631} {\path{arXiv:0708.1631}},
  \href {https://doi.org/10.1103/PhysRevD.76.105019}
  {\path{doi:10.1103/PhysRevD.76.105019}}.

\bibitem{Arnold:2002zm}
P.~B. Arnold, G.~D. Moore, L.~G. Yaffe, {Effective kinetic theory for high
  temperature gauge theories}, JHEP 0301 (2003) 030.
\newblock \href {http://arxiv.org/abs/hep-ph/0209353}
  {\path{arXiv:hep-ph/0209353}}, \href
  {https://doi.org/10.1088/1126-6708/2003/01/030}
  {\path{doi:10.1088/1126-6708/2003/01/030}}.

\bibitem{Jeon:1994if}
S.~Jeon, {Hydrodynamic transport coefficients in relativistic scalar field
  theory}, Phys. Rev. D52 (1995) 3591--3642.
\newblock \href {http://arxiv.org/abs/hep-ph/9409250}
  {\path{arXiv:hep-ph/9409250}}, \href
  {https://doi.org/10.1103/PhysRevD.52.3591}
  {\path{doi:10.1103/PhysRevD.52.3591}}.

\bibitem{Ghiglieri:2018dib}
J.~Ghiglieri, G.~D. Moore, D.~Teaney, {QCD Shear Viscosity at (almost) NLO},
  JHEP 03 (2018) 179.
\newblock \href {http://arxiv.org/abs/1802.09535} {\path{arXiv:1802.09535}},
  \href {https://doi.org/10.1007/JHEP03(2018)179}
  {\path{doi:10.1007/JHEP03(2018)179}}.

\bibitem{Laine:2005ai}
M.~Laine, Y.~{Schr\"oder}, {Two-loop QCD gauge coupling at high temperatures},
  JHEP 0503 (2005) 067.
\newblock \href {http://arxiv.org/abs/hep-ph/0503061}
  {\path{arXiv:hep-ph/0503061}}, \href
  {https://doi.org/10.1088/1126-6708/2005/03/067}
  {\path{doi:10.1088/1126-6708/2005/03/067}}.

\bibitem{Arnold:2006fz}
P.~B. Arnold, C.~Dogan, G.~D. Moore, {The Bulk Viscosity of High-Temperature
  QCD}, Phys. Rev. D74 (2006) 085021.
\newblock \href {http://arxiv.org/abs/hep-ph/0608012}
  {\path{arXiv:hep-ph/0608012}}, \href
  {https://doi.org/10.1103/PhysRevD.74.085021}
  {\path{doi:10.1103/PhysRevD.74.085021}}.

\bibitem{Laine:2014hba}
M.~Laine, K.~A. Sohrabi, {Charm contribution to bulk viscosity}, Eur. Phys. J.
  C75~(2) (2015) 80.
\newblock \href {http://arxiv.org/abs/1410.6583} {\path{arXiv:1410.6583}},
  \href {https://doi.org/10.1140/epjc/s10052-015-3297-9}
  {\path{doi:10.1140/epjc/s10052-015-3297-9}}.

\bibitem{Heiselberg:1993cr}
H.~Heiselberg, C.~Pethick, {Transport and relaxation in degenerate quark
  plasmas}, Phys.Rev. D48 (1993) 2916--2928.
\newblock \href {https://doi.org/10.1103/PhysRevD.48.2916}
  {\path{doi:10.1103/PhysRevD.48.2916}}.

\bibitem{Ghiglieri:2018ltw}
J.~Ghiglieri, H.~Kim, {Transverse momentum broadening and collinear radiation
  at NLO in the $\mathcal{N}=4$ SYM plasma}, JHEP 12 (2018) 049.
\newblock \href {http://arxiv.org/abs/1809.01349} {\path{arXiv:1809.01349}},
  \href {https://doi.org/10.1007/JHEP12(2018)049}
  {\path{doi:10.1007/JHEP12(2018)049}}.

\bibitem{Panero:2013pla}
M.~Panero, K.~Rummukainen, A.~{Sch\"afer}, {A lattice study of the jet
  quenching parameter}, Phys.Rev.Lett. 112 (2014) 162001.
\newblock \href {http://arxiv.org/abs/1307.5850} {\path{arXiv:1307.5850}},
  \href {https://doi.org/10.1103/PhysRevLett.112.162001}
  {\path{doi:10.1103/PhysRevLett.112.162001}}.

\bibitem{Moore:2019lgw}
G.~D. Moore, N.~Schlusser, {Transverse momentum broadening from the
  lattice}\href {http://arxiv.org/abs/1911.13127} {\path{arXiv:1911.13127}}.

\bibitem{Laine:2013lia}
M.~Laine, A.~Rothkopf, {Light-cone Wilson loop in classical lattice gauge
  theory}, JHEP 1307 (2013) 082.
\newblock \href {http://arxiv.org/abs/1304.4443} {\path{arXiv:1304.4443}},
  \href {https://doi.org/10.1007/JHEP07(2013)082}
  {\path{doi:10.1007/JHEP07(2013)082}}.

\bibitem{Matsui:1986dk}
T.~Matsui, H.~Satz, {$J/\psi$ Suppression by Quark-Gluon Plasma Formation},
  Phys. Lett. B178 (1986) 416--422.
\newblock \href {https://doi.org/10.1016/0370-2693(86)91404-8}
  {\path{doi:10.1016/0370-2693(86)91404-8}}.

\bibitem{Rothkopf:2019ipj}
A.~Rothkopf, {Heavy Quarkonium in Extreme Conditions}\href
  {http://arxiv.org/abs/1912.02253} {\path{arXiv:1912.02253}}.

\bibitem{Mocsy:2013syh}
A.~Mocsy, P.~Petreczky, M.~Strickland, {Quarkonia in the Quark Gluon Plasma},
  Int. J. Mod. Phys. A28 (2013) 1340012.
\newblock \href {http://arxiv.org/abs/1302.2180} {\path{arXiv:1302.2180}},
  \href {https://doi.org/10.1142/S0217751X13400125}
  {\path{doi:10.1142/S0217751X13400125}}.

\bibitem{Andronic:2015wma}
A.~Andronic, et~al., {Heavy-flavour and quarkonium production in the LHC era:
  from proton–proton to heavy-ion collisions}, Eur. Phys. J. C76~(3) (2016)
  107.
\newblock \href {http://arxiv.org/abs/1506.03981} {\path{arXiv:1506.03981}},
  \href {https://doi.org/10.1140/epjc/s10052-015-3819-5}
  {\path{doi:10.1140/epjc/s10052-015-3819-5}}.

\bibitem{Aarts:2016hap}
G.~Aarts, et~al., {Heavy-flavor production and medium properties in high-energy
  nuclear collisions - What next?}, Eur. Phys. J. A53~(5) (2017) 93.
\newblock \href {http://arxiv.org/abs/1612.08032} {\path{arXiv:1612.08032}},
  \href {https://doi.org/10.1140/epja/i2017-12282-9}
  {\path{doi:10.1140/epja/i2017-12282-9}}.

\bibitem{Svetitsky:1987gq}
B.~Svetitsky, {Diffusion of charmed quarks in the quark-gluon plasma}, Phys.
  Rev. D37 (1988) 2484--2491.
\newblock \href {https://doi.org/10.1103/PhysRevD.37.2484}
  {\path{doi:10.1103/PhysRevD.37.2484}}.

\bibitem{Braaten:1991we}
E.~Braaten, M.~H. Thoma, {Energy loss of a heavy quark in the quark - gluon
  plasma}, Phys. Rev. D44~(9) (1991) R2625.
\newblock \href {https://doi.org/10.1103/PhysRevD.44.R2625}
  {\path{doi:10.1103/PhysRevD.44.R2625}}.

\bibitem{Moore:2004tg}
G.~D. Moore, D.~Teaney, {How much do heavy quarks thermalize in a heavy ion
  collision?}, Phys. Rev. C71 (2005) 064904.
\newblock \href {http://arxiv.org/abs/hep-ph/0412346}
  {\path{arXiv:hep-ph/0412346}}, \href
  {https://doi.org/10.1103/PhysRevC.71.064904}
  {\path{doi:10.1103/PhysRevC.71.064904}}.

\bibitem{Brambilla:2016wgg}
N.~Brambilla, M.~A. Escobedo, J.~Soto, A.~Vairo, {Quarkonium suppression in
  heavy-ion collisions: an open quantum system approach}, Phys. Rev. D96~(3)
  (2017) 034021.
\newblock \href {http://arxiv.org/abs/1612.07248} {\path{arXiv:1612.07248}},
  \href {https://doi.org/10.1103/PhysRevD.96.034021}
  {\path{doi:10.1103/PhysRevD.96.034021}}.

\bibitem{Brambilla:2017zei}
N.~Brambilla, M.~A. Escobedo, J.~Soto, A.~Vairo, {Heavy quarkonium suppression
  in a fireball}, Phys. Rev. D97~(7) (2018) 074009.
\newblock \href {http://arxiv.org/abs/1711.04515} {\path{arXiv:1711.04515}},
  \href {https://doi.org/10.1103/PhysRevD.97.074009}
  {\path{doi:10.1103/PhysRevD.97.074009}}.

\bibitem{CasalderreySolana:2006rq}
J.~Casalderrey-Solana, D.~Teaney, {Heavy quark diffusion in strongly coupled
  N=4 Yang-Mills}, Phys. Rev. D74 (2006) 085012.
\newblock \href {http://arxiv.org/abs/hep-ph/0605199}
  {\path{arXiv:hep-ph/0605199}}, \href
  {https://doi.org/10.1103/PhysRevD.74.085012}
  {\path{doi:10.1103/PhysRevD.74.085012}}.

\bibitem{CaronHuot:2009uh}
S.~Caron-Huot, M.~Laine, G.~D. Moore, {A Way to estimate the heavy quark
  thermalization rate from the lattice}, JHEP 04 (2009) 053.
\newblock \href {http://arxiv.org/abs/0901.1195} {\path{arXiv:0901.1195}},
  \href {https://doi.org/10.1088/1126-6708/2009/04/053}
  {\path{doi:10.1088/1126-6708/2009/04/053}}.

\bibitem{CaronHuot:2007gq}
S.~Caron-Huot, G.~D. Moore, {Heavy quark diffusion in perturbative QCD at
  next-to-leading order}, Phys.Rev.Lett. 100 (2008) 052301.
\newblock \href {http://arxiv.org/abs/0708.4232} {\path{arXiv:0708.4232}},
  \href {https://doi.org/10.1103/PhysRevLett.100.052301}
  {\path{doi:10.1103/PhysRevLett.100.052301}}.

\bibitem{CaronHuot:2008uh}
S.~Caron-Huot, G.~D. Moore, {Heavy quark diffusion in QCD and N=4 SYM at
  next-to-leading order}, JHEP 0802 (2008) 081.
\newblock \href {http://arxiv.org/abs/0801.2173} {\path{arXiv:0801.2173}},
  \href {https://doi.org/10.1088/1126-6708/2008/02/081}
  {\path{doi:10.1088/1126-6708/2008/02/081}}.

\bibitem{Meyer:2010tt}
H.~B. Meyer, {The errant life of a heavy quark in the quark-gluon plasma}, New
  J. Phys. 13 (2011) 035008.
\newblock \href {http://arxiv.org/abs/1012.0234} {\path{arXiv:1012.0234}},
  \href {https://doi.org/10.1088/1367-2630/13/3/035008}
  {\path{doi:10.1088/1367-2630/13/3/035008}}.

\bibitem{Banerjee:2011ra}
D.~Banerjee, S.~Datta, R.~Gavai, P.~Majumdar, {Heavy Quark Momentum Diffusion
  Coefficient from Lattice QCD}, Phys. Rev. D85 (2012) 014510.
\newblock \href {http://arxiv.org/abs/1109.5738} {\path{arXiv:1109.5738}},
  \href {https://doi.org/10.1103/PhysRevD.85.014510}
  {\path{doi:10.1103/PhysRevD.85.014510}}.

\bibitem{Francis:2011gc}
A.~Francis, O.~Kaczmarek, M.~Laine, J.~Langelage, {Towards a non-perturbative
  measurement of the heavy quark momentum diffusion coefficient}, PoS
  LATTICE2011 (2011) 202.
\newblock \href {http://arxiv.org/abs/1109.3941} {\path{arXiv:1109.3941}},
  \href {https://doi.org/10.22323/1.139.0202} {\path{doi:10.22323/1.139.0202}}.

\bibitem{Francis:2015daa}
A.~Francis, O.~Kaczmarek, M.~Laine, T.~Neuhaus, H.~Ohno, {Nonperturbative
  estimate of the heavy quark momentum diffusion coefficient}, Phys. Rev.
  D92~(11) (2015) 116003.
\newblock \href {http://arxiv.org/abs/1508.04543} {\path{arXiv:1508.04543}},
  \href {https://doi.org/10.1103/PhysRevD.92.116003}
  {\path{doi:10.1103/PhysRevD.92.116003}}.

\bibitem{Brambilla:2019tpt}
N.~Brambilla, M.~A. Escobedo, A.~Vairo, P.~Vander~Griend, {Transport
  coefficients from in medium quarkonium dynamics}, Phys. Rev. D100~(5) (2019)
  054025.
\newblock \href {http://arxiv.org/abs/1903.08063} {\path{arXiv:1903.08063}},
  \href {https://doi.org/10.1103/PhysRevD.100.054025}
  {\path{doi:10.1103/PhysRevD.100.054025}}.

\bibitem{Brambilla:2019oaa}
N.~Brambilla, V.~Leino, P.~Petreczky, A.~vairo, {Heavy quark momentum diffusion
  coefficient from the lattice}, in: {37th International Symposium on Lattice
  Field Theory (Lattice 2019) Wuhan, Hubei, China, June 16-22, 2019}, 2019.
\newblock \href {http://arxiv.org/abs/1912.00689} {\path{arXiv:1912.00689}}.

\bibitem{Caswell:1985ui}
W.~E. Caswell, G.~P. Lepage, {Effective Lagrangians for Bound State Problems in
  QED, QCD, and Other Field Theories}, Phys. Lett. 167B (1986) 437--442.
\newblock \href {https://doi.org/10.1016/0370-2693(86)91297-9}
  {\path{doi:10.1016/0370-2693(86)91297-9}}.

\bibitem{Bodwin:1994jh}
G.~T. Bodwin, E.~Braaten, G.~P. Lepage, {Rigorous QCD analysis of inclusive
  annihilation and production of heavy quarkonium}, Phys. Rev. D51 (1995)
  1125--1171, [Erratum: Phys. Rev.D55,5853(1997)].
\newblock \href {http://arxiv.org/abs/hep-ph/9407339}
  {\path{arXiv:hep-ph/9407339}}, \href
  {https://doi.org/10.1103/PhysRevD.55.5853, 10.1103/PhysRevD.51.1125}
  {\path{doi:10.1103/PhysRevD.55.5853, 10.1103/PhysRevD.51.1125}}.

\bibitem{Pineda:1997bj}
A.~Pineda, J.~Soto, {Effective field theory for ultrasoft momenta in NRQCD and
  NRQED}, Nucl. Phys. Proc. Suppl. 64 (1998) 428--432, [,428(1997)].
\newblock \href {http://arxiv.org/abs/hep-ph/9707481}
  {\path{arXiv:hep-ph/9707481}}, \href
  {https://doi.org/10.1016/S0920-5632(97)01102-X}
  {\path{doi:10.1016/S0920-5632(97)01102-X}}.

\bibitem{Brambilla:1999xf}
N.~Brambilla, A.~Pineda, J.~Soto, A.~Vairo, {Potential NRQCD: An Effective
  theory for heavy quarkonium}, Nucl. Phys. B566 (2000) 275.
\newblock \href {http://arxiv.org/abs/hep-ph/9907240}
  {\path{arXiv:hep-ph/9907240}}, \href
  {https://doi.org/10.1016/S0550-3213(99)00693-8}
  {\path{doi:10.1016/S0550-3213(99)00693-8}}.

\bibitem{Brambilla:2004jw}
N.~Brambilla, A.~Pineda, J.~Soto, A.~Vairo, {Effective Field Theories for Heavy
  Quarkonium}, Rev. Mod. Phys. 77 (2005) 1423.
\newblock \href {http://arxiv.org/abs/hep-ph/0410047}
  {\path{arXiv:hep-ph/0410047}}, \href
  {https://doi.org/10.1103/RevModPhys.77.1423}
  {\path{doi:10.1103/RevModPhys.77.1423}}.

\bibitem{Laine:2006ns}
M.~Laine, O.~Philipsen, P.~Romatschke, M.~Tassler, {Real-time static potential
  in hot QCD}, JHEP 03 (2007) 054.
\newblock \href {http://arxiv.org/abs/hep-ph/0611300}
  {\path{arXiv:hep-ph/0611300}}, \href
  {https://doi.org/10.1088/1126-6708/2007/03/054}
  {\path{doi:10.1088/1126-6708/2007/03/054}}.

\bibitem{Beraudo:2007ky}
A.~Beraudo, J.~P. Blaizot, C.~Ratti, {Real and imaginary-time Q anti-Q
  correlators in a thermal medium}, Nucl. Phys. A806 (2008) 312--338.
\newblock \href {http://arxiv.org/abs/0712.4394} {\path{arXiv:0712.4394}},
  \href {https://doi.org/10.1016/j.nuclphysa.2008.03.001}
  {\path{doi:10.1016/j.nuclphysa.2008.03.001}}.

\bibitem{Brambilla:2008cx}
N.~Brambilla, J.~Ghiglieri, A.~Vairo, P.~Petreczky, {Static quark-antiquark
  pairs at finite temperature}, Phys. Rev. D78 (2008) 014017.
\newblock \href {http://arxiv.org/abs/0804.0993} {\path{arXiv:0804.0993}},
  \href {https://doi.org/10.1103/PhysRevD.78.014017}
  {\path{doi:10.1103/PhysRevD.78.014017}}.

\bibitem{Xu:1995eb}
X.-M. Xu, D.~Kharzeev, H.~Satz, X.-N. Wang, {J / psi suppression in an
  equilibrating parton plasma}, Phys. Rev. C53 (1996) 3051--3056.
\newblock \href {http://arxiv.org/abs/hep-ph/9511331}
  {\path{arXiv:hep-ph/9511331}}, \href
  {https://doi.org/10.1103/PhysRevC.53.3051}
  {\path{doi:10.1103/PhysRevC.53.3051}}.

\bibitem{Grandchamp:2001pf}
L.~Grandchamp, R.~Rapp, {Thermal versus direct J / Psi production in
  ultrarelativistic heavy ion collisions}, Phys. Lett. B523 (2001) 60--66.
\newblock \href {http://arxiv.org/abs/hep-ph/0103124}
  {\path{arXiv:hep-ph/0103124}}, \href
  {https://doi.org/10.1016/S0370-2693(01)01311-9}
  {\path{doi:10.1016/S0370-2693(01)01311-9}}.

\bibitem{Brambilla:2011sg}
N.~Brambilla, M.~A. Escobedo, J.~Ghiglieri, A.~Vairo, {Thermal width and
  gluo-dissociation of quarkonium in pNRQCD}, JHEP 12 (2011) 116.
\newblock \href {http://arxiv.org/abs/1109.5826} {\path{arXiv:1109.5826}},
  \href {https://doi.org/10.1007/JHEP12(2011)116}
  {\path{doi:10.1007/JHEP12(2011)116}}.

\bibitem{Brambilla:2013dpa}
N.~Brambilla, M.~A. Escobedo, J.~Ghiglieri, A.~Vairo, {Thermal width and
  quarkonium dissociation by inelastic parton scattering}, JHEP 05 (2013) 130.
\newblock \href {http://arxiv.org/abs/1303.6097} {\path{arXiv:1303.6097}},
  \href {https://doi.org/10.1007/JHEP05(2013)130}
  {\path{doi:10.1007/JHEP05(2013)130}}.

\bibitem{Brambilla:2010vq}
N.~Brambilla, M.~A. Escobedo, J.~Ghiglieri, J.~Soto, A.~Vairo, {Heavy
  Quarkonium in a weakly-coupled quark-gluon plasma below the melting
  temperature}, JHEP 09 (2010) 038.
\newblock \href {http://arxiv.org/abs/1007.4156} {\path{arXiv:1007.4156}},
  \href {https://doi.org/10.1007/JHEP09(2010)038}
  {\path{doi:10.1007/JHEP09(2010)038}}.

\bibitem{Maldacena:1997re}
J.~M. Maldacena, {The Large N limit of superconformal field theories and
  supergravity}, Adv.Theor.Math.Phys. 2 (1998) 231--252.
\newblock \href {http://arxiv.org/abs/hep-th/9711200}
  {\path{arXiv:hep-th/9711200}}, \href
  {https://doi.org/10.1023/A:1026654312961, 10.1023/A:1026654312961}
  {\path{doi:10.1023/A:1026654312961, 10.1023/A:1026654312961}}.

\bibitem{Witten:1998qj}
E.~Witten, {Anti-de Sitter space and holography}, Adv.Theor.Math.Phys. 2 (1998)
  253--291.
\newblock \href {http://arxiv.org/abs/hep-th/9802150}
  {\path{arXiv:hep-th/9802150}}.

\bibitem{Gubser:1998bc}
S.~Gubser, I.~R. Klebanov, A.~M. Polyakov, {Gauge theory correlators from
  noncritical string theory}, Phys.Lett. B428 (1998) 105--114.
\newblock \href {http://arxiv.org/abs/hep-th/9802109}
  {\path{arXiv:hep-th/9802109}}, \href
  {https://doi.org/10.1016/S0370-2693(98)00377-3}
  {\path{doi:10.1016/S0370-2693(98)00377-3}}.

\bibitem{CasalderreySolana:2011us}
J.~Casalderrey-Solana, H.~Liu, D.~Mateos, K.~Rajagopal, U.~A. Wiedemann,
  {Gauge/String Duality, Hot QCD and Heavy Ion Collisions}\href
  {http://arxiv.org/abs/1101.0618} {\path{arXiv:1101.0618}}, \href
  {https://doi.org/10.1017/CBO9781139136747}
  {\path{doi:10.1017/CBO9781139136747}}.

\bibitem{CaronHuot:2006te}
S.~Caron-Huot, P.~Kovtun, G.~D. Moore, A.~Starinets, L.~G. Yaffe, {Photon and
  dilepton production in supersymmetric Yang-Mills plasma}, JHEP 0612 (2006)
  015.
\newblock \href {http://arxiv.org/abs/hep-th/0607237}
  {\path{arXiv:hep-th/0607237}}, \href
  {https://doi.org/10.1088/1126-6708/2006/12/015}
  {\path{doi:10.1088/1126-6708/2006/12/015}}.

\bibitem{Huot:2006ys}
S.~C. Huot, S.~Jeon, G.~D. Moore, {Shear viscosity in weakly coupled N = 4
  super Yang-Mills theory compared to QCD}, Phys. Rev. Lett. 98 (2007) 172303.
\newblock \href {http://arxiv.org/abs/hep-ph/0608062}
  {\path{arXiv:hep-ph/0608062}}, \href
  {https://doi.org/10.1103/PhysRevLett.98.172303}
  {\path{doi:10.1103/PhysRevLett.98.172303}}.

\bibitem{Policastro:2001yc}
G.~Policastro, D.~T. Son, A.~O. Starinets, {The Shear viscosity of strongly
  coupled N=4 supersymmetric Yang-Mills plasma}, Phys. Rev. Lett. 87 (2001)
  081601.
\newblock \href {http://arxiv.org/abs/hep-th/0104066}
  {\path{arXiv:hep-th/0104066}}, \href
  {https://doi.org/10.1103/PhysRevLett.87.081601}
  {\path{doi:10.1103/PhysRevLett.87.081601}}.

\bibitem{Kovtun:2004de}
P.~Kovtun, D.~T. Son, A.~O. Starinets, {Viscosity in strongly interacting
  quantum field theories from black hole physics}, Phys. Rev. Lett. 94 (2005)
  111601.
\newblock \href {http://arxiv.org/abs/hep-th/0405231}
  {\path{arXiv:hep-th/0405231}}, \href
  {https://doi.org/10.1103/PhysRevLett.94.111601}
  {\path{doi:10.1103/PhysRevLett.94.111601}}.

\bibitem{Buchel:2008ac}
A.~Buchel, {Shear viscosity of boost invariant plasma at finite coupling},
  Nucl. Phys. B802 (2008) 281--306.
\newblock \href {http://arxiv.org/abs/0801.4421} {\path{arXiv:0801.4421}},
  \href {https://doi.org/10.1016/j.nuclphysb.2008.03.009}
  {\path{doi:10.1016/j.nuclphysb.2008.03.009}}.

\bibitem{Buchel:2008wy}
A.~Buchel, {Shear viscosity of CFT plasma at finite coupling}, Phys. Lett. B665
  (2008) 298--304.
\newblock \href {http://arxiv.org/abs/0804.3161} {\path{arXiv:0804.3161}},
  \href {https://doi.org/10.1016/j.physletb.2008.05.072}
  {\path{doi:10.1016/j.physletb.2008.05.072}}.

\bibitem{Chesler:2006gr}
P.~M. Chesler, A.~Vuorinen, {Heavy flavor diffusion in weakly coupled N=4 super
  Yang-Mills theory}, JHEP 11 (2006) 037.
\newblock \href {http://arxiv.org/abs/hep-ph/0607148}
  {\path{arXiv:hep-ph/0607148}}, \href
  {https://doi.org/10.1088/1126-6708/2006/11/037}
  {\path{doi:10.1088/1126-6708/2006/11/037}}.

\bibitem{Biondini:2017rpb}
S.~Biondini, et~al., {Status of rates and rate equations for thermal
  leptogenesis}, Int. J. Mod. Phys. A33~(05n06) (2018) 1842004.
\newblock \href {http://arxiv.org/abs/1711.02864} {\path{arXiv:1711.02864}},
  \href {https://doi.org/10.1142/S0217751X18420046}
  {\path{doi:10.1142/S0217751X18420046}}.

\bibitem{Drewes:2017zyw}
M.~Drewes, B.~Garbrecht, P.~Hernandez, M.~Kekic, J.~Lopez-Pavon, J.~Racker,
  N.~Rius, J.~Salvado, D.~Teresi, {ARS Leptogenesis}, Int. J. Mod. Phys.
  A33~(05n06) (2018) 1842002.
\newblock \href {http://arxiv.org/abs/1711.02862} {\path{arXiv:1711.02862}},
  \href {https://doi.org/10.1142/S0217751X18420022}
  {\path{doi:10.1142/S0217751X18420022}}.

\bibitem{Ghiglieri:2016xye}
J.~Ghiglieri, M.~Laine, {Neutrino dynamics below the electroweak crossover},
  JCAP 1607~(07) (2016) 015.
\newblock \href {http://arxiv.org/abs/1605.07720} {\path{arXiv:1605.07720}},
  \href {https://doi.org/10.1088/1475-7516/2016/07/015}
  {\path{doi:10.1088/1475-7516/2016/07/015}}.

\bibitem{Jackson:2019tnr}
G.~Jackson, M.~Laine, {A thermal neutrino interaction rate at NLO}, Nucl. Phys.
  B950 (2020) 114870.
\newblock \href {http://arxiv.org/abs/1910.12880} {\path{arXiv:1910.12880}},
  \href {https://doi.org/10.1016/j.nuclphysb.2019.114870}
  {\path{doi:10.1016/j.nuclphysb.2019.114870}}.

\bibitem{Ghisoiu:2014ena}
I.~Ghisoiu, M.~Laine, {Right-handed neutrino production rate at T > 160 GeV},
  JCAP 1412~(12) (2014) 032.
\newblock \href {http://arxiv.org/abs/1411.1765} {\path{arXiv:1411.1765}},
  \href {https://doi.org/10.1088/1475-7516/2014/12/032}
  {\path{doi:10.1088/1475-7516/2014/12/032}}.

\bibitem{Salvio:2011sf}
A.~Salvio, P.~Lodone, A.~Strumia, {Towards leptogenesis at NLO: the
  right-handed neutrino interaction rate}, JHEP 08 (2011) 116.
\newblock \href {http://arxiv.org/abs/1106.2814} {\path{arXiv:1106.2814}},
  \href {https://doi.org/10.1007/JHEP08(2011)116}
  {\path{doi:10.1007/JHEP08(2011)116}}.

\bibitem{Laine:2011pq}
M.~Laine, Y.~{Schr\"oder}, {Thermal right-handed neutrino production rate in
  the non-relativistic regime}, JHEP 02 (2012) 068.
\newblock \href {http://arxiv.org/abs/1112.1205} {\path{arXiv:1112.1205}},
  \href {https://doi.org/10.1007/JHEP02(2012)068}
  {\path{doi:10.1007/JHEP02(2012)068}}.

\bibitem{Biondini:2013xua}
S.~Biondini, N.~Brambilla, M.~A. Escobedo, A.~Vairo, {An effective field theory
  for non-relativistic Majorana neutrinos}, JHEP 12 (2013) 028.
\newblock \href {http://arxiv.org/abs/1307.7680} {\path{arXiv:1307.7680}},
  \href {https://doi.org/10.1007/JHEP12(2013)028}
  {\path{doi:10.1007/JHEP12(2013)028}}.

\bibitem{Graf:2010tv}
P.~Graf, F.~D. Steffen, {Thermal axion production in the primordial quark-gluon
  plasma}, Phys.Rev. D83 (2011) 075011.
\newblock \href {http://arxiv.org/abs/1008.4528} {\path{arXiv:1008.4528}},
  \href {https://doi.org/10.1103/PhysRevD.83.075011}
  {\path{doi:10.1103/PhysRevD.83.075011}}.

\bibitem{Graf:2012hb}
P.~Graf, F.~D. Steffen, {Axions and saxions from the primordial supersymmetric
  plasma and extra radiation signatures}, JCAP 1302 (2013) 018.
\newblock \href {http://arxiv.org/abs/1208.2951} {\path{arXiv:1208.2951}},
  \href {https://doi.org/10.1088/1475-7516/2013/02/018}
  {\path{doi:10.1088/1475-7516/2013/02/018}}.

\bibitem{Salvio:2013iaa}
A.~Salvio, A.~Strumia, W.~Xue, {Thermal axion production}, JCAP 1401 (2014)
  011.
\newblock \href {http://arxiv.org/abs/1310.6982} {\path{arXiv:1310.6982}},
  \href {https://doi.org/10.1088/1475-7516/2014/01/011}
  {\path{doi:10.1088/1475-7516/2014/01/011}}.

\bibitem{Pradler:2006qh}
J.~Pradler, F.~D. Steffen, {Thermal gravitino production and collider tests of
  leptogenesis}, Phys. Rev. D75 (2007) 023509.
\newblock \href {http://arxiv.org/abs/hep-ph/0608344}
  {\path{arXiv:hep-ph/0608344}}, \href
  {https://doi.org/10.1103/PhysRevD.75.023509}
  {\path{doi:10.1103/PhysRevD.75.023509}}.

\bibitem{Pradler:2006hh}
J.~Pradler, F.~D. Steffen, {Constraints on the Reheating Temperature in
  Gravitino Dark Matter Scenarios}, Phys. Lett. B648 (2007) 224--235.
\newblock \href {http://arxiv.org/abs/hep-ph/0612291}
  {\path{arXiv:hep-ph/0612291}}, \href
  {https://doi.org/10.1016/j.physletb.2007.02.072}
  {\path{doi:10.1016/j.physletb.2007.02.072}}.

\bibitem{Rychkov:2007uq}
V.~S. Rychkov, A.~Strumia, {Thermal production of gravitinos}, Phys. Rev. D75
  (2007) 075011.
\newblock \href {http://arxiv.org/abs/hep-ph/0701104}
  {\path{arXiv:hep-ph/0701104}}, \href
  {https://doi.org/10.1103/PhysRevD.75.075011}
  {\path{doi:10.1103/PhysRevD.75.075011}}.

\bibitem{Ghiglieri:2015nfa}
J.~Ghiglieri, M.~Laine, {Gravitational wave background from Standard Model
  physics: Qualitative features}, JCAP 1507~(07) (2015) 022.
\newblock \href {http://arxiv.org/abs/1504.02569} {\path{arXiv:1504.02569}},
  \href {https://doi.org/10.1088/1475-7516/2015/07/022}
  {\path{doi:10.1088/1475-7516/2015/07/022}}.

\bibitem{deForcrand:2010ys}
P.~de~Forcrand, {Simulating QCD at finite density}, PoS LAT2009 (2009) 010.
\newblock \href {http://arxiv.org/abs/1005.0539} {\path{arXiv:1005.0539}}.

\bibitem{Ipp:2003qt}
A.~Ipp, {Quantum corrections to thermodynamic properties in the large N(f)
  limit of the quark gluon plasma}, Ph.D. thesis, Vienna. Tech. U. (2003).
\newblock \href {http://arxiv.org/abs/hep-ph/0405123}
  {\path{arXiv:hep-ph/0405123}}.

\bibitem{Linde:1980ts}
A.~D. Linde, {Infrared Problem in Thermodynamics of the Yang-Mills Gas}, Phys.
  Lett. B96 (1980) 289--292.
\newblock \href {https://doi.org/10.1016/0370-2693(80)90769-8}
  {\path{doi:10.1016/0370-2693(80)90769-8}}.

\bibitem{York:2008rr}
M.~A. York, G.~D. Moore, {Second order hydrodynamic coefficients from kinetic
  theory}, Phys. Rev. D79 (2009) 054011.
\newblock \href {http://arxiv.org/abs/0811.0729} {\path{arXiv:0811.0729}},
  \href {https://doi.org/10.1103/PhysRevD.79.054011}
  {\path{doi:10.1103/PhysRevD.79.054011}}.

\bibitem{Kajantie:1997tt}
K.~Kajantie, M.~Laine, K.~Rummukainen, M.~E. Shaposhnikov, {3-D SU(N) + adjoint
  Higgs theory and finite temperature QCD}, Nucl.Phys. B503 (1997) 357--384.
\newblock \href {http://arxiv.org/abs/hep-ph/9704416}
  {\path{arXiv:hep-ph/9704416}}, \href
  {https://doi.org/10.1016/S0550-3213(97)00425-2}
  {\path{doi:10.1016/S0550-3213(97)00425-2}}.

\bibitem{Braaten:1995cm}
E.~Braaten, A.~Nieto, {Effective field theory approach to high temperature
  thermodynamics}, Phys.Rev. D51 (1995) 6990--7006.
\newblock \href {http://arxiv.org/abs/hep-ph/9501375}
  {\path{arXiv:hep-ph/9501375}}, \href
  {https://doi.org/10.1103/PhysRevD.51.6990}
  {\path{doi:10.1103/PhysRevD.51.6990}}.

\bibitem{Braaten:1995jr}
E.~Braaten, A.~Nieto, {Free energy of QCD at high temperature}, Phys.Rev. D53
  (1996) 3421--3437.
\newblock \href {http://arxiv.org/abs/hep-ph/9510408}
  {\path{arXiv:hep-ph/9510408}}, \href
  {https://doi.org/10.1103/PhysRevD.53.3421}
  {\path{doi:10.1103/PhysRevD.53.3421}}.

\bibitem{Andersen:1999fw}
J.~O. Andersen, E.~Braaten, M.~Strickland, {Hard thermal loop resummation of
  the free energy of a hot gluon plasma}, Phys. Rev. Lett. 83 (1999)
  2139--2142.
\newblock \href {http://arxiv.org/abs/hep-ph/9902327}
  {\path{arXiv:hep-ph/9902327}}, \href
  {https://doi.org/10.1103/PhysRevLett.83.2139}
  {\path{doi:10.1103/PhysRevLett.83.2139}}.

\bibitem{Andersen:1999sf}
J.~O. Andersen, E.~Braaten, M.~Strickland, {Hard thermal loop resummation of
  the thermodynamics of a hot gluon plasma}, Phys. Rev. D61 (2000) 014017.
\newblock \href {http://arxiv.org/abs/hep-ph/9905337}
  {\path{arXiv:hep-ph/9905337}}, \href
  {https://doi.org/10.1103/PhysRevD.61.014017}
  {\path{doi:10.1103/PhysRevD.61.014017}}.

\bibitem{Andersen:1999va}
J.~O. Andersen, E.~Braaten, M.~Strickland, {Hard thermal loop resummation of
  the free energy of a hot quark - gluon plasma}, Phys. Rev. D61 (2000) 074016.
\newblock \href {http://arxiv.org/abs/hep-ph/9908323}
  {\path{arXiv:hep-ph/9908323}}, \href
  {https://doi.org/10.1103/PhysRevD.61.074016}
  {\path{doi:10.1103/PhysRevD.61.074016}}.

\bibitem{Blaizot:1999ip}
J.~P. Blaizot, E.~Iancu, A.~Rebhan, {The Entropy of the QCD plasma}, Phys. Rev.
  Lett. 83 (1999) 2906--2909.
\newblock \href {http://arxiv.org/abs/hep-ph/9906340}
  {\path{arXiv:hep-ph/9906340}}, \href
  {https://doi.org/10.1103/PhysRevLett.83.2906}
  {\path{doi:10.1103/PhysRevLett.83.2906}}.

\bibitem{Blaizot:1999ap}
J.~P. Blaizot, E.~Iancu, A.~Rebhan, {Selfconsistent hard thermal loop
  thermodynamics for the quark gluon plasma}, Phys. Lett. B470 (1999) 181--188.
\newblock \href {http://arxiv.org/abs/hep-ph/9910309}
  {\path{arXiv:hep-ph/9910309}}, \href
  {https://doi.org/10.1016/S0370-2693(99)01306-4}
  {\path{doi:10.1016/S0370-2693(99)01306-4}}.

\bibitem{Blaizot:2000fc}
J.~P. Blaizot, E.~Iancu, A.~Rebhan, {Approximately selfconsistent resummations
  for the thermodynamics of the quark gluon plasma. 1. Entropy and density},
  Phys. Rev. D63 (2001) 065003.
\newblock \href {http://arxiv.org/abs/hep-ph/0005003}
  {\path{arXiv:hep-ph/0005003}}, \href
  {https://doi.org/10.1103/PhysRevD.63.065003}
  {\path{doi:10.1103/PhysRevD.63.065003}}.

\bibitem{Blaizot:2001vr}
J.~P. Blaizot, E.~Iancu, A.~Rebhan, {Quark number susceptibilities from HTL
  resummed thermodynamics}, Phys. Lett. B523 (2001) 143--150.
\newblock \href {http://arxiv.org/abs/hep-ph/0110369}
  {\path{arXiv:hep-ph/0110369}}, \href
  {https://doi.org/10.1016/S0370-2693(01)01316-8}
  {\path{doi:10.1016/S0370-2693(01)01316-8}}.

\bibitem{Ginsparg:1980ef}
P.~H. Ginsparg, {First Order and Second Order Phase Transitions in Gauge
  Theories at Finite Temperature}, Nucl. Phys. B170 (1980) 388--408.
\newblock \href {https://doi.org/10.1016/0550-3213(80)90418-6}
  {\path{doi:10.1016/0550-3213(80)90418-6}}.

\bibitem{Appelquist:1981vg}
T.~Appelquist, R.~D. Pisarski, {High-Temperature Yang-Mills Theories and
  Three-Dimensional Quantum Chromodynamics}, Phys. Rev. D23 (1981) 2305.
\newblock \href {https://doi.org/10.1103/PhysRevD.23.2305}
  {\path{doi:10.1103/PhysRevD.23.2305}}.

\bibitem{Kajantie:1995dw}
K.~Kajantie, M.~Laine, K.~Rummukainen, M.~E. Shaposhnikov, {Generic rules for
  high temperature dimensional reduction and their application to the standard
  model}, Nucl.Phys. B458 (1996) 90--136.
\newblock \href {http://arxiv.org/abs/hep-ph/9508379}
  {\path{arXiv:hep-ph/9508379}}, \href
  {https://doi.org/10.1016/0550-3213(95)00549-8}
  {\path{doi:10.1016/0550-3213(95)00549-8}}.

\bibitem{Hart:2000ha}
A.~Hart, M.~Laine, O.~Philipsen, {Static correlation lengths in QCD at high
  temperatures and finite densities}, Nucl.Phys. B586 (2000) 443--474.
\newblock \href {http://arxiv.org/abs/hep-ph/0004060}
  {\path{arXiv:hep-ph/0004060}}, \href
  {https://doi.org/10.1016/S0550-3213(00)00418-1}
  {\path{doi:10.1016/S0550-3213(00)00418-1}}.

\bibitem{Kajantie:1998yc}
K.~Kajantie, M.~Laine, A.~Rajantie, K.~Rummukainen, M.~Tsypin, {The Phase
  diagram of three-dimensional SU(3) + adjoint Higgs theory}, JHEP 11 (1998)
  011.
\newblock \href {http://arxiv.org/abs/hep-lat/9811004}
  {\path{arXiv:hep-lat/9811004}}, \href
  {https://doi.org/10.1088/1126-6708/1998/11/011}
  {\path{doi:10.1088/1126-6708/1998/11/011}}.

\bibitem{Vuorinen:2006nz}
A.~Vuorinen, L.~G. Yaffe, {Z(3)-symmetric effective theory for SU(3) Yang-Mills
  theory at high temperature}, Phys.Rev. D74 (2006) 025011.
\newblock \href {http://arxiv.org/abs/hep-ph/0604100}
  {\path{arXiv:hep-ph/0604100}}, \href
  {https://doi.org/10.1103/PhysRevD.74.025011}
  {\path{doi:10.1103/PhysRevD.74.025011}}.

\bibitem{deForcrand:2008aw}
P.~de~Forcrand, A.~Kurkela, A.~Vuorinen, {Center-Symmetric Effective Theory for
  High-Temperature SU(2) Yang-Mills Theory}, Phys.Rev. D77 (2008) 125014.
\newblock \href {http://arxiv.org/abs/0801.1566} {\path{arXiv:0801.1566}},
  \href {https://doi.org/10.1103/PhysRevD.77.125014}
  {\path{doi:10.1103/PhysRevD.77.125014}}.

\bibitem{Ghisoiu:2015uza}
I.~Ghisoiu, J.~{M\"oller}, Y.~{Schr\"oder}, {Debye screening mass of hot
  Yang-Mills theory to three-loop order}, JHEP 11 (2015) 121.
\newblock \href {http://arxiv.org/abs/1509.08727} {\path{arXiv:1509.08727}},
  \href {https://doi.org/10.1007/JHEP11(2015)121}
  {\path{doi:10.1007/JHEP11(2015)121}}.

\bibitem{Laine:2018lgj}
M.~Laine, P.~Schicho, Y.~{Schr\"oder}, {Soft thermal contributions to 3-loop
  gauge coupling}, JHEP 05 (2018) 037.
\newblock \href {http://arxiv.org/abs/1803.08689} {\path{arXiv:1803.08689}},
  \href {https://doi.org/10.1007/JHEP05(2018)037}
  {\path{doi:10.1007/JHEP05(2018)037}}.

\bibitem{Laine:2019uua}
M.~Laine, P.~Schicho, Y.~{Schr\"oder}, {A QCD Debye mass in a broad temperature
  range}, Phys. Rev. D 101~(2) (2020) 023532.
\newblock \href {http://arxiv.org/abs/1911.09123} {\path{arXiv:1911.09123}},
  \href {https://doi.org/10.1103/PhysRevD.101.023532}
  {\path{doi:10.1103/PhysRevD.101.023532}}.

\bibitem{Nishimura:2012ee}
M.~Nishimura, Y.~{Schr\"oder}, {IBP methods at finite temperature}, JHEP 1209
  (2012) 051.
\newblock \href {http://arxiv.org/abs/1207.4042} {\path{arXiv:1207.4042}},
  \href {https://doi.org/10.1007/JHEP09(2012)051}
  {\path{doi:10.1007/JHEP09(2012)051}}.

\bibitem{Ghisoiu:2012yk}
I.~Ghisoiu, Y.~{Schr\"oder}, {A New Method for Taming Tensor Sum-Integrals},
  JHEP 1211 (2012) 010.
\newblock \href {http://arxiv.org/abs/1208.0284} {\path{arXiv:1208.0284}},
  \href {https://doi.org/10.1007/JHEP11(2012)010}
  {\path{doi:10.1007/JHEP11(2012)010}}.

\bibitem{Moeller:2012da}
J.~{M\"oller}, Y.~{Schr\"oder}, {Three-loop matching coefficients for hot QCD:
  Reduction and gauge independence}, JHEP 1208 (2012) 025.
\newblock \href {http://arxiv.org/abs/1207.1309} {\path{arXiv:1207.1309}},
  \href {https://doi.org/10.1007/JHEP08(2012)025}
  {\path{doi:10.1007/JHEP08(2012)025}}.

\bibitem{Kajantie:2002wa}
K.~Kajantie, M.~Laine, K.~Rummukainen, Y.~{Schr\"oder}, {The Pressure of hot
  QCD up to g6 ln(1/g)}, Phys.Rev. D67 (2003) 105008.
\newblock \href {http://arxiv.org/abs/hep-ph/0211321}
  {\path{arXiv:hep-ph/0211321}}, \href
  {https://doi.org/10.1103/PhysRevD.67.105008}
  {\path{doi:10.1103/PhysRevD.67.105008}}.

\bibitem{Vuorinen:2003fs}
A.~Vuorinen, {The Pressure of QCD at finite temperatures and chemical
  potentials}, Phys.Rev. D68 (2003) 054017.
\newblock \href {http://arxiv.org/abs/hep-ph/0305183}
  {\path{arXiv:hep-ph/0305183}}, \href
  {https://doi.org/10.1103/PhysRevD.68.054017}
  {\path{doi:10.1103/PhysRevD.68.054017}}.

\bibitem{Gynther:2009qf}
A.~Gynther, A.~Kurkela, A.~Vuorinen, {The N(f)**3 g**6 term in the pressure of
  hot QCD}, Phys.Rev. D80 (2009) 096002.
\newblock \href {http://arxiv.org/abs/0909.3521} {\path{arXiv:0909.3521}},
  \href {https://doi.org/10.1103/PhysRevD.80.096002}
  {\path{doi:10.1103/PhysRevD.80.096002}}.

\bibitem{Kajantie:2003ax}
K.~Kajantie, M.~Laine, K.~Rummukainen, Y.~{Schr\"oder}, {Four loop vacuum
  energy density of the SU(N(c)) + adjoint Higgs theory}, JHEP 0304 (2003) 036.
\newblock \href {http://arxiv.org/abs/hep-ph/0304048}
  {\path{arXiv:hep-ph/0304048}}, \href
  {https://doi.org/10.1088/1126-6708/2003/04/036}
  {\path{doi:10.1088/1126-6708/2003/04/036}}.

\bibitem{Hietanen:2008xb}
A.~Hietanen, K.~Rummukainen, {The Diagonal and off-diagonal quark number
  susceptibility of high temperature and finite density QCD}, JHEP 04 (2008)
  078.
\newblock \href {http://arxiv.org/abs/0802.3979} {\path{arXiv:0802.3979}},
  \href {https://doi.org/10.1088/1126-6708/2008/04/078}
  {\path{doi:10.1088/1126-6708/2008/04/078}}.

\bibitem{Hietanen:2004ew}
A.~Hietanen, K.~Kajantie, M.~Laine, K.~Rummukainen, Y.~{Schr\"oder}, {Plaquette
  expectation value and gluon condensate in three dimensions}, JHEP 0501 (2005)
  013.
\newblock \href {http://arxiv.org/abs/hep-lat/0412008}
  {\path{arXiv:hep-lat/0412008}}, \href
  {https://doi.org/10.1088/1126-6708/2005/01/013}
  {\path{doi:10.1088/1126-6708/2005/01/013}}.

\bibitem{DiRenzo:2006nh}
F.~Di~Renzo, M.~Laine, V.~Miccio, Y.~{Schr\"oder}, C.~Torrero, {The Leading
  non-perturbative coefficient in the weak-coupling expansion of hot QCD
  pressure}, JHEP 0607 (2006) 026.
\newblock \href {http://arxiv.org/abs/hep-ph/0605042}
  {\path{arXiv:hep-ph/0605042}}, \href
  {https://doi.org/10.1088/1126-6708/2006/07/026}
  {\path{doi:10.1088/1126-6708/2006/07/026}}.

\bibitem{Blaizot:2003iq}
J.~Blaizot, E.~Iancu, A.~Rebhan, {On the apparent convergence of perturbative
  QCD at high temperature}, Phys.Rev. D68 (2003) 025011.
\newblock \href {http://arxiv.org/abs/hep-ph/0303045}
  {\path{arXiv:hep-ph/0303045}}, \href
  {https://doi.org/10.1103/PhysRevD.68.025011}
  {\path{doi:10.1103/PhysRevD.68.025011}}.

\bibitem{Laine:2006cp}
M.~Laine, Y.~{Schr\"oder}, {Quark mass thresholds in QCD thermodynamics},
  Phys.Rev. D73 (2006) 085009.
\newblock \href {http://arxiv.org/abs/hep-ph/0603048}
  {\path{arXiv:hep-ph/0603048}}, \href
  {https://doi.org/10.1103/PhysRevD.73.085009}
  {\path{doi:10.1103/PhysRevD.73.085009}}.

\bibitem{Andersen:2002ey}
J.~O. Andersen, E.~Braaten, E.~Petitgirard, M.~Strickland, {HTL perturbation
  theory to two loops}, Phys. Rev. D66 (2002) 085016.
\newblock \href {http://arxiv.org/abs/hep-ph/0205085}
  {\path{arXiv:hep-ph/0205085}}, \href
  {https://doi.org/10.1103/PhysRevD.66.085016}
  {\path{doi:10.1103/PhysRevD.66.085016}}.

\bibitem{Andersen:2003zk}
J.~O. Andersen, E.~Petitgirard, M.~Strickland, {Two loop HTL thermodynamics
  with quarks}, Phys. Rev. D70 (2004) 045001.
\newblock \href {http://arxiv.org/abs/hep-ph/0302069}
  {\path{arXiv:hep-ph/0302069}}, \href
  {https://doi.org/10.1103/PhysRevD.70.045001}
  {\path{doi:10.1103/PhysRevD.70.045001}}.

\bibitem{Andersen:2009tw}
J.~O. Andersen, M.~Strickland, N.~Su, {Three-loop HTL Free Energy for QED},
  Phys. Rev. D80 (2009) 085015.
\newblock \href {http://arxiv.org/abs/0906.2936} {\path{arXiv:0906.2936}},
  \href {https://doi.org/10.1103/PhysRevD.80.085015}
  {\path{doi:10.1103/PhysRevD.80.085015}}.

\bibitem{Andersen:2009tc}
J.~O. Andersen, M.~Strickland, N.~Su, {Gluon Thermodynamics at Intermediate
  Coupling}, Phys. Rev. Lett. 104 (2010) 122003.
\newblock \href {http://arxiv.org/abs/0911.0676} {\path{arXiv:0911.0676}},
  \href {https://doi.org/10.1103/PhysRevLett.104.122003}
  {\path{doi:10.1103/PhysRevLett.104.122003}}.

\bibitem{Andersen:2010ct}
J.~O. Andersen, M.~Strickland, N.~Su, {Three-loop HTL gluon thermodynamics at
  intermediate coupling}, JHEP 08 (2010) 113.
\newblock \href {http://arxiv.org/abs/1005.1603} {\path{arXiv:1005.1603}},
  \href {https://doi.org/10.1007/JHEP08(2010)113}
  {\path{doi:10.1007/JHEP08(2010)113}}.

\bibitem{Andersen:2010wu}
J.~O. Andersen, L.~E. Leganger, M.~Strickland, N.~Su, {NNLO hard-thermal-loop
  thermodynamics for QCD}, Phys. Lett. B696 (2011) 468--472.
\newblock \href {http://arxiv.org/abs/1009.4644} {\path{arXiv:1009.4644}},
  \href {https://doi.org/10.1016/j.physletb.2010.12.070}
  {\path{doi:10.1016/j.physletb.2010.12.070}}.

\bibitem{Andersen:2011sf}
J.~O. Andersen, L.~E. Leganger, M.~Strickland, N.~Su, {Three-loop HTL QCD
  thermodynamics}, JHEP 08 (2011) 053.
\newblock \href {http://arxiv.org/abs/1103.2528} {\path{arXiv:1103.2528}},
  \href {https://doi.org/10.1007/JHEP08(2011)053}
  {\path{doi:10.1007/JHEP08(2011)053}}.

\bibitem{Andersen:2011ug}
J.~O. Andersen, L.~E. Leganger, M.~Strickland, N.~Su, {The QCD trace anomaly},
  Phys. Rev. D84 (2011) 087703.
\newblock \href {http://arxiv.org/abs/1106.0514} {\path{arXiv:1106.0514}},
  \href {https://doi.org/10.1103/PhysRevD.84.087703}
  {\path{doi:10.1103/PhysRevD.84.087703}}.

\bibitem{Haque:2012my}
N.~Haque, M.~G. Mustafa, M.~Strickland, {Two-loop hard thermal loop pressure at
  finite temperature and chemical potential}, Phys. Rev. D87~(10) (2013)
  105007.
\newblock \href {http://arxiv.org/abs/1212.1797} {\path{arXiv:1212.1797}},
  \href {https://doi.org/10.1103/PhysRevD.87.105007}
  {\path{doi:10.1103/PhysRevD.87.105007}}.

\bibitem{Mogliacci:2013mca}
S.~Mogliacci, J.~O. Andersen, M.~Strickland, N.~Su, A.~Vuorinen, {Equation of
  State of hot and dense QCD: Resummed perturbation theory confronts lattice
  data}, JHEP 12 (2013) 055.
\newblock \href {http://arxiv.org/abs/1307.8098} {\path{arXiv:1307.8098}},
  \href {https://doi.org/10.1007/JHEP12(2013)055}
  {\path{doi:10.1007/JHEP12(2013)055}}.

\bibitem{Haque:2013qta}
N.~Haque, M.~G. Mustafa, M.~Strickland, {Quark Number Susceptibilities from
  Two-Loop Hard Thermal Loop Perturbation Theory}, JHEP 07 (2013) 184.
\newblock \href {http://arxiv.org/abs/1302.3228} {\path{arXiv:1302.3228}},
  \href {https://doi.org/10.1007/JHEP07(2013)184}
  {\path{doi:10.1007/JHEP07(2013)184}}.

\bibitem{Haque:2013sja}
N.~Haque, J.~O. Andersen, M.~G. Mustafa, M.~Strickland, N.~Su, {Three-loop
  pressure and susceptibility at finite temperature and density from
  hard-thermal-loop perturbation theory}, Phys. Rev. D89~(6) (2014) 061701.
\newblock \href {http://arxiv.org/abs/1309.3968} {\path{arXiv:1309.3968}},
  \href {https://doi.org/10.1103/PhysRevD.89.061701}
  {\path{doi:10.1103/PhysRevD.89.061701}}.

\bibitem{Haque:2014rua}
N.~Haque, A.~Bandyopadhyay, J.~O. Andersen, M.~G. Mustafa, M.~Strickland,
  N.~Su, {Three-loop HTLpt thermodynamics at finite temperature and chemical
  potential}, JHEP 05 (2014) 027.
\newblock \href {http://arxiv.org/abs/1402.6907} {\path{arXiv:1402.6907}},
  \href {https://doi.org/10.1007/JHEP05(2014)027}
  {\path{doi:10.1007/JHEP05(2014)027}}.

\bibitem{Andersen:2015eoa}
J.~O. Andersen, N.~Haque, M.~G. Mustafa, M.~Strickland, {Three-loop
  hard-thermal-loop perturbation theory thermodynamics at finite temperature
  and finite baryonic and isospin chemical potential}, Phys. Rev. D93~(5)
  (2016) 054045.
\newblock \href {http://arxiv.org/abs/1511.04660} {\path{arXiv:1511.04660}},
  \href {https://doi.org/10.1103/PhysRevD.93.054045}
  {\path{doi:10.1103/PhysRevD.93.054045}}.

\bibitem{Karsch:1997gj}
F.~Karsch, A.~Patkos, P.~Petreczky, {Screened perturbation theory}, Phys. Lett.
  B401 (1997) 69--73.
\newblock \href {http://arxiv.org/abs/hep-ph/9702376}
  {\path{arXiv:hep-ph/9702376}}, \href
  {https://doi.org/10.1016/S0370-2693(97)00392-4}
  {\path{doi:10.1016/S0370-2693(97)00392-4}}.

\bibitem{Chiku:1998kd}
S.~Chiku, T.~Hatsuda, {Optimized perturbation theory at finite temperature},
  Phys. Rev. D58 (1998) 076001.
\newblock \href {http://arxiv.org/abs/hep-ph/9803226}
  {\path{arXiv:hep-ph/9803226}}, \href
  {https://doi.org/10.1103/PhysRevD.58.076001}
  {\path{doi:10.1103/PhysRevD.58.076001}}.

\bibitem{Andersen:2000yj}
J.~O. Andersen, E.~Braaten, M.~Strickland, {Screened perturbation theory to
  three loops}, Phys. Rev. D63 (2001) 105008.
\newblock \href {http://arxiv.org/abs/hep-ph/0007159}
  {\path{arXiv:hep-ph/0007159}}, \href
  {https://doi.org/10.1103/PhysRevD.63.105008}
  {\path{doi:10.1103/PhysRevD.63.105008}}.

\bibitem{Andersen:2001ez}
J.~Andersen, M.~Strickland, {Mass expansions of screened perturbation theory},
  Phys. Rev. D64 (2001) 105012.
\newblock \href {http://arxiv.org/abs/hep-ph/0105214}
  {\path{arXiv:hep-ph/0105214}}, \href
  {https://doi.org/10.1103/PhysRevD.64.105012}
  {\path{doi:10.1103/PhysRevD.64.105012}}.

\bibitem{Andersen:2008bz}
J.~O. Andersen, L.~Kyllingstad, {Four-loop Screened Perturbation Theory}, Phys.
  Rev. D78 (2008) 076008.
\newblock \href {http://arxiv.org/abs/0805.4478} {\path{arXiv:0805.4478}},
  \href {https://doi.org/10.1103/PhysRevD.78.076008}
  {\path{doi:10.1103/PhysRevD.78.076008}}.

\bibitem{Lattimer:2004pg}
J.~M. Lattimer, M.~Prakash, {The physics of neutron stars}, Science 304 (2004)
  536--542.
\newblock \href {http://arxiv.org/abs/astro-ph/0405262}
  {\path{arXiv:astro-ph/0405262}}, \href
  {https://doi.org/10.1126/science.1090720}
  {\path{doi:10.1126/science.1090720}}.

\bibitem{Annala:2019puf}
E.~Annala, T.~Gorda, A.~Kurkela, J.~Nättilä, A.~Vuorinen, {Quark-matter cores
  in neutron stars}\href {http://arxiv.org/abs/1903.09121}
  {\path{arXiv:1903.09121}}.

\bibitem{Son:1998uk}
D.~T. Son, {Superconductivity by long range color magnetic interaction in high
  density quark matter}, Phys. Rev. D59 (1999) 094019.
\newblock \href {http://arxiv.org/abs/hep-ph/9812287}
  {\path{arXiv:hep-ph/9812287}}, \href
  {https://doi.org/10.1103/PhysRevD.59.094019}
  {\path{doi:10.1103/PhysRevD.59.094019}}.

\bibitem{Alford:2007xm}
M.~G. Alford, A.~Schmitt, K.~Rajagopal, T.~{Sch\"afer}, {Color
  superconductivity in dense quark matter}, Rev. Mod. Phys. 80 (2008)
  1455--1515.
\newblock \href {http://arxiv.org/abs/0709.4635} {\path{arXiv:0709.4635}},
  \href {https://doi.org/10.1103/RevModPhys.80.1455}
  {\path{doi:10.1103/RevModPhys.80.1455}}.

\bibitem{Ghisoiu:2016swa}
I.~Ghisoiu, T.~Gorda, A.~Kurkela, P.~Romatschke, M.~Säppi, A.~Vuorinen, {On
  high-order perturbative calculations at finite density}, Nucl. Phys. B915
  (2017) 102--118.
\newblock \href {http://arxiv.org/abs/1609.04339} {\path{arXiv:1609.04339}},
  \href {https://doi.org/10.1016/j.nuclphysb.2016.11.023}
  {\path{doi:10.1016/j.nuclphysb.2016.11.023}}.

\bibitem{Smirnov:2008iw}
A.~Smirnov, {Algorithm FIRE -- Feynman Integral REduction}, JHEP 0810 (2008)
  107.
\newblock \href {http://arxiv.org/abs/0807.3243} {\path{arXiv:0807.3243}},
  \href {https://doi.org/10.1088/1126-6708/2008/10/107}
  {\path{doi:10.1088/1126-6708/2008/10/107}}.

\bibitem{Bonciani:2003hc}
R.~Bonciani, P.~Mastrolia, E.~Remiddi, {Master integrals for the two loop QCD
  virtual corrections to the forward backward asymmetry}, Nucl.Phys. B690
  (2004) 138--176.
\newblock \href {http://arxiv.org/abs/hep-ph/2145} {\path{arXiv:hep-ph/2145}},
  \href {https://doi.org/10.1016/j.nuclphysb.2004.04.011}
  {\path{doi:10.1016/j.nuclphysb.2004.04.011}}.

\bibitem{Schroder:2005va}
Y.~{Schr\"oder}, A.~Vuorinen, {High-precision epsilon expansions of
  single-mass-scale four-loop vacuum bubbles}, JHEP 0506 (2005) 051.
\newblock \href {http://arxiv.org/abs/hep-ph/0503209}
  {\path{arXiv:hep-ph/0503209}}, \href
  {https://doi.org/10.1088/1126-6708/2005/06/051}
  {\path{doi:10.1088/1126-6708/2005/06/051}}.

\bibitem{Ipp:2006ij}
A.~Ipp, K.~Kajantie, A.~Rebhan, A.~Vuorinen, {The Pressure of deconfined QCD
  for all temperatures and quark chemical potentials}, Phys. Rev. D74 (2006)
  045016.
\newblock \href {http://arxiv.org/abs/hep-ph/0604060}
  {\path{arXiv:hep-ph/0604060}}, \href
  {https://doi.org/10.1103/PhysRevD.74.045016}
  {\path{doi:10.1103/PhysRevD.74.045016}}.

\bibitem{Kurkela:2016was}
A.~Kurkela, A.~Vuorinen, {Cool quark matter}, Phys. Rev. Lett. 117~(4) (2016)
  042501.
\newblock \href {http://arxiv.org/abs/1603.00750} {\path{arXiv:1603.00750}},
  \href {https://doi.org/10.1103/PhysRevLett.117.042501}
  {\path{doi:10.1103/PhysRevLett.117.042501}}.

\bibitem{Gorda:2018gpy}
T.~Gorda, A.~Kurkela, P.~Romatschke, M.~Säppi, A.~Vuorinen,
  {Next-to-Next-to-Next-to-Leading Order Pressure of Cold Quark Matter: Leading
  Logarithm}, Phys. Rev. Lett. 121~(20) (2018) 202701.
\newblock \href {http://arxiv.org/abs/1807.04120} {\path{arXiv:1807.04120}},
  \href {https://doi.org/10.1103/PhysRevLett.121.202701}
  {\path{doi:10.1103/PhysRevLett.121.202701}}.

\bibitem{Freedman:1976ub}
B.~A. Freedman, L.~D. McLerran, {Fermions and Gauge Vector Mesons at Finite
  Temperature and Density. 3. The Ground State Energy of a Relativistic Quark
  Gas}, Phys. Rev. D16 (1977) 1169.
\newblock \href {https://doi.org/10.1103/PhysRevD.16.1169}
  {\path{doi:10.1103/PhysRevD.16.1169}}.

\bibitem{Ipp:2003cj}
A.~Ipp, A.~Gerhold, A.~Rebhan, {Anomalous specific heat in high density QED and
  QCD}, Phys. Rev. D69 (2004) 011901.
\newblock \href {http://arxiv.org/abs/hep-ph/0309019}
  {\path{arXiv:hep-ph/0309019}}, \href
  {https://doi.org/10.1103/PhysRevD.69.011901}
  {\path{doi:10.1103/PhysRevD.69.011901}}.

\bibitem{Gerhold:2004tb}
A.~Gerhold, A.~Ipp, A.~Rebhan, {Non-Fermi-liquid specific heat of normal
  degenerate quark matter}, Phys. Rev. D70 (2004) 105015.
\newblock \href {http://arxiv.org/abs/hep-ph/0406087}
  {\path{arXiv:hep-ph/0406087}}, \href
  {https://doi.org/10.1103/PhysRevD.70.105015}
  {\path{doi:10.1103/PhysRevD.70.105015}}.

\bibitem{Karsch:2010hm}
F.~Karsch, B.-J. Schaefer, M.~Wagner, J.~Wambach, {Towards finite density QCD
  with Taylor expansions}, Phys. Lett. B698 (2011) 256--264.
\newblock \href {http://arxiv.org/abs/1009.5211} {\path{arXiv:1009.5211}},
  \href {https://doi.org/10.1016/j.physletb.2011.03.013}
  {\path{doi:10.1016/j.physletb.2011.03.013}}.

\bibitem{Endrodi:2011gv}
G.~{Endr\H{o}di}, Z.~Fodor, S.~D. Katz, K.~K. Szabo, {The QCD phase diagram at
  nonzero quark density}, JHEP 04 (2011) 001.
\newblock \href {http://arxiv.org/abs/1102.1356} {\path{arXiv:1102.1356}},
  \href {https://doi.org/10.1007/JHEP04(2011)001}
  {\path{doi:10.1007/JHEP04(2011)001}}.

\bibitem{Bazavov:2012ka}
A.~Bazavov, N.~Brambilla, X.~Garcia~i Tormo, P.~Petreczky, J.~Soto, A.~Vairo,
  {Determination of $\alpha_s$ from the QCD static energy}, Phys. Rev. D86
  (2012) 114031.
\newblock \href {http://arxiv.org/abs/1205.6155} {\path{arXiv:1205.6155}},
  \href {https://doi.org/10.1103/PhysRevD.86.114031}
  {\path{doi:10.1103/PhysRevD.86.114031}}.

\bibitem{Brambilla:2014jmp}
N.~Brambilla, et~al., {QCD and Strongly Coupled Gauge Theories: Challenges and
  Perspectives}, Eur. Phys. J. C74~(10) (2014) 2981.
\newblock \href {http://arxiv.org/abs/1404.3723} {\path{arXiv:1404.3723}},
  \href {https://doi.org/10.1140/epjc/s10052-014-2981-5}
  {\path{doi:10.1140/epjc/s10052-014-2981-5}}.

\bibitem{Zhai:1995ac}
C.-x. Zhai, B.~M. Kastening, {The Free energy of hot gauge theories with
  fermions through g**5}, Phys. Rev. D52 (1995) 7232--7246.
\newblock \href {http://arxiv.org/abs/hep-ph/9507380}
  {\path{arXiv:hep-ph/9507380}}, \href
  {https://doi.org/10.1103/PhysRevD.52.7232}
  {\path{doi:10.1103/PhysRevD.52.7232}}.

\bibitem{Shuryak:1977ut}
E.~V. Shuryak, {Theory of Hadronic Plasma}, Sov. Phys. JETP 47 (1978) 212--219,
  [Zh. Eksp. Teor. Fiz.74,408(1978)].

\bibitem{Kapusta:1979fh}
J.~I. Kapusta, {Quantum Chromodynamics at High Temperature}, Nucl. Phys. B148
  (1979) 461--498.
\newblock \href {https://doi.org/10.1016/0550-3213(79)90146-9}
  {\path{doi:10.1016/0550-3213(79)90146-9}}.

\bibitem{Toimela:1982hv}
T.~Toimela, {The Next Term in the Thermodynamic Potential of {QCD}}, Phys.
  Lett. B124 (1983) 407--409.
\newblock \href {https://doi.org/10.1016/0370-2693(83)91484-3}
  {\path{doi:10.1016/0370-2693(83)91484-3}}.

\bibitem{Arnold:1994ps}
P.~B. Arnold, C.-X. Zhai, {The Three loop free energy for pure gauge QCD},
  Phys.Rev. D50 (1994) 7603--7623.
\newblock \href {http://arxiv.org/abs/hep-ph/9408276}
  {\path{arXiv:hep-ph/9408276}}, \href
  {https://doi.org/10.1103/PhysRevD.50.7603}
  {\path{doi:10.1103/PhysRevD.50.7603}}.

\bibitem{Arnold:1994eb}
P.~B. Arnold, C.-x. Zhai, {The Three loop free energy for high temperature QED
  and QCD with fermions}, Phys.Rev. D51 (1995) 1906--1918.
\newblock \href {http://arxiv.org/abs/hep-ph/9410360}
  {\path{arXiv:hep-ph/9410360}}, \href
  {https://doi.org/10.1103/PhysRevD.51.1906}
  {\path{doi:10.1103/PhysRevD.51.1906}}.

\bibitem{Andersen:2012wr}
J.~O. Andersen, S.~Mogliacci, N.~Su, A.~Vuorinen, {Quark number
  susceptibilities from resummed perturbation theory}, Phys. Rev. D87~(7)
  (2013) 074003.
\newblock \href {http://arxiv.org/abs/1210.0912} {\path{arXiv:1210.0912}},
  \href {https://doi.org/10.1103/PhysRevD.87.074003}
  {\path{doi:10.1103/PhysRevD.87.074003}}.

\bibitem{Borsanyi:2010cj}
S.~Borsanyi, G.~{Endr\H{o}di}, Z.~Fodor, A.~Jakovac, S.~D. Katz, S.~Krieg,
  C.~Ratti, K.~K. Szabo, {The QCD equation of state with dynamical quarks},
  JHEP 11 (2010) 077.
\newblock \href {http://arxiv.org/abs/1007.2580} {\path{arXiv:1007.2580}},
  \href {https://doi.org/10.1007/JHEP11(2010)077}
  {\path{doi:10.1007/JHEP11(2010)077}}.

\bibitem{Borsanyi:2012rr}
S.~Borsanyi, {Thermodynamics of the QCD transition from lattice}, Nucl. Phys.
  A904-905 (2013) 270c--277c.
\newblock \href {http://arxiv.org/abs/1210.6901} {\path{arXiv:1210.6901}},
  \href {https://doi.org/10.1016/j.nuclphysa.2013.01.072}
  {\path{doi:10.1016/j.nuclphysa.2013.01.072}}.

\bibitem{Borsanyi:2013hza}
S.~Borsanyi, Z.~Fodor, S.~D. Katz, S.~Krieg, C.~Ratti, K.~K. Szabo, {Freeze-out
  parameters: lattice meets experiment}, Phys. Rev. Lett. 111 (2013) 062005.
\newblock \href {http://arxiv.org/abs/1305.5161} {\path{arXiv:1305.5161}},
  \href {https://doi.org/10.1103/PhysRevLett.111.062005}
  {\path{doi:10.1103/PhysRevLett.111.062005}}.

\bibitem{Bazavov:2013uja}
A.~Bazavov, H.-T. Ding, P.~Hegde, F.~Karsch, C.~Miao, S.~Mukherjee,
  P.~Petreczky, C.~Schmidt, A.~Velytsky, Quark number susceptibilities at high
  temperatures, Phys.Rev.D 88~(9) (2013) 094021.
\newblock \href {http://arxiv.org/abs/1309.2317} {\path{arXiv:1309.2317}},
  \href {https://doi.org/10.1103/PhysRevD.88.094021}
  {\path{doi:10.1103/PhysRevD.88.094021}}.

\bibitem{Vuorinen:2002ue}
A.~Vuorinen, {Quark number susceptibilities of hot QCD up to g**6 ln g}, Phys.
  Rev. D67 (2003) 074032.
\newblock \href {http://arxiv.org/abs/hep-ph/0212283}
  {\path{arXiv:hep-ph/0212283}}, \href
  {https://doi.org/10.1103/PhysRevD.67.074032}
  {\path{doi:10.1103/PhysRevD.67.074032}}.

\bibitem{Blaizot:2002xz}
J.~P. Blaizot, E.~Iancu, A.~Rebhan, {Comparing different hard thermal loop
  approaches to quark number susceptibilities}, Eur. Phys. J. C27 (2003)
  433--438.
\newblock \href {http://arxiv.org/abs/hep-ph/0206280}
  {\path{arXiv:hep-ph/0206280}}, \href
  {https://doi.org/10.1140/epjc/s2002-01103-5}
  {\path{doi:10.1140/epjc/s2002-01103-5}}.

\bibitem{Ding:2015fca}
H.~T. Ding, S.~Mukherjee, H.~Ohno, P.~Petreczky, H.~P. Schadler, Diagonal and
  off-diagonal quark number susceptibilities at high temperatures, Phys.Rev.D
  92~(7) (2015) 074043.
\newblock \href {http://arxiv.org/abs/1507.06637} {\path{arXiv:1507.06637}},
  \href {https://doi.org/10.1103/PhysRevD.92.074043}
  {\path{doi:10.1103/PhysRevD.92.074043}}.

\bibitem{Borsanyi:2012cr}
S.~Borsanyi, G.~{Endr\H{o}di}, Z.~Fodor, S.~D. Katz, S.~Krieg, C.~Ratti, K.~K.
  Szabo, {QCD equation of state at nonzero chemical potential: continuum
  results with physical quark masses at order $mu^2$}, JHEP 08 (2012) 053.
\newblock \href {http://arxiv.org/abs/1204.6710} {\path{arXiv:1204.6710}},
  \href {https://doi.org/10.1007/JHEP08(2012)053}
  {\path{doi:10.1007/JHEP08(2012)053}}.

\bibitem{Kurkela:2009gj}
A.~Kurkela, P.~Romatschke, A.~Vuorinen, {Cold Quark Matter}, Phys.Rev. D81
  (2010) 105021.
\newblock \href {http://arxiv.org/abs/0912.1856} {\path{arXiv:0912.1856}},
  \href {https://doi.org/10.1103/PhysRevD.81.105021}
  {\path{doi:10.1103/PhysRevD.81.105021}}.

\bibitem{Kneur:2019tao}
J.-L. Kneur, M.~B. Pinto, T.~E. Restrepo, {Renormalization group improved
  pressure for cold and dense QCD}, Phys.\ Rev.\ D 100~(11) (2019) 114006.
\newblock \href {http://arxiv.org/abs/1908.08363} {\path{arXiv:1908.08363}},
  \href {https://doi.org/10.1103/PhysRevD.100.114006}
  {\path{doi:10.1103/PhysRevD.100.114006}}.

\bibitem{Weissenborn:2011qu}
S.~Weissenborn, I.~Sagert, G.~Pagliara, M.~Hempel, J.~Schaffner-Bielich, {Quark
  Matter In Massive Neutron Stars}, Astrophys. J. 740 (2011) L14.
\newblock \href {http://arxiv.org/abs/1102.2869} {\path{arXiv:1102.2869}},
  \href {https://doi.org/10.1088/2041-8205/740/1/L14}
  {\path{doi:10.1088/2041-8205/740/1/L14}}.

\bibitem{Fraga:2013qra}
E.~S. Fraga, A.~Kurkela, A.~Vuorinen, {Interacting quark matter equation of
  state for compact stars}, Astrophys. J. 781~(2) (2014) L25.
\newblock \href {http://arxiv.org/abs/1311.5154} {\path{arXiv:1311.5154}},
  \href {https://doi.org/10.1088/2041-8205/781/2/L25}
  {\path{doi:10.1088/2041-8205/781/2/L25}}.

\bibitem{Kurkela:2014vha}
{Kurkela, Aleksi and Fraga, Eduardo S. and Schaffner-Bielich, J\"urgen and
  Vuorinen, Aleksi}, {Constraining neutron star matter with Quantum
  Chromodynamics}, Astrophys. J. 789 (2014) 127.
\newblock \href {http://arxiv.org/abs/1402.6618} {\path{arXiv:1402.6618}},
  \href {https://doi.org/10.1088/0004-637X/789/2/127}
  {\path{doi:10.1088/0004-637X/789/2/127}}.

\bibitem{Kojo:2014rca}
T.~Kojo, P.~D. Powell, Y.~Song, G.~Baym, {Phenomenological QCD equation of
  state for massive neutron stars}, Phys. Rev. D91~(4) (2015) 045003.
\newblock \href {http://arxiv.org/abs/1412.1108} {\path{arXiv:1412.1108}},
  \href {https://doi.org/10.1103/PhysRevD.91.045003}
  {\path{doi:10.1103/PhysRevD.91.045003}}.

\bibitem{Heinimann:2016zbx}
O.~Heinimann, M.~Hempel, F.-K. Thielemann, {Towards generating a new supernova
  equation of state: A systematic analysis of cold hybrid stars}, Phys. Rev.
  D94~(10) (2016) 103008.
\newblock \href {http://arxiv.org/abs/1608.08862} {\path{arXiv:1608.08862}},
  \href {https://doi.org/10.1103/PhysRevD.94.103008}
  {\path{doi:10.1103/PhysRevD.94.103008}}.

\bibitem{Gorda:2016uag}
T.~Gorda, {Global properties of rotating neutron stars with QCD equations of
  state}, Astrophys. J. 832~(1) (2016) 28.
\newblock \href {http://arxiv.org/abs/1605.08067} {\path{arXiv:1605.08067}},
  \href {https://doi.org/10.3847/0004-637X/832/1/28}
  {\path{doi:10.3847/0004-637X/832/1/28}}.

\bibitem{Annala:2017llu}
E.~Annala, T.~Gorda, A.~Kurkela, A.~Vuorinen, {Gravitational-wave constraints
  on the neutron-star-matter Equation of State}, Phys. Rev. Lett. 120~(17)
  (2018) 172703.
\newblock \href {http://arxiv.org/abs/1711.02644} {\path{arXiv:1711.02644}},
  \href {https://doi.org/10.1103/PhysRevLett.120.172703}
  {\path{doi:10.1103/PhysRevLett.120.172703}}.

\bibitem{Schafer:2004zf}
{Sch\"afer, Thomas and Schwenzer, Kai}, {Non-Fermi liquid effects in QCD at
  high density}, Phys. Rev. D70 (2004) 054007.
\newblock \href {http://arxiv.org/abs/hep-ph/0405053}
  {\path{arXiv:hep-ph/0405053}}, \href
  {https://doi.org/10.1103/PhysRevD.70.054007}
  {\path{doi:10.1103/PhysRevD.70.054007}}.

\bibitem{Gerhold:2005uu}
A.~Gerhold, A.~Rebhan, {Fermionic dispersion relations in ultradegenerate
  relativistic plasmas beyond leading logarithmic order}, Phys. Rev. D71 (2005)
  085010.
\newblock \href {http://arxiv.org/abs/hep-ph/0501089}
  {\path{arXiv:hep-ph/0501089}}, \href
  {https://doi.org/10.1103/PhysRevD.71.085010}
  {\path{doi:10.1103/PhysRevD.71.085010}}.

\bibitem{Laine:2003bd}
M.~Laine, M.~{Veps\"al\"ainen}, {Mesonic correlation lengths in high
  temperature QCD}, JHEP 02 (2004) 004.
\newblock \href {http://arxiv.org/abs/hep-ph/0311268}
  {\path{arXiv:hep-ph/0311268}}, \href
  {https://doi.org/10.1088/1126-6708/2004/02/004}
  {\path{doi:10.1088/1126-6708/2004/02/004}}.

\bibitem{Laine:2009dh}
M.~Laine, M.~{Veps\"al\"ainen}, {On the smallest screening masses in hot QCD},
  JHEP 09 (2009) 023.
\newblock \href {http://arxiv.org/abs/0906.4450} {\path{arXiv:0906.4450}},
  \href {https://doi.org/10.1088/1126-6708/2009/09/023}
  {\path{doi:10.1088/1126-6708/2009/09/023}}.

\bibitem{Brandt:2014uda}
B.~Brandt, A.~Francis, M.~Laine, H.~Meyer, {A relation between screening masses
  and real-time rates}, JHEP 1405 (2014) 117.
\newblock \href {http://arxiv.org/abs/1404.2404} {\path{arXiv:1404.2404}},
  \href {https://doi.org/10.1007/JHEP05(2014)117}
  {\path{doi:10.1007/JHEP05(2014)117}}.

\bibitem{Burnier:2009bk}
Y.~Burnier, M.~Laine, M.~{Veps\"al\"ainen}, {Dimensionally regularized Polyakov
  loop correlators in hot QCD}, JHEP 01 (2010) 054, [Erratum:
  JHEP01,180(2013)].
\newblock \href {http://arxiv.org/abs/0911.3480} {\path{arXiv:0911.3480}},
  \href {https://doi.org/10.1007/JHEP01(2010)054, 10.1007/JHEP01(2013)180}
  {\path{doi:10.1007/JHEP01(2010)054, 10.1007/JHEP01(2013)180}}.

\bibitem{Brambilla:2010xn}
N.~Brambilla, J.~Ghiglieri, P.~Petreczky, A.~Vairo, {The Polyakov loop and
  correlator of Polyakov loops at next-to-next-to-leading order}, Phys. Rev.
  D82 (2010) 074019.
\newblock \href {http://arxiv.org/abs/1007.5172} {\path{arXiv:1007.5172}},
  \href {https://doi.org/10.1103/PhysRevD.82.074019}
  {\path{doi:10.1103/PhysRevD.82.074019}}.

\bibitem{Laine:2010tc}
M.~Laine, M.~{Veps\"al\"ainen}, A.~Vuorinen, {Ultraviolet asymptotics of scalar
  and pseudoscalar correlators in hot Yang-Mills theory}, JHEP 10 (2010) 010.
\newblock \href {http://arxiv.org/abs/1008.3263} {\path{arXiv:1008.3263}},
  \href {https://doi.org/10.1007/JHEP10(2010)010}
  {\path{doi:10.1007/JHEP10(2010)010}}.

\bibitem{Laine:2010fe}
M.~Laine, M.~{Veps\"al\"ainen}, A.~Vuorinen, {Intermediate distance correlators
  in hot Yang-Mills theory}, JHEP 12 (2010) 078.
\newblock \href {http://arxiv.org/abs/1011.4439} {\path{arXiv:1011.4439}},
  \href {https://doi.org/10.1007/JHEP12(2010)078}
  {\path{doi:10.1007/JHEP12(2010)078}}.

\bibitem{Burnier:2012ze}
Y.~Burnier, M.~Laine, {Massive vector current correlator in thermal QCD}, JHEP
  11 (2012) 086.
\newblock \href {http://arxiv.org/abs/1210.1064} {\path{arXiv:1210.1064}},
  \href {https://doi.org/10.1007/JHEP11(2012)086}
  {\path{doi:10.1007/JHEP11(2012)086}}.

\bibitem{Berwein:2012mw}
M.~Berwein, N.~Brambilla, J.~Ghiglieri, A.~Vairo, {Renormalization of the
  cyclic Wilson loop}, JHEP 03 (2013) 069.
\newblock \href {http://arxiv.org/abs/1212.4413} {\path{arXiv:1212.4413}},
  \href {https://doi.org/10.1007/JHEP03(2013)069}
  {\path{doi:10.1007/JHEP03(2013)069}}.

\bibitem{Burnier:2013vsa}
Y.~Burnier, M.~Laine, {Charm mass effects in bulk channel correlations}, JHEP
  11 (2013) 012.
\newblock \href {http://arxiv.org/abs/1309.1573} {\path{arXiv:1309.1573}},
  \href {https://doi.org/10.1007/JHEP11(2013)012}
  {\path{doi:10.1007/JHEP11(2013)012}}.

\bibitem{Berwein:2015ayt}
M.~Berwein, N.~Brambilla, P.~Petreczky, A.~Vairo, {Polyakov loop at
  next-to-next-to-leading order}, Phys. Rev. D93~(3) (2016) 034010.
\newblock \href {http://arxiv.org/abs/1512.08443} {\path{arXiv:1512.08443}},
  \href {https://doi.org/10.1103/PhysRevD.93.034010}
  {\path{doi:10.1103/PhysRevD.93.034010}}.

\bibitem{Berwein:2017thy}
M.~Berwein, N.~Brambilla, P.~Petreczky, A.~Vairo, {Polyakov loop correlator in
  perturbation theory}, Phys. Rev. D96~(1) (2017) 014025.
\newblock \href {http://arxiv.org/abs/1704.07266} {\path{arXiv:1704.07266}},
  \href {https://doi.org/10.1103/PhysRevD.96.014025}
  {\path{doi:10.1103/PhysRevD.96.014025}}.

\bibitem{Bazavov:2016uvm}
A.~Bazavov, N.~Brambilla, H.~T. Ding, P.~Petreczky, H.~P. Schadler, A.~Vairo,
  J.~H. Weber, {Polyakov loop in 2+1 flavor QCD from low to high temperatures},
  Phys. Rev. D93~(11) (2016) 114502.
\newblock \href {http://arxiv.org/abs/1603.06637} {\path{arXiv:1603.06637}},
  \href {https://doi.org/10.1103/PhysRevD.93.114502}
  {\path{doi:10.1103/PhysRevD.93.114502}}.

\bibitem{Bazavov:2018wmo}
A.~Bazavov, N.~Brambilla, P.~Petreczky, A.~Vairo, J.~H. Weber, {Color screening
  in (2+1)-flavor QCD}, Phys. Rev. D98~(5) (2018) 054511.
\newblock \href {http://arxiv.org/abs/1804.10600} {\path{arXiv:1804.10600}},
  \href {https://doi.org/10.1103/PhysRevD.98.054511}
  {\path{doi:10.1103/PhysRevD.98.054511}}.

\bibitem{Kajantie:2013gab}
K.~Kajantie, M.~Krssak, A.~Vuorinen, {Energy momentum tensor correlators in hot
  Yang-Mills theory: holography confronts lattice and perturbation theory},
  JHEP 05 (2013) 140.
\newblock \href {http://arxiv.org/abs/1302.1432} {\path{arXiv:1302.1432}},
  \href {https://doi.org/10.1007/JHEP05(2013)140}
  {\path{doi:10.1007/JHEP05(2013)140}}.

\bibitem{Bazavov:2019qoo}
A.~Bazavov, N.~Brambilla, X.~G. Tormo, I, P.~Petreczky, J.~Soto, A.~Vairo,
  J.~H. Weber, {Determination of the QCD coupling from the static energy and
  the free energy}, Phys. Rev. D100~(11) (2019) 114511.
\newblock \href {http://arxiv.org/abs/1907.11747} {\path{arXiv:1907.11747}},
  \href {https://doi.org/10.1103/PhysRevD.100.114511}
  {\path{doi:10.1103/PhysRevD.100.114511}}.

\bibitem{Burnier:2008ia}
Y.~Burnier, M.~Laine, M.~{Veps\"al\"ainen}, {Heavy quark medium polarization at
  next-to-leading order}, JHEP 02 (2009) 008.
\newblock \href {http://arxiv.org/abs/0812.2105} {\path{arXiv:0812.2105}},
  \href {https://doi.org/10.1088/1126-6708/2009/02/008}
  {\path{doi:10.1088/1126-6708/2009/02/008}}.

\bibitem{Burnier:2010rp}
Y.~Burnier, M.~Laine, J.~Langelage, L.~Mether, {Colour-electric spectral
  function at next-to-leading order}, JHEP 08 (2010) 094.
\newblock \href {http://arxiv.org/abs/1006.0867} {\path{arXiv:1006.0867}},
  \href {https://doi.org/10.1007/JHEP08(2010)094}
  {\path{doi:10.1007/JHEP08(2010)094}}.

\bibitem{Laine:2011xm}
M.~Laine, A.~Vuorinen, Y.~Zhu, {Next-to-leading order thermal spectral
  functions in the perturbative domain}, JHEP 09 (2011) 084.
\newblock \href {http://arxiv.org/abs/1108.1259} {\path{arXiv:1108.1259}},
  \href {https://doi.org/10.1007/JHEP09(2011)084}
  {\path{doi:10.1007/JHEP09(2011)084}}.

\bibitem{Zhu:2012be}
Y.~Zhu, A.~Vuorinen, {The shear channel spectral function in hot Yang-Mills
  theory}, JHEP 03 (2013) 002.
\newblock \href {http://arxiv.org/abs/1212.3818} {\path{arXiv:1212.3818}},
  \href {https://doi.org/10.1007/JHEP03(2013)002}
  {\path{doi:10.1007/JHEP03(2013)002}}.

\bibitem{Vuorinen:2015wla}
A.~Vuorinen, Y.~Zhu, {On the infrared behavior of the shear spectral function
  in hot Yang-Mills theory}, JHEP 03 (2015) 138.
\newblock \href {http://arxiv.org/abs/1502.02556} {\path{arXiv:1502.02556}},
  \href {https://doi.org/10.1007/JHEP03(2015)138}
  {\path{doi:10.1007/JHEP03(2015)138}}.

\bibitem{CaronHuot:2009ns}
S.~Caron-Huot, {Asymptotics of thermal spectral functions}, Phys. Rev. D79
  (2009) 125009.
\newblock \href {http://arxiv.org/abs/0903.3958} {\path{arXiv:0903.3958}},
  \href {https://doi.org/10.1103/PhysRevD.79.125009}
  {\path{doi:10.1103/PhysRevD.79.125009}}.

\bibitem{Ghiglieri:2019lzz}
J.~Ghiglieri, U.~A. Wiedemann, {Thermal width of the Higgs boson in hot QCD
  matter}, Phys. Rev. D99 (2019) 054002.
\newblock \href {http://arxiv.org/abs/1901.04503} {\path{arXiv:1901.04503}},
  \href {https://doi.org/10.1103/PhysRevD.99.054002}
  {\path{doi:10.1103/PhysRevD.99.054002}}.

\bibitem{Gynther:2007bw}
A.~Gynther, M.~Laine, Y.~{Schr\"oder}, C.~Torrero, A.~Vuorinen, {Four-loop
  pressure of massless O(N) scalar field theory}, JHEP 04 (2007) 094.
\newblock \href {http://arxiv.org/abs/hep-ph/0703307}
  {\path{arXiv:hep-ph/0703307}}, \href
  {https://doi.org/10.1088/1126-6708/2007/04/094}
  {\path{doi:10.1088/1126-6708/2007/04/094}}.

\bibitem{Andersen:2009ct}
J.~O. Andersen, L.~Kyllingstad, L.~E. Leganger, {Pressure to order g**8 log g
  of massless phi**4 theory at weak coupling}, JHEP 08 (2009) 066.
\newblock \href {http://arxiv.org/abs/0903.4596} {\path{arXiv:0903.4596}},
  \href {https://doi.org/10.1088/1126-6708/2009/08/066}
  {\path{doi:10.1088/1126-6708/2009/08/066}}.

\bibitem{Moore:2002md}
G.~D. Moore, {Pressure of hot QCD at large N(f)}, JHEP 10 (2002) 055.
\newblock \href {http://arxiv.org/abs/hep-ph/0209190}
  {\path{arXiv:hep-ph/0209190}}, \href
  {https://doi.org/10.1088/1126-6708/2002/10/055}
  {\path{doi:10.1088/1126-6708/2002/10/055}}.

\bibitem{Ipp:2003jy}
A.~Ipp, A.~Rebhan, {Thermodynamics of large N(f) QCD at finite chemical
  potential}, JHEP 06 (2003) 032.
\newblock \href {http://arxiv.org/abs/hep-ph/0305030}
  {\path{arXiv:hep-ph/0305030}}, \href
  {https://doi.org/10.1088/1126-6708/2003/06/032}
  {\path{doi:10.1088/1126-6708/2003/06/032}}.

\bibitem{Ipp:2003yz}
A.~Ipp, A.~Rebhan, A.~Vuorinen, {Perturbative QCD at nonzero chemical
  potential: Comparison with the large N(f) limit and apparent convergence},
  Phys. Rev. D69 (2004) 077901.
\newblock \href {http://arxiv.org/abs/hep-ph/0311200}
  {\path{arXiv:hep-ph/0311200}}, \href
  {https://doi.org/10.1103/PhysRevD.69.077901}
  {\path{doi:10.1103/PhysRevD.69.077901}}.

\bibitem{Blaizot:2005fd}
J.-P. Blaizot, A.~Ipp, A.~Rebhan, {Study of the gluon propagator in the
  large-N(f) limit at finite temperature and chemical potential for weak and
  strong couplings}, Annals Phys. 321 (2006) 2128--2155.
\newblock \href {http://arxiv.org/abs/hep-ph/0508317}
  {\path{arXiv:hep-ph/0508317}}, \href
  {https://doi.org/10.1016/j.aop.2005.11.012}
  {\path{doi:10.1016/j.aop.2005.11.012}}.

\bibitem{Blaizot:2005wr}
J.-P. Blaizot, A.~Ipp, A.~Rebhan, U.~Reinosa, {Asymptotic thermal quark masses
  and the entropy of QCD in the large-N(f) limit}, Phys.Rev. D72 (2005) 125005.
\newblock \href {http://arxiv.org/abs/hep-ph/0509052}
  {\path{arXiv:hep-ph/0509052}}, \href
  {https://doi.org/10.1103/PhysRevD.72.125005}
  {\path{doi:10.1103/PhysRevD.72.125005}}.

\bibitem{Romatschke:2019rjk}
P.~Romatschke, {Simple non-perturbative resummation schemes beyond mean-field:
  case study for scalar $\phi^4$ theory in 1+1 dimensions}, JHEP 03 (2019) 149.
\newblock \href {http://arxiv.org/abs/1901.05483} {\path{arXiv:1901.05483}},
  \href {https://doi.org/10.1007/JHEP03(2019)149}
  {\path{doi:10.1007/JHEP03(2019)149}}.

\bibitem{Romatschke:2019ybu}
P.~Romatschke, {Finite-Temperature Conformal Field Theory Results for All
  Couplings: O(N) Model in 2+1 Dimensions}, Phys. Rev. Lett. 122~(23) (2019)
  231603.
\newblock \href {http://arxiv.org/abs/1904.09995} {\path{arXiv:1904.09995}},
  \href {https://doi.org/10.1103/PhysRevLett.123.209901,
  10.1103/PhysRevLett.122.231603} {\path{doi:10.1103/PhysRevLett.123.209901,
  10.1103/PhysRevLett.122.231603}}.

\bibitem{Romatschke:2019qbx}
P.~Romatschke, M.~Säppi, {Thermal free energy of large Nf QED in 2+1
  dimensions from weak to strong coupling}, Phys. Rev. D100 (2019) 073009.
\newblock \href {http://arxiv.org/abs/1908.09835} {\path{arXiv:1908.09835}},
  \href {https://doi.org/10.1103/PhysRevD.100.073009}
  {\path{doi:10.1103/PhysRevD.100.073009}}.

\bibitem{Gynther:2005dj}
A.~Gynther, M.~{Veps\"al\"ainen}, {Pressure of the standard model at high
  temperatures}, JHEP 01 (2006) 060.
\newblock \href {http://arxiv.org/abs/hep-ph/0510375}
  {\path{arXiv:hep-ph/0510375}}, \href
  {https://doi.org/10.1088/1126-6708/2006/01/060}
  {\path{doi:10.1088/1126-6708/2006/01/060}}.

\bibitem{Gynther:2005av}
A.~Gynther, M.~{Veps\"al\"ainen}, {Pressure of the standard model near the
  electroweak phase transition}, JHEP 03 (2006) 011.
\newblock \href {http://arxiv.org/abs/hep-ph/0512177}
  {\path{arXiv:hep-ph/0512177}}, \href
  {https://doi.org/10.1088/1126-6708/2006/03/011}
  {\path{doi:10.1088/1126-6708/2006/03/011}}.

\bibitem{Laine:2015kra}
M.~Laine, M.~Meyer, {Standard Model thermodynamics across the electroweak
  crossover}, JCAP 1507~(07) (2015) 035.
\newblock \href {http://arxiv.org/abs/1503.04935} {\path{arXiv:1503.04935}},
  \href {https://doi.org/10.1088/1475-7516/2015/07/035}
  {\path{doi:10.1088/1475-7516/2015/07/035}}.

\bibitem{Blaizot:2006tk}
J.~P. Blaizot, E.~Iancu, U.~Kraemmer, A.~Rebhan, {Hard thermal loops and the
  entropy of supersymmetric Yang-Mills theories}, JHEP 06 (2007) 035.
\newblock \href {http://arxiv.org/abs/hep-ph/0611393}
  {\path{arXiv:hep-ph/0611393}}, \href
  {https://doi.org/10.1088/1126-6708/2007/06/035}
  {\path{doi:10.1088/1126-6708/2007/06/035}}.

\bibitem{Nieto:1999kc}
A.~Nieto, M.~H.~G. Tytgat, {Effective field theory approach to N=4
  supersymmetric Yang-Mills at finite temperature}\href
  {http://arxiv.org/abs/hep-th/9906147} {\path{arXiv:hep-th/9906147}}.

\bibitem{Kim:1999sg}
C.-j. Kim, S.-J. Rey, {Thermodynamics of large N superYang-Mills theory and AdS
  / CFT correspondence}, Nucl. Phys. B564 (2000) 430--440.
\newblock \href {http://arxiv.org/abs/hep-th/9905205}
  {\path{arXiv:hep-th/9905205}}, \href
  {https://doi.org/10.1016/S0550-3213(99)00532-5}
  {\path{doi:10.1016/S0550-3213(99)00532-5}}.

\bibitem{VazquezMozo:1999ic}
M.~A. Vazquez-Mozo, {A Note on supersymmetric Yang-Mills thermodynamics}, Phys.
  Rev. D60 (1999) 106010.
\newblock \href {http://arxiv.org/abs/hep-th/9905030}
  {\path{arXiv:hep-th/9905030}}, \href
  {https://doi.org/10.1103/PhysRevD.60.106010}
  {\path{doi:10.1103/PhysRevD.60.106010}}.

\bibitem{Chesler:2009yg}
P.~M. Chesler, A.~Gynther, A.~Vuorinen, {On the dispersion of fundamental
  particles in QCD and N=4 Super Yang-Mills theory}, JHEP 09 (2009) 003.
\newblock \href {http://arxiv.org/abs/0906.3052} {\path{arXiv:0906.3052}},
  \href {https://doi.org/10.1088/1126-6708/2009/09/003}
  {\path{doi:10.1088/1126-6708/2009/09/003}}.

\bibitem{Brauner:2016fla}
{Brauner, Tom\'{a}\v{s} and Tenkanen, Tuomas V. I. and Tranberg, Anders and
  Vuorinen, Aleksi and Weir, David J.}, {Dimensional reduction of the Standard
  Model coupled to a new singlet scalar field}, JHEP 03 (2017) 007.
\newblock \href {http://arxiv.org/abs/1609.06230} {\path{arXiv:1609.06230}},
  \href {https://doi.org/10.1007/JHEP03(2017)007}
  {\path{doi:10.1007/JHEP03(2017)007}}.

\bibitem{Andersen:2017ika}
J.~O. Andersen, T.~Gorda, A.~Helset, L.~Niemi, T.~V.~I. Tenkanen, A.~Tranberg,
  A.~Vuorinen, D.~J. Weir, {Nonperturbative Analysis of the Electroweak Phase
  Transition in the Two Higgs Doublet Model}, Phys. Rev. Lett. 121~(19) (2018)
  191802.
\newblock \href {http://arxiv.org/abs/1711.09849} {\path{arXiv:1711.09849}},
  \href {https://doi.org/10.1103/PhysRevLett.121.191802}
  {\path{doi:10.1103/PhysRevLett.121.191802}}.

\bibitem{Niemi:2018asa}
L.~Niemi, H.~H. Patel, M.~J. Ramsey-Musolf, T.~V.~I. Tenkanen, D.~J. Weir,
  {Electroweak phase transition in the real triplet extension of the SM:
  Dimensional reduction}, Phys. Rev. D100~(3) (2019) 035002.
\newblock \href {http://arxiv.org/abs/1802.10500} {\path{arXiv:1802.10500}},
  \href {https://doi.org/10.1103/PhysRevD.100.035002}
  {\path{doi:10.1103/PhysRevD.100.035002}}.

\bibitem{Asaka:2006rw}
T.~Asaka, M.~Laine, M.~Shaposhnikov, {On the hadronic contribution to sterile
  neutrino production}, JHEP 06 (2006) 053.
\newblock \href {http://arxiv.org/abs/hep-ph/0605209}
  {\path{arXiv:hep-ph/0605209}}, \href
  {https://doi.org/10.1088/1126-6708/2006/06/053}
  {\path{doi:10.1088/1126-6708/2006/06/053}}.

\bibitem{Asaka:2006nq}
T.~Asaka, M.~Laine, M.~Shaposhnikov, {Lightest sterile neutrino abundance
  within the nuMSM}, JHEP 01 (2007) 091, [Erratum: JHEP02,028(2015)].
\newblock \href {http://arxiv.org/abs/hep-ph/0612182}
  {\path{arXiv:hep-ph/0612182}}, \href
  {https://doi.org/10.1088/1126-6708/2007/01/091, 10.1007/JHEP02(2015)028}
  {\path{doi:10.1088/1126-6708/2007/01/091, 10.1007/JHEP02(2015)028}}.

\bibitem{Bellwied:2015lba}
R.~Bellwied, S.~Borsanyi, Z.~Fodor, S.~D. Katz, A.~Pasztor, C.~Ratti, K.~K.
  Szabo, {Fluctuations and correlations in high temperature QCD}, Phys. Rev.
  D92~(11) (2015) 114505.
\newblock \href {http://arxiv.org/abs/1507.04627} {\path{arXiv:1507.04627}},
  \href {https://doi.org/10.1103/PhysRevD.92.114505}
  {\path{doi:10.1103/PhysRevD.92.114505}}.

\bibitem{Kurkela:2011ti}
A.~Kurkela, G.~D. Moore, Thermalization in weakly coupled nonabelian plasmas,
  JHEP 12 (2011) 044.
\newblock \href {http://arxiv.org/abs/1107.5050} {\path{arXiv:1107.5050}},
  \href {https://doi.org/10.1007/JHEP12(2011)044}
  {\path{doi:10.1007/JHEP12(2011)044}}.

\bibitem{Kurkela:2014tea}
A.~Kurkela, E.~Lu, {Approach to Equilibrium in Weakly Coupled Non-Abelian
  Plasmas}, Phys. Rev. Lett. 113~(18) (2014) 182301.
\newblock \href {http://arxiv.org/abs/1405.6318} {\path{arXiv:1405.6318}},
  \href {https://doi.org/10.1103/PhysRevLett.113.182301}
  {\path{doi:10.1103/PhysRevLett.113.182301}}.

\bibitem{Kurkela:2015qoa}
A.~Kurkela, Y.~Zhu, {Isotropization and hydrodynamization in weakly coupled
  heavy-ion collisions}, Phys. Rev. Lett. 115~(18) (2015) 182301.
\newblock \href {http://arxiv.org/abs/1506.06647} {\path{arXiv:1506.06647}},
  \href {https://doi.org/10.1103/PhysRevLett.115.182301}
  {\path{doi:10.1103/PhysRevLett.115.182301}}.

\bibitem{Kurkela:2018wud}
A.~Kurkela, A.~Mazeliauskas, J.-F. Paquet, S.~Schlichting, D.~Teaney, {Matching
  the Nonequilibrium Initial Stage of Heavy Ion Collisions to Hydrodynamics
  with QCD Kinetic Theory}, Phys. Rev. Lett. 122~(12) (2019) 122302.
\newblock \href {http://arxiv.org/abs/1805.01604} {\path{arXiv:1805.01604}},
  \href {https://doi.org/10.1103/PhysRevLett.122.122302}
  {\path{doi:10.1103/PhysRevLett.122.122302}}.

\bibitem{Kurkela:2018xxd}
A.~Kurkela, A.~Mazeliauskas, {Chemical Equilibration in Hadronic Collisions},
  Phys. Rev. Lett. 122 (2019) 142301.
\newblock \href {http://arxiv.org/abs/1811.03040} {\path{arXiv:1811.03040}},
  \href {https://doi.org/10.1103/PhysRevLett.122.142301}
  {\path{doi:10.1103/PhysRevLett.122.142301}}.

\bibitem{Mrowczynski:1993qm}
S.~Mrowczynski, Plasma instability at the initial stage of ultrarelativistic
  heavy ion collisions, Phys.Lett.B 314 (1993) 118--121.
\newblock \href {https://doi.org/10.1016/0370-2693(93)91330-P}
  {\path{doi:10.1016/0370-2693(93)91330-P}}.

\bibitem{Mrowczynski:1996vh}
S.~Mrowczynski, Color filamentation in ultrarelativistic heavy ion collisions,
  Phys.Lett.B 393 (1997) 26--30.
\newblock \href {http://arxiv.org/abs/hep-ph/9606442}
  {\path{arXiv:hep-ph/9606442}}, \href
  {https://doi.org/10.1016/S0370-2693(96)01621-8}
  {\path{doi:10.1016/S0370-2693(96)01621-8}}.

\bibitem{Romatschke:2003ms}
P.~Romatschke, M.~Strickland, {Collective modes of an anisotropic quark gluon
  plasma}, Phys.Rev. D68 (2003) 036004.
\newblock \href {http://arxiv.org/abs/hep-ph/0304092}
  {\path{arXiv:hep-ph/0304092}}, \href
  {https://doi.org/10.1103/PhysRevD.68.036004}
  {\path{doi:10.1103/PhysRevD.68.036004}}.

\bibitem{Arnold:2003rq}
P.~B. Arnold, J.~Lenaghan, G.~D. Moore, {QCD plasma instabilities and bottom up
  thermalization}, JHEP 0308 (2003) 002.
\newblock \href {http://arxiv.org/abs/hep-ph/0307325}
  {\path{arXiv:hep-ph/0307325}}, \href
  {https://doi.org/10.1088/1126-6708/2003/08/002}
  {\path{doi:10.1088/1126-6708/2003/08/002}}.

\bibitem{Mrowczynski:2004kv}
S.~Mrowczynski, A.~Rebhan, M.~Strickland, {Hard loop effective action for
  anisotropic plasmas}, Phys. Rev. D 70 (2004) 025004.
\newblock \href {http://arxiv.org/abs/hep-ph/0403256}
  {\path{arXiv:hep-ph/0403256}}, \href
  {https://doi.org/10.1103/PhysRevD.70.025004}
  {\path{doi:10.1103/PhysRevD.70.025004}}.

\bibitem{Romatschke:2004jh}
P.~Romatschke, M.~Strickland, Collective modes of an anisotropic quark-gluon
  plasma ii, Phys.Rev.D 70 (2004) 116006.
\newblock \href {http://arxiv.org/abs/hep-ph/0406188}
  {\path{arXiv:hep-ph/0406188}}, \href
  {https://doi.org/10.1103/PhysRevD.70.116006}
  {\path{doi:10.1103/PhysRevD.70.116006}}.

\bibitem{Arnold:2004ti}
P.~B. Arnold, J.~Lenaghan, G.~D. Moore, L.~G. Yaffe, Apparent thermalization
  due to plasma instabilities in quark-gluon plasma, Phys.Rev.Lett. 94 (2005)
  072302.
\newblock \href {http://arxiv.org/abs/nucl-th/0409068}
  {\path{arXiv:nucl-th/0409068}}, \href
  {https://doi.org/10.1103/PhysRevLett.94.072302}
  {\path{doi:10.1103/PhysRevLett.94.072302}}.

\bibitem{Rebhan:2004ur}
A.~Rebhan, P.~Romatschke, M.~Strickland, {Hard-loop dynamics of non-Abelian
  plasma instabilities}, Phys. Rev. Lett. 94 (2005) 102303.
\newblock \href {http://arxiv.org/abs/hep-ph/0412016}
  {\path{arXiv:hep-ph/0412016}}, \href
  {https://doi.org/10.1103/PhysRevLett.94.102303}
  {\path{doi:10.1103/PhysRevLett.94.102303}}.

\bibitem{Rebhan:2005re}
A.~Rebhan, P.~Romatschke, M.~Strickland, {Dynamics of quark-gluon-plasma
  instabilities in discretized hard-loop approximation}, JHEP 09 (2005) 041.
\newblock \href {http://arxiv.org/abs/hep-ph/0505261}
  {\path{arXiv:hep-ph/0505261}}, \href
  {https://doi.org/10.1088/1126-6708/2005/09/041}
  {\path{doi:10.1088/1126-6708/2005/09/041}}.

\bibitem{Arnold:2005qs}
P.~B. Arnold, G.~D. Moore, {The Turbulent spectrum created by non-Abelian
  plasma instabilities}, Phys. Rev. D 73 (2006) 025013.
\newblock \href {http://arxiv.org/abs/hep-ph/0509226}
  {\path{arXiv:hep-ph/0509226}}, \href
  {https://doi.org/10.1103/PhysRevD.73.025013}
  {\path{doi:10.1103/PhysRevD.73.025013}}.

\bibitem{Arnold:2005ef}
P.~B. Arnold, G.~D. Moore, {QCD plasma instabilities: The NonAbelian cascade},
  Phys. Rev. D 73 (2006) 025006.
\newblock \href {http://arxiv.org/abs/hep-ph/0509206}
  {\path{arXiv:hep-ph/0509206}}, \href
  {https://doi.org/10.1103/PhysRevD.73.025006}
  {\path{doi:10.1103/PhysRevD.73.025006}}.

\bibitem{Arnold:2005vb}
P.~B. Arnold, G.~D. Moore, L.~G. Yaffe, {The Fate of non-Abelian plasma
  instabilities in 3+1 dimensions}, Phys. Rev. D 72 (2005) 054003.
\newblock \href {http://arxiv.org/abs/hep-ph/0505212}
  {\path{arXiv:hep-ph/0505212}}, \href
  {https://doi.org/10.1103/PhysRevD.72.054003}
  {\path{doi:10.1103/PhysRevD.72.054003}}.

\bibitem{Romatschke:2006wg}
P.~Romatschke, A.~Rebhan, {Plasma Instabilities in an Anisotropically Expanding
  Geometry}, Phys. Rev. Lett. 97 (2006) 252301.
\newblock \href {http://arxiv.org/abs/hep-ph/0605064}
  {\path{arXiv:hep-ph/0605064}}, \href
  {https://doi.org/10.1103/PhysRevLett.97.252301}
  {\path{doi:10.1103/PhysRevLett.97.252301}}.

\bibitem{Rebhan:2008uj}
A.~Rebhan, M.~Strickland, M.~Attems, {Instabilities of an anisotropically
  expanding non-Abelian plasma: 1D+3V discretized hard-loop simulations}, Phys.
  Rev. D 78 (2008) 045023.
\newblock \href {http://arxiv.org/abs/0802.1714} {\path{arXiv:0802.1714}},
  \href {https://doi.org/10.1103/PhysRevD.78.045023}
  {\path{doi:10.1103/PhysRevD.78.045023}}.

\bibitem{Rebhan:2009ku}
A.~Rebhan, D.~Steineder, {Collective modes and instabilities in anisotropically
  expanding ultrarelativistic plasmas}, Phys. Rev. D 81 (2010) 085044.
\newblock \href {http://arxiv.org/abs/0912.5383} {\path{arXiv:0912.5383}},
  \href {https://doi.org/10.1103/PhysRevD.81.085044}
  {\path{doi:10.1103/PhysRevD.81.085044}}.

\bibitem{Ipp:2010uy}
A.~Ipp, A.~Rebhan, M.~Strickland, {Non-Abelian plasma instabilities: SU(3) vs.
  SU(2)}, Phys. Rev. D 84 (2011) 056003.
\newblock \href {http://arxiv.org/abs/1012.0298} {\path{arXiv:1012.0298}},
  \href {https://doi.org/10.1103/PhysRevD.84.056003}
  {\path{doi:10.1103/PhysRevD.84.056003}}.

\bibitem{Attems:2012js}
M.~Attems, A.~Rebhan, M.~Strickland, {Instabilities of an anisotropically
  expanding non-Abelian plasma: 3D+3V discretized hard-loop simulations}, Phys.
  Rev. D 87~(2) (2013) 025010.
\newblock \href {http://arxiv.org/abs/1207.5795} {\path{arXiv:1207.5795}},
  \href {https://doi.org/10.1103/PhysRevD.87.025010}
  {\path{doi:10.1103/PhysRevD.87.025010}}.

\bibitem{Mrowczynski:2016etf}
S.~Mrowczynski, B.~Schenke, M.~Strickland, Color instabilities in the
  quark–gluon plasma, Phys.Rept. 682 (2017) 1--97.
\newblock \href {http://arxiv.org/abs/1603.08946} {\path{arXiv:1603.08946}},
  \href {https://doi.org/10.1016/j.physrep.2017.03.003}
  {\path{doi:10.1016/j.physrep.2017.03.003}}.

\bibitem{Burnier:2009yu}
Y.~Burnier, M.~Laine, M.~{Veps\"al\"ainen}, {Quarkonium dissociation in the
  presence of a small momentum space anisotropy}, Phys. Lett. B678 (2009)
  86--89.
\newblock \href {http://arxiv.org/abs/0903.3467} {\path{arXiv:0903.3467}},
  \href {https://doi.org/10.1016/j.physletb.2009.05.067}
  {\path{doi:10.1016/j.physletb.2009.05.067}}.

\bibitem{Dumitru:2009fy}
A.~Dumitru, Y.~Guo, M.~Strickland, {The Imaginary part of the static gluon
  propagator in an anisotropic (viscous) QCD plasma}, Phys. Rev. D79 (2009)
  114003.
\newblock \href {http://arxiv.org/abs/0903.4703} {\path{arXiv:0903.4703}},
  \href {https://doi.org/10.1103/PhysRevD.79.114003}
  {\path{doi:10.1103/PhysRevD.79.114003}}.

\bibitem{Nopoush:2017zbu}
M.~Nopoush, Y.~Guo, M.~Strickland, {The static hard-loop gluon propagator to
  all orders in anisotropy}, JHEP 09 (2017) 063.
\newblock \href {http://arxiv.org/abs/1706.08091} {\path{arXiv:1706.08091}},
  \href {https://doi.org/10.1007/JHEP09(2017)063}
  {\path{doi:10.1007/JHEP09(2017)063}}.

\bibitem{Hauksson:2020etn}
S.~Hauksson, S.~Jeon, C.~Gale, {Hard probes of non-equilibrium quark-gluon
  plasma}, in: {28th International Conference on Ultrarelativistic
  Nucleus-Nucleus Collisions (Quark Matter 2019) Wuhan, China, November 4-9,
  2019}, 2020.
\newblock \href {http://arxiv.org/abs/2001.10046} {\path{arXiv:2001.10046}}.

\bibitem{Landshoff:1992ne}
P.~Landshoff, A.~Rebhan, {Covariant gauges at finite temperature}, Nucl.Phys.
  B383 (1992) 607--621.
\newblock \href {http://arxiv.org/abs/hep-ph/9205235}
  {\path{arXiv:hep-ph/9205235}}, \href
  {https://doi.org/10.1016/0550-3213(92)90089-T}
  {\path{doi:10.1016/0550-3213(92)90089-T}}.

\bibitem{Landshoff:1993ag}
P.~Landshoff, A.~Rebhan, {Thermalization of longitudinal gluons}, Nucl.Phys.
  B410 (1993) 23--36.
\newblock \href {http://arxiv.org/abs/hep-ph/9303276}
  {\path{arXiv:hep-ph/9303276}}, \href
  {https://doi.org/10.1016/0550-3213(93)90571-6}
  {\path{doi:10.1016/0550-3213(93)90571-6}}.

\end{thebibliography}
